%% file: thesis.tex
\documentclass[12pt,lot, lof, singlespace]{puthesis}

\title{Gyrokinetic Continuum Simulation of Turbulence in Open-Field-Line Plasmas}

\submitted{September 2017}  
\copyrightyear{2017}  
\author{Eric Leon Shi}
\adviser{Gregory W. Hammett}  
\department{Astrophysical Sciences\\Program in Plasma Physics}

\usepackage{listings,inconsolata}
\usepackage[cmyk, dvipsnames]{xcolor}

\usepackage{amsfonts}
\usepackage{amsmath}
\usepackage{amssymb}
\usepackage[round]{natbib}

\usepackage{textcomp}
\usepackage{xspace}
\usepackage{epigraph}

\DeclareMathOperator\erf{erf}

\usepackage{graphicx}
\usepackage[caption=false]{subfig} 
\usepackage{verbatim}

\usepackage{multirow}
\usepackage{longtable}

\usepackage{booktabs}

\def\CC{{C\nolinebreak[4]\hspace{-.05em}\raisebox{.4ex}{\tiny\bf ++}}}

\ifdefined\printmode

\usepackage{url}

\else

\ifdefined\proquestmode

\usepackage{hyperref}
\hypersetup{bookmarksnumbered}

\makeatletter
\hypersetup{pdftitle=\@title,pdfauthor=\@author}
\makeatother

\else

\usepackage{hyperref}
\hypersetup{colorlinks,bookmarksnumbered,urlcolor=Cyan,citecolor=Cyan,linkcolor=Magenta}

\makeatletter
\hypersetup{pdftitle=\@title,pdfauthor=\@author}
\makeatother

\fi 
\fi 

\ifodd 0
\renewcommand{\maketitlepage}{}
\renewcommand*{\makecopyrightpage}{}
\renewcommand*{\makeabstract}{}

\else

\abstract{
\input{abstract}
}

\acknowledgements{
\input{acknowledgements}
}

\dedication{To my parents.}

\fi  

\begin{document}

\makefrontmatter


\include{ch-intro/chapter-intro}
\include{ch-models-and-numerical-methods/chapter-models-and-numerical-methods}
\include{ch-1d-sol/chapter-1d-sol}
\include{ch-simulations-of-lapd/chapter-simulations-of-lapd}
\include{ch-simulations-of-helical-sol/chapter-simulations-of-helical-sol}
\include{ch-exponentially-weighted-basis-functions/chapter-exponentially-weighted-basis-functions}
\include{ch-conclusion/chapter-conclusion}
\appendix 
\include{ch-appendicies/notes-on-plotting}
\include{ch-appendicies/1d-sol-differences}
\include{ch-appendicies/lapd-discrete-energy}
\include{ch-appendicies/helical-sol-initial-conditions}

\include{ch-appendicies/interchange}
\singlespacing

\cleardoublepage
\ifdefined\phantomsection
  \phantomsection  
\else
\fi
\addcontentsline{toc}{chapter}{Bibliography}

\bibliographystyle{thesis-style}
\bibliography{thesis}

\end{document}

%% file: abstract.tex
The properties of the boundary plasma in a tokamak are now recognized to
play a key role in determining the achievable fusion power and the
lifetimes of plasma-facing components.
Accurate quantitative modeling and improved qualitative understanding 
of the boundary plasma ultimately require five-dimensional gyrokinetic turbulence simulations,
which have been successful in predicting turbulence and transport in the core.
Gyrokinetic codes for the boundary plasma must be able to handle large-amplitude fluctuations,
electromagnetic effects, open and closed magnetic field lines, magnetic X-points,
and the dynamics of impurities and neutrals.
The additional challenges of boundary-plasma simulation
necessitate the development of new gyrokinetic codes or major modifications 
to existing core gyrokinetic codes.

In this thesis, we develop the first gyrokinetic continuum code capable of simulating
plasma turbulence on open magnetic field lines, which is a key feature of a tokamak scrape-off layer.
In contrast to prior attempts at this problem, we use an energy-conserving discontinuous Galerkin discretization
in space.
To model the interaction between the plasma and the wall, we design conducting-sheath boundary conditions
that permit local currents into and out of the wall.
We start by designing spatially one-dimensional kinetic models of parallel SOL dynamics 
and solve these systems using novel continuum algorithms.
By generalizing these algorithms to higher dimensions and adding a model for collisions,
we present results from the first gyrokinetic continuum simulations of turbulence on two
types of open-field-line systems.
The first simulation features uniform and straight field lines, such as found in some linear plasma devices.
The second simulation is of a hypothetical model we developed of the NSTX scrape-off layer 
featuring helical field lines.
In the context of discontinuous Galerkin methods, we also explore the use of exponentially
weighted polynomials for a more efficient velocity-space discretization of the distribution function
when compared to standard polynomials.
We show that standard implementations do not conserve any moments of the distribution function, and
we develop a modified algorithm that does.
These developments comprise a major step towards a gyrokinetic continuum code for
quantitative predictions of turbulence and transport in the boundary plasma of magnetic fusion devices.

%% file: acknowledgements.tex
\epigraph{Many fall in the face of chaos, but not this one... not today.}
{\textit{Narrator, Darkest Dungeon}}

\noindent After more than six years at Princeton, I am finally ready to move on.
Although I am eagerly anticipating what the future holds,
I will also miss the tightly structured life that I have become accustomed to over the years.
I would not have been able to complete my thesis without the direct
support of many people.

Of course, I must thank Greg Hammett for being my adviser.
Although I did not decide to come to Princeton to work with him (or anyone else) in particular,
I realized later on in my graduate career how lucky I was to find an adviser
who was an expert in both turbulence and numerical methods,
had a code-development project written in a modern programming language,
and had the patience to teach a student who knew so little.
It is fitting that we both have a tendency to send e-mails well-past midnight.
I will be happy if I can one day attain even a small fraction of his level of
knowledge, creativity, and intuition.

I thank Tim Stoltzfus-Dueck for always making the time to share his
incredible knowledge and powerful intuition about plasma physics with me.
Tim has been a great mentor and teacher, often meeting with me several times a week,
and I owe much of what I learned about SOL physics to him.
Tim played an essential role in interpreting the physics of the simulations and suggesting
systematic ways to test the code when the results appeared to be incorrect.
Tim's commitment to do physics in an objective and transparent way has greatly influenced
the way I conduct and present my own work.
I hope that Tim will one day use his remarkable aptitude for teaching to train future generations
of plasma physicists in edge and SOL physics.

I am very grateful to Ammar Hakim for creating the Gkeyll plasma-physics framework
and for developing some of the additional capabilities required for my thesis.
I owe a large part of the timely completion of my thesis (by
Princeton standards, at least) to Ammar, who has designed the code with modern
and sound software-engineering concepts.
I would also like to thank Ammar for teaching me the basics about DG methods
and for closely working with me for my second-year project.

I thank John Krommes for being an excellent teacher, academic adviser,
and even boss. Although he must have been disappointed in how little I learned from his
`Irreversible Processes in Plasmas' class when I took it as a second-year student,
I hope he recognized that I finally started learning something in
the two semesters during which I worked as a grader for his class.
As my academic adviser, John always listened intently to whatever dilemmas I brought to him
and drew upon his experience to provide profoundly wise advice.
I also credit John with instilling a strong belief in adhering to established rules and conventions
for technical writing.
Unfortunately, I will not return your \$31 if you catch five writing mistakes in my thesis.

I thank my thesis readers, Stewart Zweben and Matt Kunz, for carefully reading my thesis
and providing valuable feedback.
Greg Hammett and Tim Stoltzfus-Dueck also provided valuable comments for my thesis.
I would also like to thank Stewart Zweben (figures~\ref{fig:intro-shot-138844} and \ref{fig:intro-blob-sequence}),
the IAEA and Thomas Eich (figure~\ref{fig:intro-eich}),
and Troy Carter and David Schaffner (figures~\ref{fig:lapd-diagram}, \ref{fig:lapd-experiment-fluctuations},
and \ref{fig:lapd-visible-light}) for permission to use their figures 
in my thesis to illustrate important results.

I thank my other research collaborators for taking the time to
learn about my research or code: Qingjiang Pan, Tess Bernard, and Jimmy Juno.
I credit Noah Mandell with the formidible task of building the Gkeyll code on various systems.
I also thank Dan Ruiz and Jeff Parker for enlisting my help on a side
project about DW--ZF interactions.

I am very grateful to Curt Hillegas and Matt Kunz for their support for my use of the 
Perseus cluster at Princeton, which enormously sped up my time to graduate.
I also thank Frank Jenko and Jason TenBarge for providing access
to additional computational resources.
I thank Troy Carter, Amitava Bhattacharjee, Paolo Ricci, Stewart Zweben, 
Frank Jenko, Bill Tang, Hantao Ji, and Bob Kaita for their encouragement and support.
I also acknowledge useful discussions with John Krommes, Troy Carter, Paolo Ricci,
Stewart Zweben, Denis St-Onge, Seung-Ho Ku, and Michael Churchill.

I thank Dan Ruiz for his consistent friendship over the year and being a great office mate.
He probably would have graduated even sooner if he did not have me as an office mate to
provide constant distractions.
I am in awe of his personal integrity and ability to conjure original research ideas.
I am also grateful to the friends I made at Princeton: Mike Hay, Jeff Parker,
Seth Davidovits, Lee Ellison, Jake Nichols, Yao Zhou, Kenan Qu, Chang Liu, and Yuan Shi.

I thank Beth Leman, Barbara Sarfaty, and Jen Jones for shielding me from the host of
complicated administrative details associated with being a graduate student at PPPL.

Lastly, I thank my family for their unconditional support.
Rita, I would not have been so determined in my goals if I did not have an older sibling
to show me what could be achieved.
Mom and Dad, thank you for all the sacrifices you have made to help your children pursue their dreams.
You have given me so much and have never asked for a single thing in return.

This work was funded by the U.S. Department of Energy under Contract No.~DE-AC02-09CH11466, through
the Max-Planck/Princeton Center for Plasma Physics and the Princeton Plasma Physics Laboratory.
Some simulations in Chapters~\ref{ch:lapd} and \ref{ch:helical-sol}
were performed on the Perseus cluster at the TIGRESS high-performance
computer center at Princeton University, which is jointly
supported by the Princeton Institute for Computational Science and Engineering and the Princeton
University Office of Information Technology's Research Computing department.
Initial development on the simulations in Chapter~\ref{ch:lapd}
used the Edison system at the National Energy Research Scientific Computing Center,
a DOE Office of Science User Facility supported by the Office of Science of the
U.S. Department of Energy under Contract No.~DE-AC02-05CH11231.
The simulations in Chapters~\ref{ch:lapd} and \ref{ch:helical-sol} also used the Extreme Science and Engineering Discovery Environment (XSEDE),
which is supported by National Science Foundation grant number~ACI-1548562.

%% file: ch-intro/chapter-intro.tex
\chapter{Introduction\label{ch:intro}}

\section{The Boundary Plasma}
\begin{figure}
  \centering
  \includegraphics[width=0.35\linewidth]{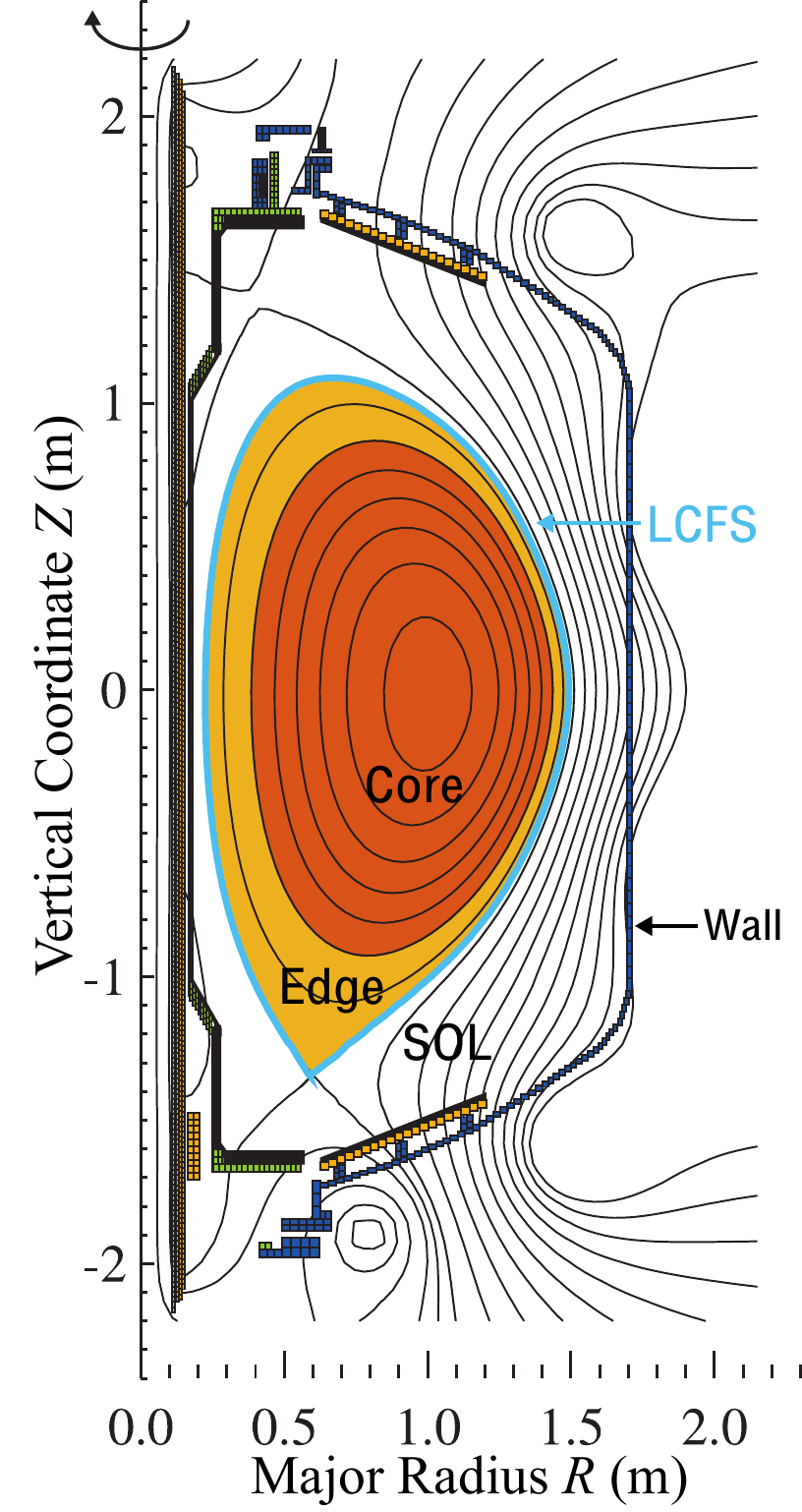}
    \caption[Illustration of the core, edge, and SOL regions in a diverted tokamak.]
    {Illustration of the core, edge, and SOL regions in a diverted tokamak.
    The LCFS separating the edge and SOL is indicated by the blue line.
    The separation between the core and edge regions is less well defined.
    The black curves are magnetic flux surfaces from the EFIT magnetic reconstruction of
    an NSTX shot.
    The flux surfaces in the confined core and edge are closed, while the flux surfaces
    in the SOL are open.
    This figure was adapted with permission from \citet[figure 1.1]{StoltzfusDueck2009}.}
    \label{fig:tokamak_cartoon}
\end{figure}

On a basic level, the plasma in a tokamak can be separated into core, edge,
and scrape-off-layer (SOL) regions.\footnote{Note that authors sometimes collectively 
refer to the edge and SOL as indicated in figure \ref{fig:tokamak_cartoon} as the edge.
For this thesis, it is important to distinguish between these two regions, so the term `boundary plasma'
is used to refer to the collective edge and SOL plasma.}
These regions are indicated in figure \ref{fig:tokamak_cartoon}, which shows a poloidal cross section of a diverted
tokamak plasma.
The core is the hot, innermost region at the center of the plasma where fusion power is produced.
Profile scale lengths in the core are on the order of the plasma minor radius, and
a temperature ${\sim}10$~keV is required to reach the break-even condition for D--T fusion \citep{Kaw2012}.
The edge is a thin layer surrounding the core often characterized by steep pressure gradients
\citep{StoltzfusDueck2009,Zweben2007,Boedo2009}
and temperatures ${\sim}100$~eV.
In the high-confinement mode (H-mode), first discovered by \citet{Wagner1982}, a
strong sheared poloidal flow correlated with a reduction in edge turbulence
levels is generically observed in the edge.
The physics behind the generation of this sheared flow layer and the subsequent reduction of turbulent fluxes
have received much attention over the years \citep{Wagner2007,Connor2000,Terry2000}.
In the core and edge, the magnetic field lines trace out nested, closed flux surfaces.
Charged particles, which rapidly flow along the field lines, are therefore confined in these regions.
Turbulence or collisions with other particles cause transport across these flux surfaces.

While a significant fraction of the tokamak plasma lies in the core and edge,
there exists a region outside the edge called the SOL where the magnetic field lines
no longer trace out closed flux surfaces and instead intersect material walls after
winding around toroidally a number of times
(in practice it is not possible to create a magnetic field with field lines that are perfectly
tangential to the wall at every point \citep{Stangeby2000,Ricci2015b}).
Charged particles in the SOL are therefore rapidly lost to material surfaces in contact with the magnetic field lines,
where recombination occurs.
On a basic level, the SOL properties are set through a balance between plasma outflow from the edge,
cross-field turbulent transport, and strong parallel losses at the divertor or limiter plates \citep{Ricci2015,
Mosetto2014},
where a Debye sheath layer forms to keep electron and ion particle fluxes to the wall approximately equal.
Plasma--surface interactions (PSIs) such as recycling and impurity influx are also important
in setting the particle and power balances in the SOL.
Due to plasma--wall interactions, the SOL plasma is much colder
($T_e \sim 10$--$100$~eV \citep{Zweben2007,Stangeby2000,StoltzfusDueck2009})
than the core and edge plasmas.
The SOL and edge are separated by a boundary referred to as the \textit{last-closed flux surface} (LCFS).

The properties of the boundary plasma constrain the performance and component lifetime of tokamak
fusion reactor by affecting both the details of how heat is exhausted in the SOL as well as
how much fusion power can be generated in the core, assuming that core profiles are stiff 
\citep{Kotschenreuther1995,Doyle2007,Kardaun2008},.
A thorough understanding of the physics of boundary plasma is therefore key to improving the overall
viability of the tokamak concept, but there are still many gaps in our present understanding
that need to be filled in before ITER begins the very long-awaited D--T operations in ${\sim}2035$.
Given the complexities of edge and SOL turbulence, numerical simulations have become 
important in furthering our theoretical understanding of the boundary plasma.
This thesis attempts to add to this line of research by developing gyrokinetic simulations of turbulence in
open-field-line (as in the SOL) plasmas using a class of grid-based (`continuum' or `Eulerian') numerical methods.
In this introduction, we briefly review some basic features of the SOL and
motivate the usefulness of gyrokinetic simulations as a tool to address important boundary-plasma physics questions
in present-day and future tokamaks.

\subsection{Basic Features of the SOL}
\begin{figure}
  \centering
  \includegraphics[width=\linewidth]{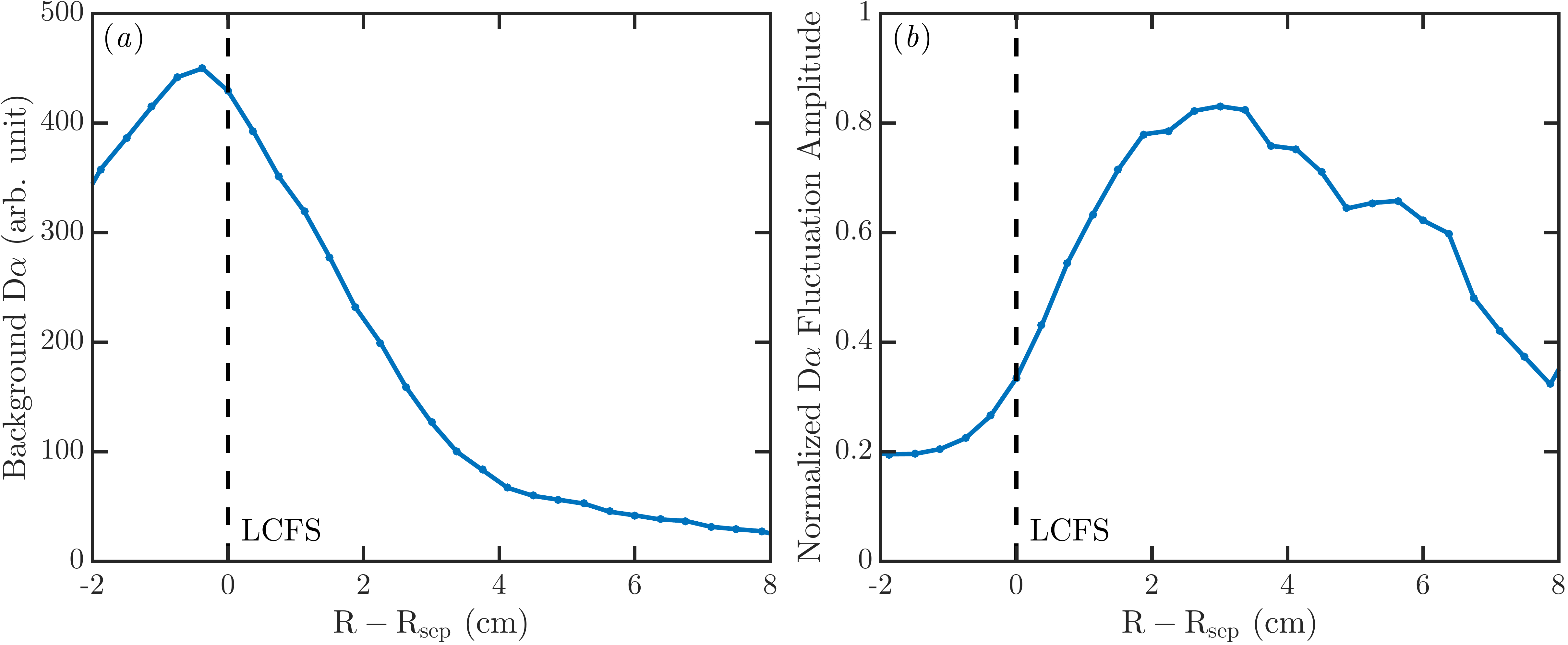}
    \caption[Profiles of the steady-state D$\alpha$ light emission and corresponding fluctuation levels
    measured using gas puff imaging in the boundary of an H-mode NSTX plasma.]
    {($a$) Radial profile of the steady-state D$\alpha$ light emission measured in the outer midplane of an
    H-mode NSTX plasma (Discharge 138844).
    ($b$) Radial profile of the locally
    normalized root mean square D$\alpha$-light-emission fluctuation amplitude
    showing large relative fluctuation levels in the edge and SOL.
    The LCFS inferred from the EFIT magnetic reconstruction is
    indicated by the dashed line.
    These profiles are obtained by averaging over a vertical band in the GPI images
    over a 10 ms period. The GPI camera records images at a rate of ${\sim}4 \times 10^5$ frames/s.
    S.\,Zweben provided the data for these plots.}
    \label{fig:intro-shot-138844}
\end{figure}

The SOL plasma is defined as the plasma in the region extending from the LCFS to the material wall.
Since the SOL plasma is in direct contact with solid surfaces, PSIs that lower plasma temperatures through
radiative processes are inevitable \citep{Stangeby2000}.
Just outside the LCFS, the SOL features steep exponential profiles and near-Gaussian probability distribution functions (PDFs).
As one goes farther out radially in the SOL from the LCFS, the profiles become slowly decreasing or flat, and
the PDFs become increasingly non-Gaussian \citep{Zweben2007}.
Relative electron density fluctuation levels in the SOL are also fairly large,
increasing from ${\sim}5\%$ near the LCFS to ${\sim}100\%$ near the first wall \citep{Zweben2007}.
Figure~\ref{fig:intro-shot-138844} shows typical profiles of the D$\alpha$ light emission measured in the boundary
plasma on NSTX using a gas puff imaging (GPI) diagnostic \citep{Zweben2017}.
The D$\alpha$ light emission is influenced by the neutral D$_2$ gas-puff density,
the local electron density, and the local electron temperature \citep{Zweben2017},
but the relative fluctuation levels should mostly be set by plasma fluctuations.

On the open magnetic field lines in the SOL, charged particles rapidly flow along the field lines 
towards solid surfaces (e.g., walls, limiters, divertors), where they recombine and are lost \citep{Chen1984}.
Since electrons are much more mobile than ions, they initially are lost to the surfaces
in contact with the magnetic field lines faster than the ions, leaving
the plasma with a thin layer of net positive charge, which is confined to a layer at the plasma--material boundary
a few Debye lengths wide due to Debye shielding.
This layer is called the \textit{electrostatic Debye sheath} \citep{Stangeby2000,Chen1984} and 
its primary purpose is to establish a potential barrier that repels incident electrons and accelerates
incident ions into the surface, keeping the particle fluxes of
electrons and ions lost to the surfaces approximately equal.
The plasma itself maintains quasineutrality.
Only electrons with sufficient parallel velocity can surmount the sheath potential drop to reach
the surface.
The sheath plays an important role in governing both the properties of the SOL plasma and
how particles and energy are lost to solid surfaces.

As charged particles in the SOL rapidly flow along field lines to solid surfaces through parallel motion,
they also undergo a much slower but non-negligible motion across the magnetic field.
As in the core, this cross-field transport is dominated by turbulence, but unlike the core,
a significant fraction of the cross-field transport in the SOL \citep{Boedo2009a,Boedo2003} is due to
the radial convection of coherent structures called \textit{blobs} or \textit{filaments} \citep{Zweben2004,Terry2007,Boedo2014,Zweben1985b,Zweben1985a}.
Plasma blobs are highly elongated along the field line, with typical parallel scales ${\sim}$1--10~m
due to the rapid parallel motion of charged particles and much smaller cross-field scales ${\sim}$1--10~cm \citep{Zweben2017}.
Blobs have elevated densities and/or temperatures compared to the background plasma.
Although the blob-formation mechanisms are not as well-understood as their
transport mechanisms \citep{Krasheninnikov2016},
they are commonly observed to form in the edge, from where they are ejected across the
LCFS into the SOL \citep{Boedo2003,Terry2003,Zweben2004}.
Cross-field transport in the SOL is highly intermittent due to blob propagation \citep{Zweben2007} and is consequently
poorly described in terms of effective diffusion coefficients and convective velocities \citep{Naulin2007}.

\begin{figure}
  \centering
  \includegraphics[width=\linewidth]{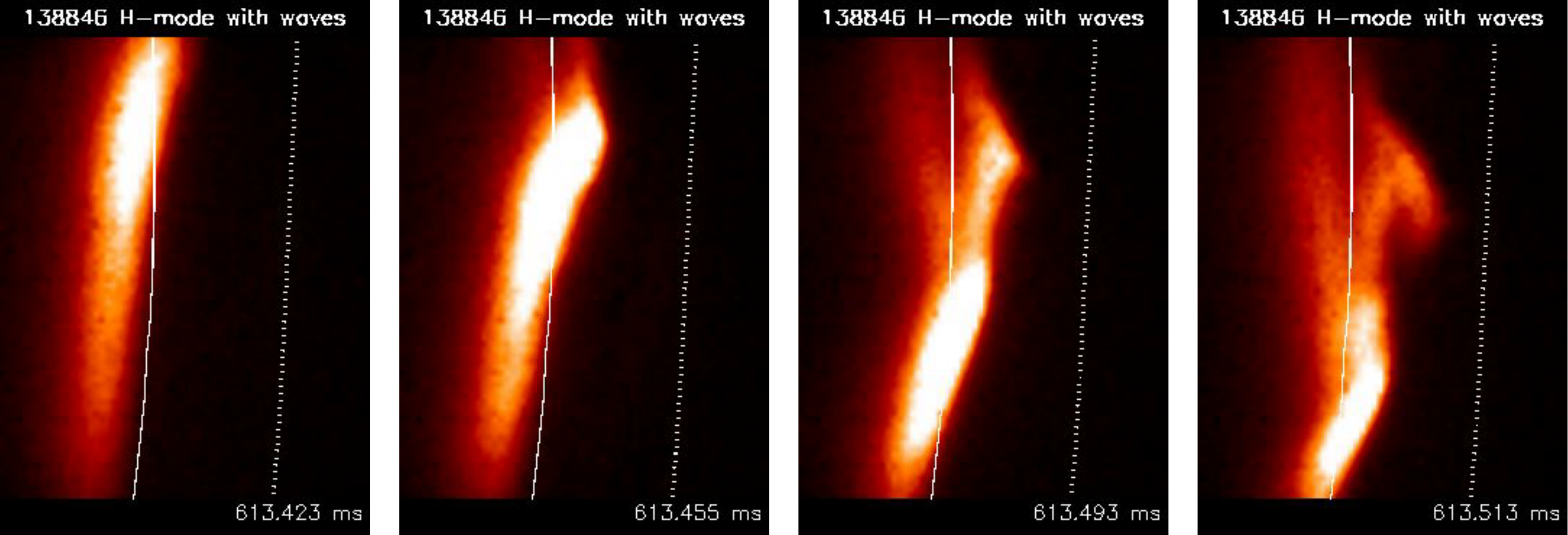}
    \caption[Sequence of GPI images from an H-mode NSTX plasma showing the ejection of a plasma blob across
    the LCFS into the SOL.]
    {Sequence of GPI images from an H-mode NSTX plasma (Discharge 138846) showing the ejection of a plasma blob across
    the LCFS into the SOL. These images show the raw camera data taken near the outer midplane,
    covering a $24~\mathrm{cm} \times 30~\mathrm{cm}$ region. 
    The LCFS is indicated by the solid line and is inferred from the EFIT magnetic reconstruction,
    while the dotted lines indicate the limiter shadow.
    The horizontal axis is the radially outward direction, and the vertical axis is the upward direction.
    These images are used with permission from S.\,Zweben (\url{http://w3.pppl.gov/~szweben/NSTX2013/NSTX2013.html}).}
    \label{fig:intro-blob-sequence}
\end{figure}

Figure~\ref{fig:intro-blob-sequence} shows raw camera images from a GPI diagnostic
viewed along the local magnetic field direction near the outer midplane of the NSTX tokamak.
These images show the 2D cross-field blob structure of a 3D plasma filament as it is ejected across the LCFS into the SOL.
The standard picture of blob propagation involves the polarization of a blob by the curvature drift,
which self-generates a vertical electric field across the blob that results in a radially outward
$E \times B$ drift
\citep{Krasheninnikov2001,Krasheninnikov2008,Grulke2006}.
The efficient and fast convective radial transport of blobs towards the main
chamber walls (instead of along the field lines to the divertor) can result in damage
to first-wall components and the contamination of the core plasma by wall impurities \citep{Rudakov2005,Terry2007},
which lowers the plasma temperature through radiative cooling.
Both the generation of blobs in the edge and
the structure and motion of blobs in the SOL \citep{Zweben2016,Zweben2006}
are ongoing research topics of practical interest.

\subsubsection{Particle and Heat Loss in the SOL}
The loss of particles and heat in the SOL to plasma-facing components (PFCs) is a major concern for
future high-power devices like ITER \citep{Lipschultz2012} and Demo because a significant fraction (${\sim}$20\%) of the
heat produced in the core is transported across the LCFS into the SOL by convection and conduction
\citep{Loarte2007}, where the heat must be exhausted somehow.
The unmitigated steady-state parallel heat flux in the SOL is expected to
be ${\sim}1$~GW~m$^{-2}$ for ITER \citep{Loarte2007} and ${\sim}20$~GW~m$^{-2}$ for Demo \citep{Goldston2015},
while material limitations set the maximum tolerable heat flux normal to the divertor plates at ${\sim}10$~MW~m$^{-2}$
for steady-state and ${\sim}20$~MW~m$^{-2}$ for transients \citep{Loarte2007}, such as from disruptions
and edge-localized modes.
An often-quoted comparison in the fusion community is the transient heat flux of approximately 6~MW~m$^{-2}$
experienced by some components of a space vehicle during atmospheric reentry.
The use of an extremely shallow incidence angle between the divertor plates and field lines (${\sim}2$--$5^\circ$)
significantly reduces the heat flux normal to the divertor plates, 
but up to ${\sim}95\%$ of the power \citep{Goldston2015} may need to be dissipated in the SOL through various means
before reaching the divertor plates to bring normal heat fluxes down to tolerable levels.
Strategies to reduce the heat load on the divertor plates are still under development
and include radiative divertor detachment \citep{Soukhanovskii2017}, advanced divertor geometries \citep{Kotschenreuther2016,Ryutov2008,Valanju2009},
and applied resonant magnetic perturbations \citep{Ahn2010,Evans2005,Jakubowski2009}.

The heat fluxes on the ITER divertor targets are believed to be enormous because experiments
on present-day machines
indicate that the heat exhausted into the SOL flows towards the divertor plates in an
extremely narrow channel whose width at the outer midplane (quantified as an exponential
decay length) is insensitive to the machine size \citep{Loarte2007,Makowski2012,Eich2013}.
There exists major uncertainty about the validity of
empirical extrapolations to ITER, however.
The amount of power spreading along the ITER divertor legs is also not well understood
(empirically or theoretically), and it is possible that power-spreading effects on ITER will
be principal in setting the heat-flux width at the divertor plates,
making such considerations of the midplane heat-flux width unimportant
for predicting divertor-plate heat loads.
Presumably, the location and strength of heat loads deposited on PFCs is set through a balance
between confined plasma outflow across the LCFS, parallel losses at the sheaths,
and cross-field turbulent or neoclassical transport in the SOL.
Therefore, credible numerical investigations of the SOL heat-flux width
require the use of sophisticated turbulence codes.

\begin{figure}
  \centering
  \includegraphics[width=0.75\linewidth]{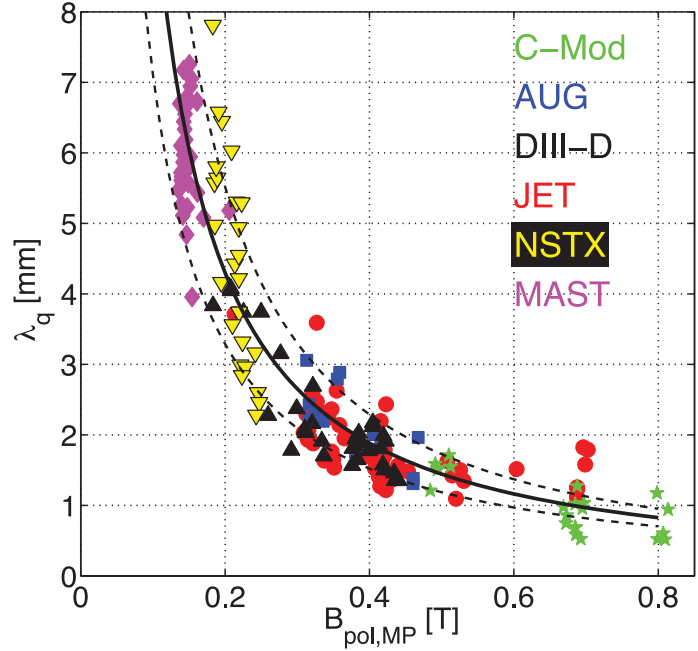}
    \caption[Outer-midplane SOL heat-flux width versus outer-midplane poloidal magnetic field
    for a multi-machine database of low-recycling H-mode discharges.]
    {Outer-midplane SOL heat-flux width $\lambda_q$ versus outer-midplane poloidal magnetic field $B_\mathrm{pol}$
    for a multi-machine database of low-recycling H-mode discharges.
    The value of $\lambda_q$ is inferred from infrared thermography measurements made at the outer divertor target.
    The regression $\lambda_q \text{ (mm)} = (0.63 \pm 0.08) \times B_{\mathrm{pol}}^{-1.19 \pm 0.08}$
    is indicated by the solid and dashed lines.
    Regression studies indicate that $\lambda_q$ decreases approximately linearly with
    the value of $B_\mathrm{pol}$ and has much weaker dependencies on other machine parameters,
    including machine size.
    Figure reprinted from \citet[figure 3]{Eich2013} with permission. Copyright \textcopyright\xspace 2013 by the IAEA.}
    \label{fig:intro-eich}
\end{figure}

Figure~\ref{fig:intro-eich} shows a plot from \cite{Eich2013} of the outer-midplane SOL heat-flux width $\lambda_q$
versus the outer-midplane poloidal magnetic field $B_\mathrm{pol}$ computed from a multi-machine database of
low-recycling H-mode tokamak discharges.
In that study, the outer-midplane heat-flux width for each discharge was determined as a fit parameter
to match infrared thermography measurements of the heat-flux profile at the outer divertor target \citep{Eich2011}.
\citet{Eich2013} found that ``the strongest and essentially only dependence amongst the regression variables
tested, at least for the conventional aspect ratio tokamaks, is an inverse scaling with plasma current (or
equivalently [an inverse] dependence on outboard midplane poloidal magnetic field)''.
\citet{Eich2013} and \citet{Makowski2012} also extrapolated the scaling to the ITER H-mode diverted plasma, finding an
outer-midplane SOL heat-flux width of ${\approx}1$ mm, much smaller than the 5~mm value used in ITER design specifications
and the 3.6 mm value from transport modeling \citep{Kukushkin2013}.
\citet{Goldston2012} developed a heuristic drift-based (HD) model of the SOL heat-flux width that
was consistent with the results of \citet{Eich2013} and \citet{Makowski2012}.
The HD model notably does not attempt to include turbulence,
which could spoil its extrapolation to new parameter regimes (such as the ITER SOL).

It is important to study the heat-flux-width problem using turbulence simulations
to provide a first-principles-based check of the empirical predictions and to investigate
ways to broaden the heat-flux width, such as by increasing turbulent heat transport
along the divertor legs.
A recent electrostatic gyrokinetic simulation predicted the outer-midplane heat-flux width on ITER to be
approximately 5.6~mm \citep{Chang2017b},
but no other gyrokinetic codes currently have the capability to cross-check this result.
More research is required to understand the physics setting the SOL heat-flux width (at both the outer midplane
and at the divertor plates), the validity of such empirical extrapolations to ITER, and
the implications of such a narrow width on divertor-plate heat loads \citep{Goldston2015}.

\subsection{Basic Plasma Physics Experiments}
While experimental measurements of the SOL plasma in tokamaks
have been highly successful in improving our understanding of SOL turbulence \citep{Zweben2007,Tynan2009},
the complicated physics situation in a tokamak can make detailed comparisons of analytical theories
and numerical simulations with experimental data extremely challenging.
Fortunately, there are a number of basic plasma physics experiments
that produce highly reproducible, well-diagnosed plasmas in simple, open-magnetic-field-line configurations.
Measurements of turbulence and transport in these laboratory plasma devices
can shed light on topics of relevance to tokamak plasmas, such as sheared-flow suppression of turbulence 
\citep{Carter2009,Schaffner2013,Gentle2010}, turbulence intermittency \citep{Carter2006,Windisch2011},
and turbulence saturation.
The plasmas are usually relatively cold ($T_e \sim 10$~eV) and low pressure ($\beta \sim 10^{-4}$),
so electrostatic 3D fluid simulations are commonly used to study the turbulence in basic
plasma physics experiments
\citep{Friedman2013,Ricci2009a,Rogers2010,Naulin2008}.
Although the plasmas created in these devices are not in the same parameter regime as fusion plasmas,
the free energy sources, dissipation mechanisms, and nonlinear transfer processes
identified in these devices may be generic and thus relevant to tokamaks \citep{Tynan2009,Fasoli2006}.

Linear devices create a column of plasma in a uniform, axial background magnetic field.
Some examples of linear devices include LAPD \citep{Gekelman1991,Gekelman2016},
CSDX \citep{Burin2005,Thakur2014}, VINETA \citep{Franck2002},
HelCat \citep{Gilmore2015}, and Mirabelle \citep{Pierre1987,Brandt2011}.
Turbulence in linear devices is typically driven by temperature or density profile gradients (drift-wave instabilities).
Toroidal basic plasma physics devices, called \textit{simple magnetized tori} (SMTs), create
a magnetic geometry composed of helical magnetic field lines (a superposition of vertical
and toroidal magnetic fields) that intersect the vessel on each end.
Compared to linear devices, the turbulence and transport in SMTs are more
relevant to a tokamak SOL because they can also be driven by magnetic-field-line-curvature effects,
which are important in the generation of blobs (as we will see in Chapter \ref{ch:helical-sol}).
Examples of SMTs include TORPEX \citep{Fasoli2006} and Helimak \citep{Gentle2008}.

The relative simplicity, availability of detailed diagnostics, and ease of
parameter scans in basic plasma physics experiments also facilitates comparisons
between simulations and experiments.
The Global Braginskii Code (GBS) has been used to perform rigorous validation studies with
data from TORPEX \citep{Ricci2015,Ricci2011,Ricci2009b}.
In the development of complex numerical codes, comparisons with basic plasma physics experiments
can be made relatively early on with a minimal set of features implemented in the code
to benchmark and test the models and numerical algorithms in a physically relevant setting.
We adopt this strategy in this thesis, where we first simulate turbulence in LAPD to
test 5D gyrokinetic continuum simulations before adding additional levels of complexity.

\section{Boundary-Plasma Modeling}
Due to the complications of the edge and SOL, numerical simulations have become an important tool for
elucidating the physics of the boundary plasma.
The use of a particular simulation can be classified as interpretive, predictive,
or generic \citep{Mitchell2000}.
Since these classifications are seldom explained in plasma-physics literature, and sometimes
incorrectly applied, we first provide a brief description of the different types of modeling.

\noindent
\hangindent=15pt \textbf{Interpretive modeling.}
To aid in experimental analysis and physical understanding,
interpretive modeling uses simulations with data collected from an experiment as inputs
to estimate other variables described by reduced models.
These estimated variables are often difficult or impossible to measure in the actual experiment
due to diagnostic limitations.
Some interpretive models have several free parameters that can be tuned automatically so that a subset of the output
data matches the corresponding experimentally measured data (e.g., density and temperature profiles),
making it straightforward (but somewhat questionable) to run additional simulations in a predictive manner
by using hypothetical initial conditions and plasma parameters \citep{Ricci2015b}.
These tools have become standard in analyzing data from tokamak discharges to infer transport levels
and energy confinement.
Since interpretive simulations are limited by the fidelity of the underlying reduced models (which
can be empirical rather than physics based) and
by the availability and quality of experimental data,
caution must be used when drawing physics conclusions based on
interpretive simulations and when extrapolating the models to parameter regimes
that are outside the range of demonstrated validity.

\noindent
\hangindent=15pt \textbf{Predictive modeling.}
As their name suggests, the goal of predictive models is to make quantitative predictions about
the physics given a collection of initial conditions and assumptions
(e.g., plasma sources, boundary conditions, fluid vs. kinetic description).
They are particularly valuable for testing and improving interpretive models
and predicting the plasma behavior in situations that cannot be investigated by experiments
(such as in future fusion reactors like ITER).
The use of free parameters in predictive models should be minimized,
and the insensitivity of results to specific values of the model free parameters must be verified.
Ideally, predictive simulations should solve first-principles equations without \textit{ad hoc} terms 
and empirical models \citep{Barnes2008,Candy2009,Gorler2014}.
However, such first-principles simulations for a given application (e.g., turbulence on transport time scales)
can be computationally infeasible even on state-of-the-art high-performance computers,
in which case reduced models must be used in some capacity
\citep{Kotschenreuther1995,Staebler2007,Kinsey2011}.
A code that can make quantitatively accurate predictions of the boundary plasma is likely
to require a high degree of sophistication, having models of the numerous PSIs that occur in the SOL. 
An example of a boundary-plasma predictive simulation is the use of the XGC1 gyrokinetic PIC code to calculate
the SOL heat-flux width in ITER \citep{Chang2017b},
although this is a prediction that cannot be validated for many years.
Predictive simulations for high-fusion-power ITER regimes are particularly valuable because mistakes
in the design of ITER based on our understanding of existing experiments
can result in irreparable damage to the machine and to the overall ITER project.

\noindent
\hangindent=15pt \textbf{Generic modeling.}
Generic modeling concerns the modeling of hypothetical situations, often with a number of highly simplifying
assumptions.
This type of modeling is useful for helping assemble a theoretical picture to explain observed phenomena
and for testing the theoretical models of others using first-principles simulations.
Since the purpose is neither to analyze a particular experiment nor to
make definite predictions, sophisticated physics models that are essential for accurate predictive
modeling can be neglected, which reduces the computing requirements and code complexity.
Some uses of generic modeling for the boundary plasma include understanding the low-to-high confinement-mode
(L--H) transition in tokamaks \citep{Connor2000}, which is a rapid bifurcation between two plasma states associated
with the formation of an edge transport barrier.
Many studies of the mechanisms driving turbulence in the edge and SOL also employ generic modeling
\citep{Scott2005,Ribeiro2008,Halpern2016}.
The simulations that are presented in this thesis also fall into this category.

\subsection{Fluid Modeling}
Fluid-based codes have been and still are commonly used to model the boundary plasma.
One class of these codes solves simplified transport equations 
based on the Braginskii fluid equations in two dimensions assuming axisymmetry.
Plasma turbulence is not captured in these models, so the use of \textit{ad hoc} anomalous diffusion
terms \citep{Dekeyser2011} or coupling to a turbulence code \citep{Schneider2006,Rognlien2004} is
required to model the turbulent transport across the magnetic field.
Perhaps the most widely used of these fluid plasma transport codes is the SOLPS package \citep{Schneider1992,Rozhansky2009}, 
which contains a 2D boundary-plasma transport component coupled to a kinetic Monte Carlo model for neutral transport \citep{Reiter2005}.
SOLPS has become the standard tool used for ITER divertor and boundary-plasma modeling \citep{Wiesen2015}.
Some physics models implemented in SOLPS include impurities, charge-exchange, ionization, radiation, and sputtering.
Other boundary-plasma transport codes which also solve similar equations include UEDGE \citep{Rognlien1999,Rognlien1994}
and EDGE2D \citep{Simonini1994}.
Although interpretive transport codes have become packed with features over decades of development,
incorporating many sophisticated models for physics believed to be relevant in the boundary plasma \citep{Schneider2007},
the basic fact that they often do not capture plasma turbulence self-consistently makes
it questionable to use them in a predictive capacity,
especially as a principal tool for quantitative modeling of the challenging SOL in ITER, a machine which 
costs ${\sim}\mathrm{US}\$20$ billion just to construct \citep{Kramer2016}.
Some shortcomings of fluid plasma transport codes for boundary-plasma modeling are discussed by \citet{Boedo2009}.

Boundary-plasma turbulence codes solving drift-ordered fluid equations
\citep{Zeiler1997,Simakov2003,Scott1997,Xu1998} in two or three dimensions
have also been developed \citep{Ricci2008,Dudson2009,Naulin2008,Xu2000,Xu2008,Tamain2010},
although many codes employ approximations or omit physics that might have significant
effects on the results, such as neglecting adiabatic coupling\footnote{
Adiabatic coupling refers to the coupling of density and pressure through the parallel current \citep{StoltzfusDueck2009}.
The importance of retaining the adiabatic response for capturing the correct
qualitative character of boundary-plasma turbulence has been emphasized by \citet{Scott2007,Scott2006,Scott2005}
and for L--H-transition physics by \citet{StoltzfusDueck2016}}
by dropping electron pressure in the Ohm's law \citep{Park2015,Li2016,Li2017,Garcia2006,Beyer2011},
neglecting geodesic coupling\footnote{The importance of retaining the geodesic coupling effect
for zonal-flow dynamics in a toroidal geometry
has been emphasized by \citet{Scott2005a}.} \citep{Bisai2005a,Li2016,Li2017,Shurygin2001},
or assuming cold ions ($T_i/T_e \approx 1$--4 is typically
observed just outside the LCFS and increases with radius in the SOL \citep{Kocan2011}).
Their underlying models are usually derived by applying a low-frequency approximation to the
Braginskii equations in which the gyrofrequencies are ordered fast compared to the frequencies
of interest \citep{Scott2003,Zeiler1997}.
Since plasma turbulence is captured in these codes, they avoid one of the major drawbacks of transport codes
at the cost of increased computational expense.

There are also boundary-plasma turbulence codes that solve 3D electromagnetic gyrofluid equations
\citep{Ribeiro2005,Ribeiro2008,Xu2013,Kendl2010},
which are more robust than drift-ordered fluid equations and model finite-Larmor-radius
and Landau-damping effects \citep{Dorland1993,Snyder1997}.
Gyrofluid codes have models for the treatment of dynamics on the ion-gyroradius scale and smaller, a regime
in which drift-ordered fluid equations break down \citep{Scott2007a,Scott2007}.
Both fluid and gyrofluid turbulence codes have yielded many insights into edge and SOL turbulence, but
attempts to make quantitative comparisons with experimental data from tokamaks are 
rare and have produced mixed results \citep{Zweben2009,Halpern2017,Halpern2015,Cohen2013}.
These codes keep just a few moments and cannot fully capture potentially important kinetic effects, such as
trapped particles \citep{Lackner2012}, nonlinear wave--particle interactions, and suprathermal electrons,
and their model assumptions can be violated in edge and SOL plasmas \citep{Batishchev1997,Takizuka2017}.
While fluid and gyrofluid models have been useful in revealing the qualitative physics of
the boundary plasma, satisfactory and reliable quantitative prediction of boundary-plasma
properties are believed to require the use of kinetic simulations in some capacity
\citep{Cohen2008,Scott2010b,Scott2003},
including the direct or indirect coupling of a fluid transport code to kinetic turbulence
code \citep{Schneider2007}.

For these reasons, there are efforts to develop first-principles gyrokinetic codes for boundary-plasma turbulence
simulation \citep{Chang2009,Shi2017,Dorf2016,Korpilo2016}.
Unlike drift-reduced Braginskii-fluid approaches, gyrokinetic approaches use equations
that are valid across a wide range of collisionality regimes, even if the collisional mean free
path is not small compared to the parallel scale length
or if the ion drift-orbit excursions are not small compared to radial gradient length scales \citep{Cohen2008}.
Gyrokinetic simulations, however, are much more computationally expensive than fluid simulations, so
fluid-based transport and turbulence codes for the boundary plasma will remain useful
for modeling the boundary plasma.
The results from gyrokinetic simulations are also expected to
aid in improving the fidelity of boundary-plasma fluid simulations \citep{Ricci2015b}.

\subsection{Gyrokinetic Modeling}
Full six-dimensional kinetic modeling of plasma turbulence in tokamaks on macroscopic time
and length scales by solving the Vlasov--Maxwell or Vlasov--Poisson equations have memory and processing power
requirements several orders beyond what is currently possible on present-day and near-term
supercomputers.
Fortunately, there is a way to reduce the often insurmountable full 6D problem to a tractable 5D one, given
that certain assumptions are well satisfied.
Gyrokinetic theory is a reduced five-dimensional description of low-frequency plasma dynamics constructed
by systematically removing the details of the charged particles' rapid gyromotion due to a magnetic field
and other high-frequency phenomena \citep{Krommes2012,Tronko2016,Krommes2010,Sugama2000,Brizard2007,Brizard2000a}.
This time-scale separation is well justified for particles in a strong background magnetic field with
weak spatial inhomogeneity,
such as those present in tokamaks and stellarators, in which case the frequencies of turbulent fluctuations
are much smaller than the ion gyrofrequency.
The gyrokinetic system, which describes the evolution of a gyrocenter (the gyro-averaged particle position)
distribution function over a 5D phase space, is much easier to simulate when compared to 6D kinetic descriptions of
particle distribution functions because of the reduced dimensionality
and from relaxing the restriction on the time step from the plasma period to turbulence time scales
and the restriction on the grid spacing from the Debye length to the gyroradius \citep{Garbet2010}.

While 5D gyrokinetic simulations require much more computational resources
than comparable 3D fluid simulations due to the high dimensionality
of the gyrokinetic system, their use to study turbulent transport in the tokamak core
has now become routine \citep{Garbet2010}.
A number of important verification studies and cross-code benchmarks on
core gyrokinetic codes have been performed \citep{Dimits2000,Tronko2017,Lapillonne2010,McMillan2010,Rewoldt2007}.
More recently, gyrokinetic models have also been used to study astrophysical
turbulence \citep{Schekochihin2009,Numata2010}.
For many of the same reasons why quantitative modeling in the boundary plasma is difficult to approach 
analytically, gyrokinetic codes for boundary-plasma simulation are much less mature than than their
core counterparts. 

Some complications that must be faced by boundary-plasma codes include the need to
handle large-amplitude fluctuations (invalidating
conventional $\delta f$ approaches \citep{Hu1994}),
open and closed magnetic field lines with a LCFS and X-point
(which can cause difficulties with coordinates),
electromagnetic fluctuations \citep{Scott2007a,Scott2010b},
a wide range of space and time scales, a wide range of collisionality regimes,
sheath boundary conditions, plasma--wall interactions, atomic physics,
and the existence of a high-frequency electrostatic shear Alfv\'{e}n ($\omega_H$) mode
in electrostatic simulations \citep{Lee1987,Belli2005}
or sheath-interaction modes that one does not want to artificially excite.
Major extensions to existing core gyrokinetic codes or new codes are required to handle
the additional challenges of the edge and SOL regions.

\subsubsection{Numerical Implementations of the Gyrokinetic Equations \label{sec:intro-gk-codes}}
A variety of numerical methods have been developed for the computationally challenging
solution to the gyrokinetic equations.
The two main types of numerical methods for solving the gyrokinetic equations are \textit{continuum} methods 
\citep{Jenko2001} and the \textit{particle-in-cell} (PIC) method \citep{Lee1983,Bottino2015,Birdsall2004}.
There is also a third approach called \textit{semi-Lagrangian} methods, which are a hybrid
between continuum and PIC methods, but they have been seldom-used for gyrokinetic simulation
so far \citep{Grandgirard2006,Grandgirard2007}.
While gyrokinetic simulations have been useful in elucidating the physical mechanisms behind
tokamak and stellarator microturbulence, the ultimate goal of gyrokinetic simulations
is to produce quantitatively reliable predictions of core and boundary plasma properties.
As the history of core gyrokinetic codes has demonstrated, it is important to explore both 
PIC and continuum (and semi-Lagrangian) approaches as independent cross-checks against each other and to
continuously shore up the specific weaknesses of each method.

Continuum methods are Eulerian approaches to solve a kinetic equation (e.g., the 5D gyrokinetic equation)
by discretizing the equation on a fixed phase-space mesh.
Standard numerical methods developed for the solution of partial differential equations are used,
including finite-difference, finite-volume, spectral, pseudospectral, finite-element, and
discontinuous Galerkin (DG) methods.
The PIC method is a Lagrangian approach that 
solves the kinetic equation using a finite set of particles called markers.
Since it is often computationally infeasible to use a number of markers on the same order
as the number of physical particles in a real plasma, a much lower number of markers are used in practice \citep{Tskhakaya2007}.
Each marker in the simulation then represents a `macroparticle' or `superparticle' 
encapsulating many physical particles.
Starting with a set of markers that sample the initial positions in phase space,
the marker positions are advanced over a small time step according to the characteristics
of the kinetic equation.
The source terms for the 3D field equations are then computed on a fixed grid from the markers, and the
resulting fields are interpolated back to the marker positions so that the markers can be advanced again
for next time step.
PIC methods can be considered as a kind of Monte Carlo method that uses a finite set of markers to
approximate integrals involving the distribution function \citep{Bottino2015,Krommes2012}.
The 3D field equations in both continuum and PIC approaches are solved using standard grid-based algorithms.

Both classes of numerical methods are associated with a unique set of advantages, disadvantages, and challenges.
PIC methods automatically maintain the positivity of the distribution function and are relatively straightforward
to implement \citep{Krommes2012}.
They are also not subject to the Courant--Friedrichs--Lewy (CFL) condition when 
explicit time stepping is used \citep{Garbet2010},
which results in instability if violated in a continuum simulation.\footnote{Explicit time steps
that violate the CFL condition in PIC simulations can result in inaccuracies such as
numerical heating or numerical diffusion.}
PIC methods have been used for plasma simulation for several decades, and the first
gyrokinetic simulation used a PIC method \citep{Lee1983}.
Computational plasma physicists can consequently draw upon a larger body of knowledge in the field
when developing PIC codes, while the first continuum gyrokinetic code GS2 was developed much later
\citep{Dorland2000}.\footnote{The development of GS2 was motivated in part by the difficulties
gyrokinetic PIC codes were having with handling electromagnetic fluctuations.}
As a Monte Carlo sampling technique, PIC methods have statistical noise in moments of the distribution
function, and this error scales with the number $N$ of markers as $1/\sqrt{N}$.
Although there are a variety of techniques to reduce sampling noise \citep{Garbet2010},
the effect of noise in PIC simulations can be subtle \citep{Krommes1994,Krommes2007}
and can even dominate the results of a simulation, which has
led to misunderstandings in the past \citep{Nevins2005,Bottino2007}.
\citet{Wilkie2016} recently documented two time-discretization-independent numerical instabilities
in the electrostatic $\delta f$-PIC algorithm, one of which is even converged on particle number.
Statistical noise in PIC codes has also made electromagnetic simulations challenging \citep{Hatzky2007},
which is a problem in the literature referred as the \textit{Amp\'{e}re's law cancellation problem}.
This problem refers to the inaccuracy in the numerical cancellation of two large terms
in Amp\'ere's law, which becomes severe at moderate to high plasma $\beta$ \citep{Chen2001,Mishchenko2004}
and small perpendicular wave numbers \citep{Hatzky2007}.
A variety of techniques for PIC simulations have been proposed over the years
\citep{Chen2003,Mishchenko2014b,Kleiber2016,Hatzky2007} to mitigate this problem, and recent
developments have been promising.

Continuum methods are not Monte Carlo methods, so they avoid having to deal with the challenging
statistical-noise issues that PIC methods face.
These methods are however subject to a time-step size restriction from
the CFL condition if explicit time stepping is used, which can be highly restrictive
in electrostatic simulations \citep{Lee1987,Belli2005} and
for the treatment of diffusive and hyperdiffusive terms.
Therefore, the fastest dynamics of a system must always be resolved when explicit methods are used,
even if they do not affect the results.
Implicit or semi-implicit time stepping methods are required to avoid the CFL restriction, which can
require computationally expensive global matrix-inversion operations and processor communication.
Continuum methods usually do not automatically guarantee the positivity of the distribution function,
which can be a particularly difficult issue to address for high-order schemes \citep{Garbet2010}.
Continuum methods often use upwind techniques that introduce some numerical dissipation
that result in some smoothing of the particle distribution function at small scales, which
can be beneficial in addressing the `entropy paradox' \citep{Krommes1994} and numerical-recurrence issues
\cite{Garbet2010}, which can affect the results if the grid is too coarse \citep{Candy2006}.
The inclusion of electromagnetic effects in continuum codes has been a long-solved issue\footnote{A key
realization was to compute all terms appearing in Amp\'ere's law numerically and in a consistent manner.
This idea was first presented by G.\,Hammett and F.\,Jenko in presentations at the Plasma Microturbulence
Project meeting at General Atomics on July 25, 2001 \citep[see][footnote~13]{Chen2003}.}
\citep{Kotschenreuther1995,Dorland2000,Jenko2001,Candy2003a,Candy2003b,Dannert2004},
although electromagnetic simulations for certain types of problems can still be challenging for
physics and not numerical reasons.
High-order continuum methods, which perform more calculations per grid point than low-order methods,
provide a path for faster convergence with solution size than PIC methods.
Continuum methods for gyrokinetic and fully kinetic simulation still appear to be associated with misconceptions
in the literature regarding larger memory requirements or worse computational efficiency 
\citep{Takizuka2017,Doyle2007,Kardaun2008} when compared to PIC methods, despite attempts to show that
they are similarly efficient for relevant problems \citep{Jenko2000b,Candy2006}.
We note that comparisons between PIC and continuum codes may also be highly problem dependent.
Nevertheless, it is essential to have independent PIC and continuum codes for the cross-verification of
results, especially of those that cannot yet be experimentally verified.

Another difference between PIC and continuum codes is that PIC codes have increasingly good velocity
space resolution for longer-wavelength modes, while continuum codes have a
velocity-space resolution that is independent of spatial scale.
The required velocity-space resolution is highly problem dependent, and the resonance
broadening by nonlinear scattering (island overlap) and collisions should
be accounted for.
The effects described in \citet{Su1968} show that collisional diffusion sets a limit on
velocity-space resolution requirements that scales as $\nu^{1/3}$,
where $\nu$ is the collision frequency, so the moderate collisionality
of the boundary plasma often means that the velocity-space resolution does not need
to be very high \citep[see also][]{Smith1997}.

\subsubsection{Gyrokinetic Codes for the Boundary Plasma}
Currently, the most sophisticated gyrokinetic code for the boundary plasma is the XGC1 gyrokinetic
particle code \citep{Chang2009},
which presently uses a `hybrid-Lagrangian scheme' \citep{Ku2016} and includes
a realistic diverted-plasma geometry, neutral particles, charge-exchange 
and ionization interactions, and radiation cooling \citep{Chang2017b}.
Recent electrostatic XGC1 simulations have predicted the midplane heat-flux width
on an attached ITER plasma to be 5.6~mm, which can be compared to the 5~mm
ITER design specification and the 1~mm
empirical scaling based on present experiments \citep{Chang2017b}.
The authors were able to reproduce the measured midplane heat-flux widths in three major
tokamaks to build confidence in their `gyrokinetic projection' to ITER.
The heat-flux width in XGC1 simulations of present-day tokamaks was found to be dominated
by ion-drift-orbit excursions, while the heat-flux width in the ITER simulation was
found to be dominated by turbulent electron-heat-flux spreading.
It remains to be seen what challenges the inclusion of electromagnetic effects in XGC1
simulations with kinetic electrons will present 
and whether similar calculations can be obtained at much cheaper computational cost through
other codes or models (the ITER simulation
reported in \citet{Chang2017b} used 300 billion markers and ran on 90\% of the Titan
computer for a few days).
ELMFIRE \citep{Korpilo2016,Heikkinen2008} is another 5D gyrokinetic PIC code being
extended to handle the boundary plasma, although it is still in an early stage of development.

On the other hand, the development of gyrokinetic continuum codes for the boundary plasma has
lagged significantly behind gyrokinetic PIC codes, despite the promising review of three main efforts
at the time by \citet{Cohen2008}.
Unfortunately, negative results are seldom reported in science, and so
it is unclear what issues have held up the development of these gyrokinetic continuum codes.
Some hints are found in the literature, however.
Recent papers reporting G5D simulations \citep{Kawai2017,Idomura2014,Idomura2012} focus only on core turbulence
and no longer have indications that the code is being extended to the boundary plasma.
The FEFI code \citep{Scott2010b,Scott2006a} tried to proceed directly
to electromagnetic simulations in the SOL, but ran into difficulties arising from sheath-model stability
and shear-Alfv\'en dynamics in low-density regions \citep{Zweben2009}.
The development of TEMPEST \citep{Xu2007} was apparently halted sometime after results from 4D
gyrokinetic transport simulations were presented in \citet{Xu2010}, perhaps due to issues stemming from
the non-conservation properties of the underlying numerical scheme \citep{Cohen2008}.
Some members of the TEMPEST team eventually began the development of the COGENT code \citep{Dorf2012},
which features a conservative fourth-order finite-volume discretization \citep{Colella2011}.
COGENT was recently used to perform axisymmetric 4D gyrokinetic transport simulations in a realistic geometry
with an anomalous radial-diffusion term to model radial transport due to turbulence \citep{Dorf2016}.
A modified version of the GENE code (widely used for core turbulence simulation)
has recently begun development for boundary-plasma applications \citep{Pan2016}.
A gyrokinetic continuum code with similar capabilities of XGC1 would be extremely valuable to the plasma physics
field, since XGC1 is the only gyrokinetic code capable of performing ITER boundary-plasma simulations 
at present.
Having the same prediction made by two (or several) independent gyrokinetic codes for the boundary plasma 
using different numerical approaches would be highly reassuring.

In continuum codes, spectral techniques are commonly used in some directions, which can have problems
with Gibbs phenomena that result in negative overshoots.
Most algorithms used in magnetic fusion research are designed for
cases in which viscous or dissipative scales are fully resolved and do not use limiters,
and thus can have problems with small negative oscillations.
Negative densities may result in unphysical behavior in the solution
(for example, a negative density in the tail of the electron distribution
function can reverse the slope of the sheath current versus sheath potential
relation), and inaccuracies in the sheath boundary conditions can also lead qualitatively
incorrect results.
Some finite-difference algorithms make it easier to calculate derivatives
across the LCFS with field-aligned coordinates, but may have problems
with particle conservation, and small imbalances in electron and ion gyrocenter
densities may drive large errors in the electric field.

For these reasons, we were motivated to investigate discontinuous Galerkin methods for
gyrokinetic continuum simulation in boundary plasmas.
DG methods are a class of finite-element methods
that use discontinuous basis functions (typically piecewise polynomials) to represent the solution in each cell.
The most popular version of DG is the Runge--Kutta discontinuous Galerkin method
\citep{Cockburn1998b,CockburnShu2001,Shu2009}, which uses a high-order DG method for space discretization and
explicit, strong stability-preserving (SSP) high-order Runge--Kutta methods \citep{Gottlieb2001} for time
discretization.
The RKDG method was originally developed for the solution of nonlinear, time-dependent hyperbolic systems
and has found use in numerous applications \citep{Cockburn2000}, especially for Euler and
Navier--Stokes equations.
Since continuity in the solution is not required across cell interfaces,
DG methods gain a number of important benefits that are not available to traditional finite-element methods.
The RKDG method is attractive because it is highly local,
highly parallelizable, able to handle complex geometries,
allows high-order accuracy, and enforces local conservation laws.
For an introduction to DG methods, the reader is referred to \citet{Shu2009,Durran2010,Hesthaven2010}.

\section{Thesis Overview}
This thesis focuses on efforts towards the development of a gyrokinetic continuum code
for the simulation of boundary-plasma turbulence.
Specifically, we investigate the application of discontinuous Galerkin algorithms to handle the difficulties
in open-field-line plasmas, which is the situation found in the SOL.
The algorithms that we use have been implemented in the Gkeyll code,
which is a framework for kinetic and fluid plasma simulations using a variety of grid-based numerical algorithms.
The Gkeyll code is primarily developed at the Princeton Plasma Physics Laboratory, with
contributors from a variety of institutions around the United States.
We note that we have not yet performed a `gyrokinetic continuum simulation of a tokamak SOL'.
At a minimum, a tokamak SOL simulation needs to include a confined edge region where plasma is sourced
and a realistic diverted geometry including a LCFS and X-point.
Additionally, kinetic modeling of wall-recycled neutrals
and models of radiative power losses, charge-exchange interactions, and ionization are required
for quantitative prediction, since these processes play important energy-dissipation roles in the SOL.
Nevertheless, major steps towards this goal have been completed in the course of this thesis.

In Chapter \ref{ch:models-and-numerical-methods}, we discuss the important models and algorithms used
in other chapters of the thesis.
We first describe the gyrokinetic model that has been implemented in Gkeyll,
which at present employs a number of simplifications (electrostatics, long-wavelength, linear polarization)
to make the problem tractable in the scope of a PhD thesis.
Next, we discuss discontinuous Galerkin algorithms, starting from a 1D example before
covering the specific energy-conserving version of DG that we use in our simulations.
We also discuss important aspects of a simplified Lenard--Bernstein collision operator and
how positivity issues in the distribution function are addressed.
A key component of simulating plasma dynamics on open magnetic field lines is the sheath-model boundary
condition applied at the material interfaces, and we describe two kinds of sheath models that can be used
and note an important shortcoming of these models concerning the Bohm sheath criterion
that is not usually acknowledged.

In Chapter \ref{ch:1d-sol}, we present results from our initial efforts to investigate the feasibility
of using DG methods for gyrokinetic continuum simulation in the boundary plasma in spatially 1D
kinetic simulations.
We describe the construction of a simplified 1D1V (one position dimension, one velocity dimension)
gyrokinetic model that incorporates ion polarization effects 
through a specified perpendicular wavenumber in a modified gyrokinetic Poisson equation.
Combined with logical-sheath boundary conditions, this model is then used to simulate the parallel propagation of an
ELM heat pulse in the SOL, which is a problem that has been studied before using kinetic simulations that
fully resolved the sheath and fluid simulations that used sheath boundary conditions.
This model is then extended to 1D2V and some collisional effects are added through a Lenard--Bernstein
collision operator.
Despite not directly resolving the sheath, our 1D1V and 1D2V gyrokinetic simulations 
agree quantitatively well with comparable fully kinetic simulations that are much more
computationally expensive due to restrictive spatial and temporal resolution requirements.

Chapter \ref{ch:lapd} presents the key accomplishment of the thesis, which are the first
5D gyrokinetic continuum simulations of turbulence in a straight-magnetic-field-line geometry.
Specifically, we present simulations of the Large Plasma Device (LAPD), which is a basic plasma physics experiment
at the University of California, Los Angeles.
Compared to a realistic scrape-off layer, the LAPD plasma is at a much colder temperature and
is not subject to magnetic-curvature effects.
The LAPD plasma is very well diagnosed, so we compare turbulence
characteristics from our simulations to previous LAPD measurements published elsewhere.
We also describe a simple modification to the sheath-model boundary conditions that allows us to simulate
a set of LAPD experiments to investigate sheared-flow-suppression of turbulence using bias-induced flows.

In Chapter \ref{ch:helical-sol}, we add some additional complexity to the open-field-line simulations
by adding magnetic-curvature effects to simulate turbulence on helical field lines.
While the new magnetic geometry makes the simulations particularly suitable for simulating the plasma
turbulence in SMTs, we design a test case for a helical SOL using parameters relevant for the NSTX SOL.
The helical-SOL simulations are qualitatively different from the LAPD simulations, with
the generation and radial propagation of blobs playing an important role in transporting plasma across the magnetic
field lines.
We also show how the magnetic-field-line incidence angle affects plasma profiles and turbulence characteristics
in these simulations.
In this simple model, we show that the heat-flux width is strongly affected by the strength of the 
vertical (poloidal) magnetic field.

In Chapter~\ref{ch:exp-basis}, we present a numerical method that
uses exponentially weighted polynomials to represent the solution while maintaining important conservation
properties.
The use of non-polynomial basis functions is motivated by the need to use as few pieces of data
to represent the solution as possible, and exponentially weighted polynomials appear to be a reasonable
choice to represent the distribution function in problems in which collisions are strong.
Previous work \citep{Yuan2006} in using non-polynomial basis functions in standard DG methods does not conserve
number, momentum, and energy.
Using 1D numerical tests of collisional relaxation,
we show how the new method conserves important quantities to machine precision.
Results from a non-trivial calculation of the parallel heat flux in a simplified
Spitzer--H\"arm test problem are then presented to compare the accuracy and efficiency of the new method
with standard DG methods using polynomials.
The generalization of this method to high dimensions and the implementation of this method in Gkeyll is left for
future work.

In Chapter \ref{ch:conclusion}, we summarize the main results of this thesis and discuss what are
high-priority directions for near-term future work.

\subsection{A Note on Color Maps}
Due to the continued popularity of the rainbow color map for representing
interval data in plasma-physics papers,
we briefly explain why authors should avoid using the rainbow color map for such purposes.
We decided to place this discussion here in the introduction chapter instead of
in the appendices to help spread general awareness of this issue.
In fact, even we are guilty of using a rainbow color map for the 2D plots
in a recent publication \citep{Shi2017} before this issue was brought to our attention.

As done by most authors, we present 2D data using pseudocoloring,
in which data is displayed by mapping scalar data values to colors
according to a \textit{color map}.
While many authors still use a color map that is ordered according to the visible light spectrum,
known as the \textit{rainbow color map},
data visualization experts have long recognized that the rainbow color map
is confusing and misleading \citep{Eddins2014,Borland2007,Ware1988,Rogowitz1998}.
Many of these issues arise from the lack of
\textit{perceptual ordering} and \textit{perceptual uniformity} in the rainbow color map.
Perceptual ordering refers to a color map that uses a sequence of colors with a
consistent, inherent ordering:
Given a set of distinct colors, will most people order the color in the same sequence
based on their color perception (and not a mnemonic)?
Many issues in the rainbow color map come from our perception of yellow as the brightest color,
while the rainbow color map assigns the largest data values to red.
Perceptual uniformity refers how the same difference between two data values 
corresponds to the same perceived difference in color on the entire color scale.
The perceived difference between the colors representing the values 1 and 2 should also
be the same between the colors representing the values 9 and 10. 

\begin{figure}
  \centerline{\includegraphics[width=\textwidth]{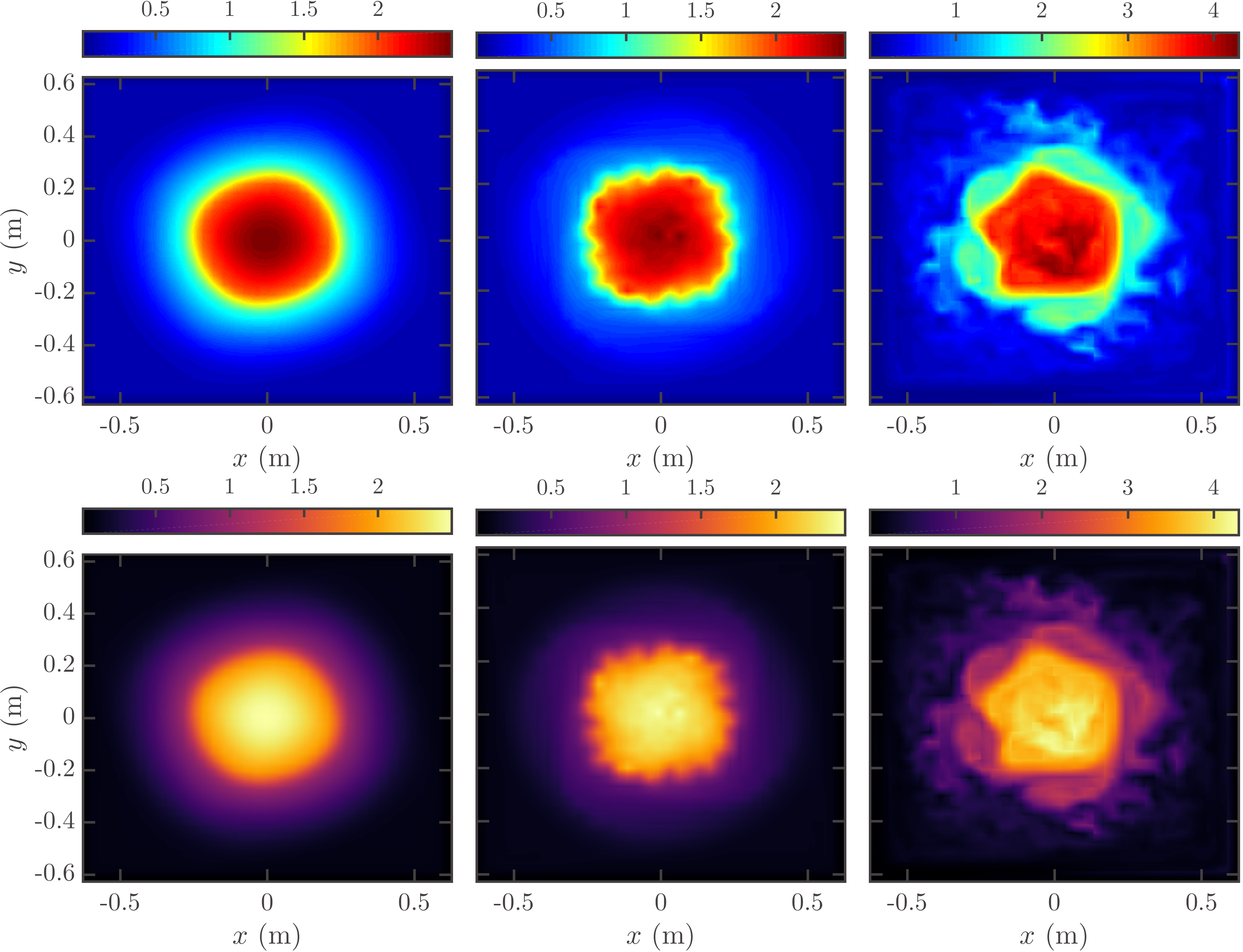}}
  \caption[Comparison of electron-density snapshots from gyrokinetic simulations of LAPD 
  using the rainbow color map and a perceptual color map.]
  {Comparison of electron-density snapshots (in 10$^{18}$~m$^{-3}$) from gyrokinetic simulations of LAPD 
  using (top row) the rainbow color map and  (bottom row) a perceptual color map.
  Several visual artifacts appear in the plots using the rainbow color map
  but not in the corresponding set of plots using a perceptual color map.
  In the upper set of plots, yellow edges and cyan `halos' are prominent,
  and large-scale dark-blue artifacts are visible in the outer regions in all three plots.
  The yellow outlines imply distinct, sharp-gradient regions in the data
  that are not actually present.
  Since green and cyan are extremely similar in hue,
  green and cyan regions tend to blend together
  and smear out fine structures, while such details are not obscured in the corresponding
  perceptual plots.
  Before the reader consults the color key, the dark-red features near the center
  of each upper plot also falsely suggest a depression in the data,
  since it resembles a shadow.}
  \label{fig:intro-lapd-color-map}
\end{figure}

The color map used for 2D data visualization in this thesis is the `inferno' color map \citep{Smith2016},
which is available Matlab, Matplotlib, and R.
The inferno color map is perceptually uniform and perceptually ordered.
Figure~\ref{fig:intro-lapd-color-map} shows a comparison of
electron-density snapshots from a gyrokinetic simulation of LAPD (discussed in Chapter~\ref{ch:lapd}) using
the rainbow color map (called the \textit{jet} color map in Matlab) and the inferno color map.
By comparing the two sets of plots, several visual artifacts can be identified in the plots
with the rainbow color map.
Additionally, the detail in green and cyan regions is wiped out due to the perceptual similarity
of these two colors.

%% file: ch-models-and-numerical-methods/chapter-models-and-numerical-methods.tex
\chapter{Models and Numerical Methods \label{ch:models-and-numerical-methods}}
In this chapter, we discuss some of the common models and numerical algorithms 
used elsewhere in this thesis.
Although it is well known that scientific results must be reproducible,
authors sometimes do not describe their simulations in published papers
in enough detail for other researchers to replicate their results.
One should strive to publicly document the model equations, initial conditions,
boundary conditions, grid resolution, and any special techniques that were necessary
to obtain stable simulations.
We hope to facilitate any future attempts to reproduce our results and to break away
from the tendency to be vague about the `ugly parts' of a simulation
by providing a clear description of our approach
for the open-field-line simulations discussed in Chapters~\ref{ch:lapd}
and \ref{ch:helical-sol}, including the difficulties we encountered
and what was done to address them.

In Section~\ref{sec:model}, we describe the gyrokinetic model we
solve in Chapters~\ref{ch:lapd} and \ref{ch:helical-sol}.
We then provide an overview of discontinuous Galerkin (DG) methods 
and discuss how we apply an energy-conserving DG method \citep{Liu2000} to solve
the gyrokinetic system in Section~\ref{sec:algorithms}.
We provide details about the numerical implementation of the Lenard--Bernstein collision operator
in Section~\ref{sec:collision_operator}, which also motivates the need for
a positivity-adjustment procedure described in Section~\ref{sec:positivity}.
Lastly, we specify the sheath-model boundary conditions that are used in our simulations
in Section~\ref{sec:sheath}.

These models and algorithms are implemented in the Gkeyll code,
which is a framework for kinetic and fluid plasma-physics problems.
Recently, Gkeyll has been used for fluid studies of magnetic reconnection \citep{Wang2015,Ng2015},
kinetic simulations of the Vlasov--Maxwell system \citep{Juno2017},
and kinetic and multi-fluid sheath modeling \citep{Cagas2017}.
Gkeyll is composed of a core component written
in \CC{} and problem-specific configuration components written in the Lua scripting language
\citep{Ierusalimschy1996}.

Gkeyll was originally designed and developed at the Princeton
Plasma Physics Laboratory (PPPL) by A.\,Hakim, who created the core framework
components (e.g., Lua integration, data structures, and grid classes)
used in the various applications.
For the simulations presented in this thesis, A.\,Hakim added additional
domain-decomposition capabilities, developed the gyrokinetic-Poisson-equation solver,
and contributed to the extension of the energy-conserving Liu~\&~Shu algorithm \citep{Liu2000}
to generic Hamiltonian systems (Section~\ref{sec:liu-algorithm}).
There are currently several code contributors from various academic institutions
across the United States.
At the time of writing, the Gkeyll repository is stored on Bitbucket and
access must be requested.\footnote{The author is well aware of the irony in
discussing the need for transparency in the plasma-simulation literature
while working on a closed-source code.
Access to the code is available upon request, however, and there are currently plans to make Gkeyll publically
available.}

The core \CC{} component contains implementations of the basic self-contained classes,
such as data structures to store solutions, rectangular finite-element meshes,
and solvers, which generically perform operations on input data to produce output data.
Currently, Gkeyll has classes that implement discontinuous Galerkin, finite-element,
finite-volume, and finite-difference algorithms.
Most of the work in this thesis required the addition of new capabilities to the Gkeyll code
in the form of solvers.

The configuration component uses Lua for problem-specific applications of the Gkeyll code.
The core Gkeyll code only needs to be compiled by the user once, while a Lua script is provided as an
input argument to the executable and is automatically compiled at run time.
Lua is used for much more than supplying parameter values for a simulation.
In a Lua script, a user defines which objects (implemented in the core code) are
to be created for a simulation and how these objects interact with each other.
To elaborate, there is no implementation of an `open-field-line simulation' in the core \CC{} code.
There are several classes relevant to this problem, such as implementations
of sheath boundary conditions, basis functions, a Poisson-equation solver,
a kinetic-equation solver, and more, but a Lua script is required to connect all these pieces together
and determine the sequence in which various tasks are performed.
Lua scripts for the 5D gyrokinetic continuum simulations have approximately 3000 lines of code.

\section{Gyrokinetic Model \label{sec:model}}
The gyrokinetic model discussed in this section is used in simulations of the
Large Plasma Device (LAPD) (Chapter~\ref{ch:lapd}) and a helical scrape-off-layer (SOL)
model (Chapter~\ref{ch:helical-sol}).
Several versions of full-$f$ gyrokinetic equations have been derived with
various formulations, ordering assumptions, and levels of accuracy
\citep{Brizard2007,Sugama2000,Hahm2009,Parra2011,Parra2014,Dimits2012,McMillan2016},
and can generically be written in the form of an evolution equation for the
gyrocenter distribution function, with a Poisson bracket for the phase-space
velocities and expressions for the Lagrangian/Hamiltonian, coupled to field equations to
determine the potentials.
Here, we solve a long-wavelength (drift-kinetic) limit of electrostatic full-$f$ gyrokinetic equations with
a linearized polarization term for simplicity, as summarized by \citet{Idomura2009}.
As the code is further developed, it can be extended to more accurate and more
general equations, though the equations will still have this generic structure.

The fundamental assumption of standard gyrokinetics is that there is a coordinate system
in which things change slowly compared to the gyrofrequency.
In some gyrokinetic derivations, the ordering assumptions are written in a more restrictive
form requiring that the fluctuation amplitudes must be small.
But as discussed in various places \citep[such as][]{Hahm2009,Dimits2012,McMillan2016},
more general derivations that are appropriate for the edge region of fusion devices are possible,
such as using a small vorticity ordering \citep{McMillan2016},
which allows large flows and large-amplitude fluctuations at long wavelengths.

We solve a full-$f$ gyrokinetic equation written in the conservative form \citep{Brizard2007,Sugama2000,Idomura2009}
\begin{equation}
  \frac{\partial \mathcal{J} f_s}{\partial t} + \nabla_{\boldsymbol{R}} \cdot (\mathcal{J} \dot{\boldsymbol{R}} f_s) +
  \frac{\partial}{\partial v_\parallel}(\mathcal{J} \dot{v}_\parallel f_s) = \mathcal{J} C[f_s] + \mathcal{J} S_s, \label{eq:gke}
\end{equation}
where $f_s = f_s(\boldsymbol{R}, v_\parallel, \mu, t)$ is the gyrocenter distribution function for species $s$,
$\mathcal{J} = B_\parallel^*$ is the Jacobian of the gyrocenter coordinates,
$B_\parallel^* = \boldsymbol{b} \cdot \boldsymbol{B^*}$, $\boldsymbol{B^*} = \boldsymbol{B} + (B v_\parallel/\Omega_s) \nabla \times \boldsymbol{b}$,
$C[f_s]$ represents the effects of collisions, $\Omega_s = q_s B/m_s$, and $S_s = S_s(\boldsymbol{R}, v_\parallel,\mu, t)$ represents plasma sources (e.g.,
neutral ionization or core plasma outflow).
In a straight-field-line geometry, $B_\parallel^*$ simplifies to $B$.
The phase-space advection velocities are defined as
$\dot{\boldsymbol{R}} = \{\boldsymbol{R}, H\}$ and $\dot{v_\parallel} = \{v_\parallel, H\}$,
where the gyrokinetic Poisson bracket is
\begin{equation}
\{F,G\} = \frac{\boldsymbol{B^*}}{m_s B_\parallel^*} \cdot \left( \nabla_{\boldsymbol{R}} F \frac{\partial G}{\partial v_\parallel} - \frac{\partial F}{\partial v_\parallel} \nabla_{\boldsymbol{R}} G \right)
  - \frac{1}{q_s B_\parallel^*} \boldsymbol{b} \cdot \nabla_{\boldsymbol{R}} F \times \nabla_{\boldsymbol{R}} G. \label{eq:gkPB}
\end{equation}

The gyrocenter Hamiltonian is
\begin{equation}
H_s = \frac{1}{2} m_s v_\parallel^2 + \mu B + q_s \langle \phi \rangle_\alpha, \label{eq:hamiltonian}
\end{equation}
where $\langle \phi \rangle_\alpha$ is the gyro-averaged potential (with the gyro-angle denoted by $\alpha$).
In the simulations discussed in Chapters~\ref{ch:lapd} and \ref{ch:helical-sol}, we consider a long-wavelength limit of the gyrokinetic system and
neglect gyroaveraging in the Hamiltonian to take $\langle \phi \rangle_\alpha = \phi$.
This system has similarities to some versions of drift kinetics (and is sometimes referred to as the drift-kinetic
limit of gyrokinetics \citep{Dorf2016,Dorf2013,Cohen2008}), but is unlike versions
that include the polarization drift in the kinetic equation or determine
the potential from some other equation.
In a straight-magnetic-field geometry, (\ref{eq:gke})--(\ref{eq:hamiltonian}) reduce to the description
of parallel streaming, an $E \times B$ drift, and acceleration along the field line due to $E_\parallel$
(see (\ref{eq:gksimple}), which is solved in our LAPD simulations).

The potential is solved for using the long-wavelength gyrokinetic Poisson equation with a linearized ion polarization density
\begin{equation}
-\nabla_\perp \cdot \left( \frac{n_{i0}^g q_i^2 \rho_{\mathrm{s}0}^2}{T_{e0} } \nabla_\perp \phi \right) =  \sigma_g = q_i n_i^g(\boldsymbol{R}) - e n_e(\boldsymbol{R}), \label{eq:gkp}
\end{equation}
where $\rho_{\mathrm{s}0} = c_{\mathrm{s}0} / \Omega_i$, $c_{\mathrm{s}0} = \sqrt{T_{e0}/m_i}$, and $n_{i0}^g$ is the background
ion gyrocenter density that we will take to be a constant in space and in time.
Gyroaveraging in the gyrocenter densities is neglected in this long-wavelength limit.
The replacement of $n_{i}^g(\boldsymbol{R})$ by $n_{i0}^g$ on the left-hand side of (\ref{eq:gkp})
is analogous to the Boussinesq approximation employed in some Braginskii fluid codes \citep{Dudson2015,Halpern2016,Angus2014}.
We note that the use of a linearized ion polarization charge density is formally valid when ion density fluctuations
are small.

Note that (\ref{eq:gkp}) is a statement of quasineutrality, where the right-hand side is the
gyrocenter component of the charge density $\sigma_g$, and the left-hand side is the negative of the
ion polarization charge density, $-\sigma_{\rm pol}$ (due to the plasma response to a cross-field electric field),
so this equation is equivalent to $0 = \sigma = \sigma_g + \sigma_{\rm pol}$.
The simulations are done in a Cartesian geometry with $x$ and $y$ being
used as coordinates perpendicular to the magnetic field, which lies solely in the $z$ direction.
Therefore, $\nabla_\perp = \boldsymbol{\hat{x}} \partial_x  + \boldsymbol{\hat{y}} \partial_y$.

\section{\label{sec:algorithms}Discontinuous Galerkin Algorithms}
Consider a numerical method to solve the following time-dependent partial differential equation
over a domain $\Omega$:
\begin{equation}
  \frac{\partial f}{\partial t} + G(f)= 0 \label{eq:general_1d_pde},
\end{equation}
where $G$ is an operator that involves spatial derivatives in $x$
and an initial condition and boundary conditions are prescribed.
One class of numerical methods, called series-expansion methods,
solves (\ref{eq:general_1d_pde}) by approximating $f(x,t)$ as a linear combination of a finite number of 
predetermined basis functions:
\begin{equation}
  f \approx f_h = \sum_{k=1}^N f_k(t) \psi_k(x). \label{eq:basis_expansion}
\end{equation}
In almost all practical cases of interest, it is not possible to find a solution that satisfies
(\ref{eq:general_1d_pde}) by this expansion because the $\psi_k$'s are generally not eigenfunctions of $G$
\citep{Durran2010}, so the $\approx$ sign is used to relate $f$ to the numerical solution $f_h$.
Therefore, we settle for finding the degrees of freedom $f_1(t),\dots,f_N(t)$ that minimize some kind of error
that quantifies the degree to which the numerical solution fails to satisfy (\ref{eq:general_1d_pde}).
For this purpose, it is convenient to define the residual $R(f_h)$
\begin{equation}
  R(f_h) = \frac{\partial f_h}{\partial t} + G(f_h) 
\end{equation}
and try to minimize a function involving the residual in an integral sense over $\Omega$ or at
a set of $N$ points in $\Omega$.

In Galerkin methods, a system of equations for the time evolution of the degrees of freedom,
$\partial f_1/\partial t,\dots,\partial f_N/ \partial t$,
is obtained by requiring that the residual be orthogonal to each basis function:
\begin{equation}
  \int_\Omega \mathrm{d}x \, R(f_h) \psi_k(x) = 0, \qquad 1 \le k \le N. \label{eq:galerkin_approx}
\end{equation}
The solution to (\ref{eq:galerkin_approx}) determines the time evolution of the degrees of freedom
such that the squared-$L^2$-norm error
\begin{equation}
  \left( \| R(f_h) \|_2 \right)^2  = \int_\Omega \mathrm{d}x \, R(f_h)^2 =
  \int_\Omega \mathrm{d}x \left( \frac{\partial f_h}{\partial t} + G(f_h) \right)^2 \label{eq:residual_sq}
\end{equation}
is minimized.
To see this, we substitute the expansion (\ref{eq:basis_expansion}) into (\ref{eq:residual_sq}),
take a derivative with respect to $\dot{f}_j = \mathrm{d} f_j / \mathrm{d}t$, 
and look for critical points:
\begin{align}
  0 =& \frac{\partial}{\partial \dot{f}_j} \int_\Omega \mathrm{d}x \left[ \left( \sum_{k=1}^N \dot{f}_k \psi_k \right)
    + G\left(\sum_{k=1}^N f_k \psi_k \right)^2 \right]^2 \\
    =& 2 \int_\Omega \mathrm{d}x \left[ \left( \sum_{k=1}^N \dot{f}_k \psi_k \right)
    + G\left(\sum_{k=1}^N f_k \psi_k \right)^2 \right] \psi_j \\
    =& 2 \int_\Omega \mathrm{d}x \, R(f_h) \psi_j(x).
\end{align}
This choice of $\dot{f}_j$ minimizes the squared-$L^2$-norm error because the 
second derivative of with respect to $\dot{f}_j$ is positive:
\begin{align}
  \frac{\partial^2}{\partial \dot{f}_j^2} \int_\Omega \mathrm{d}x R(f_h)^2 =
  2 \int_\Omega \mathrm{d}x\, \psi_j(x)^2.
\end{align}

The Galerkin approximation (\ref{eq:galerkin_approx}) is used in many series-expansion methods,
including the spectral method, some finite-element methods, and the discontinuous Galerkin method.
The Runge--Kutta discontinuous Galerkin (RKDG) method \citep{Cockburn1998b,CockburnShu2001,Shu2009}
is a semi-discrete numerical method that uses a discontinuous Galerkin discretization for the spatial variables
and explicit high-order-accurate Runge--Kutta methods (made of convex combinations
of first-order Euler steps) for time discretization.
The method is particularly well-suited for the solution of nonlinear, time-dependent hyperbolic conservation laws
and has found use in numerous applications \citep{Cockburn2000}.
As its name implies, the discontinuous Galerkin method is a Galerkin method that uses discontinuous basis
functions.
In contrast to other Galerkin methods, the condition (\ref{eq:galerkin_approx}) is enforced element-by-element 
instead of globally over the entire domain.

The use of discontinuous basis functions that are smoothly varying within a cell but zero everywhere outside
of it enables several benefits for DG methods.
Computations are highly localized in the sense that data only needs to be shared with immediate neighbors
regardless of the basis function degree,\footnote{This statement comes with some exceptions. In the
energy-conserving algorithm we use for the gyrokinetic system, a non-local solve is required for the
electrostatic potential.}
which is a desirable property for scalability and parallel efficiency
on massively parallel architectures. 
For other high-resolution methods, a wider stencil needs to be used to achieve high-order accuracy \citep{CockburnShu2001}.
The locality of the DG algorithm also makes it well-suited for adaptive $h$ (element size)
and $p$ (basis function degree) refinement and coarsening \citep{Remacle2003}.
We also note that a DG method that expands a scalar solution
as only a constant in each cell is a finite-volume, monotone scheme,
so DG methods can be considered as a higher-order generalization
of finite-volume methods \citep{Cockburn1998b,CockburnShu2001}.

\subsection{DG for 1D Conservation Laws}
We review the RKDG method for a 1D
nonlinear conservation law
\begin{equation}
  \frac{\partial f}{\partial t} + \frac{\partial g(f)}{\partial x} = 0. \label{eq:1d_conservation}
\end{equation}
As in finite-element methods, we first partition the domain $\Omega$ into a number of cells
$I_j = \left[x_{j-\frac{1}{2}},x_{j+\frac{1}{2}} \right]$,
for $1 \le j \le N$ and define an approximation space for the discrete solution $f_h$ 
\begin{equation}
  V_h = \{ v:v|_{I_j} \in V(I_j); 1\le j \le N \},
\end{equation}
where the local space $V(I_j)$ is usually taken to be $P^k(I_j)$,
the space of polynomials up to degree $k$ for $x \in I_j$.
Local basis functions that span $V(I_j)$ are required for numerical implementation,
and the solution in $I_j$ is expressed as
\begin{equation}
  f_h(x) = \sum_{l=1}^k f_j^l \psi_j^l, \qquad x \in I_j, \label{eq:basis_expansion_dg}
\end{equation}
so the RKDG method prescribes a way to solve for the evolution of the degrees of freedom
\begin{equation}
  \boldsymbol{f}_j = \begin{pmatrix}
    f^1_j\\
    \vdots \\
    f^k_j
  \end{pmatrix}, \qquad 1 \le j \le N.
\end{equation}

Legendre and Lagrange polynomials are typically used as basis functions in 
modal and nodal DG representations, respectively.
We have also implemented Serendipity basis functions \citep{Arnold2011},
which are an attractive alternative to Lagrange polynomials.
The Serendipity finite-element space has fewer basis functions (smaller dimension) than
the Lagrange finite-element space, while achieving the same convergence rate.
The Serendipity finite-element space can also represent solutions that are continuous
across elements.
The disparity between the size of these two finite element spaces grows
with the degree of the finite element space and the space dimension.
For example, $k=(2+1)^5=243$ degrees of freedom are required to represent the solution in 5D
for a second-degree Lagrange finite element, while only $k=112$ degrees of freedom are needed
for a second-degree Serendipity finite element, and one could expect a factor of five speedup
when using Serendipity finite elements in this case due to prevalence of matrix operations of size $k\times k$
in DG methods.
We have not yet explored the use of higher-order Serendipity elements for 5D gyrokinetic simulations
and use first-order elements (where the two finite-element spaces are identical) in those
simulations for simplicity.
DG methods also allow for the use of non-polynomial basis functions \citep{Yuan2006},
which can also result in significant savings over polynomial basis functions.
These ideas are explored in Chapter~\ref{ch:exp-basis}.

Recalling that Galerkin methods minimize the squared-$L^2$-norm error by requiring that the residual be orthogonal
to the basis functions (\ref{eq:galerkin_approx}), we multiply (\ref{eq:1d_conservation})
by an arbitrary test function $v(x)$ and integrate over a cell $I_j$
\begin{multline}
  \int_{I_j} \mathrm{d}x\, v(x) \frac{\partial f(x,t)}{\partial t} =
  \int_{I_j} \mathrm{d}x\, \frac{\partial v}{\partial x} g \left( f(x,t) \right)
  -g\left( f \left(x_{j+\frac{1}{2}},t \right)\right) v\left(x_{j+\frac{1}{2}}\right) \\
  +g\left( f \left(x_{j-\frac{1}{2}},t \right)\right) v\left(x_{j-\frac{1}{2}}\right),
\end{multline}
where an integration by parts was performed to move the spatial derivative on $g$ onto the test function $v$.
Next, the exact solution $f$ is replaced by the numerical solution $f_h$,
the flux $g\left( f(x,t) \right)$ evaluated at the element interfaces is replaced
by the numerical flux $\hat{g} \left( f_h(x,t) \right)$, and the test function $v$
is replaced by $v_h \in V(I_j)$:
\begin{multline}
  \int_{I_j} \mathrm{d}x\, v_h(x) \frac{\partial f_h(x,t)}{\partial t} =
  \int_{I_j} \mathrm{d}x\, \frac{\partial v_h}{\partial x} g \left ( f_h(x,t)\right)
  -\hat{g}\left( f_h\left(x_{j+\frac{1}{2}},t\right)\right) v_h(x_{j+\frac{1}{2}}^-) \\
  + \hat{g} \left( f_h\left(x_{j-\frac{1}{2}},t\right)\ \right) v_h(x_{j-\frac{1}{2}}^+). \label{eq:1d_conservation_full}
\end{multline}
The $-$ and $+$ superscripts indicate that a discontinuous function is evaluated at the interface using
the left and right limits of the discontinuous numerical solution, respectively.
In (\ref{eq:1d_conservation_full}), the test functions are evaluated inside $I_j$.

The numerical flux $\hat{g}$ is an approximation to the exact flux $g$ and depends on the value
of $f_h$ on each side of a boundary:
\begin{equation}
  \hat{g} \left( f_h \left(x_{j+\frac{1}{2}},t \right) \right)
  = \hat{g} \left( f_h\left(x^-_{j+\frac{1}{2}},t\right), f_h\left(x^+_{j+\frac{1}{2}},t\right)\right).
\end{equation}
The single-valued numerical flux is only defined at cell interfaces, where $f_h$ is discontinuous.
The numerical flux does not appear in the volume integrals because there is no ambiguity in the value of 
$f_h$ to use in the interior of a cell.
The choice of numerical flux must be consistent with the physical flux $g(a)$ that it approximates:
$\hat{g}(a,a) = g(a)$.
The numerical flux must also be a non-decreasing function of its first argument and a non-increasing
function of its second argument, which are required for the numerical scheme to reduce to a monotone
finite-volume scheme when the solution is approximated by a constant in each cell \citep{CockburnShu2001}.
A standard choice for the numerical flux is to use the \textit{upwind numerical flux}.
If $g(f) = cf$, then the upwind flux is
\begin{equation}
  \hat{g}(a,b) =  \begin{cases}
    ca & \text{if } c \ge 0, \\
    cb & \text{if } c < 0.
  \end{cases}
  \label{eq:upwind_flux}
\end{equation}
It is important to use a numerical flux suitable for the problem at hand because
the choice affects the approximation quality \citep{CockburnShu2001}.

We can now obtain a system of $k$ coupled equations that can be solved for the $k$ unknowns $\partial f_j^l/\partial t$
in each cell $I_j$ by substituting the basis function expansion for $f$ (\ref{eq:basis_expansion_dg})
and taking $v_h = \psi_j^m$ for $1 \le m \le k$ in (\ref{eq:1d_conservation_full}):
\begin{multline}
  \sum_{l=1}^k \mathbb{M}(m,l) \frac{\partial f_j^l}{\partial t} =
  \int_{I_j} \mathrm{d}x\, \frac{\partial \psi_j^m}{\partial x} g \left( f_h(x,t) \right)
  -\hat{g}\left( f_h\left(x_{j+\frac{1}{2}},t\right)\right) \psi_j^m \left(x_{j+\frac{1}{2}}^-\right) \\
  + \hat{g}\left(f_h\left(x_{j-\frac{1}{2}},t\right)\right) \psi_j^m \left(x_{j-\frac{1}{2}}^+\right),
  \label{eq:1d_conservation_num}
\end{multline}
where $\mathbb{M}(m,l) = \int_{I_j} \mathrm{d}x\, \psi_j^m(x) \psi_j^l(x)$ are the components
of the $k\times k$ mass matrix for $I_j$. Equation (\ref{eq:1d_conservation_num})
can be written as a matrix equation $\mathbb{M} \dot{\boldsymbol{f}}_j = \boldsymbol{c}$,
where the element $c_m$ of the vector $\boldsymbol{c}$ is the right hand side of (\ref{eq:1d_conservation_num})
evaluated for $1 \le m \le k$.
In this example, $\mathbb{M}$ is independent of time and is the same for all $I_j$,
and so it is advantageous to compute
$\mathbb{M}$ once and store its inverse $\mathbb{M}^{-1}$ at the beginning of a simulation.

Integrals are numerically evaluated using Gaussian quadrature methods.
In the Gkeyll code, Gauss--Legendre quadrature is typically used to approximate the definite integral
of a function $f(x)$ from $x=a$ to $x=b$ by taking a weighted sum of
the evaluation of the function to be integrated at a set of $N_q$ points:
\begin{equation}
  \int_{a}^{b} \mathrm{d}x \, f(x) \approx \sum_{j=1}^{N_q} w_j f(x_j),
\end{equation}
where the approximation is exact when $f(x)$ is a polynomial of degree $2N_q-1$ or less.
On the interval $[-1,1]$, the Gauss--Legendre quadrature nodes $x_j$ are the
zeros of $P_{N_q}(x)$, the $N_q$th Legendre polynomial for $x \in [-1,1]$,
and the associated weights $w_j$ are computed as \citep{Press2007}
\begin{equation}
  w_j = \frac{2}{(1-x_j)^2 [P_{N_q}'(x_j)]^2}.
\end{equation}
The quadrature rule on the interval $[-1,1]$ can be scaled to an arbitrary interval $[a,b]$
using the relation
\begin{equation}
  \int_a^b \mathrm{d}x \, f(x) = \frac{b-a}{2} \int_{-1}^1 \mathrm{d}x\,
  f\left( \frac{b-a}{2} x + \frac{a+b}{2} \right),
\end{equation}
so the quadrature rule becomes \citep{Press2007}
\begin{equation}
  \int_a^b \mathrm{d}x \, f(x) \approx \frac{b-a}{2}
  \sum_{j=1}^{N_q} w_j f\left(\frac{b-a}{2}x_j + \frac{a+b}{2} \right).
\end{equation}
We note that other quadrature rules with different weights and nodes can be constructed
for general integrals of the form $\int_a^b \mathrm{d}x \, W(x) f(x)$, where $W(x)$
is an arbitrary (possibly non-polynomial) weight function \citep[for an investigation of several choices, see][]{Landreman2013}.
Gaussian quadrature rules for integrals involving multiple directions are simply computed
by taking tensor products of the 1D quadrature rule.

In order to apply Gaussian quadrature methods to evaluate integrals as they appear in (\ref{eq:1d_conservation_num}),
auxiliary matrices\footnote{When DG is applied to systems with multiple spatial dimensions,
separate quadrature rules are needed for the evaluation of integrals on each surface of the element
in addition to the quadrature rule for volume integrals.}
are needed to compute the set of function evaluations at quadrature nodes
$\{f(x_1),\dots,f(x_{N_q})\}$ from the vector $\boldsymbol{f}_j$.
For this reason, it is necessary to compute and store the $N_q \times k$ quadrature matrix
\begin{equation}
  \mathbb{Q} =
    \begin{pmatrix}
      \psi_j^1(x_1) & \psi_j^2(x_1) & \cdots & \psi_j^k(x_1) \\
      \psi_j^1(x_2) & \psi_j^2(x_2) & \cdots & \psi_j^k(x_2) \\
      \vdots & \vdots & \ddots & \vdots \\
      \psi_j^1(x_{N_q}) & \psi_j^2(x_{N_q}) & \cdots & \psi_j^k(x_{N_q})
    \end{pmatrix}
\end{equation}
from the basis functions whose analytical form is known.

We also compute and store a matrix that calculates $\partial f_h /\partial x$ in terms of
the same basis functions used to represent the spatial dependence of $f_h$,
which is used when the evaluation of spatial derivatives is required (such as in the volume integral 
on the right-hand side of (\ref{eq:1d_conservation_num})).
That is, we seek the weights $f_{x,j}^l$ in element $I_j$ such that
\begin{equation}
  \frac{\partial f_h}{\partial x} = \sum_{l=1}^k f_{x,j}^l \psi_j^l, \qquad x \in I_j. \label{eq:1d_deriv}
\end{equation}
We multiply $(\ref{eq:1d_deriv})$ by each basis function $\psi_j^m$,
substitute the basis function expansion for $f_h$, and integrate
over $I_j$ to get the system of equations that can be solved for
the $f_{x,j}^i$:
\begin{equation}
  \int_{I_j} \mathrm{d}x\, \psi_j^m \sum_{l=1}^k f_j^l \frac{\partial \psi_j^l}{\partial x} =
  \int_{I_j} \mathrm{d}x\, \psi_j^m \sum_{l=1}^k f_{x,j}^l \psi_j^l, \qquad 1 \le m \le k.
\end{equation}
By defining an additional auxiliary $k \times k$ matrix $\mathbb{S}_x$ whose elements are
\begin{equation}
  \mathbb{S}_x(m,n) = \int_{I_j} \mathrm{d} x \, \frac{\partial \psi_j^m}{\partial x} \psi_j^n,
\end{equation}
we see that $\boldsymbol{f}_{x,j} = \mathbb{M}^{-1} \mathbb{S}_x \boldsymbol{f}_j$.

Finally, once (\ref{eq:1d_conservation_num}) has been solved for the weights
$\partial f_j^l / \partial t$, we use an explicit high-order strong-stability-preserving (SSP)
time stepping method to advance the solution in time.
These methods achieve high-order accuracy by taking convex combinations of first-order
forward Euler steps.
Specifically, the three-stage, third-order SSP Runge--Kutta method \citep{Gottlieb2001}
is used in the simulations discussed in this thesis
and is the most popular choice \citep{Shu2009} for solving equations of the form
\begin{equation}
  \frac{\partial f}{\partial t} = L(f), \label{eq:euler_example}
\end{equation}
where $L(f)$ is the space discretization of the $-\partial g(f) / \partial x$
operator in (\ref{eq:1d_conservation}).
If the first-order Euler time discretization of (\ref{eq:euler_example}) is
stable under a certain norm for a sufficiently small time step (e.g., satisfying the
Courant--Friedrichs--Levy condition),
high-order SSP methods are designed to automatically maintain the strong stability property for certain
higher-order time discretizations \citep{Gottlieb2001} under a possibly more restrictive time step.
To advance $f(x,t^{n}) \equiv f^n$ to $f(x,t^{n}+\Delta t) \equiv f^{n+1}$,
the third-order SSP Runge--Kutta method is
\begin{align}
  f^{(1)} =& f^n + \Delta t L\left(f^n\right), \label{eq:ssp-rk3-1}\\
  f^{(2)} =& \frac{3}{4} f^n + \frac{1}{4} \left[ f^{(1)} + \Delta t L\left(f^{(1)} \right) \right],  \label{eq:ssp-rk3-2} \\
  f^{n+1} =& \frac{1}{3} f^n + \frac{2}{3} \left[ f^{(2)} + \Delta t L\left(f^{(2)} \right) \right]  \label{eq:ssp-rk3-3}.
\end{align}
In this method, the time step restriction is the same as the time step needed for the forward-Euler
method to be strongly stable.
One downside to this method is the considerable storage requirement.
At any given instant of the algorithm, three versions of $f$ need to be stored.
For two plasma species (electrons and one ion species), the storage requirement is doubled.
It might be useful to consider the use of the low-storage, third-order SSP Runge--Kutta method,
which only needs to store two versions of $f$, but requires a 3.125 times smaller time step
\citep{Gottlieb2001}.

\subsection{DG for the Gyrokinetic System\label{sec:liu-algorithm}}
An energy-conserving (in the continuous-time limit) discontinuous Galerkin algorithm \citep{Liu2000}
is used to discretize the equations in space.
Although \citet{Liu2000} presented their algorithm for the 2D incompressible Euler and Navier--Stokes equations,
G.\,Hammett recognized the general applicability of their algorithm for Hamiltonian systems,
and A.\,Hakim contributed to the generalization.
Upwind interface fluxes (\ref{eq:upwind_flux}) are used in the discretization of the 
gyrokinetic equation (\ref{eq:gke}).
This algorithm requires that the Hamiltonian be represented on a continuous subset
of the basis set used to represent the distribution function.
Therefore, the distribution function is represented using discontinuous ($C^{-1}$) polynomials,
while the electrostatic potential is represented using continuous ($C^0$) polynomials (equivalent
to continuous finite elements).
In addition to contributing to the formulation of the generalized algorithm,
A.\,Hakim also implemented and performed a number of two-dimensional tests
(e.g. incompressible Euler equations, 1D1V Vlasov--Poisson system)
to benchmark the convergence and conservation
properties of this algorithm.
Additional details regarding the generalization of the \citet{Liu2000} algorithm
will be provided in future paper by A.\,Hakim et al.

\citet{Liu2000} presented their algorithm for 2D incompressible Euler equations in a
vorticity--stream function formulation: $\partial \rho(x,y,t) / \partial t + \nabla \cdot ({\bf u} \rho) = 0$,
with the stream function $\psi$ given by $\nabla^2 \psi = \rho$,
and the velocity ${\bf u} = \nabla^{\perp} \psi = (-\partial_y \psi, \partial_x \psi)$.
They showed analytically that the DG space discretization of these equations conserves
energy if the basis functions for $\psi$ are in a continuous subspace of the basis
functions used for the vorticity $\rho$.\footnote{The numerical tests of \citet{Liu2000} 
did not actually use basis sets that satisfied these properties.}
This problem can also be written in a Hamiltonian form
$\partial \rho / \partial t = - \{\psi,\rho\}$, where $\psi$
is the Hamiltonian and the Poisson bracket in this case is
$\{\psi,\rho\} = \partial_x \psi \partial_y \rho - \partial_y \psi \partial_x \rho$.

Here, we consider how the algorithm of \citet{Liu2000} is applied to the gyrokinetic equation
system (\ref{eq:gke})--(\ref{eq:gkp}), although we will neglect the collisions and source terms in
the gyrokinetic equation (\ref{eq:gke}) to focus on the Hamiltonian part of the system.
First, we discretize the $N$-dimensional phase-space domain $\Omega$ by dividing it into a number of
elements $I_j$.
We also define the configuration-space domain $\Omega_x$ and its respective partition $\mathcal{T}^x$.

Next, we define the approximation spaces \citep[following the notation of][]{Liu2000}
\begin{align}
  V^k =& \left\{ v : v|_{I_j} \in Q^k (I_j), \forall I_j \in \mathcal{T} \right\}, \\
  W^k_0 =& V^k \cap C_0 (\Omega),
\end{align}
where $Q^k (I_j)$ is the space of polynomials in $N$ variables with each variable degree at most $k$ for $z \in I_j$.

We note that the gyrokinetic equation (\ref{eq:gke}) can be written in the general form
\begin{equation}
  \frac{\partial}{\partial t}\left( \mathcal{J} f \right) + \frac{\partial}{\partial z^j} \left(\mathcal{J} \dot{z}^j f \right) = 0,
\end{equation}
where the coordinates $\boldsymbol{z} = (z^1,z^2,z^3,z^4,z^5) = (x,y,z,v_\parallel,\mu)$ 
and $\dot{z}^i = \mathrm{d} z^i / \mathrm{d}t = \{z^i, H \}$,
The Poisson bracket can be written as
\begin{equation}
  \{f,g\} = \frac{\partial f}{\partial z^i} \Pi^{ij} \frac{\partial g}{\partial z^j},
\end{equation}
so $\dot{z}^i = \Pi^{ij} \partial H/\partial z^j$.
The Poisson matrix $\boldsymbol{\Pi}$ is assumed to be antisymmetric.
As a consequence of this antisymmetry, the component of
the characteristic velocities normal to a surface is continuous on that surface.
This means that $\boldsymbol{n} \cdot \dot{\boldsymbol{z}}$ evaluated at an interface
between two cells has the same value when approached from either side of the interface.
Equivalently, this means that $\boldsymbol{n} \cdot \dot{\boldsymbol{z}}$ evaluated on the interface between
two cells can be correctly evaluated using the solution from only one of the cells.
To see this, we compute
\begin{equation}
  \boldsymbol{n} \cdot \dot{\boldsymbol{z}} = n_i \Pi^{ij} \partial H/\partial z^j =
  \tau^i \partial H/\partial z^j = \boldsymbol{\tau} \cdot \nabla H \label{eq:nDotZDot},
\end{equation}
which is a term that will appear in surface integrals when we perform an integration by parts on the 
weak form of the advection equation. 
We also see that $\boldsymbol{\tau}$ is orthogonal to $\boldsymbol{n}$:
\begin{equation}
  \boldsymbol{\tau} \cdot \boldsymbol{n} = n_i \Pi^{ij} n_j = \left\{ \boldsymbol{n},\boldsymbol{n} \right\} = 0 \label{eq:tauDotN}.
\end{equation}
Since we have required that the Hamiltonian be continuous, we see from (\ref{eq:nDotZDot}) and (\ref{eq:tauDotN}) 
that $\boldsymbol{n} \cdot \dot{\boldsymbol{z}}$ is continuous on cell surfaces.
While $\boldsymbol{n}\cdot\nabla H$ can be discontinuous at cell surfaces because $H$ is only required to be $C^0$ continuous,
$\boldsymbol{\tau} \cdot \nabla H$ will be continuous on a surface if $H$
is also continuous on that surface. 

As in standard DG methods, we multiply the gyrokinetic equation (\ref{eq:gke}) by a test function $v$ and
integrate over each cell $I_j$:
\begin{equation}
  \int_{I_j} \mathrm{d}\Lambda \, v \frac{\partial}{\partial t}\left( \mathcal{J} f \right) +
  \oint_{\partial I_j} \mathrm{d}S \, v \mathcal{J} \boldsymbol{n} \cdot \dot{\boldsymbol{z}} \hat{f}
  - \int_{I_j} \mathrm{d}\Lambda \, \mathcal{J} \nabla v \cdot \dot{\boldsymbol{z}} f = 0. \label{eq:weak_gke}
\end{equation}
By taking $v = 1$ and summing over all cells, we see that particle number is conserved by
the discrete scheme (assuming that the constant 1 is in the approximation space):
\begin{align}
  \sum_{I_j} \int_{I_j} \mathrm{d}\Lambda \, \frac{\partial}{\partial t}\left( \mathcal{J} f \right) + 
  \oint_{\partial I_j} \mathrm{d}S \, \mathcal{J} \boldsymbol{n} \cdot \dot{\boldsymbol{z}} \hat{f} =& 0 \nonumber \\
  \frac{\partial}{\partial t} \left( \sum_{I_j} \int_{I_j} \mathrm{d}\Lambda \, \mathcal{J} f \right) =& 0.
\end{align}

For the 5D gyrokinetic system, $\mathcal{J}\boldsymbol{\Pi}$ has the following form (see (\ref{eq:gkPB})):
\begin{align}
\mathcal{J}\boldsymbol{\Pi} &= 
	\begin{pmatrix}
		0 & - b_z/q_s & b_y/q_s & B_x^*/m_s & 0 \\
		 b_z/q_s & 0 & -b_x/q_s & B_y^*/m_s & 0\\
		-b_y/q_s & b_x/q_s & 0 & B_z^*/m_s & 0 \\
		-B_x^*/m_s & -B_y^*/m_s & -B_z^*/m_s & 0 & 0 \\
		0 & 0 & 0 & 0 & 0
	\end{pmatrix}.
  \label{eq:gkPB_full}
\end{align}

Following a standard finite-element approach,
in order to satisfy continuity constraints on the potential,
we project the gyrokinetic Poisson equation onto the space of basis
functions for the potential, yielding a single non-local 3D solve to find
the potential.
Multiplying (\ref{eq:gkp}) by a test function $w$
and integrating over the entire configuration-space domain $\Omega_x$ (not just over an individual cell, since
the basis functions for $\phi$ couple the cells together):
\begin{align}
  - \int_{\Omega_x} \mathrm{d}^3 x \, w \nabla_\perp\left(\epsilon \nabla_\perp \phi \right) =
  \int_{\Omega_x} \mathrm{d}^3 x \, w \sigma_g, \label{eq:discrete_gkp}
\end{align}
where $\epsilon = n_{i0}^g q_i^2 \rho_{\mathrm{s}0}^2/T_{e0}$.
There are ways to solve (\ref{eq:discrete_gkp}) by performing a set of
independent 2D solves, followed by a local self-adjoint smoothing/interpolation
operation, but we have not yet implemented this potentially more efficient approach.
Although a non-local solve in three dimensions is required for the potential,
the 5D gyrokinetic equation itself can be solved in a highly local manner.

We can see that there will be a conserved energy by taking $v$ to be the discrete Hamiltonian:
\begin{equation}
\int_{I_j} \mathrm{d}\Lambda \, H \frac{\partial}{\partial t}\left( \mathcal{J} f \right) +
  \oint_{\partial I_j} \mathrm{d}S H \mathcal{J} \boldsymbol{n} \cdot \dot{\boldsymbol{z}} \hat{f}
  - \int_{I_j} \mathrm{d}\Lambda \, \mathcal{J} \nabla H \cdot \dot{\boldsymbol{z}} f = 0. \label{eq:h_times_gke}
\end{equation}
The third term is zero because $\nabla H \cdot \dot{\boldsymbol{z}} = \{ H, H \} =0$.
By summing (\ref{eq:h_times_gke}) over all elements, the second term vanishes because
the surface integrals not on the domain boundaries appear in equal and opposite pairs due to the
$C^0$ continuity of $H$, while the surface integrals on the domain boundaries are
zero by boundary conditions (e.g., zero flux or $\phi = 0$ ).
Therefore, we have
\begin{equation}
  \sum_{I_j} \int_{I_j} \mathrm{d}\Lambda \, H \frac{\partial}{\partial t}\left( \mathcal{J} f \right) = 0. \label{eq:h_times_gke_simple}
\end{equation}
To identify the conserved energy for the gyrokinetic system, we insert the discrete Hamiltonian
into (\ref{eq:h_times_gke_simple}) and sum over all species:
\begin{equation}
  \sum_s \sum_{I_j} \int_{I_j} \mathrm{d}\Lambda \, \left(\frac{1}{2}m_s v_\parallel^2 + \mu B + q_s \phi \right)
  \frac{\partial}{\partial t}\left( \mathcal{J} f_s \right) = 0. \label{eq:dg_W_tot}
\end{equation}
Carrying out the velocity-space integrals and identifying the first two terms as a thermal energy,
we have
\begin{equation}
  \frac{\partial W_k}{\partial t} + \sum_{I_j} \int_{I_j} \mathrm{d}^3 x \, \phi \frac{\partial \sigma}{\partial t} = 0,
\end{equation}
where $W_k = \sum_s \sum_{I_j}  \int_{I_j} \mathrm{d}\Lambda \, (m_s v_\parallel^2/2 + \mu B) \mathcal{J} f_s$ and 
the species summation on the last term was carried out to write it in terms of the gyrocenter charge density $\sigma$.
Next, we take a time derivative of (\ref{eq:discrete_gkp}) and set $w = \phi$ to get
\begin{align}
  \int_\Omega \mathrm{d}^3 x \, \phi \frac{\partial \sigma_g}{\partial t}
   =& - \int_\Omega \mathrm{d}^3 x \, \phi \nabla_\perp\left(\epsilon \nabla_\perp \frac{\partial \phi}{\partial t} \right) \nonumber \\
   =& \int_\Omega \mathrm{d}^3 x \, \epsilon \nabla_\perp \phi \frac{\partial \nabla_\perp \phi}{\partial t} \nonumber \\
   =& \frac{\partial}{\partial t} \int_\Omega \mathrm{d}^3 x \, \frac{\epsilon}{2} \left(\nabla_\perp \phi \right)^2, \label{eq:dg_W_phi}
\end{align}
where the surface terms are zero from the boundary conditions (e.g., periodic or $\phi = 0$ on the side walls).
Therefore, we have the energy conservation law for the discrete system
\begin{align}
  \frac{\partial}{\partial t} \left( W_k + W_\phi \right) = 0,
\end{align}
where the $E \times B$ energy $W_\phi = \int_\Omega \mathrm{d}^3 x \, \epsilon \left(\nabla_\perp \phi \right)^2/2$.

For simplicity, we use nodal, linear basis functions to approximate the solution in each element
for the 5D gyrokinetic simulations in Chapters~\ref{ch:lapd} and \ref{ch:helical-sol}.
This choice leads to $32$ degrees of freedom per cell in the 5D phase-space mesh
($8$ degrees of freedom in the 3D configuration-space mesh).
With the $32$ degrees of freedom specified in a cell, $f$ can be evaluated anywhere within the cell without
additional approximation.
With knowledge of the basis functions, we can generate data for plotting without having to rely on
approximate interpolation methods.
This means that the data we show in various plots in this thesis have not been filtered or smoothed
in post processing to make a more visually attractive image.
Some more details about the plotting procedures used to create figures for this thesis are provided
in Appendix~\ref{ch:plot-creation}.

The choice of a linear-polynomial basis set means that $v_\parallel^2$, which appears in the Hamiltonian (\ref{eq:hamiltonian}),
cannot be exactly represented.
We therefore approximate $v_\parallel^2$ in the linear basis set by requiring that the
piecewise-linear approximation equal $v_\parallel^2$ at the DG nodes.
These nodes are located on the vertices of uniform rectangular cells, so the nodes of cell $j$
either have $v_\parallel = v_{c,j} - \Delta v_\parallel/2$ or $v_\parallel = v_{c,j} + \Delta v_\parallel/2$,
where $v_{c,j}$ is the $v_\parallel$ coordinate of the center of cell $j$.
By approximating $v_\parallel^2$ in such a manner, $v_\parallel^2$ will vary linearly in cell $j$ from
$\left( v_{c,j} - \Delta v_\parallel/2 \right)^2$ to $\left( v_{c,j} + \Delta v_\parallel/2 \right)^2$.
Note that the approximated $v_\parallel^2$ is continuous across elements, as required for numerical energy conservation,
and its first derivative is discontinuous across elements, i.e. $\partial_{v_\parallel} H = v_{c,j}$.
We use rectangular meshes with uniform cell spacing, but we note that most of the DG algorithms discussed in this
thesis are trivially generalizable to non-uniform and non-rectangular meshes.

\section{Collision Operator \label{sec:collision_operator}}
Electron--electron and ion--ion collisions are implemented using a Lenard--Bernstein
model collision operator \citep{Lenard1958}
\begin{align}
C_{ss}[f_s] =& \nu_{ss} \frac{\partial}{\partial \boldsymbol{v}} \cdot \left[ (\boldsymbol{v} - \boldsymbol{u}_s) f_s + v_{t,ss}^2 \frac{\partial f_s}{\partial \boldsymbol{v}}\right] \nonumber\\
=& \nu_{ss} \frac{\partial}{\partial v_\parallel} \left[ (v_\parallel - u_{\parallel,s}) f_s + v_{t,ss}^2 \frac{\partial f_s}{\partial v_\parallel}\right]
  + \nu_{ss} \frac{\partial}{\partial \mu}\left[2\mu f_s + 2 \frac{m_s v_{t,ss}^2}{B} \mu \frac{\partial f_s}{\partial \mu} \right], \label{eq:lbCollisionOp}
\end{align}
where standard expressions are used for collision frequency $\nu_{ss}$ \citep[p. 37]{Huba2013}, $n_s v_{t,ss}^2 = \int \mathrm{d}^3 v \, \left( \boldsymbol{v} - \boldsymbol{u_s}\right)^2 f_s /3$,
and $n_s u_{\parallel,s} = \int \mathrm{d}^3v \, v_\parallel f_s$.
The exact expressions used in our simulations for $\nu_{ee}$ and $\nu_{ii}$ are \citep{Fitzpatrick2011}
\begin{align}
  \nu_{ee} &= \frac{n_e(\boldsymbol{R},t) e^4 \ln \Lambda}{6 \sqrt{2} \pi^{3/2} \epsilon_0^2 \sqrt{m_e}
  T_e(\boldsymbol{R},t)^{3/2}} ,\\ 
  \nu_{ii} &= \frac{n_i(\boldsymbol{R},t) e^4 \ln \Lambda}{12 \pi^{3/2} \epsilon_0^2 \sqrt{m_i}
  T_i(\boldsymbol{R},t)^{3/2}} ,
\end{align}
where $\ln \Lambda = 6.6 - 0.5 \ln \left( n_0/10^{20} \right) + 1.5 \ln T_{e0}$ is Coulomb logarithm
for $n_0$ expressed in $m^{-3}$ and $T_e$ expressed in eV.
As implied by these expressions, the variation of densities and temperatures
in time and in space is taken into account for the collision frequencies 
used in the code, except in the Coulomb logarithm.

This collision operator relaxes to a local Maxwellian, contains pitch-angle scattering, and analytically conserves number, momentum, and energy.
Note that the collision frequency is independent of velocity; the $v^{-3}$ dependence of the collision frequency expected for Coulomb collisions
is neglected.
This collision operator is long wavelength and ignores finite-gyroradius
corrections, which lead to classical cross-field diffusion.
This model operator represents many of the key features of the full Landau operator,
including velocity-space diffusion that preferentially damps small velocity-space scales, but
is much simpler to implement in the code.
Collisions with neutrals are neglected at present.

For simplicity, electron--ion collisions are also modeled using Lenard--Bernstein collision operator rather than an operator that only causes pitch-angle scattering:
\begin{equation}
C_{ei}[f_e] = \nu_{ei} \frac{\partial}{\partial v_\parallel} \left[ (v_\parallel - u_{\parallel,i}) f_e + v_{t,ei}^2 \frac{\partial f_e}{\partial v_\parallel}\right]
  + \nu_{ei} \frac{\partial}{\partial \mu} \left[2\mu f_e + 2 \frac{m_e v_{t,ei}^2}{B} \mu \frac{\partial f_e}{\partial \mu} \right] \label{eq:lbCollisionOpElcIon},
\end{equation}
where $\nu_{ei} = \nu_{ee}/1.96$ is used in the simulations and $n_e v_{t,ei}^2 = \int \mathrm{d}^3 v \, \left( \boldsymbol{v} - \boldsymbol{u_i}\right)^2 f_e /3$.
The coefficients are chosen so that the electrons relax to become isotropic in the frame of the mean ion velocity,
and conserves energy while losing mean momentum to the ions.
Because electron--electron collisions cause both pitch-angle scattering and energy diffusion,
we set $\nu_{ei}$ to a smaller value than $\nu_{ee}$.
The corresponding small change in the ion velocity is neglected, leading to a small
$\mathcal{O}(m_e/m_i)$ violation of momentum conservation.
The very slow energy exchange due to the $C_{ie}$ operator is also neglected.

The collision operators that we have implemented are constructed to numerically conserve
number and energy, but do not conserve momentum.
Analytical expressions for the forms of $u_{\parallel,s'}$ and $v_{t,ss'}^2$ necessary for exact numerical
conservation momentum and energy are difficult to derive due to the complicated
nature of the calculation of diffusion terms (see the following section).
Number conservation comes from the fact that the DG method solves the collision operators
in a weak form and the polynomial space spanned by the basis functions includes the constant 1.
Energy conservation is achieved by choosing the numerical value of $v_{t,ss'}^2$ by first calculating
the power of the drag and diffusion terms, e.g.,
\begin{align}
  P_{\mathrm{drag}} =& \nu_{ss'} \int \mathrm{d} \Lambda\, H \left( \frac{\partial}{\partial v_\parallel} \left[ (v_\parallel - u_{\parallel,s'}) f_s \right]
  + \frac{\partial}{\partial \mu}\left(2\mu f_s\right) \right) \\
  P_{\mathrm{diff}} =& \nu_{ss'} \int \mathrm{d} \Lambda\, H \left[ v_{t,ss'}^2 \frac{\partial^2 f_s}{\partial v_\parallel^2}
  + \frac{\partial}{\partial \mu}\left(2 \frac{m_s v_{t,ss'}^2}{B} \mu \frac{\partial f_s}{\partial \mu} \right) \right],
\end{align}
and then multiplying the diffusion terms in (\ref{eq:lbCollisionOp}) or (\ref{eq:lbCollisionOpElcIon})
by a near-unity constant such that energy is exactly conserved by the collision operator:
\begin{multline}
  C_{ss'}[f_s] = \nu_{ss'} \frac{\partial}{\partial v_\parallel} \left[ (v_\parallel - u_{\parallel,s'}) f_s + c_E v_{t,ss'}^2 \frac{\partial f_s}{\partial v_\parallel}\right] \\
  + \nu_{ss'} \frac{\partial}{\partial \mu} \left[2\mu f_s + 2 c_E \frac{m_s v_{t,ss'}^2}{B} \mu \frac{\partial f_s}{\partial \mu} \right],
\end{multline}
where $c_E = -P_{\mathrm{drag}}/P_{\mathrm{diff}} \approx 1$.
This choice to conserve energy is not unique.
In principle, a similar procedure can be used to construct a same-species collision operator that also
numerically conserves momentum, but we have not yet done this.

\subsection{Second-Order Derivatives}
Second-order derivatives in the collision operator
are calculated using the recovery-based DG method \citep{VanLeer2005},
which has the desirable property of producing symmetric solutions.
Originally, we used the popular local discontinuous Galerkin method \citep{Cockburn1998} for
the second-order derivatives, but the asymmetries inherent to the use of alternating fluxes
combined with the positivity-adjustment procedure described in Section \ref{sec:positivity} led
to large, unphysical asymmetries (such as in the flow) in our LAPD simulations.
We note that we do not need to follow the approach of \citet{VanLeer2007} for 2D diffusion problems
because the grids we used are composed of quadrilaterals.

\begin{figure}
  \centerline{\includegraphics[width=0.6\textwidth]{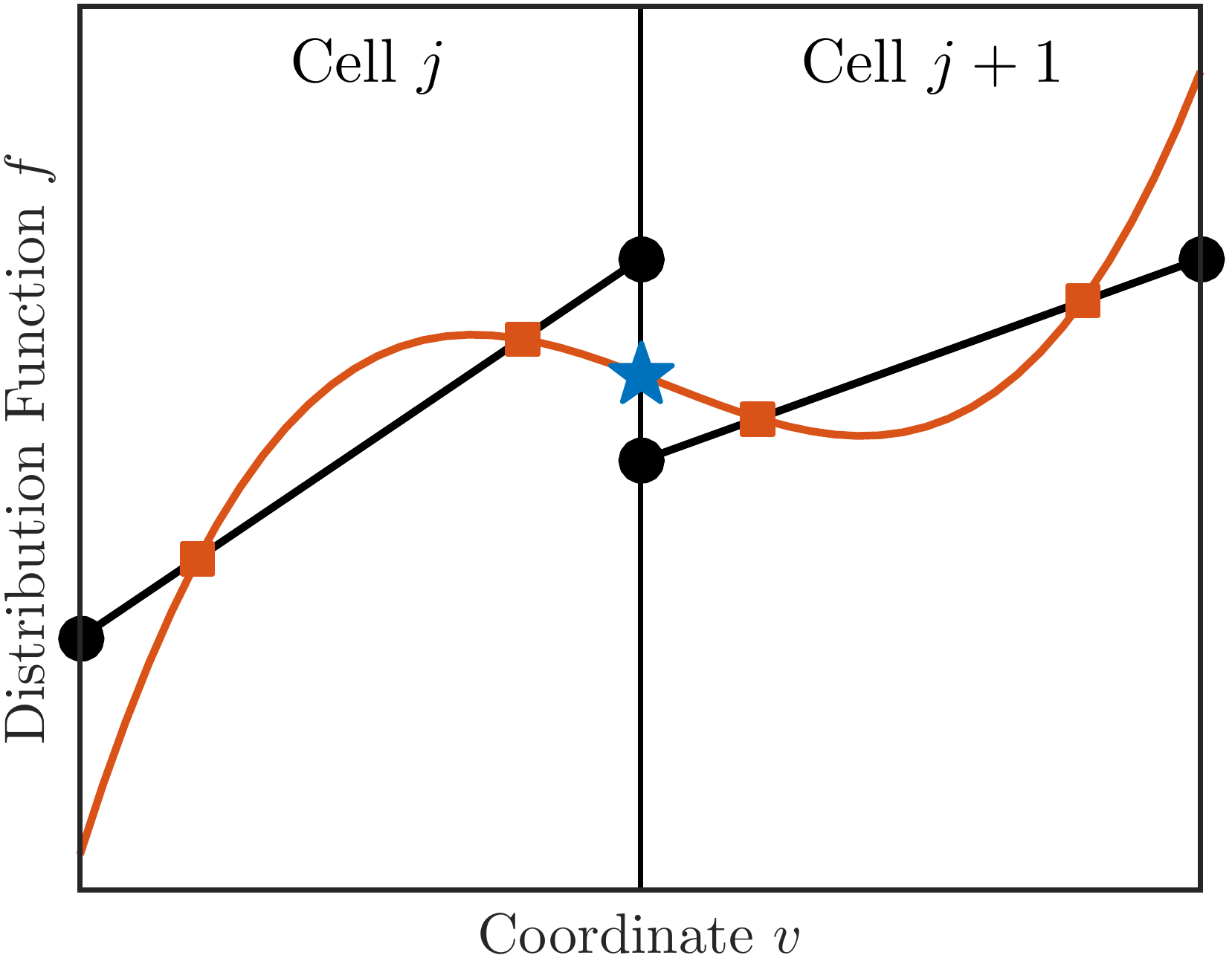}}
  \caption[Illustration of how a recovery polynomial that spans two neighboring cells is constructed
  using the DG solution from both cells.]{Illustration of how a recovery polynomial (red) that spans
  two neighboring cells is constructed using the DG solution from both cells (black).
  The recovery polynomial is identical to the DG solution in the weak sense, and it is equivalent to the
  Lagrange interpolation polynomial that passes through $f$ at the $p+1$ Gauss--Legendre quadrature points
  in each cell (red squares).
  The locally recovered smooth solution and its derivatives are used in the evaluation of fluxes
  on the shared boundary (blue).}
  \label{fig:recovery_1d}
\end{figure}

Before we discuss how we apply the recovery-based DG method in our 5D simulations,
we will first review how this method is used to solve diffusion equations in 1D.
Consider the diffusion equation
\begin{equation}
  \frac{\partial f}{\partial t} = D \frac{\partial^2 f}{\partial v^2}, \label{eq:1d_diffusion}
\end{equation}
where $D$ is a positive diffusion coefficient.
To enforce the Galerkin condition (\ref{eq:galerkin_approx}) element by element,
we multiply (\ref{eq:1d_diffusion}) by each basis function $\psi_k$ and then integrate by parts:
\begin{align}
  \int_{I_j} \mathrm{d}v \, \psi_k \frac{\partial f}{\partial t} = \left.
  D \psi_k \frac{\partial f}{\partial v} \right|_{v_{j-\frac{1}{2}}}^{v_{j+\frac{1}{2}}}
  - D \int_{I_j} \mathrm{d}v \, \frac{\partial \psi_k}{\partial v}\frac{\partial f}{\partial v}.
\end{align}
In the recovery-based DG method, one more integration by parts is performed, and we have
\begin{equation}
  \int_{I_j} \mathrm{d}v \, \psi_k \frac{\partial f}{\partial t} = \left. D \left( \psi_k \frac{\partial f_r}{\partial v} 
  -\frac{\partial \psi_k}{\partial v} f_r \right)\right|_{v_{j-\frac{1}{2}}}^{v_{j+\frac{1}{2}}}
  + D \int_{I_j} \mathrm{d}v \, \frac{\partial^2 \psi_k}{\partial v^2}f \label{eq:recovery_evo}
\end{equation}
after replacing $f$ and $\partial_{v} f$ in the interface terms by a recovered smooth polynomial $f_r$
and its first derivative.
The basis function and its derivative evaluated at the interfaces is evaluated from inside $I_j$.
The volume-integral term in (\ref{eq:recovery_evo}) is zero unless basis functions of degree $p \ge 2$ are used.
The recovery polynomial $f_r$ is constructed using the DG solution in two cells sharing a common boundary
for the evaluation of $f_r$ and $\partial_{v} f_r$ on the shared boundary.
The recovery polynomial $f_r$ is chosen to be identical in a weak sense to the DG solution $f$, so
it satisfies the following system of equations \citep[see][p. 13]{VanLeer2005}:
\begin{equation}
  \left.\begin{aligned}
    \int_{I_j} \mathrm{d} v\, f_r \psi_k &= \int_{I_j} \mathrm{d} v\, f \psi_k \\
    \int_{I_{j+1}} \mathrm{d} v\, f_r \psi_k &= \int_{I_{j+1}} \mathrm{d} v\, f \psi_k
  \end{aligned}\right\}, \quad k=0,\dots,p. \label{eq:recovery_system_1d}
\end{equation}
Since there are $2p+2$ pieces of data ($p+1$ pieces of data from each cell) available for the
reconstruction, $f_r$ is determined as a polynomial of degree $2p+1$ on $v\in \left[v_{j-\frac{1}{2}},v_{j+\frac{3}{2}}\right]$:
\begin{equation}
  f_r(\xi) = f_0 + \xi f_0' + \frac{1}{2} \xi^2 f_0'' + \dots + \frac{1}{(2p+1)!} \xi^{2p+1}f_0^{(2p+1)} ,
\end{equation}
where $\xi = v - v_{j+\frac{1}{2}}$.
\citet{VanLeer2005} notes that the 1D recovery scheme for the diffusion operator using a piecewise-linear basis 
is fourth-order accurate.
Figure~\ref{fig:recovery_1d} illustrates the recovery procedure for a case in which the
DG solution is represented using piecewise-linear polynomials (two degrees of freedom per cell).
In this figure, a 3rd-order recovery polynomial that spans cells $j$ and $j+1$ is constructed
for the evaluation of $f_r$ and $\partial_v f_r$ at $v_{j+\frac{1}{2}}$,
which appear in the boundary-flux terms in (\ref{eq:recovery_evo}).

Commenting on the code implementation of this procedure, we note that we do not need to directly
solve (\ref{eq:recovery_system_1d}) for $f_r$.
The following discussion is not made in \citet{VanLeer2005}.
Instead, $f_r$ can be constructed as a Lagrange interpolation polynomial \citep{LagrangePolynomialWeb},
requiring that $f_r$ be equivalent to $f$ at a set of $2p+2$ points.
From the requirement that $f$ and $f_r$ be identical in the weak sense (\ref{eq:recovery_evo}),
we see that points must be chosen to be the Gauss--Legendre quadrature nodes in each cell
for a $p+1$-point quadrature rule:
\begin{equation}
  f_r(\xi) = \sum_{l = 0}^{2p+1} f(\xi_l) L_l(\xi),
\end{equation}
where the $\xi_l$ and $\xi_k$ are the Gauss--Legendre quadrature nodes and
\begin{equation}
  L_l(\xi) = \prod_{\substack{k=0 \\ k \neq l}}^{2p+1} \frac{\xi-\xi_k}{\xi_l - \xi_k}.
\end{equation}
We emphasize that the equivalence of the two approaches to obtain $f_r$ only holds
when nodes for the $p+1$ point Gauss--Legendre quadrature rule are used in the Lagrange interpolation polynomial.
Almost always, we use a Gauss--Legendre quadrature rule with more points everywhere else in the code,\footnote{
The reason for using a quadrature rule with more than $p+1$ points is so that certain integrals can be evaluated
exactly.} and so it is important not to mix up the use of these different quadrature rules in the code.

For completeness, we also note that
\begin{equation}
  \frac{\partial L_l}{\partial \xi} = L_l(\xi) \sum_{\substack{k=0\\k\neq l}}^{2p+1} \frac{1}{\xi-\xi_k},
\end{equation}
which is useful for computing $f_r'$.
Therefore, we can easily pre-compute matrices that calculate the vector $[f_r(\xi=0), f_r'(\xi=0)]$ when multiplied by
a vector containing the degrees of freedom in two neighboring elements.

Now, we discuss the extension of the 1D recovery procedure to higher dimensions, specifically 2D.
In the simulations presented in this thesis, diffusion terms appear in the collision operators
like (\ref{eq:lbCollisionOp}), so an implementation of diffusion in $(v_\parallel,\mu)$ space is required.
At present, the $(v_\parallel,\mu)$ grid is composed of rectangular elements, which greatly simplifies the
implementation of the recovery-based DG method.
Let us consider the following diffusion equation on the $(v_\parallel,\mu)$ grid:
\begin{equation}
  \frac{\partial f}{\partial t} = D \frac{\partial}{\partial \mu} \left( \mu \frac{\partial f}{\partial \mu} \right).
\end{equation}
As before, we multiply this equation by each basis function $\psi_k$ and integrate over all space,
performing two integration by parts:
\begin{multline}
  \iint_{I_j} \mathrm{d}v_\parallel \, \mathrm{d} \mu \, \psi_k \frac{\partial f}{\partial t} =
  D \int_{\partial I_j} \mathrm{d}v_\parallel \, \left.\left[ \psi_k \mu \frac{\partial f_r}{\partial \mu} 
  - \frac{\partial \psi_k}{\partial \mu} \mu f_r \right] \right|_{\mu_{j-\frac{1}{2}},R}^{\mu_{j+\frac{1}{2}},L} \\ 
  + D \iint_{I_j} \mathrm{d}v_\parallel \, \mathrm{d} \mu \, \left( \frac{\partial^2 \psi_k}{\partial \mu^2} \mu + \frac{\partial \psi_k}{\partial \mu} \right) f.
\end{multline}
The boundary fluxes are now computed as surface integrals in $v_\parallel$ involving $f_r$ and
$\partial_\mu f_r$, and $f_r$ is a function of both $v_\parallel$ and $\mu$.
Specifically, $f_r$ is degree $p$ in $v_\parallel$ and degree $2p+1$ in $\mu$, which results in polynomial
expansion involving $2(p+1)^2$ monomials and matches the total available degrees of freedom
from the two neighboring cells.
The recovery polynomial satisfies
\begin{equation}
  \left.\begin{aligned}
    \iint_{I_j} \mathrm{d} v\, \mathrm{d}\mu\, f_r \psi_k &= \iint_{I_j} \mathrm{d} v\, \mathrm{d}\mu\, f \psi_k \\
    \iint_{I_{j+1}} \mathrm{d} v\, \mathrm{d}\mu\, f_r \psi_k &= \iint_{I_{j+1}} \mathrm{d} v\, \mathrm{d}\mu\, f \psi_k
  \end{aligned}\right\}, \quad k=0,\dots,p. \label{eq:recovery_system_2d}
\end{equation}
As before, we do not need to directly solve (\ref{eq:recovery_system_2d}) for the $2(p+1)^2$ monomial coefficients
of $f_r$, although this approach is certainly a valid one.
Instead, we can directly compute what we ultimately need from the recovery-based DG method, i.e. the
values of $f_r$ and $\partial_\mu f_r$ evaluated at the surface quadrature points.
To do this, we recover a 1D polynomial in $\mu$ that spans both cells at \textit{at every surface quadrature node.}
The linear variation in $v_\parallel$ of $f_r$ is automatically contained in the basis functions $\psi_k$,
and the results of performing a separate 1D recovery calculation at every surface quadrature node are
equivalent to those obtained from solving (\ref{eq:recovery_system_2d}) for $f_r$ and interpolating $f_r$
to the surface quadrature nodes for evaluation of the boundary fluxes.

\begin{figure}
  \centerline{\includegraphics[width=0.75\textwidth]{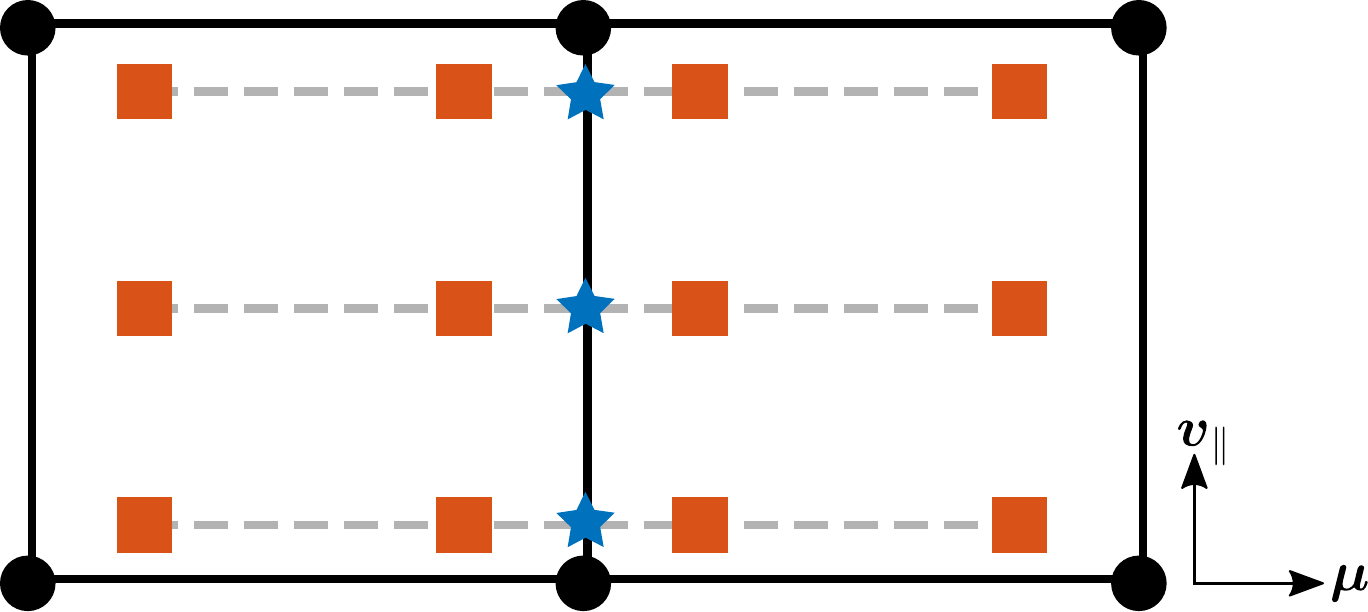}}
  \caption[Illustration of how 1D recovery polynomials are constructed for the evaluation of diffusion terms in 2D.]{
  Illustration of how 1D recovery polynomials are constructed for the evaluation of diffusion terms in 2D.
  Two neighboring elements are shown here ($p=1$), with the solution nodes located at the black circles.
  At each surface quadrature point (blue stars), a high-order, 1D Lagrange polynomial
  of degree $2p+1$ that passes through the solution at a set of $2p+2$ points (red squares) is constructed,
  which is the locally recovered smooth solution. The recovered polynomial that passes through
  each surface quadrature point is then used to compute the local value of $f_r$ and $\partial_\mu f_r$.}
  \label{fig:recovery_sketch}
\end{figure}

The 2D recovery-based DG method currently implemented in the code is illustrated in figure~\ref{fig:recovery_sketch},
where a 3-point quadrature rule is used for the numerical evaluation of the
$\int \mathrm{d}v_\parallel$ surface integral as an example of a case in which a higher-order quadrature
rule is needed to evaluate integrals exactly.
At each surface quadrature node, a 1D recovery procedure is performed to calculate the value of $f_r$ and $\partial_\mu f_r$ there.
Therefore, $f$ needs to be evaluated at certain points in the cell (the red squares in figure~\ref{fig:recovery_sketch})
to facilitate these calculations.
Since the collision operator also contains a term for diffusion in $v_\parallel$, an analogous procedure
is performed on the shared cell boundaries in $v_\parallel$.

\subsection{Numerical Tests}
Here, we present a benchmark of the self-species collision operator (\ref{eq:lbCollisionOp})
by solving the equation $\partial_t f_e = C_{ee}[f_e]$ from $t = 0$ to $t = 10$ $\mu$s.
For this test, the 5D distribution function for electrons is initialized to a top-hat distribution
with temperature $T_{\mathrm{tar}}$:
\begin{equation}
  f(v_\parallel,\mu) =
  \begin{cases}
    c_{n} & \mu \le \mu_0 \text{ and } u_0 - v_0 \le v_\parallel \le u_0 + v_0, \\
    0 & \text{otherwise},
  \end{cases}
  \label{eq:collisionTestInit}
\end{equation}
where $c_n$ is chosen so that the density of discretized distribution function is exactly $2 \times 10^{18}$ m$^{-3}$,
$u_0$ is arbitrarily set to $\sqrt{T_{\mathrm{tar}}/m_e}$, $v_0 = \sqrt{3 T_{\mathrm{tar}}/m_e}$,
$\mu_0 = 2 T_{\mathrm{tar}}/B_0$, and $B_0 = 0.0398$ T.
The distribution function is initialized according to (\ref{eq:collisionTestInit}) at nodes,
i.e. (\ref{eq:collisionTestInit}) is evaluated at each of the 32 nodes in a 5D element
to set the value of $f$ for the initial condition.

\begin{figure}
  \centerline{\includegraphics[width=\textwidth]{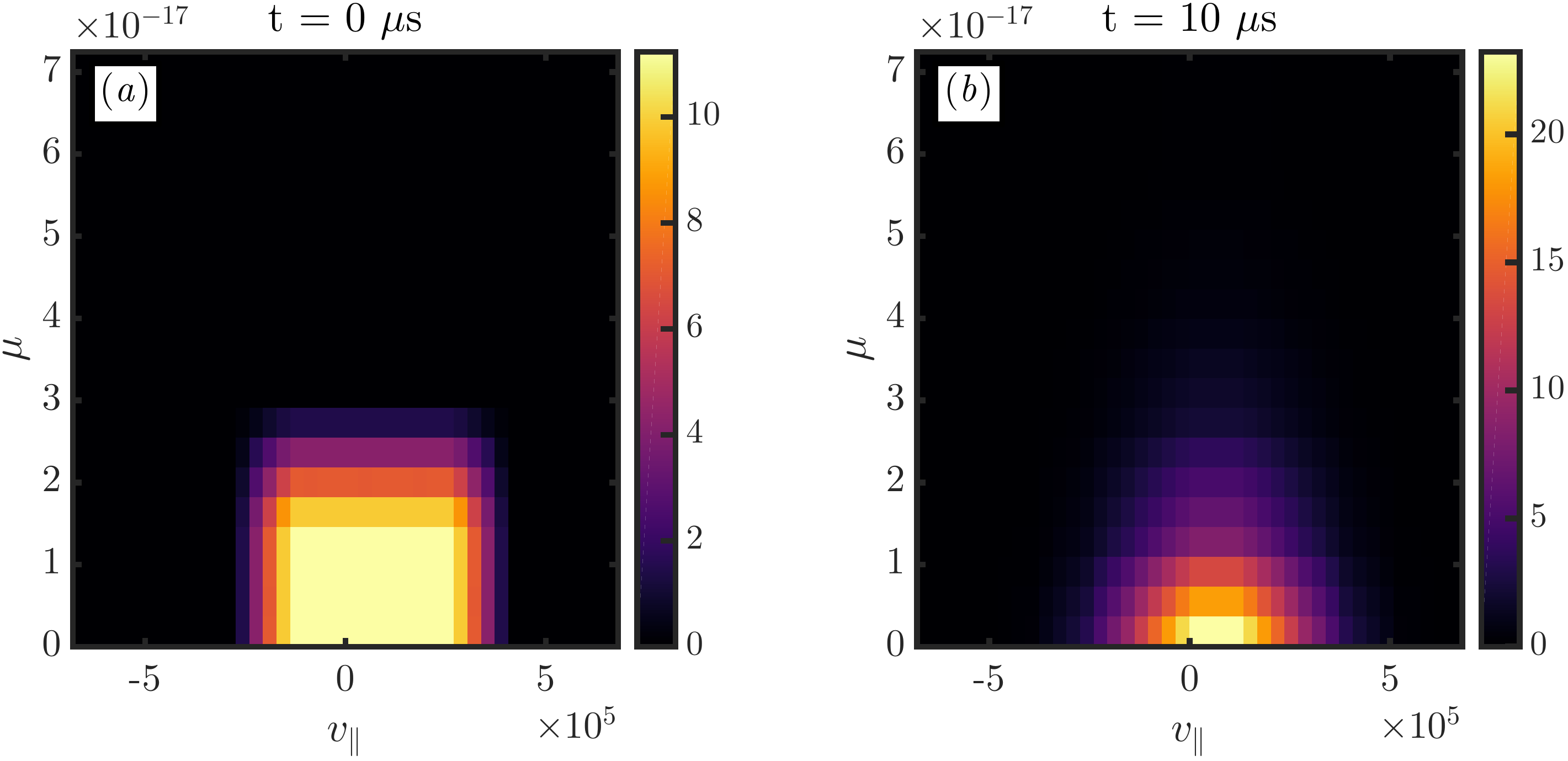}}
  \caption[Initial and final distribution function for the collision operator benchmark
  with $T_{\mathrm{tar}} = 3$~eV.]{
    Comparison of ($a$) initial and ($b$) final distribution function for the collision operator benchmark
    with $T_{\mathrm{tar}} = 3$~eV.}
  \label{fig:collision_distribution_3_eV}
\end{figure}

\begin{figure}
  \centerline{\includegraphics[width=\textwidth]{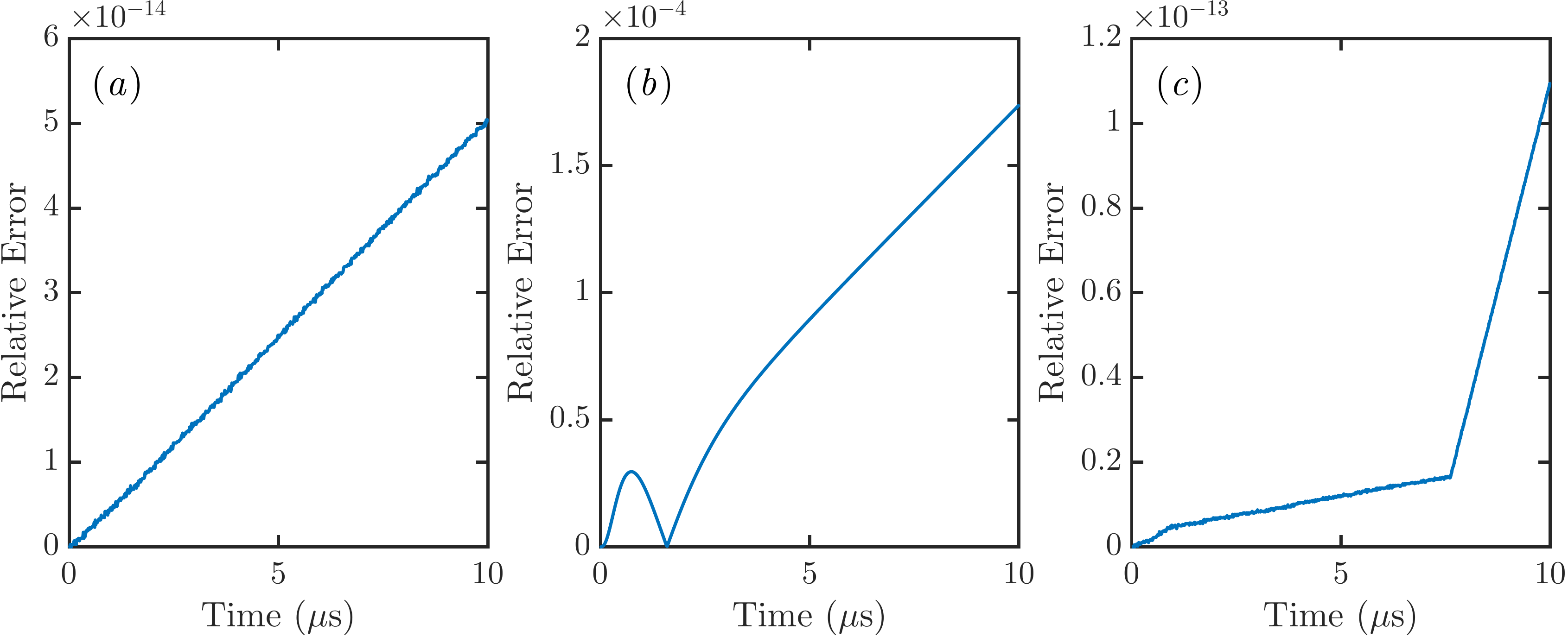}}
  \caption[Time evolution of the relative error in number, momentum, and energy for a collision operator test.]{
    Time evolution of the relative error in ($a$) number, ($b$) momentum, and ($c$)
    energy for a collision operator test.}
  \label{fig:collision_conservation}
\end{figure}

The velocity-space grid covers $v_\parallel \in \left [-4 \sqrt{T_{e,\mathrm{grid}}/m_s},4 \sqrt{T_{e,\mathrm{grid}}/m_s} \right]$,
where $T_{e,\mathrm{grid}} = 3$~eV,
and $\mu \in \left [0, 0.75 m_e v_{\parallel,\mathrm{max}}^2/(2 B_0) \right]$.
Ten cells are used in $v_\parallel$, and five cells are used in $\mu$.
A single cell is used in the position-space dimensions for this 5D test, which allows
this test to be executed on a single processor.
A realistic (not reduced) collision frequency is used and no positivity-correction procedures
(discussed in Section \ref{sec:positivity}) are applied.
Figure~\ref{fig:collision_distribution_3_eV} shows the initial and final states of the distribution function
for this test with $T_{\mathrm{tar}} = 3$~eV.
The distribution function visually appears to relax to a Maxwellian distribution function, as expected.
Figure~\ref{fig:collision_conservation} shows the number, momentum, and energy-conservation
properties of the collision operator, as measured in this benchmark.
The use of a CFL number of 0.1 results in time steps of size $\Delta t = 1.29552 \times 10^{-8}$ s,
so 772 RK3 time steps are taken to advance the distribution function from $t=0$ to $t=10$ $\mu$s
\footnote{A final time step of size $\Delta t = 1.15754\times 10^{-8}$ s is taken so that the end time is exactly 10 $\mu$s.}.

\begin{figure}
  \centerline{\includegraphics[width=\textwidth]{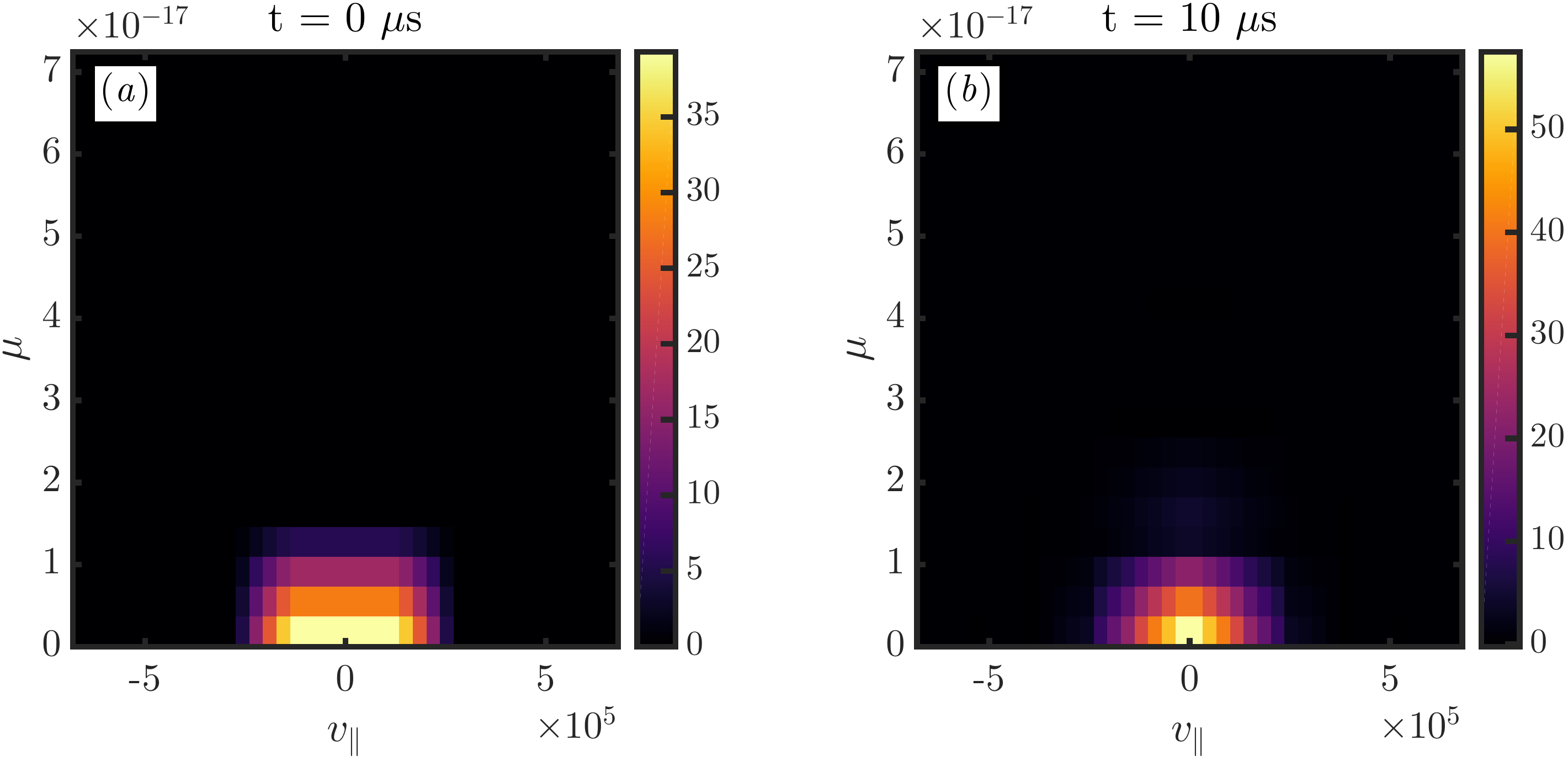}}
  \caption[Comparison of initial and final distribution function for the collision operator test
  with $T_{\mathrm{tar}} = 1$~eV.]{
    Comparison of ($a$) initial and ($b$) final distribution function for the collision operator test
    with $T_{\mathrm{tar}} = 1$~eV.}
  \label{fig:collision_distribution_1_eV}
\end{figure}

\begin{figure}
  \centerline{\includegraphics[width=\textwidth]{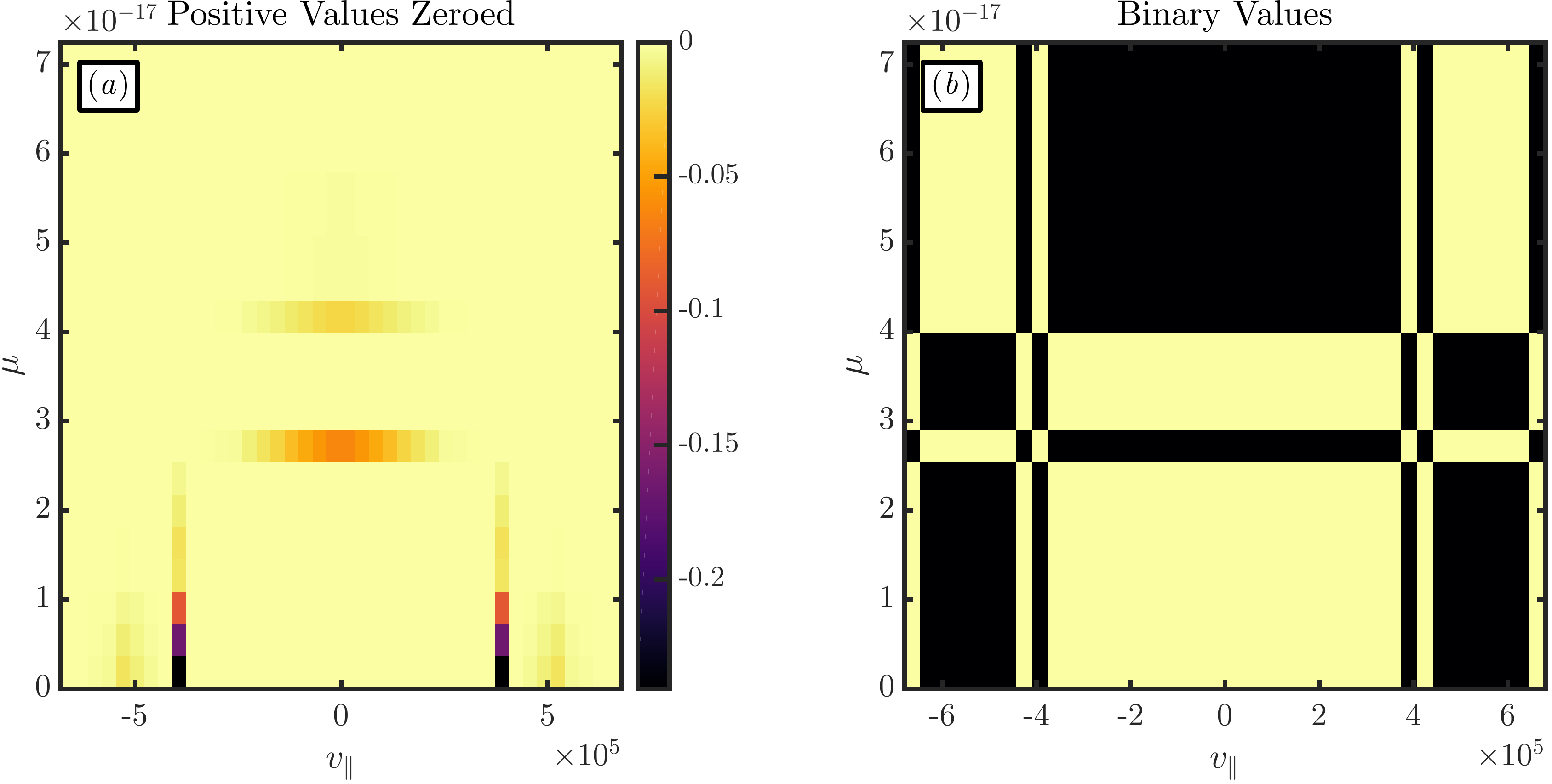}}
  \caption[Negative regions in the distribution function after
  evolution to a steady state by the self-species collision operator.]
  {Identification of negative regions in the distribution function after
  evolution to a steady state by the self-species collision operator, showing
  ($a$) the distribution function with all positive values set to 0, leaving
  the negative values unmodified and
  ($b$) the negative regions of the distribution function in black and
  the positive regions in yellow.}
  \label{fig:collision_distribution_1_eV_negative}
\end{figure}

To motivate the discussion in the next section about positivity issues,
we run the same collision-operator benchmark, but initialize the distribution function with a much colder
temperature.
The grid is still based on $T_{e,\mathrm{grid}} = 3$~eV, but the distribution function has temperature
$T_{\mathrm{tar}} = 1$~eV, which is below the minimum $T_\perp$ that can be represented on
the grid. By initializing the distribution function in a nodal manner,
the initial numerical value of $T_\perp$ for the distribution function will be the minimum $T_\perp$ and not 1~eV.
Figure~\ref{fig:collision_distribution_1_eV} shows the initial and final states of the distribution function
for this test, which still appear reasonable.
Upon closer inspection of the distribution function, however,
we see that negative regions now appear at large $v_\parallel$ and large $\mu$.
Figure~\ref{fig:collision_distribution_1_eV_negative} shows plots that focus on the
negative regions in the same distribution function from figure~\ref{fig:collision_distribution_1_eV}.
These negative regions of the distribution function are unphysical and can lead
to issues with code stability.
For example, these negative regions can grow over time and eventually cause the calculation of the
density or temperature at a location $\boldsymbol{R}$ to be negative.

\section{\label{sec:positivity}Positivity of the Distribution Function}
One challenge with continuum methods is making sure that the distribution function does not go negative.
Particle-in-cell methods do not have to deal with this issue.
We found it necessary to adjust the distribution function of each species at every time step so that $f_s \ge 0$ at
every node to avoid stability issues.
After much investigation, the main source of negativity in the distribution function
in our LAPD simulations (Chapter~\ref{ch:lapd}) appears to be the collision operator at locations 
where the perpendicular temperature of the distribution function is
close to the lowest perpendicular temperature that can be represented on the grid.

If one considers a velocity-space grid composed of uniform cells with widths $\Delta v_\parallel$
and $\Delta \mu$ in the
parallel and perpendicular coordinates, the minimum temperatures for a realizable distribution
are computed by assuming that the distribution function
is non-zero at the node located at $(v_\parallel = 0, \mu = 0)$ and 0 at all other nodes.
Using piecewise-linear basis functions,
\begin{align}
T_{\parallel,\mathrm{min}} =& \frac{m}{6} \left(\Delta v_\parallel \right)^2, \\
T_{\perp,\mathrm{min}} =& \frac{B}{3} \Delta \mu, \\
T_{\mathrm{min}} =& \frac{1}{3} \left(T_{\parallel,\mathrm{min}} + 2 T_{\perp,\mathrm{min}} \right).
\end{align}
Typical values of $\Delta v_\parallel$ and $\Delta \mu$ for a uniformly spaced grid that contains a few $v_{t} = \sqrt{T/m}$
usually result in $T_{\parallel,\mathrm{min}} < T_{\perp,\mathrm{min}}$.
A situation can occur in which the collision operator will try to relax the $T_\perp$
of the distribution function at a location $\boldsymbol{R}$ to a value below $T_{\perp,\mathrm{min}}$,
resulting in negative regions appearing in the distribution function.

This positivity issue might be avoided by choosing a velocity-space grid that has
$T_{\parallel,\mathrm{min}} = T_{\perp,\mathrm{min}} = T_{\mathrm{min}}$, either by increasing the resolution in $\mu$
relative to the resolution in $v_\parallel$,
using non-polynomial basis functions \citep{Yuan2006} that guarantee the positivity of the distribution function,
or using $\sqrt{\mu}$ as a coordinate instead of $\mu$.
Initial investigation in the use of a non-uniformly spaced grid in $\mu$
showed that the positivity issues resulting from the collision operator were nearly eliminated,
but more severe positivity issues arose in the kinetic-equation solver.
A possible explanation for why the positivity issues in the kinetic-equation solver
are not an issue in the case with a uniformly spaced grid in $\mu$ is because
a much lower $T_{\perp,\mathrm{min}}$
also allows regions with much lower $T_\parallel$ to exist in the simulation
(since collisions keep $T_\perp \approx T_\parallel$,
and the LAPD case is highly collisional), and the advection of a distribution function with
$T_\parallel$ close to $T_{\parallel,\mathrm{min}}$
might result in positivity issues for the kinetic-equation solver.

For now, we use a simpler correction procedure described in this section,
which has a philosophy similar to the correction operator used by \citet{Taitano2015}.
The magnitude of the correction operator scales with the truncation error of the method,
and so it vanishes as the grid is refined and does not affect the order of accuracy of the
algorithm while making the simulation more robust on coarse grids by preserving key conservation laws.
Our relatively simple positivity-adjustment procedure
is to eliminate the negative nodes of the distribution functions
while keeping the number density and thermal energy unchanged.
This procedure is conceptually similar to `filling algorithms', which attempt
to remove negative regions in the solution by moving mass in from nearby positive regions
\citep{Durran2010,Rood1987}.
First, the number density, parallel energy, perpendicular energy, and parallel momentum for each species are computed.
Next, all negative nodes of the distribution functions are set to zero, resulting in changes
to the thermal energy and density at locations where the distribution functions have been modified.
To compensate for the increased density, the distribution function is scaled uniformly in velocity space at each configuration-space node
to restore the original density.
In cases which the original density at a location $\boldsymbol{R}$ is negative to begin with,
a Maxwellian distribution with zero flow velocity and a temperature profile identical to the
initial condition is added to both species such that the density of both species at that location is above
some floor value (we used $n_\mathrm{floor} = 10^{-5} n_0$ in our tests in Chapter~\ref{ch:lapd})
and the charge density $\sigma_g$ is unchanged.

The remaining task is to modify the distribution function so that no additional energy is added through the positivity-adjustment procedure.
To remove parallel thermal energy $ \int \mathrm{d}^3 v\, m_s v_\parallel^2 f_s/2$ added through the positivity-adjustment
procedure, we use a numerical drag term of the form
\begin{equation}
\frac{\partial f}{\partial t} = \frac{\partial}{\partial v_\parallel} \left[\alpha_{\mathrm{corr},v_\parallel} \left(v_\parallel - u_\parallel \right) f \right],
\end{equation}
where $\alpha_{\mathrm{corr},v_\parallel}$ is a small numerical correction drag rate that is chosen
each time step to remove the extra parallel energy added.
To guarantee that the numerical drag term will not cause any nodes to go negative,
this operator is implemented in a finite-volume sense, adjusting the mean values:
\begin{equation}
\frac{\bar{f}_j^{n+1} - \bar{f}_j^{n}}{\Delta t} = \frac{\alpha_{\mathrm{corr},v_\parallel}}{\Delta v_\parallel}
  \left( (v_\parallel-u_\parallel)_{j+1/2} \hat{f}^n_{j+1/2} - (v_\parallel-u_\parallel)_{j-1/2} \hat{f}^n_{j-1/2}
  \right),
\end{equation}
where the interface flux $\hat{f}^n_{j+1/2} = g(\bar{f}^n_{j}, \bar{f}^n_{j+1})$ is chosen in an upwind sense according to the sign of $v_\parallel-u_\parallel$
and $\bar{f}_j$ denotes the cell-averaged value of $f_j$.
To ensure that the parallel drag term does not modify the perpendicular energy $ \int \mathrm{d}^3 v\, \frac{1}{2} m_s v_\perp^2 f_s $,
this operator is applied at fixed $(\boldsymbol{R},\mu)$.
In our tests, we found that $\alpha_{\mathrm{corr},v_\parallel}$ cannot be generally chosen to restore the parallel thermal energy
at every position-space node, since there is a limit on how large $\alpha_{\mathrm{corr},v_\parallel}$
can be while keeping $\bar{f}_j \geq 0$ in every cell.
Instead, we choose $\alpha_{\mathrm{corr},v_\parallel}$ to restore the cell-averaged parallel energy
\begin{equation}
\bar{W}_{\parallel,j} = \int_{x_j - \Delta x/2}^{x_j + \Delta x/2} \mathrm{d}x
\int_{y_j - \Delta y/2}^{y_j + \Delta y/2} \mathrm{d}y
\int_{z_j - \Delta z/2}^{z_j + \Delta z/2} \mathrm{d}z \int \mathrm{d}^3 v\, \frac{1}{2} m_s v_\parallel^2 f_s,
\end{equation}
which results in some position-space diffusion of energy.

We employ a similar procedure to remove the unphysical perpendicular energy added through positivity:
\begin{equation}
\frac{\partial f}{\partial t} = \frac{\partial}{\partial \mu} \left( 2 \alpha_{\mathrm{corr},\mu} \mu f \right).
\end{equation}
Here, the factor $\alpha_{\mathrm{corr},\mu}$ is chosen to restore the cell-averaged perpendicular energy.
Similarly, this operation modifies the perpendicular energy without changing the parallel energy.
Generally speaking, all of the parallel energy added through positivity can usually be removed
through the numerical drag operator while a small amount ($<10\%$) of perpendicular energy added
through positivity remains even after applying the numerical drag operator, a consequence from the choice of a
uniformly spaced grid in $\mu$ (energy is typically added in the distribution function tails, so a uniformly spaced energy grid will
be more constrained than a quadratically spaced energy grid in removing positivity-added energy using a numerical drag operator).
We observe that much of the extra energy added through the positivity-adjustment procedure
are in cells located in the outer region $r > 0.4$ m and near the boundaries in
the parallel direction where much of it may be quickly lost in the
outflows through the sheaths, so we think the extra energy added will not have a
significant impact on the turbulence characteristics in the main part of the simulation.
(Our present model for the physical source in the LAPD simulations is uniform in $z$, but
there may be a localized source from recycling near the real end plates, which may offset the need for
the numerical positivity source there.)
Ultimately, we hope to develop a more satisfactory solution to deal with these
positivity issues.

\section{Sheath Boundary Conditions \label{sec:sheath}}
\begin{figure}
  \centerline{\includegraphics[width=\textwidth]{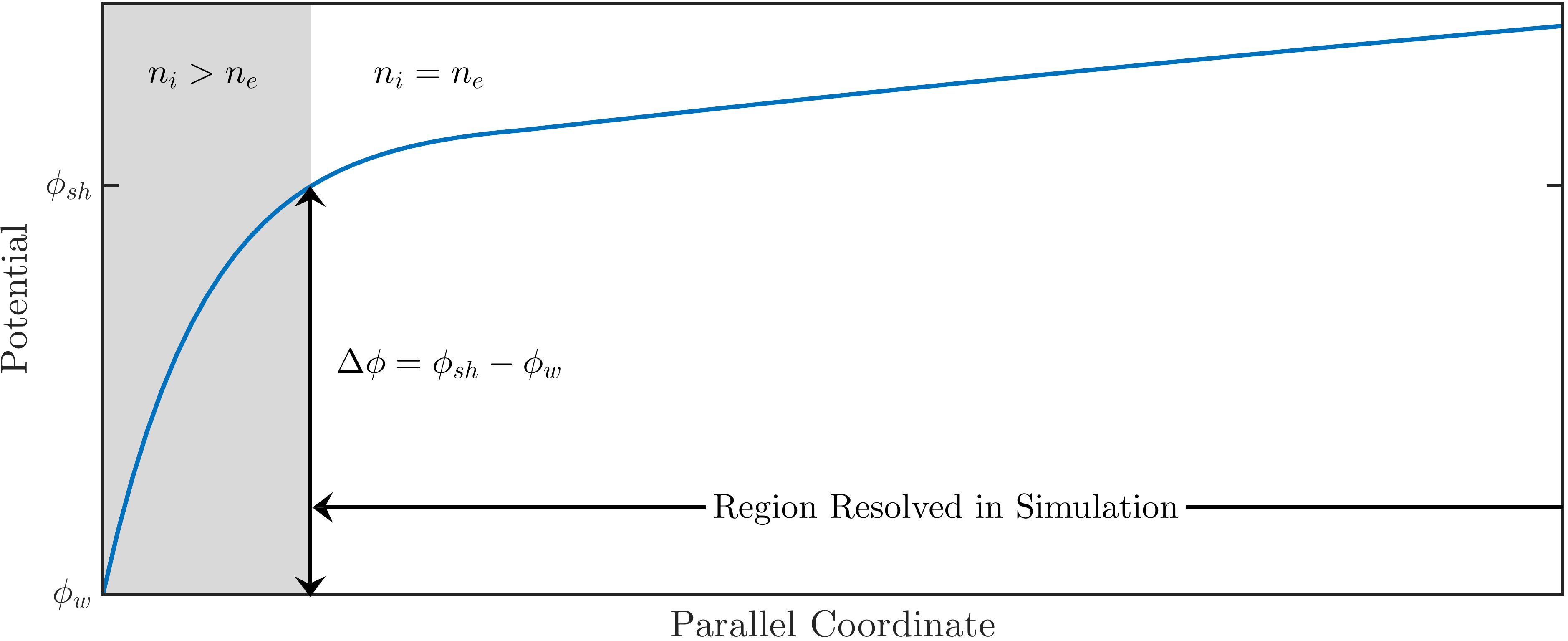}}
  \caption[Illustration of the potential structure near a plasma--material interface.]{
  Illustration of the potential structure near a plasma--material interface.
  The shaded region is the Debye sheath, which is a region of net positive charge that is a few
  electron-Debye-lengths wide. In gyrokinetic simulations, the sheath cannot be resolved since
  the plasma is assumed to be quasineutral ($\sum_i Z_i n_i=n_e$) and its effects must
  be modeled using boundary conditions. Additionally, spatial and temporal resolution
  of the sheath in turbulence simulations is computationally impractical.}
  \label{fig:sheath_simple}
\end{figure}

A layer of net positive charge called the \textit{electrostatic Debye sheath} forms at the plasma--material interface, such as
where open magnetic field lines intersect a grounded conducting divertor or limiter in a tokamak.
As shown in figure~\ref{fig:sheath_simple}, the sheath sets up a potential drop that
accelerates ions into the wall and repels low-energy electrons, keeping the particle flux of electrons and ions
into the wall in approximate balance.
Accurate modeling of sheath effects is important for the calculation
of particle and heat fluxes to plasma-facing components and plasma--surface interactions.
The sheath is a few electron-Debye-lengths wide and forms on a time scale of order the electron plasma period,
which are both very disparate scales compared to the turbulence scales of interest in gyrokinetics,
so it is natural and desirable to treat the sheath through model boundary conditions
to avoid the need to directly resolve it.
The implementation of sheath-boundary conditions should be considered a distinguishing feature of
a gyrokinetic code for open-field-line simulations.

The electron Debye length in LAPD is ($\lambda_{De} \sim$10$^{-6}$ m), which is very small compared
to the ion gyroradius ($\sim$10$^{-2}$ m) and even smaller compared to the parallel scales of
the turbulence ($\sim$10 m).
The electron plasma frequency is $\omega_{pe} \sim$10$^{9}$ s$^{-1}$,
which is much larger compared to the ion gyrofrequency
($\sim$10$^{6}$ s$^{-1}$), and even larger than the turbulence frequencies of interest
($ \omega_* \sim$10$^{4}$ s$^{-1}$ at $k_\theta \rho_{\mathrm{s}0} \sim$0.3).
Since the quasineutrality and low-frequency assumptions of gyrokinetics break
down in the sheath, gyrokinetic models cannot directly handle sheaths.
There is also a transition region between the collisional upstream region
and the collisionless sheath, with a width of order the mean free path.
This region is not resolved in the simulations we have performed.
We also note that the end walls (divertor plates) in tokamaks and in basic plasma physics
experiments are typically grounded, so sheath-model boundary conditions should be
reflective of this situation.

\subsection{Logical-Sheath Model\label{sec:logical-sheath}}
The 1D gyrokinetic model with kinetic electrons described in Chapter~\ref{ch:1d-sol} uses
logical-sheath boundary conditions, which were originally developed for fully kinetic 1D2V PIC
simulations by \citet{Parker1993} to address inaccuracies in sheath effects
that were observed when using a direct implicit PIC method with $\Delta t \omega_{pe} > 1$.
In that work, the authors needed to use a coarser spatial resolution $\Delta z$
as the time-step size $\Delta t$ was increased in order to avoid numerical heating and cooling,
so the sheath became increasingly poorly resolved in the implicit limit $\Delta t \omega_{pe} > 1$.
Logical-sheath boundary conditions were developed to model the essential sheath effects
in simulations with coarse spatial resolution of the sheath.

Logical-sheath boundary conditions impose the steady state behavior $j_\parallel = 0$ to the wall
at each instant, and they are sometimes
referred to as insulting-wall boundary conditions in fluid models (in contrast to conducting-wall
boundary conditions).
For a normal, positively charged sheath, all incident ions flow into the wall because they are accelerated
by the sheath potential drop, while incident electrons are partially reflected.
Only electrons with a high enough parallel velocity can surpass the sheath potential drop and reach
the wall, and the rest are reflected back into the plasma.
In a logical sheath, the sheath potential is determined from the
$j_\parallel = 0$ condition, so the lowest-parallel-velocity electrons are reflected to
satisfy $j_\parallel = 0$, and then the sheath potential is calculated using the $v_\parallel$
of the slowest electrons that have enough parallel energy to overcome the sheath potential drop
and flow into the wall.
No boundaries conditions are applied to the ions in the logical-sheath model.

\begin{figure}
  \centerline{\includegraphics[width=\textwidth]{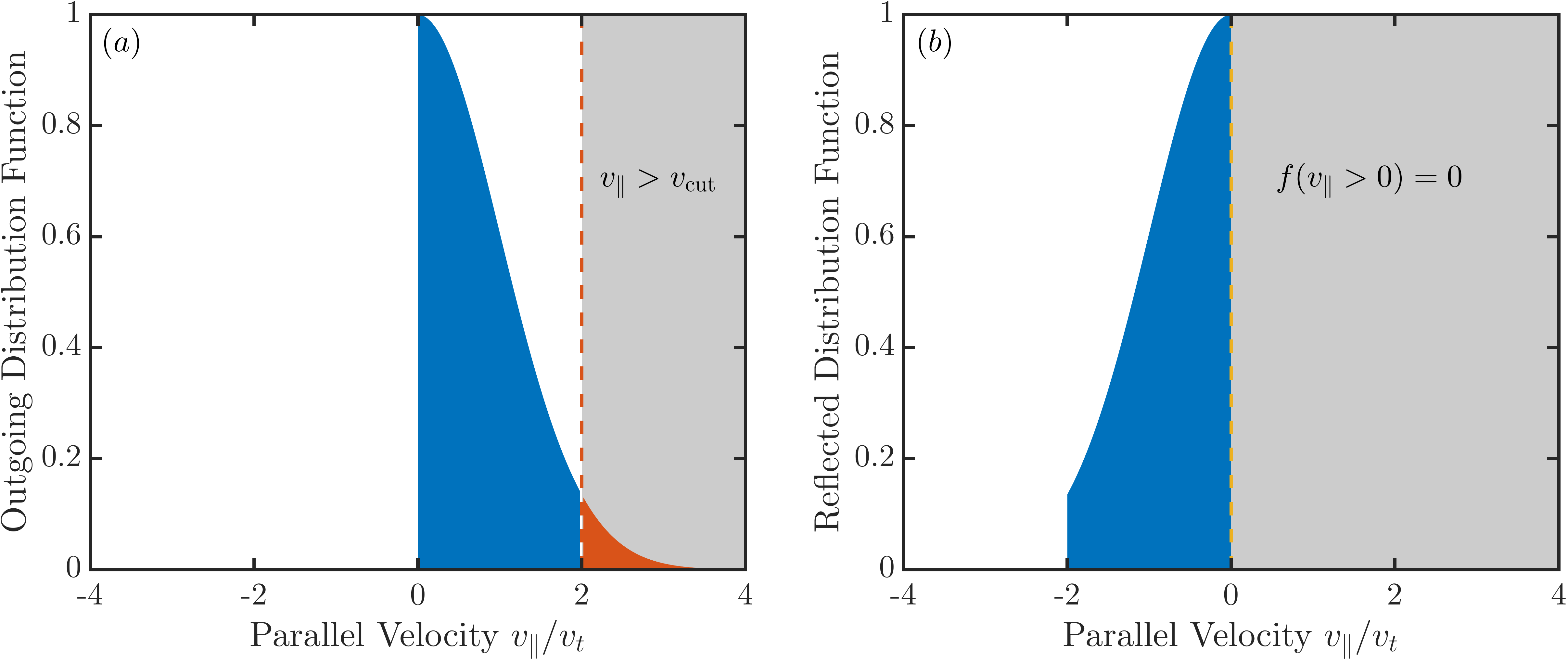}}
  \caption[Application of sheath boundary conditions to an incident electron distribution function.]
  {Application of sheath boundary conditions to an incident electron distribution function.
  ($a$) The distribution function incident to the sheath. The cutoff velocity, determined by the 
  $j_\parallel = 0$ condition in the logical-sheath model or by the gyrokinetic Poisson equation in the
  conducting-sheath model, is indicated by the dashed red line. Electrons with $v_\parallel > v_{\mathrm{cut}}$
  are in red and have enough energy to surpass the sheath potential, while incident electrons with
  $v_\parallel \le v_\mathrm{cut}$ are reflected back into the plasma.
  ($b$) The reflected distribution function is the portion of the incident distribution function
  that does not have sufficient parallel velocity to overcome the sheath potential, reflected about the
  $v_\parallel = 0$ axis. These particles come back into the plasma.}
  \label{fig:sheath-illustration}
\end{figure}

If logical-sheath boundary conditions are applied at $z=z_R$, the typical $j_\parallel = 0$
condition for a 1D1V (one) problem is expressed as
\begin{align}
\int_0^\infty \mathrm{d}v_\parallel \, v_\parallel f_i\left(z_R,v_\parallel,t\right)
  =& \int_{v_{\mathrm{cut}}}^\infty \mathrm{d}v_\parallel \, v_\parallel f_e\left(z_R,v_\parallel,t\right)
  \label{eq:sheath_condition}.
\end{align}
The cutoff velocity $v_{\mathrm{cut}}>0$ is found numerically by
first finding the velocity-space cell in which the cutoff velocity lies
($v_{c,j} - \Delta v_\parallel/2 \le v_\parallel \le v_{c,j} + \Delta v_\parallel/2$,
where $v_{c,j}$ is the $v_\parallel$ in the center of cell $j$) and then
using a bisection method such that (\ref{eq:sheath_condition}) is satisfied
to a desired tolerance level.
The sheath potential $\phi_{sh}$ is then determined using the relation
$e \Delta \phi = e(\phi_{sh}-\phi_w) = m_e v_\mathrm{cut}^2/2$,
where the wall potential $\phi_w$ is usually taken to be 0~V for a grounded wall.
Figure~\ref{fig:sheath-illustration} illustrates how the incident electrons are partially reflected
by the sheath-model boundary conditions.

In order to reflect all electrons incident on the sheath with parallel outgoing velocity in the range
$0 < v_\parallel < v_{\mathrm{cut}}$, the electron distribution function in this interval is copied into
ghost cells according to
\begin{align}
  f_e(z_R,-v_\parallel,t) &= f_e(z_R,v_\parallel,t), \quad 0<v_\parallel<v_\mathrm{cut},
\end{align}
and $f_e(z_R,-v_\parallel,t) = 0$ for $v_\parallel>v_\mathrm{cut}$. This condition can also be written as
$f_e(z_R,-v_\parallel,t) = f_e(z_R,v_\parallel,t) H(v_\mathrm{cut} -v_\parallel)$ for $v_\parallel > 0$.
This condition results in the reflection of electrons with velocity in the range
$0 < v_\parallel < v_\mathrm{cut}$ back into the domain with the opposite velocity, while
the electrons with energy sufficient to overcome the sheath potential
will flow out of the system to the divertor plates.

To summarize, our implementation of the logical sheath in 1D kinetic continuum models with a single ion species
has the following steps:
\begin{enumerate}
  \item Calculate the outward ion and electron fluxes at the sheath entrance,
    $\Gamma_i = \int_0^\infty \mathrm{d}v_\parallel \, v_\parallel f_i$ and
    $\Gamma_e = \int_0^\infty \mathrm{d}v_\parallel \, v_\parallel f_e$
    respectively.
  \item Compare $\Gamma_i$ to $\Gamma_e$ to identify the reflected species.
    \begin{itemize}
      \item $\Gamma_e > \Gamma_i$ (Typical Case): Reflect slowest outgoing \textit{electrons} in next step.
      \item $\Gamma_e > \Gamma_i$ (Rare Case): Reflect slowest outgoing \textit{ions} in next step.
    \end{itemize}
  \item For the reflected species, find $v_\mathrm{cut}$, the $v_\parallel$ above which
    particles can overcome the sheath potential drop $\Delta \phi$ and leave the system,
    such that $\int_{v_\mathrm{cut}}^\infty \mathrm{d} v_\parallel \, v_\parallel f_e = \Gamma_i$ if
    electrons are reflected or $\int_{v_\mathrm{cut}}^\infty \mathrm{d} v_\parallel \, v_\parallel f_i = \Gamma_e$
    if ions are reflected. Numerically, the cutoff velocity is found using the bisection method.
  \item For the reflected species, reflect the outgoing distribution function in the interval
    $0 \le v_\parallel \le v_\mathrm{cut}$ about the $v_\parallel = 0$ axis, and
    copy the result to the boundaries of the appropriate ghost cells.
  \item Determine the sheath potential $\phi_{sh}$ from the cutoff velocity as
    $e \Delta \phi = e(\phi_{sh}-\phi_w) = m_e v_\mathrm{cut}^2/2$,
    which will be used as boundary conditions for the gyrokinetic Poisson equation solve.
\end{enumerate}

The implementation of logical-sheath boundary conditions needs a slight modification
for use in a continuum code.
Typically, the cutoff velocity will fall within a cell and not exactly on a cell edge.
A direct projection of the discontinuous reflected distribution onto
the basis functions used in a cell could lead to negative values of the
distribution function at some velocities in the cell. Future work could
consider methods of doing higher-order projections that incorporate
positivity constraints, but for now we have used a simple scaling
method, in which the entire distribution function inside the `cutoff
cell' is copied into the ghost cell and then scaled by the fraction
required to ensure that the electron flux at the domain edge equals the ion flux.
For scaling the reflected distribution function in the cutoff cell on
the right boundary, this fraction is
\begin{align}
  c &= \frac{\int_{v_{c,j} - \Delta v_\parallel/2}^{v_\mathrm{cut}}\mathrm{d}v_\parallel\,
  v_\parallel f_e(z_R,v_\parallel,t)}{\int_{v_{c,j}-\Delta v_\parallel/2}^{v_{c,j}+\Delta v_\parallel/2}
  \mathrm{d}v_\parallel \, v_\parallel f_e(z_R,v_\parallel,t)},
\end{align}
where $\Delta v_\parallel$ is the cell width in $v_\parallel$ (assumed to be uniform),
and $v_{c,j}$ denotes the $v_\parallel$ coordinate of the center of the cutoff cell.

\begin{figure}
  \centerline{\includegraphics[width=\textwidth]{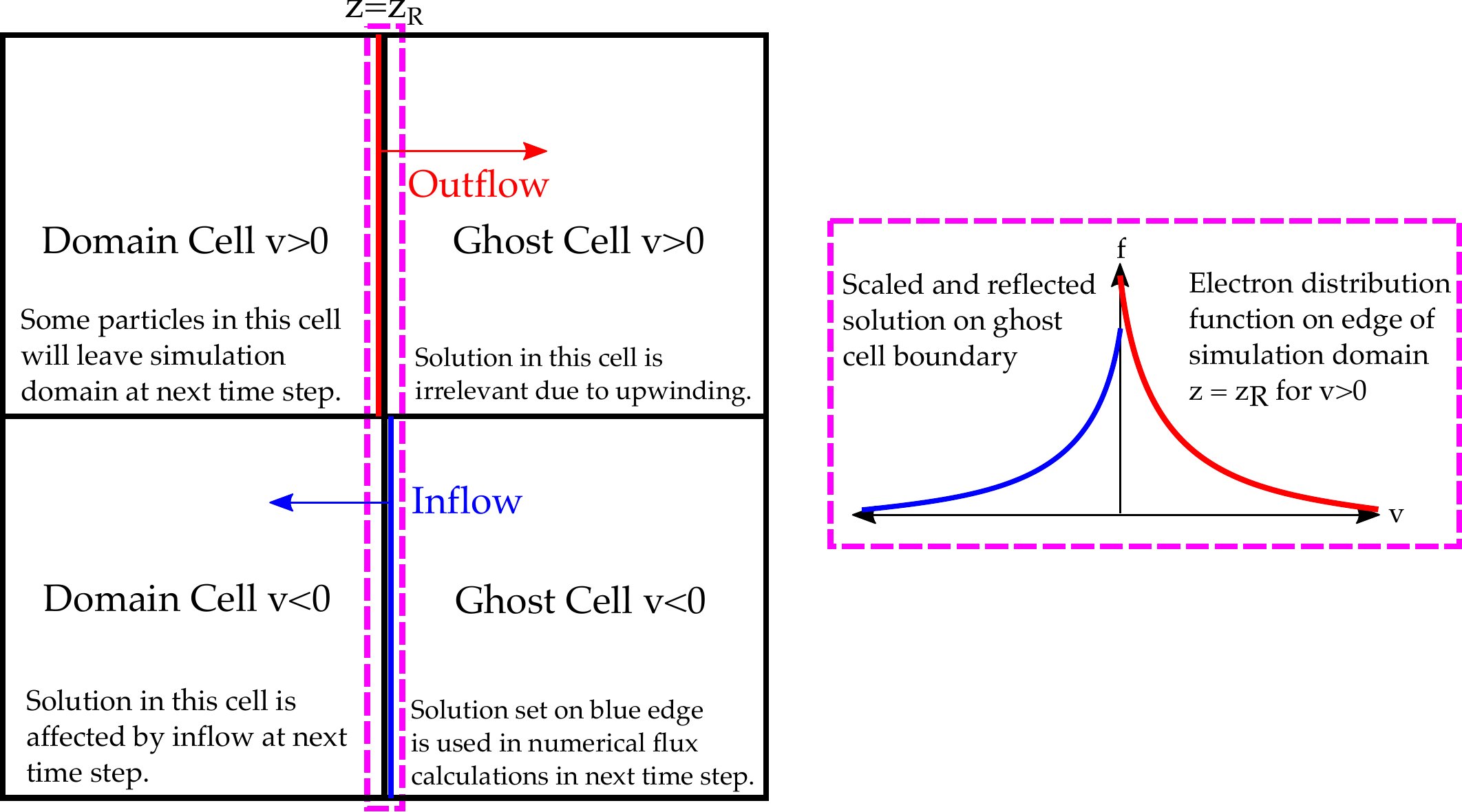}}
  \caption[Illustration of how the sheath model boundary conditions use ghost cells to
  set the reflected distribution function.]{An illustration of how the sheath model boundary conditions use
  ghost cells to set the reflected distribution function. Left: sheath boundary conditions are
  applied to the distribution function at $z=z_R$, which results in the determination of the 
  reflected distribution function (blue edge) in the ghost cell based on the outgoing distribution
  function (red edge) in the domain cell.
  Right: the distribution function on the $z=z_R$ boundary after application of sheath boundary conditions,
  which corresponds to the dashed-box region on the left diagram.}
  \label{fig:ghost-cells}
\end{figure}

For cells whose parallel velocity extents do not bound $v_{\mathrm{cut}}$, the reflection procedure
is straightforward: find the corresponding ghost cell $j'$ with $v_{c,j'} = -v_{c,j}$ and copy the
solution after reflection about the $v_\parallel$ axis.
Figure~\ref{fig:ghost-cells} illustrates how ghost cells are used to provide inflow 
characteristics in sheath-model boundary conditions.
In 5D gyrokinetic simulations, this reflection procedure is applied at each position-space node on the upper and
lower surfaces in $z$ at the end of the simulation domain, which are assumed to be end plates.
Sheath boundary conditions are not applied on the $x$ and $y$ boundaries.

\subsection{Conducting-Sheath Boundary Conditions\label{sec:lapd_sheath}}
For 5D gyrokinetic simulations, we use conducting-sheath boundary conditions
instead of logical-sheath boundary conditions.
Based on our success with the logical-sheath boundary conditions in 1D models (Chapter~\ref{ch:1d-sol}),
we originally implemented logical-sheath boundary conditions for LAPD simulations, but we encountered stability
issues.
We found better robustness by implementing a set of sheath boundary conditions motivated
by how some fluid (or gyrofluid) codes determine $\phi$ everywhere (including the sheath potential)
from the fluid vorticity equation (or polarization equation)
and then use the sheath potential to set the boundary condition
on the parallel electron velocity \citep{Xu1998,Rogers2010,Friedman2013,Ribeiro2005}.

In conducting-sheath boundary conditions, we use the gyrokinetic Poisson equation
(\ref{eq:gkp}) to solve for the potential $\phi(x,y,z)$ everywhere in the simulation domain.
The sheath potential $\phi_{sh}(x,y)$ on each boundary in $z$ (where the field lines intersect the wall)
is obtained by simply evaluating $\phi$ on that boundary, so at the upper boundary in $z$,
$\phi_{sh}(x,y) = \phi(x,y,L_z/2)$.
The wall is taken to be just outside the simulation domain and the wall potential $\phi_w$ is
0 for a grounded wall.
Outgoing particles with $\frac{1}{2} m_s v_\parallel^2 < -q_s \left( \phi_{sh}-\phi_w \right)$ are reflected
(e.g., when $\phi_{sh}-\phi_w$ is positive, some electrons will be reflected),
while the rest of the outgoing particles leave the simulation domain.

In the conducting-sheath approach, the sheath potential is determined by other
effects (the gyrokinetic Poisson equation or the related fluid vorticity equation in fluid
simulations)
and is used to determine what fraction of electrons are reflected and thus the
resulting currents to the wall.
In contrast to the logical-sheath model, the conducting-sheath model allows parallel current
fluctuations into the wall, since the $\Gamma_i = \Gamma_e$ condition is never directly imposed,
with steady-state current paths closing through the walls.
As we show in Chapters~\ref{ch:lapd} and \ref{ch:helical-sol}, we observe large local current
fluctuations and local steady-state currents in some simulations, so qualitative differences might be expected
between logical-sheath or conducting-sheath boundary conditions.
If one starts with an initial condition where $\sigma_g=0$ in (\ref{eq:gkp}) so $\phi=0$,
then electrons will rapidly leave the plasma, causing the gyrocenter charge $\sigma_g$
to rise to be positive, and thus the sheath potential will quickly rise to reflect most of the electrons,
and bring the sheath currents down to a much smaller level,
while allowing the sheath currents to self-consistently fluctuate in interactions with the turbulence.

The boundary condition we use for ions is the same as the one used in the logical-sheath model \citep{Parker1993}
(a variant of which are used in the XGCa and XGC1 gyrokinetic PIC codes \citep{Churchill2016}):
the ions just pass out freely at whatever velocity they have been accelerated to
by the potential drop from the upstream source region to the sheath entrance.
The only boundary condition that the sheath model imposes on the ions is that there are no incoming ions,
i.e. at the incoming lower sheath boundary we have the boundary condition that
$f_i(x,y,z=-L_z/2,v_\parallel, \mu) = 0$ for all $v_{\parallel} \ge 0$.
While this leads to a well-posed set of boundary conditions and appears to give
physically reasonable results for the simulations carried out in this thesis,
it might need improvements in some parameter regimes.

\subsection{Future Considerations for Sheath Models}
Sheaths have long been studied in plasma physics, including kinetic effects
and angled magnetic fields, and there is a vast literature on them.
The standard treatments look at steady-state results in one dimension, in which the potential
is determined by solving the Poisson equation along a field line
(for the case here in which the magnetic field is perpendicular to the surface),
but for gyrokinetic turbulence, we need to consider time-varying
fluctuations in which the sheath region needs to couple to an upstream
gyrokinetic region where the potential is determined in 2D planes
perpendicular to the magnetic field by solving the gyrokinetic quasineutrality
equation (\ref{eq:gkp}).
The details of how this matching or coupling is carried out may depend on the particular
numerical algorithm used and how it represents electric fields near a boundary.

There are a range of possible sheath models of different levels of
complexity and accuracy that could be considered in future work.
The present model does not guarantee that the Bohm sheath criterion is met,
which requires that the ion outflow velocity exceed the sound speed, $u_{\parallel i} \ge c_\mathrm{s}$,
for a steady-state sheath and in the sheath-entrance region.
However, the present simulations of LAPD start at a low density and ramp up
the density to an approximate steady state over a period of a few sound transit times,
and during this phase, the pressure and potential drop from the central
source region to the edges is large enough to accelerate ions to near-sonic velocities.
We show in Chapter~\ref{ch:lapd} that the steady-state outflow velocities are very close to or slightly exceed
the sound speed in our LAPD simulations.

There could be other cases where the acceleration of ions in the upstream region
is not strong enough to enforce the Bohm sheath criterion for a steady-state result.
In such a case, some kind of rarefaction fan may propagate from near the sheath,
accelerating ions back up to a sonic level.
This situation is very similar to the Riemann problem for the expansion of a gas into a vacuum \citep{Munz1994}
or into a perfectly absorbing surface, which leads to a rarefaction wave that always maintains
$u_{\parallel i} \ge c_\mathrm{s}$ at the boundary (but also modifies the density and temperature at the outflow
boundary because of the rarefaction in the expanding flow).
A Riemann solver has been implemented in the two-fluid version of Gkeyll for 1D simulations that resolve the sheath 
\citep{Cagas2017}, and the results were compared with a fully kinetic solver.
Exact and approximate Riemann solvers are often used in computational fluid dynamics to
determine upwind fluxes at an interface \citep{LeVeque2002,Durran2010}.
It could be useful to work out a kinetic analogue of this process, or a kinetic model based
on the approximate fluid result, but those are beyond the scope of this thesis.

There is ongoing research to develop improved sheath models for fluid codes.
In some past fluid simulations of LAPD, the parallel ion dynamics were neglected
and modeled by sink terms to maintain a desired steady state on average \citep{Popovich2010b,Friedman2012,Friedman2013}.
\citet{Rogers2010} included parallel ion dynamics in their fluid simulations
and imposed the boundary condition $u_{\parallel i} = c_\mathrm{s}$, thus avoiding the
problem of $u_{\parallel i} < c_\mathrm{s}$.
This boundary condition could be generalized to allow $u_{\parallel i} > c_\mathrm{s}$ at the sheath entrance to
handle cases in which turbulent fluctuations or other effects give more upstream acceleration \citep{Togo2016,Dudson2017}.
\citet{Loizu2012} carried out a kinetic study to develop improved sheath model boundary conditions
for fluid codes that include various effects (including the magnetic pre-sheath \citep{Chodura1982} in an oblique
magnetic field and the breakdown of the ion drift approximation) that have been incorporated
into later versions of the GBS code \citep{Halpern2016}.

%% file: ch-1d-sol/chapter-1d-sol.tex
\chapter{1D Scrape-Off-Layer Models\label{ch:1d-sol}}
The development of a complex gyrokinetic code begins with the investigation
of tests in a spatially one-dimensional geometry.
In this chapter, some simple 1D kinetic models of parallel propagation in
an open-field-line region are explored.
We apply an electrostatic, gyrokinetic-based model to simulate
the parallel plasma transport of an edge-localized-mode (ELM) heat pulse
in the scrape-off layer (SOL) to a
divertor plate. We focus on a test problem that has been studied
previously, using parameters chosen to model a heat
pulse driven by an ELM in JET. Previous work used direct particle-in-cell (PIC)
equations with full dynamics and collisions, or Vlasov or fluid equations
with only parallel dynamics. With the use of the gyrokinetic
quasineutrality equation and logical-sheath
boundary conditions in our model, spatial and temporal resolution requirements
are no longer set by the electron Debye length $\lambda_{De}$ and plasma period $\omega_{pe}^{-1}$,
respectively.
This test problem also helps illustrate some of the
physics contained in the Hamiltonian form of the gyrokinetic
equations and some of the numerical challenges in developing a
gyrokinetic code for the boundary plasma.
The discussion in this chapter on 1D1V models is based on \citet{Shi2015},
while the results from 1D2V simulations have not been published before.
We note that the work presented here was later used as a test case
for a version of the GENE gyrokinetic code that was being extended to handle the SOL using
finite-volume methods \citep{Pan2016}.

\section{Introduction}
One of the major issues for the operation of ITER and subsequent higher-power tokamaks
in the high-confinement mode (H-mode)
is the power load on plasma-facing components (PFCs) from periodic bursts of energy
expelled into the scrape-off layer by Type I ELMs.
ELMs are an MHD instability triggered by a steep pressure gradient in the edge plasma
(e.g., the H-mode pedestal) that is followed by a loss of plasma stored energy and profile
relaxation \citep{Leonard2014}.
Excessive total and peak power loads from ELM heat pulses can cause
melting or ablation of solid surfaces, such as the divertor targets
and the main chamber wall \citep{Pitts2005}.
In a single ITER discharge, several hundred ELMs are expected \citep{Loarte2007}.
The ability to suppress ELMs or at least mitigate the damage
they cause to PFCs is crucial for the viability of
reactor-scale tokamaks.
An accurate prediction of how heat is transported
in the plasma boundary in future devices is important for the development of
mitigation concepts.

Numerical simulations of ELM heat-pulse propagation are valuable in understanding
the time-dependence of power loads on divertor targets due to different sizes of ELMs.
Such simulations can also be used to predict the peak divertor surface temperature
due to an ELM, which is important for material-erosion considerations.
\citet{Pitts2007} studied the parallel propagation of an ELM heat pulse
in the SOL to a divertor plate using 1D3V PIC simulations, including the effects
of collisions.
The authors recognized the inadequacy of fluid codes for this task, since they
usually employ approximate flux limiters on parallel heat fluxes and assume
constant sheath heat-transmission coefficients.
The authors simulated ELM crashes with different energies,
temperatures, densities, and durations.

Some quantities of interest in the simulations of \citet{Pitts2007}
included the time variation of sheath heat-transmission factors,
how much of the energy deposited on the divertor plate is from electrons, and how much of
ELM energy is deposited before the peak heat flux at the divertor plate,
which affects the peak surface temperature \citep{Loarte2007,Leonard2014}.
These simulations were able to reproduce the characteristic heat-flux rise time 
on the order of the fast-ion (at the pedestal temperature) sound-transit time seen on several machines.
The authors found that the fraction of energy deposited before the peak heat flux
to be in the range 0.25--0.35, which was consistent with experimental measurements on JET.
This agreement with experimental data for low ELM energies
provided some confidence in their simulation results for a hypothetical 2.46~MJ ELM (approximating
a small Type I ELM on ITER).

\citet{Manfredi2011} later developed a Vlasov-Poisson code 
to study this problem. \citet{Havlickova2012} carried out a benchmark of
fluid, Vlasov, and PIC approaches to this problem.
An implementation of this test case in BOUT++ was used to compare non-local
and diffusive heat-flux models for SOL modeling \citep{Omotani2013}.
With the exception of initial conditions, the parameters we have adopted for these simulations are
described in \citet{Havlickova2012}.
This test case involves just one spatial dimension (along the field line),
treating an ELM as an intense source near the midplane without trying to
directly calculate the magnetohydrodynamic instability and reconnection processes that
drive the ELM. Nevertheless, this test problem is useful for testing codes
and understanding some of the physics involved in parallel propagation
and divertor heat fluxes.

Unlike previous approaches, we have developed and studied
gyrokinetic-based models with logical-sheath boundary conditions \citep{Parker1993} using
kinetic electrons or by assuming an adiabatic response for the electrons.
As is often done in gyrokinetics, a gyrokinetic quasineutrality equation
(which includes a polarization-shielding term) is used, so the Debye length does not need to be resolved.
To handle the sheath, logical-sheath boundary conditions \citep{Parker1993} are used,
which maintain zero local net current to the wall at each time step.
Logical-sheath boundary conditions are described in detail in Section~\ref{sec:logical-sheath}.
Although our simulations are
one-dimensional, perpendicular effects can be incorporated by assuming
axisymmetry. In an axisymmetric system, poloidal gradients have
components that are both parallel and perpendicular to the magnetic
field. The perpendicular ion polarization dynamics then enter the field equation by accounting for
the finite pitch of the magnetic field.

A major advantage of the models we have developed is their low
computational cost. Earlier kinetic models with explicit time stepping have been described as
computationally intensive \citep{Pitts2007} due to restrictions in the time
step to ${\sim}\omega_{pe}^{-1}$ and in the spatial resolution to ${\sim}\lambda_{De}$.
A 1D Vlasov model using an asymptotic-preserving
implicit numerical scheme described in \citet{Manfredi2011} was only able to
relax these restrictions somewhat for this problem, using
$\Delta x \sim 2 \lambda_{De}$ and $\Delta t \sim 4 / \omega_{pe}$ because
their simulation still included the sheath directly.
By using a gyrokinetic-based model with logical-sheath boundary
conditions, we verify that our simulations can use grid sizes and time steps that are several
orders of magnitude larger than this and still accurately model the sheath potential drop
and the consequent outflow of particles and heat from the SOL.
We find that the divertor-plate heat fluxes (electron, ion, and total)
obtained in our gyrokinetic simulation are within ${\sim}5\%$
of those from a Vlasov--Poisson simulation \citep[see][figure~2]{Havlickova2012}.
For simplicity, the simulations are fully explicit at present.
While fluid models require much fewer computational resources when compared to kinetic models,
they miss some kinetic effects,
including the effect of hot tail electrons on the heat flux on the
divertor plate and the subsequent rise of sheath potential.

In this chapter, we focus on simulations in one spatial dimension using the parameters
of for a 0.4~MJ ELM on JET \citep[see][table~1]{Pitts2007}.
Section~\ref{sec:esmodel} describes an electrostatic 1D gyrokinetic-based
model with a modification to the ion polarization term to set a
minimum value for the wavenumber.
Numerical implementation details and initial conditions are described in Section~\ref{sec:1d_sol_numerics}.
Results from 1D1V numerical simulations are presented in Section~\ref{sec:1d_sol_results}.
We later extended our simulations to 1D2V and added a model for self-species collisions.
These results, which result in improved agreement with PIC simulations, are presented
in Section~\ref{sec:1d-sol-collisions}.
We summarize the main results of this chapter in Section~\ref{sec:1d-sol-conclusions}.

\section{Electrostatic 1D Gyrokinetic Model with Kinetic Electrons\label{sec:esmodel}}
In this chapter, we focus on the long-wavelength, drift-kinetic limit of gyrokinetics and ignore
finite-Larmor-radius effects for simplicity. Polarization effects are kept in the
gyrokinetic Poisson equation, and the model has the general form
of gyrokinetics.

\begin{figure}
\centering
 \subfloat[]{\includegraphics[height=17em]{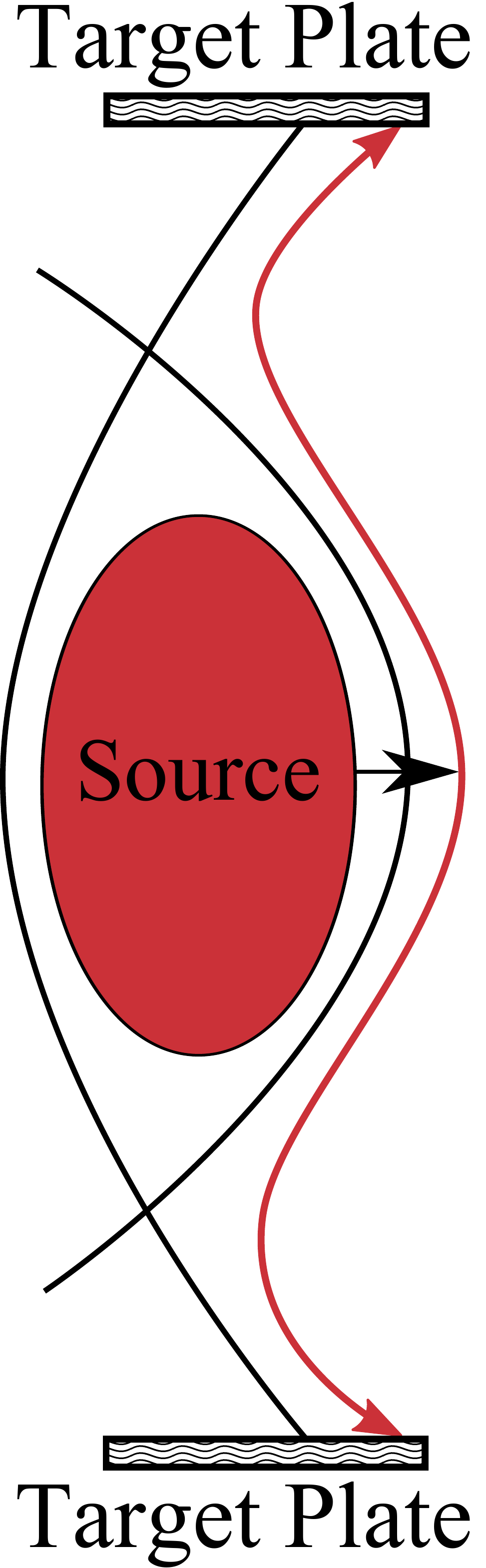}}
 \qquad
 \subfloat[]{\includegraphics[height=17em]{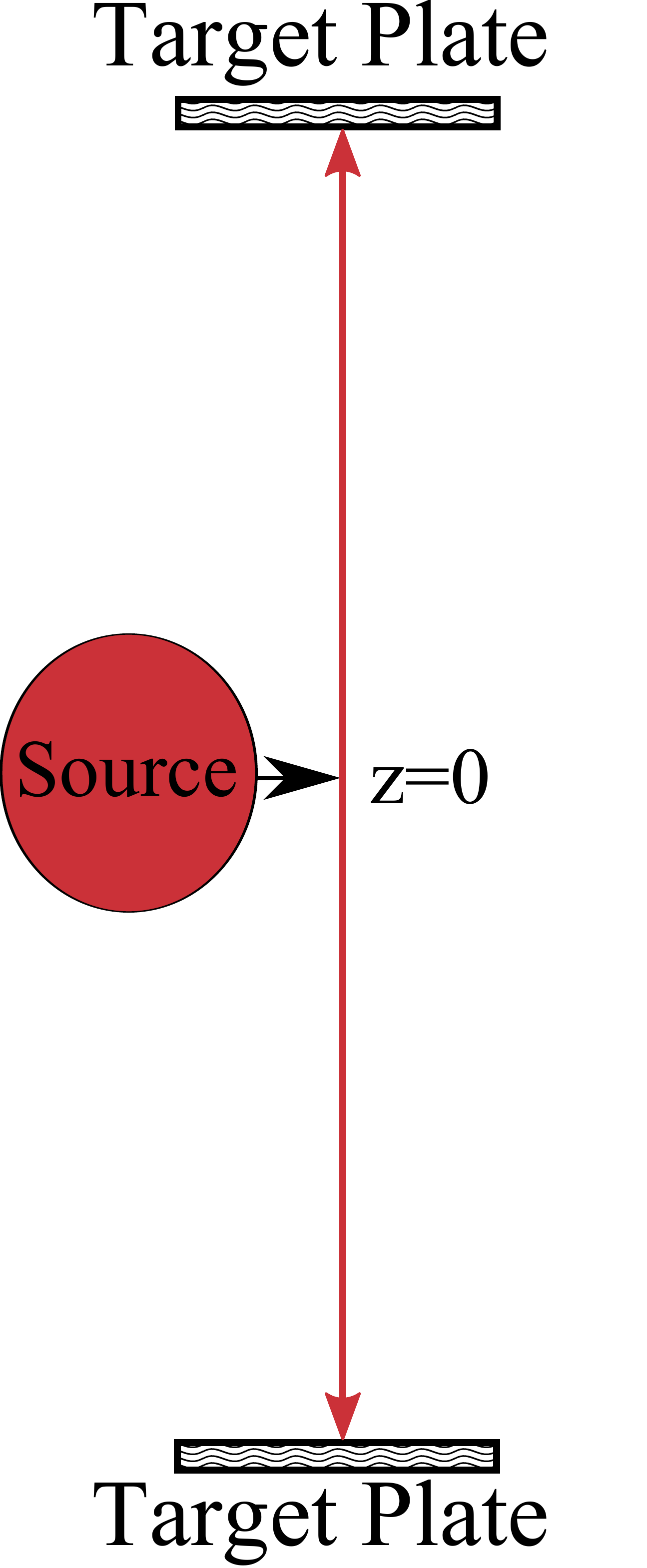}}
 \subfloat[]{\includegraphics[width=0.5\textwidth]{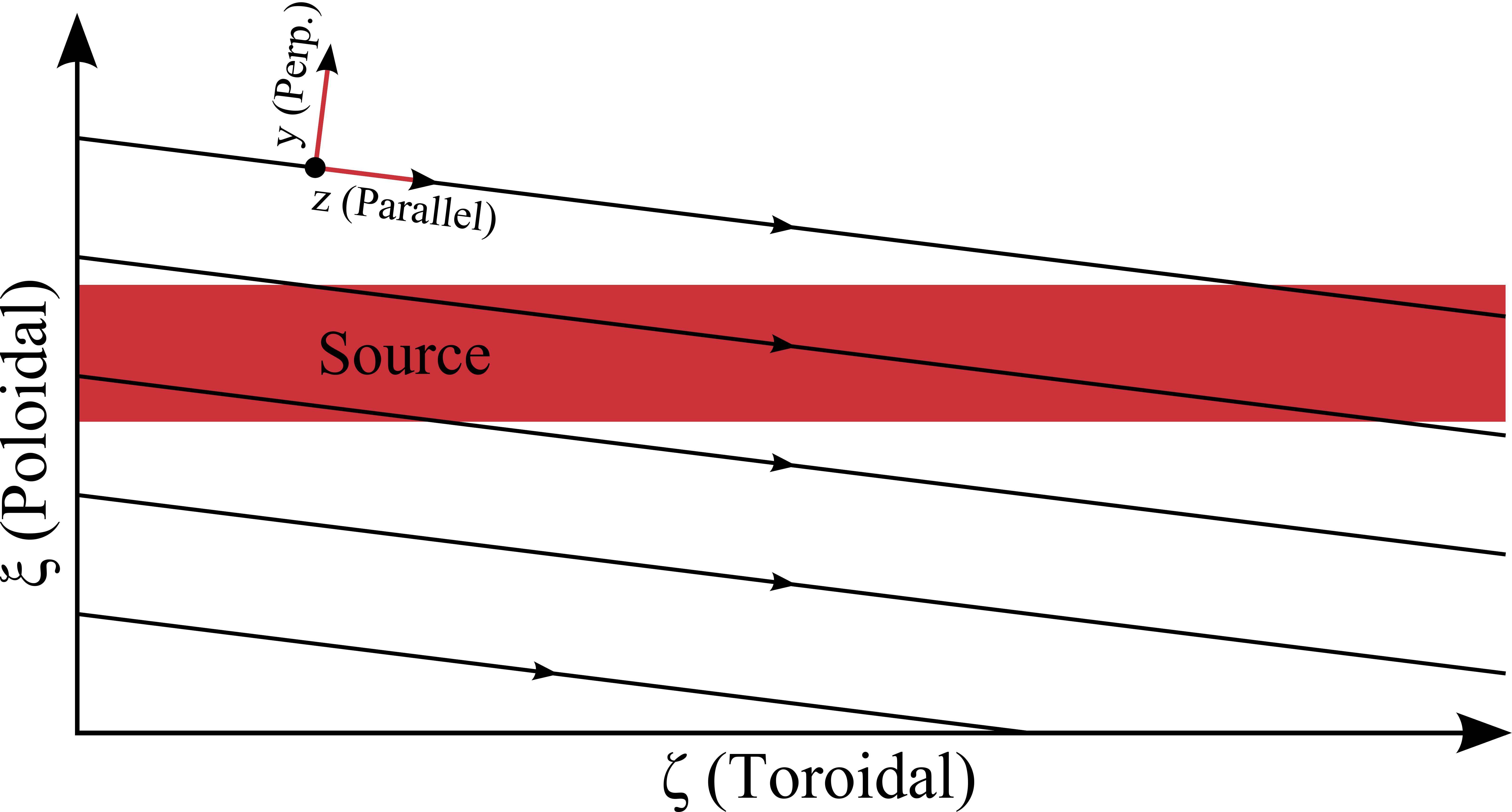}}
 \caption[Geometry used in the ELM heat-pulse test problem.]
 {Illustration of the geometry used in the ELM
 heat-pulse test problem. The scrape-off layer region in the
  poloidal cross section ($a$) is treated as straight ($b$) in this test,
 with the ELM represented by an intense source near
 the midplane region. The time history of the resulting heat flux to
  the target plate is calculated in the simulation. The side view ($c$) illustrates
 that although there is no toroidal variation in this axisymmetric problem, poloidal variations
 lead to both parallel and perpendicular gradient components.}
  \label{fig:sol_geometry} 
\end{figure}

The geometry used in the ELM heat-pulse test problem is illustrated in
figure~\ref{fig:sol_geometry}. The Vlasov and fluid codes used by \citet{Havlickova2012}
consider only the parallel dynamics, while the 1D3V PIC code used by \citet{Havlickova2012}
includes full-orbit (not gyro-averaged) particle dynamics
in an axisymmetric system and so would automatically include
polarization effects on time scales longer than an ion gyroperiod.

The gyrokinetic equation can be written as a Hamiltonian evolution
equation for species $s$ of a plasma
\begin{align}
 \frac{\partial f_s}{\partial t} = \{H_s, f_s \} \label{eqn:gk},
\end{align}
where $H_s =
p_\parallel^2 / 2 m_s + q_s\phi - m_s V_E^2 / 2$ is the Hamiltonian
for the 1D electrostatic case considered here,
$p_\parallel = m_s v_\parallel$ is the parallel momentum, and
$\{f,g\} = (\partial f / \partial z) (\partial g / \partial p_\parallel) -
(\partial f / \partial p_\parallel) (\partial g / \partial z)$ is
the Poisson bracket operator for any two functions $f$ and $g$. The
potential is determined by a
gyrokinetic Poisson equation (in the long-wavelength quasineutral limit):
\begin{align}
-\partial_\perp \left( \epsilon_\perp \partial_\perp \phi \right)
 = \frac{\sigma_g}{\epsilon_0} = \frac{1}{\epsilon_0} \sum_s q_s \int d v_\parallel \, f_s \label{eqn:gk-Poisson}.
\end{align}

Here, $\sigma_g$ is the {\em guiding-center} charge density, while the
left-hand side is the negative of the polarization contribution to the
density, where the plasma perpendicular dielectric is
\begin{align}
\epsilon_\perp = \frac{c^2}{v_A^2} = \sum_s \frac{n_s m_s}{\epsilon_0 B^2}.
\label{eqn:epsilon_perp}
\end{align}
The ion polarization dominates this term, but a sum
over all species has been included for generality.

In the Hamiltonian, $V_E = - (1/B) \partial_\perp \phi$ is the $E \times B$
drift in the radial direction (out of the plane in
figure~\ref{fig:sol_geometry}$(c)$). Since there
is no variation in the radial direction, there is no explicit
$\boldsymbol{V}_E \cdot \nabla$ term, and $V_E$ only enters through the
second-order contribution to the Hamiltonian, $-m V_E^2 / 2$.
\citet{Krommes2013,Krommes2012} provides some physical interpretations of
this term, and \citet{Krommes2013} gives a derivation of it in the cold-ion limit.

The conserved energy is given by
\begin{align}
W_{\rm tot} &= \int \mathrm{d}z \sum_s \int \mathrm{d}v_\parallel \, f_s H_s \nonumber\\
 &= W_K + \int \mathrm{d}z \, \sigma_g \phi - \frac{1}{2}\int \mathrm{d}z \, \rho V_E^2,
\end{align}
where $W_K = \int \mathrm{d}z \sum_s\int \mathrm{d}v_\parallel \, f_s m_s v_\parallel^2 /
2$ is the kinetic energy, and $\rho$ is the total mass density. Using the
gyrokinetic Poisson equation (\ref{eqn:gk-Poisson}) to substitute
for $\sigma_g$ in this equation and doing an integration by parts,
one finds that the total conserved energy can be written as
\begin{align}
  W_{\rm tot} &= \frac{1}{2}\int \mathrm{d}z \sum_s \int \mathrm{d}v_\parallel \, f_s \left(m_s v_\parallel^2 + m_s V_E^2\right) \nonumber\\
 &= W_K + \frac{1}{2}\int \mathrm{d}z \, \rho V_E^2,
\end{align}
where the global neutrality condition $\int \mathrm{d}z \, \sigma_g = 0$ was used to eliminate boundary terms.

To verify energy conservation, first note that $\int \mathrm{d}z \int d
v_\parallel H_s \partial f_s / \partial t = 0$ by multiplying the gyrokinetic equation (\ref{eqn:gk}) by
the Hamiltonian and integrating over all of phase-space.
Here, periodic boundary conditions are used for simplicity; there are
losses to the wall in a bounded system.
The rate of change of the total conserved energy is then written as
\begin{align}
\frac{d W_{\rm tot}}{dt} &= \int \mathrm{d}z \sum_s \int \mathrm{d}v_\parallel \, f_s \left(
q_s \frac{\partial \phi}{\partial t} - \frac{m_s}{2} \frac{\partial V_E^2}{\partial t}
\right) \nonumber\\
&= \int \mathrm{d}z \, \left( \sigma_g \frac{\partial \phi}{\partial t} 
 - \frac{1}{2} \sum_s n_s m_s \frac{\partial V_E^2}{\partial t}\right).
\end{align}
Using the gyrokinetic Poisson equation (\ref{eqn:gk-Poisson}) to substitute for $\sigma_g$
and integrating by parts, one finds that these two terms cancel, so $d W_{\rm tot}/dt = 0$.
Note that the small second-order Hamiltonian term $H_2 = -(m/2) V_E^2$
was needed to get exact energy conservation.
In many circumstances, the $E \times B$ energy $m V_E^2$ is only a very small
correction to the parallel kinetic energy $m v_\parallel^2/2$, but it is
still assuring to know that exact energy conservation is possible.
This automatically occurs in the Lagrangian field theory approach to
full-$f$ gyrokinetics \citep{Sugama2000,Brizard2000,Krommes2012}, in which the
gyrokinetic Poisson equation results from a functional derivative of the action with
respect to the potential $\phi$, so a term that is linear in $\phi$ in
the gyrokinetic Poisson equation comes from a term that is quadratic in $\phi$ in the Hamiltonian.

\subsection{\label{subsec:1dsolenergy}Electrostatic Model with a Modified Ion Polarization Term}
One can obtain a wave dispersion relation by linearizing
({\ref{eqn:gk}}) and (\ref{eqn:gk-Poisson}) and Fourier
transforming in time and space.
With the additional assumption that $q_e = -q_i$ and neglecting ion
perturbations (except for the ion polarization density), one has
\begin{align}
  k_\perp^2 \rho_{\mathrm{s}}^2 + \left[1 + \xi Z\left(\xi\right)\right]&= 0.
\end{align}
Here, $\rho_\mathrm{s}^2 = T_e/(m_i \Omega_{ci}^2)$, $\xi = \omega/(\sqrt{2}
k_\parallel v_{\rm te})$, $v_{\rm te} = \sqrt{T_e/m_e}$, and the plasma dispersion function is
$Z(\xi)=\pi^{-1/2}\int \mathrm{d}t 
\exp(-t^2)/(t-\xi)$ (or the analytic continuation of this for ${\rm Im}(\xi)
\le 0$).
In the limit $\xi \gg 1$, the solution to the dispersion relation is a wave with frequency
\begin{align}
  \omega_H &= \frac{k_\parallel v_{\rm te}}{|k_\perp| \rho_\mathrm{s}}\label{eqn:omegaH}.
\end{align}
For $k_\perp \rho_\mathrm{s} \ll 1$, this is a high-frequency wave that must be handled
carefully to remain numerically stable. Note that this wave does not affect parallel
transport in the SOL because the main heat pulse propagates at the ion sound speed,
and this wave is even faster than the electrons for $k_\perp \rho_\mathrm{s} \ll 1$.

This wave is the electrostatic limit of the shear Alfv\'en 
wave \citep{Lee1987,Belli2005}, which lies in the regime of inertial Alfv\'en waves \citep{Lysak1996,Vincena2004}.
The difficulties introduced by such a wave could be eased by including
magnetic perturbations from $A_\parallel$, in which case the dispersion relation (in the
fluid electron regime $\xi \gg 1$) becomes $\omega^2 = k_\parallel^2
v_{\rm te}^2 / (\hat{\beta}_e + k_\perp^2 \rho_\mathrm{s}^2)$, where $\hat{\beta}_e = 
(\beta_e/2) (m_i/m_e)$ and $\beta_e = 2 \mu_0 n_e T_e / B^2$ \citep{Belli2005}.
In the electrostatic limit $\hat{\beta}_e = 0$, we
recover (\ref{eqn:omegaH}), but retaining a finite $\hat{\beta}_e$ would set a
maximum frequency at low $k_\perp$ of $\omega = k_\parallel v_{\rm te} / 
\hat{\beta}_e^{1/2} = k_\parallel v_A$, where $v_A$ is the Alfv\'en velocity, 
avoiding the $k_\perp \rho_\mathrm{s} \rightarrow 0$ singularity of the electrostatic case.

For electrostatic simulations, a modified ion polarization term can be
introduced to effectively set a minimum value for the perpendicular wave
number $k_\perp$.
This modification can be used to slow down the electrostatic shear Alfv\'en wave to
make it more numerically tractable.
Even when magnetic fluctuations are included, one still might
want to consider an option of introducing a long-wavelength modification
for numerical convenience or efficiency.

When choosing how to select the minimum value for $k_\perp \rho_\mathrm{s}$, it is
useful to consider the set of $k_\perp$'s represented on the
grid for particular simulation parameters.
Consider an axisymmetric system (as in figure~\ref{fig:sol_geometry}$(c$)) with
constant $B/B_\xi$, where $B$ is the total magnetic field, and
$B_\xi$ and $B_\zeta$ are the components of $\boldsymbol{B}$ in the poloidal and
toroidal directions.
It follows that $\partial_\perp = (B_\zeta/B_\xi) \partial_\parallel$, so
\begin{align}
k_{\perp,\mathrm{max}} &= \frac{B_\zeta}{B_\xi} k_{\parallel,\mathrm{max}}.
\end{align}
The maximum parallel wavenumber can be estimated as
$k_{\parallel,\mathrm{max}} \Delta z \sim \pi N_{nc}$, 
where $\Delta z$ is the width of a single cell in position space, and
$N_{nc}$ is the total degrees of freedom per cell used in the
DG representation of the position coordinate.

Therefore, one has
\begin{align}
k_{\perp,\mathrm{max}} &= \frac{B_\zeta}{B_\xi} \frac{\pi N_{nc}}{\Delta z}.
\end{align}
In our simulations, $N_{nc}=3$ and $\Delta z = 10$ m using $8$ cells in the
spatial direction to represent an $80$ m parallel length.
Assuming that $B_\xi/B = \sin(6^\circ)$, 
one estimates that $k_{\perp,\mathrm{max}}\rho_\mathrm{s} \approx 2.5 \times
10^{-2}$ for 1.5~keV deuterium ions with $B=2$~T. Thus, the
perpendicular wave wavenumbers represented by a typical grid are fairly
small.

The general modified gyrokinetic Poisson equation we consider is of
the form
\begin{align}
-\partial_\perp ( C_\epsilon \epsilon_\perp \partial_\perp \phi)
 + s_\perp(z,t)
 (\phi - \langle \phi \rangle) = \frac{\sigma_g(z)}{\epsilon_0},
\label{eqn:mod-gk-Poisson}
\end{align}
where $s_\perp(z,t)=k^2_{\rm min}(z) {\epsilon_\perp(z,t)}$ is a
shielding factor (we allow $k_{\rm min}$ to depend on position but not
on time in order to preserve energy conservation, as described later in this section)
and $\langle \phi \rangle$ is a
dielectric-weighted, flux-surface-averaged 
potential defined as 
\begin{align}
\langle \phi \rangle &= \frac{\int \mathrm{d}z \, s_\perp \phi}{\int \mathrm{d}z \, s_\perp} \label{eqn:phi-avg}.
\end{align}
The fixed coefficient $C_\epsilon$ is for generality, making it easier to consider various limits later.

The sound gyroradius is chosen to be defined by $\rho_{\mathrm{s}}^2(z,t) = c_s^2(z,t)/ \Omega_{ci}^2 =
T_e(z,t)/(m_i \Omega_{ci}^2)$, using the mass and cyclotron frequency of
a main ion species. A time-independent sound gyroradius (using a
typical or initial value for the electron temperature $T_{e0}$) is defined by
 $\rho_{\mathrm{s}0}^2(z) = c_{s0}^2(z)/ \Omega_{ci}^2 = T_{e0}(z)/(m_i \Omega_{ci}^2)$. Note
that the shielding factor can also be written as $s_\perp(z,t)=[k_{\rm
min}(z) \rho_{\mathrm{s}0}(z)]^2 {\epsilon_\perp(z,t)} / \rho^2_{\mathrm{s}0}(z)$.

For simplicity, $k_{\rm min} \rho_{\mathrm{s}0}$ is chosen to be a constant independent of
position. 
Its value should be small enough that the wave in (\ref{eqn:omegaH}) is high enough in
frequency that it does not interact with other dynamics of
interest, but not so high in frequency that it forces the explicit time
step to be excessively small. For some of our simulations, we use
$k_{\rm min} \rho_\mathrm{s} = 0.2$,
which leads to only a 2\% correction to the ion acoustic 
wave frequency $\omega = k_\parallel c_s / \sqrt{1 + k_\perp^2 \rho_\mathrm{s}^2}$ at
long wavelengths. Convergence can be checked by taking the
limit $k_{\rm min} \rho_{\mathrm{s}0} \rightarrow 0$.

As a simple limit, one can even set $C_\epsilon = 0$ and keep just the $s_\perp$ term, which replaces the
usual differential gyrokinetic Poisson equation with a simpler algebraic model.
This approach should work fairly well for low frequency dynamics.
The basic idea is that for long-wavelength ion-acoustic
dynamics, the left-hand side of (\ref{eqn:mod-gk-Poisson})
is small, so the potential is primarily determined by 
the requirement that it adjust to keep the electron density on the
right-hand side almost equal to the ion guiding center density.
At low frequencies, the electrons are near-adiabatic (in parallel force balance),
so the density depends in part on the potential. In future work, one could consider
using an implicit method, perhaps using the method here as a 
preconditioner. Alternatively, electromagnetic effects will slow down
the high-frequency wave so that explicit methods may be sufficient.

The flux-surface-averaged potential $\langle \phi \rangle$ is subtracted
off in (\ref{eqn:mod-gk-Poisson}) so that the model polarization term is gauge invariant like the
usual polarization term.
This choice is also related to our form of the logical-sheath boundary
condition (Section~\ref{sec:logical-sheath}), which enforces
zero current to the wall.
Since the electron and ion guiding-center fluxes
to the wall are the equal, the net guiding center charge vanishes, $\int \mathrm{d}z \, \sigma_g = 0$.
Just as the net guiding-center charge vanishes, our model polarization charge
density, $s_\perp (\phi - \langle \phi \rangle)$, also averages to zero. 
This approach neglects ion polarization losses to the wall, which is consistent in this
model because integrating (\ref{eqn:gk-Poisson}) over all space then
gives $\partial_\perp \phi = 0$ at the plasma edge.
One could consider future modifications to account for polarization drift
losses to the wall, but the present model is found to agree fairly well
with full-orbit PIC results.

With this approach, it is also necessary to modify the Hamiltonian in order to preserve
energy consistency with this modified gyrokinetic Poisson equation.
The modified Hamiltonian is written in the form
\begin{align}
H_s = \frac{1}{2} m_s v_\parallel^2 + q_s (\phi - \langle \phi \rangle) - \frac{1}{2} m_s \hat{V}_E^2,
\end{align}
where $\hat{V}_E^2$ is a modified $E \times B$ velocity that is
chosen to conserve energy. 
The constant $\langle \phi \rangle$ term in $H_s$ has no effect on the
gyrokinetic equation because only gradients of $\phi$ matter, but it
simplifies the energy conservation calculation.
The total energy is still $W_{\rm tot} = \int \mathrm{d}z \sum_s \int
\mathrm{d}v_\parallel \, f_s H_s$, and its time derivative (neglecting boundary
terms that are straightforward to evaluate) can be written as
\begin{align}
\frac{d W_{\rm tot}}{dt} &= \int \mathrm{d}z \sum_s \int \mathrm{d}v_\parallel \, f_s \frac{\partial H}{\partial t} \nonumber \\
 &= \int \mathrm{d}z \left( \sigma_g \frac{\partial}{\partial t} \left( \phi -
\langle \phi \rangle \right) 
 - \sum_s \frac{1}{2} n_s m_s \frac{\partial}{\partial t} \hat{V}_E^2 \right).
\end{align}
Using the modified gyrokinetic Poisson equation (\ref{eqn:mod-gk-Poisson}) and
integrating the first term by parts gives
\begin{align}
\frac{d W_{\rm tot}}{dt} 
 &= \int \mathrm{d}z \Bigg( \sum_s \frac{1}{2} n_s m_s
C_\epsilon \frac{\partial}{\partial t} V_E^2 +\frac{\epsilon_0}{2} s_\perp \frac{\partial}{\partial t} 
 (\phi - \langle \phi \rangle)^2 - \sum_s \frac{1}{2} n_s m_s \frac{\partial}{\partial t} \hat{V}_E^2 \Bigg),
\end{align}
so energy is conserved if one chooses
\begin{align}
\hat{V}_E^2 &= C_\epsilon V_E^2 + \frac{\epsilon_0 s_\perp}{\sum_s n_s m_s} (\phi - \langle
\phi \rangle)^2
\end{align}
and require that the coefficient $\epsilon_0 s_\perp/ (\sum_s n_s m_s)$ be
independent of time so that it comes outside of a time derivative.
Using (\ref{eqn:epsilon_perp}) and the definition of $s_\perp$ after
(\ref{eqn:mod-gk-Poisson}), one sees that $\epsilon_0 s_\perp/ (\sum_s n_s m_s) =
k_{\rm min}^2(z) /B^2$, which is indeed independent of time because $k_{\rm
min}$ was chosen not to have any time dependence.

In the limit that one uses only the algebraic model polarization term with $C_\epsilon=0$, one finds that
\begin{align}
  \hat{V}_E^2 = (k_{\rm min} \rho_{\mathrm{s}0})^2 \left( \frac{e \delta \phi}{T_{e0}} \right)^2 c_{s0}^2,
\end{align}
where $\delta \phi = \phi - \langle \phi \rangle$.
For $k_{\rm min} \rho_{\mathrm{s}0} = 0.2$ and $e \delta \phi / T_{e0} \sim 1$, this $E \times B$
energy could be order 4\% of the total energy.
We use the $C_\epsilon = 0$ limit of this model for the simulations discussed in the next section.

\section{Numerical Implementation Details\label{sec:1d_sol_numerics}}
One detail of solving the modified gyrokinetic Poisson equation (\ref{eqn:mod-gk-Poisson}) is how to
determine the flux-surface-averaged component, which is related to the boundary conditions.
Consider the case in which $\epsilon_\perp = 0$, and expand $\phi =
\langle \phi \rangle + \delta \phi$. Then $\delta \phi$ is determined by
the algebraic equation
\begin{align}
s_\perp(z) \, \delta \phi(z) = \frac{\sigma_g(z)}{\epsilon_0}\label{eqn:algebraicPoisson}.
\end{align}
Imposing the boundary condition that the value of $\phi$ at the plasma edge be
equal to the sheath potential gives $\phi(z_R) = \phi_{sh} = \langle \phi \rangle + \delta \phi(z_R)$
(the left and right boundaries have been assumed to be symmetric here), which gives an additional equation
to determine $\langle \phi \rangle$.
The final expression is 
\begin{align}
  \phi(z) = \delta \phi(z) - \delta \phi(z_R) + \phi_{sh}.
\end{align}

In order to maintain energy conservation, it is important that the algorithm preserve the numerical
equivalent of certain steps in the analytic derivation.
In our algorithm, based on Liu and Shu's \citep{Liu2000} algorithm for
the incompressible Euler equation, $\phi$ must be obtained using
continuous finite elements, although the charge density $\sigma_g$ is discontinuous in our Poisson equation.

To preserve the integrations involved in energy conservation, it is important
to ensure that one can multiply (\ref{eqn:algebraicPoisson}) by the fluctuating potential,
integrate over all space, and preserve
\begin{align}
\int \mathrm{d}z \, \delta \phi \, s_\perp \, \delta \phi = \frac{1}{\epsilon_0} \int \mathrm{d}z \, \delta \phi \, \sigma_g.
\end{align}
This requirement ensures that a potential part of the energy on the right-hand side is exactly related to a
field-like-energy on the left-hand side.
This quantity will be preserved if one projects the modified Poisson equation onto all of the continuous basis functions
$\psi_j$ that are used for $\phi$ (i.e., $\phi(z) = \sum_j \phi_j
\psi_j(z)$) to ensure that
\begin{align}
\langle \psi_j s_\perp \phi \rangle = \langle \psi_j \sigma_g \rangle.
\end{align}

For piecewise-linear basis functions, this leads to a tridiagonal equation for $\phi_j$
that has to be inverted to determine $\phi$. Because $s_\perp \propto n(z,t)$ varies in time,
this will take a little bit of work, but as one goes to higher dimensions
in velocity space, the Poisson solve (which is only in the
lower-dimensional configuration space)
will be a negligible fraction of the computational time.

\section{Simulation Results\label{sec:1d_sol_results}}
The main parameters used for our simulations were described in
\citet{Havlickova2012,Pitts2007} and were chosen to model an ELM on the JET
tokamak for a case in which the density and temperature at the top of the
pedestal were $n_{\mathrm{ped}} = 5 \times 10^{19}$~m$^{-3}$ and $T_{\mathrm{ped}} =
1.5$~keV. The ELM is modeled as an intense particle and heat source in
the SOL that lasts for 200~$\mu$s, spread over a poloidal extent $L_\mathrm{pol}=2.6$~m
around the midplane and a radial width $\Delta R=10$~cm.
The model SOL has a major radius $R=3$~m, and this source
corresponds to a total ELM energy of about 0.4~MJ.
Note that $R$ and $\Delta R$ are simply volume-scaling parameters
and do not affect the simulation results.
The simulation domain has a length along field lines
$2 L_\parallel = 80$~m, which models a magnetic field line
on JET with an incidence angle $\theta$ of $6^\circ$.

\begin{table}
\begin{center}
  \caption[Summary of parameters for the ELM-heat-pulse simulations.]
  {Summary of parameters for the ELM-heat-pulse simulations. These parameters
  are based on a simulation for a 0.4~MJ ELM released into the JET SOL,
  originally described in \citet{Pitts2007}. This case was later used as a benchmark
  problem in \citet{Havlickova2012}.}
  \bigskip
  \begin{tabular}{ccl}
  \toprule
  \textbf{Symbol} & \textbf{Value} & \textbf{Description} \\
  \midrule
    $t_\mathrm{ELM}$ & 200~$\mu$s & ELM pulse duration \\
    $n_\mathrm{ped}$ & $5 \times 10^{19}$~m$^{-3}$ & Pedestal density \\
    $T_\mathrm{ped}$ & 1.5~keV & Temperature of ELM-pulse ions and electrons \\
    $W_\mathrm{ELM}$ & 0.4~MJ & Total ELM energy \\
    $L_s$ & 25~m & Parallel length of source region \\
    $2 L_\parallel$ & 80~m & Parallel length of simulation domain \\
    $\tau_e$ & 2.5~$\mu$s & Electron transit time ($L_\parallel/v_{te,\mathrm{ped}}$) \\
    $\tau_i$ & 149~$\mu$s & Ion transit time ($L_\parallel/c_{\mathrm{s},\mathrm{ped}}$) \\
    $k_\perp \rho_{\mathrm{s}0}$ & 0.2 & Perpendicular wavenumber \\
    $S_0$ & $9.07 \times 10^{23}$~m$^{-3}$~s$^{-1}$ & Density source rate \\
    $\theta$ & $6^\circ$ & Magnetic-field-line incidence angle \\
  \bottomrule
  \end{tabular}
  \label{tab:1d_sol_parameters}
\end{center}
\end{table}

The kinetic equation with the source term on the right-hand side is
\begin{align}
\frac{\partial f}{\partial t} -\{H,f\}
= g(t) \, S(z) \, F_M\boldsymbol{(}v_\parallel, T_{S}(t)\boldsymbol{)}\label{eq:1d-sol-source},
\end{align}
where $F_M\boldsymbol{(}v_\parallel, T_S(t)\boldsymbol{)}$ is a unit Gaussian in variable $v_\parallel$
with a time-dependent temperature $T_S(t)$.
The function $S(z)$ is the same for both particle species, and is represented as
\begin{align}
S(z) &= \begin{cases}
  S_0 \cos\left(\frac{\pi z}{L_s}\right) & |z| < \frac{L_s}{2},\\
  0 & \mathrm{else},
\end{cases}
\end{align}
where $L_s = 25$~m is the length of the source along the magnetic field line.
The value of $S_0$ was computed using the scaling \citep{Havlickova2012}
\begin{align}
  S_0 = A \, n_{\mathrm{ped}} \, c_{\mathrm{s},\mathrm{ped}} / L_s,
\end{align}
where the constant of proportionality $A$ was chosen to be $1.2 \sqrt{2} \approx 1.7$ for comparison with \citet{Havlickova2012}.
In our simulations, $S_0 \approx 9.07\times 10^{23}$ m$^{-3}$\thinspace s$^{-1}$.
\begin{figure}
\centering
\includegraphics[width=\textwidth]{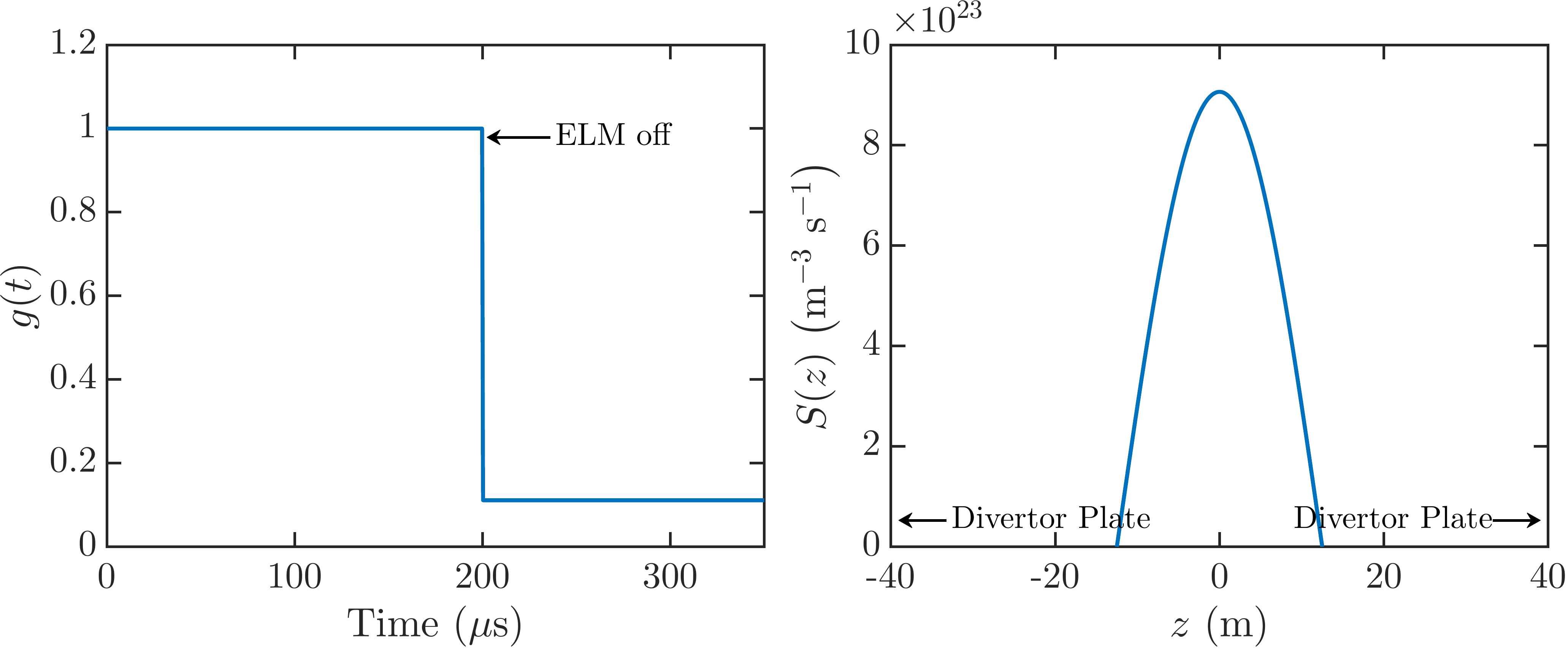}
  \caption[Spatial and temporal profiles of the source term used in
  the ELM heat-pulse test problem.]
  {\label{fig:1d_sol_source} Spatial and temporal profiles of the source term used in
  the ELM heat-pulse test problem. After the intense-ELM source is turned off at $t=200$~$\mu$s,
  the particle sources are reduced to 1/9th their original value, and the electron and ion source
  temperatures are also reduced from 1.5~keV to 210~eV and 260~eV, respectively.}
\end{figure}

The function $g(t)$ in (\ref{eq:1d-sol-source}) is used to model the time-dependence of the particle source:
\begin{align}
g(t) &= \begin{cases} 1 & 0 < t < 200 \, \mu\mathrm{s}, \\
1/9 & t > 200 \, \mu\mathrm{s}.
\end{cases}
  \label{eq:1d-sol-g-factor}
\end{align}

The post-ELM source also has reduced electron and ion temperature, represented
by the $T_S(t)$ parameter in the Maxwellian term $F_M$ in (\ref{eq:1d-sol-source}), which
has the value $1.5$~keV from $0 < t < 200$~$\mu$s for both ions and electrons.
The electron temperature for $t > 200$~$\mu$s is 210~eV, and the ion temperature
is reduced to 260~eV. The end time for the simulation is $t=350$~$\mu$s.
The functions $g(t)$ and $S(z)$ are shown in figure~\ref{fig:1d_sol_source}.

We performed our simulations using second-order Serendipity basis
functions \citep{Arnold2011} on a grid with eight cells in the spatial direction and 32
cells in the velocity direction. In one dimension, second-order basis functions correspond
to piecewise-quadratic basis functions, or three degrees of freedom within
each cell. The case with kinetic electrons and kinetic ions takes about three
minutes to run on a personal computer.

\subsection{Initial Conditions}
In previous papers that looked at this problem, the codes were typically
run for a while with the same weak source that would be used in the
post-ELM phase to reach a quasi-steady state before the intense ELM
source was turned on. The authors found that the final results were not very
sensitive to the duration of the pre-ELM phase or the initial conditions
used for it. However, there is formally no normal steady state for this
problem in the collisionless limit (low-energy particles build up over time
without collisions). To remove a possible source of ambiguity for
future benchmarking, here we specify more precise initial
conditions chosen to approximately match initial conditions at the
beginning of the ELM phase used in previous work.

We model the initial electron distribution function as
\begin{align}
f_{e0}(z,v_\parallel) &= n_{e0}(z) F_M(v_\parallel, T_{e0}),
\end{align}
with $T_{e0} = 75$~eV.
The electron density profile (in $10^{19}$ m$^{-3}$) is chosen to be
\begin{align}
n_{e0}(z) &= 0.7 + 0.3\left(1-\left|\frac{z}{L_\parallel}\right|\right) + 0.5\cos\left(\frac{\pi z}{L_s}\right)H\left(\frac{L_s}{2} - |z|\right).
\end{align}
The initial ion distribution function is modeled as
\begin{equation}
f_{i0}(z,v_\parallel) = \begin{cases}
  F_L & z < -\frac{L_s}{2}, \\
\left(\frac{1}{2}- \frac{z}{L_s}\right) F_L + \left(\frac{1}{2} + \frac{z}{L_s}\right) F_R
& -\frac{L_s}{2} < z < \frac{L_s}{2}, \\
F_R & z > \frac{L_s}{2}. \end{cases}
\end{equation}
Here, $F_L$ and $F_R$ are left and right half-Maxwellians defined as
\begin{align}
F_R(z,v_\parallel; T_{i0}) &= \hat{n}(z)F_M(v_\parallel, T_{i0})H(v_\parallel),\\
F_L(z,v_\parallel; T_{i0}) &= \hat{n}(z)F_M(v_\parallel, T_{i0})H(-v_\parallel),
\end{align}
where $\hat{n}(z) = 2n_{i0}(z)$, $H$ is the Heaviside step function, and
the initial ion temperature profile (in eV) is defined as
\begin{align}
T_{i0}(z) &= 100 + 45\left(1-\left|\frac{z}{L_\parallel}\right|\right)
  + 30\cos\left(\frac{\pi z}{L_s}\right)H\left(\frac{L_s}{2} - |z|\right).
\end{align}

The expressions for the $n_{e0}$ and $T_{i0}$ profiles were chosen to
approximate those described in private communication with the author of
\citet{Havlickova2012}, which were originally
obtained from simulations that had run for a while with a weaker source
to achieve a quasi-steady state before the strong ELM source was turned
on, as described at the beginning of this subsection.

Given an initial electron density profile, we then calculate an initial
ion-guiding-center-density profile to minimize the excitation of
high-frequency shear Alfv\'{e}n waves.
We do this by choosing the initial ion-guiding-center-density $n_i(z)$
so that it gives a potential $\phi(z)$ that results in the electron
density's being consistent with a Boltzmann equilibrium, i.e., the
electrons are initially in parallel force balance (adiabatic) and do not excite
high-frequency shear Alfv\'en waves.
An adiabatic-electron response is
\begin{align}
n_e(z) &= C \exp \left(\frac{e \phi(z)}{T_e}\right).
\end{align}
Taking the log of the above equation and then an $n_e$-weighted average, one has
\begin{align}
\langle \log n_e \rangle_{n_e} &= \log C + \frac{e\langle\phi\rangle_{n_e}}{T_{e0}},
\end{align}
where $T_e$ has been assumed to be a constant $T_{e0}$.

Note that one is free to add an arbitrary constant to $\phi$ since only
gradients of $\phi$ affect the dynamics.
Choosing the additional constraint that $\langle \phi \rangle_{n_e} =
0$, one can express the constant $C$ in terms of $n_e$. This convention
for $\langle \phi \rangle_{n_e}$ is only for convenience, as any
constant can be added to $\phi$ in the plasma interior without affecting
the results. After the first time step, the sheath boundary condition
will be imposed, which will give a non-zero value for the average
potential.

One then has the following equation for $\phi$:
\begin{align}
\frac{e\phi}{T_{e0}} &= \log n_e - \langle \log n_e \rangle_{n_e}.
\end{align}
This $\phi$ can be used with the gyrokinetic Poisson equation to solve for $n_i(z)$ by iteration.
With a small $m_e/m_i$ ratio, the gyrokinetic Poisson equation can be written as
\begin{align}
  n_i(z) \left(1 - k_\perp^2 \rho_{\mathrm{s}0}^2 \frac{e(\phi - \langle \phi \rangle_{n_i})}{T_{e0}}\right) &= n_e(z),
\end{align}
where with the small $m_e/m_i$ ratio approximation, the dielectric-weighted
average is equivalent to an ion density-weighted average.
The left-hand side of this equation is a nonlinear function of $n_i$
(because it appears as a leading coefficient and in the density-weighted
average $\langle \phi \rangle_{n_i}$), which is solved for by using iteration:
\begin{align}
  n_i^{j+1}(z) &= \frac{n_e(z)}{1 - k_\perp^2 \rho_{\mathrm{s}0}^2 \frac{e}{T_{e0}}
 \left(\phi - \langle \phi \rangle_{n_i^{j}} \right)}.
\end{align}
Note that the averaged $\phi$ on the right-hand side is weighted by $n_i^j$,
the previous iteration's ion density.
Convergence can be improved by adding a constant to $n_i(z)$ each
iteration to enforce global neutrality $\langle n_i \rangle = \langle
n_e \rangle$.
In our tests, the initial ion density profile was calculated to $10^{-15}$ relative error in five iterations.

\subsection{Divertor Heat Flux with Drift-Kinetic Electrons}
\begin{figure}
\centering
\includegraphics[width=0.75\textwidth]{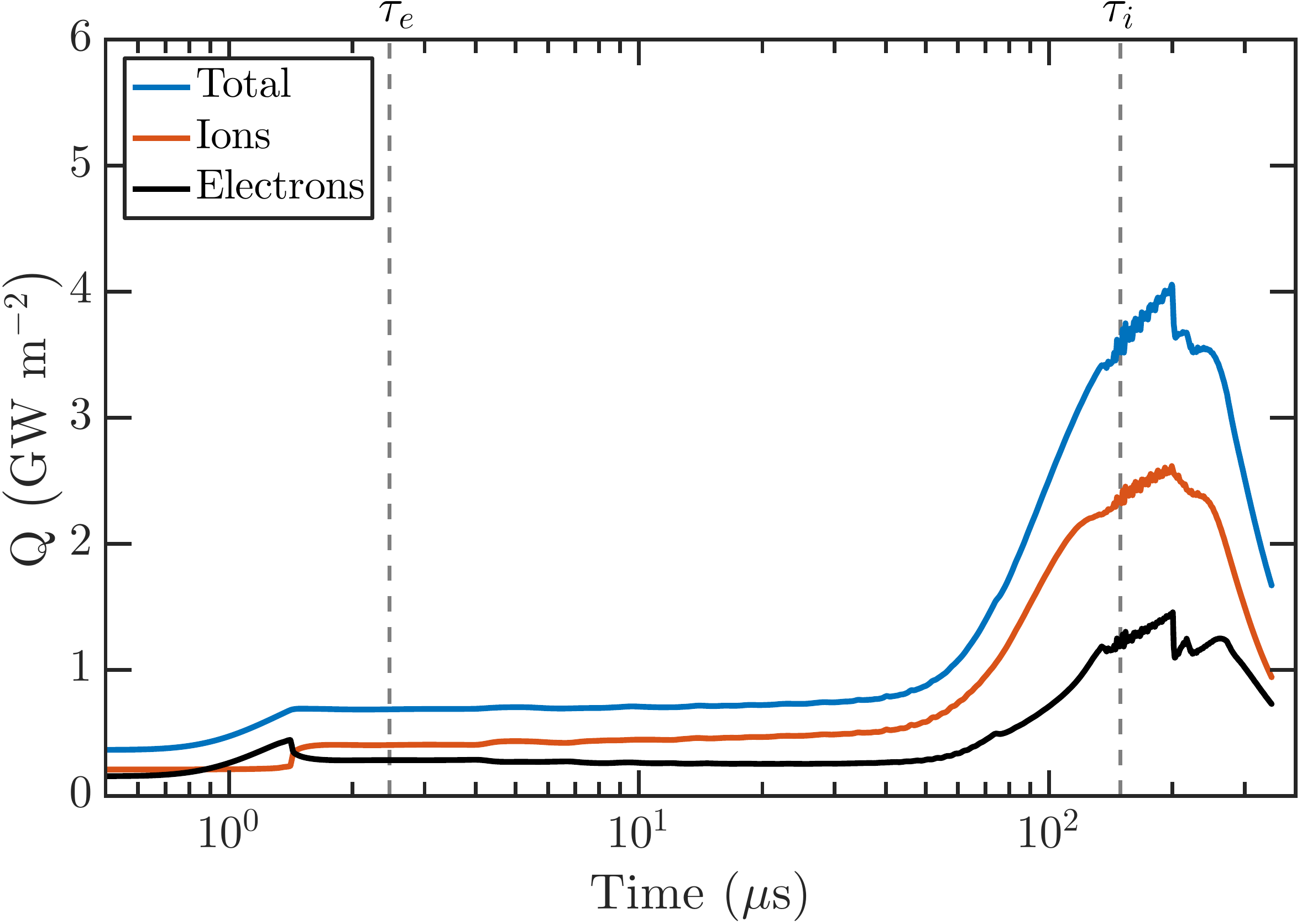}
\caption[Parallel heat flux at the divertor
 plate versus time with drift-kinetic electrons in the ELM heat-pulse problem.]
  {\label{fig:heat-flux-es} Parallel heat flux at the divertor
 plate versus time with drift-kinetic electrons. The electron and ion
 transit times $\tau_e$ and $\tau_i$ are indicated by the
 vertical dashed lines. The sharp drop in the electron heat flux occurs
  in the microseconds after the intense upstream source is turned off (see (\ref{eq:1d-sol-g-factor})
  and nearby discussion).}
\end{figure}

Figure~\ref{fig:heat-flux-es} shows the parallel heat flux on the target plate versus time
using the 1D electrostatic model with a fixed $k_\perp \rho_{\mathrm{s}0}=0.2$.
A rapid response in the electron heat flux is observed at early times,
on the order of the electron transit timescale $\tau_e \sim
L_\parallel/v_{te, \mathrm{ped}} \approx 2.5$~$\mu$s.
This response is due to fast electrons reaching the target plate, 
which initially cause a modest rise in the electron heat flux from $t \sim 1$~$\mu$s to $t \approx 1.5$~$\mu$s.
This build-up of fast electrons results in a rise in the sheath potential at $t \approx 1.5$~$\mu$s, which
causes a modest rise in the ion heat flux
and a modest drop in the electron heat flux until the arrival of the
bulk ion heat flux at a later time.
We did a scan in $k_\perp^2 \rho_{\mathrm{s}0}^2$ over a factor of 25 (0.02--0.5)
and found only a few percent variation in the resulting plot of heat flux versus
time, verifying that the results are not sensitive to the exact value of this
parameter (as long as it is small).

The sheath is seen to play an essential role in mediating the power loads to the divertor.
As discussed by \citet{Leonard2014}, one of the
original motivations for these calculations
was to address concerns that ELM energy deposition on time scales short compared to the ion-sound-transit time
could lead to melting of the divertor due to excessive surface temperatures
\citep{Pitts2007,Leonard2014,Loarte2007}.
Our results confirm the previous 1D3V PIC calculations of \citet{Pitts2007} that
found that the bulk of the ELM energy arrives
at the target plate only on the slower ion-sound-transit time scale
$\tau_i \sim L_\parallel/c_{\mathrm{s},\mathrm{ped}} \approx 149$~$\mu$s,
although there is a modest rise in the heat flux on ${\sim}\tau_e$.
This observation is consistent with measurements from several machines \citep{Loarte2007}.

The bulk of the ELM energy is carried by the ions, which arrive at the
target plate on the order of the ion-sound transit time $\tau_i$.
Note that although the source power is equally divided between electrons and ions,
the electric field transfers some thermal energy from electrons to the ions as the ions
flow to the divertor plates.
The reduction of source strength and temperature after $200$~$\mu$s results
in the abrupt drop seen in the heat fluxes.
The parallel heat flux (parallel to the magnetic field) on the right
target plate for each species is calculated as
\begin{align}
  Q_s &= \frac{1}{2}m_s \int_{v_{c,s}}^{\infty} \mathrm{d}v_\parallel \, f_s v_\parallel^3 +
  (T_\perp + q_s\phi_{sh}) \int_{v_{c,s}}^{\infty} \mathrm{d}v_\parallel \, f_s v_\parallel,
\end{align}
where $v_{c,s} = \sqrt{\mathrm{max}(-2 q_s \phi_{sh}/m_s, 0)}$ accounts
for the reflection of electrons by the sheath.
The $q_s \phi_{sh}$ term in the second integral models the acceleration
of ions and deceleration of electrons as they pass through the sheath
to the divertor plate, a region that is not resolved in our models.
We have assumed that each species has a constant perpendicular
temperature $T_\perp = T_{\mathrm{ped}}$ for comparison with the 1D Vlasov
results in \citet{Havlickova2012}.
Note that the incidence angle of the magnetic field ($\theta \approx 6^\circ$) is not factored into
this measure of heat flux on the target plate.
The heat flux normal to the target plate is $Q_{s,n} = Q_s\sin(\theta)$,
where $\theta$ is the (usually very small) angle between the magnetic field
and the target-plate surface.

Figure~\ref{fig:heat-flux-es} indicates good quantitative agreement with the collisionless 
1D1V Vlasov results in \citet[figure~2]{Havlickova2012}, supporting the accuracy of the logical-sheath boundary
conditions and the gyrokinetic-based model used here.
Specifically, our model quantitatively reproduces the peak total heat flux of ${\approx}4$~GW~m$^{-2}$,
the individual contributions to the total heat flux by electrons and ions within ${\sim}5\%$,
and the features on the
electron and ion transit scales found in the 1D1V Vlasov simulation.
As with the Vlasov simulation of \citet{Havlickova2012}, we obtain a lower peak total heat flux
and lower fraction of energy carried by ions in our model compared to the
PIC simulation of \citet{Havlickova2012}, which obtained
peak heat fluxes of ${\approx}1.1$~GW~m$^{-2}$ for electrons and
${\approx}3.9$~GW~m$^{-2}$ for ions.

Small differences between our 1D1V results and the Vlasov results could be attributed
to the use of different initial conditions, while discrepancies with the PIC simulation
can be attributed to the lack of collisions, as discussed in \citet{Havlickova2012}.
An extension to our model with self-species collisions is discussed in Section~\ref{sec:1d-sol-collisions},
where we confirm that the inclusion of collisions eliminates the major discrepancies in the 
heat fluxes.

In a previous study with a 1D1V PIC code \citep{Pitts2007},
the fraction of total energy deposited on the divertor plate by each species was also of interest.
We find that ${\approx}35\%$ of the energy load
on the divertor target is deposited by the electrons in the time interval $0 < t < 350$~$\mu$s
and that ${\approx}51\%$ of the energy in the same time time interval arrives before the peak
in the heat flux at $t \approx 200$~$\mu$s.
Both of these fractions appear to be
larger than what was reported in the PIC simulation, which found that
${\approx}23\%$ of the total energy load is deposited by the electrons and that
${\approx}27\%$ of the total ELM energy arrives before the peak in the heat flux.

\begin{figure}
\centering
\includegraphics[width=\textwidth]{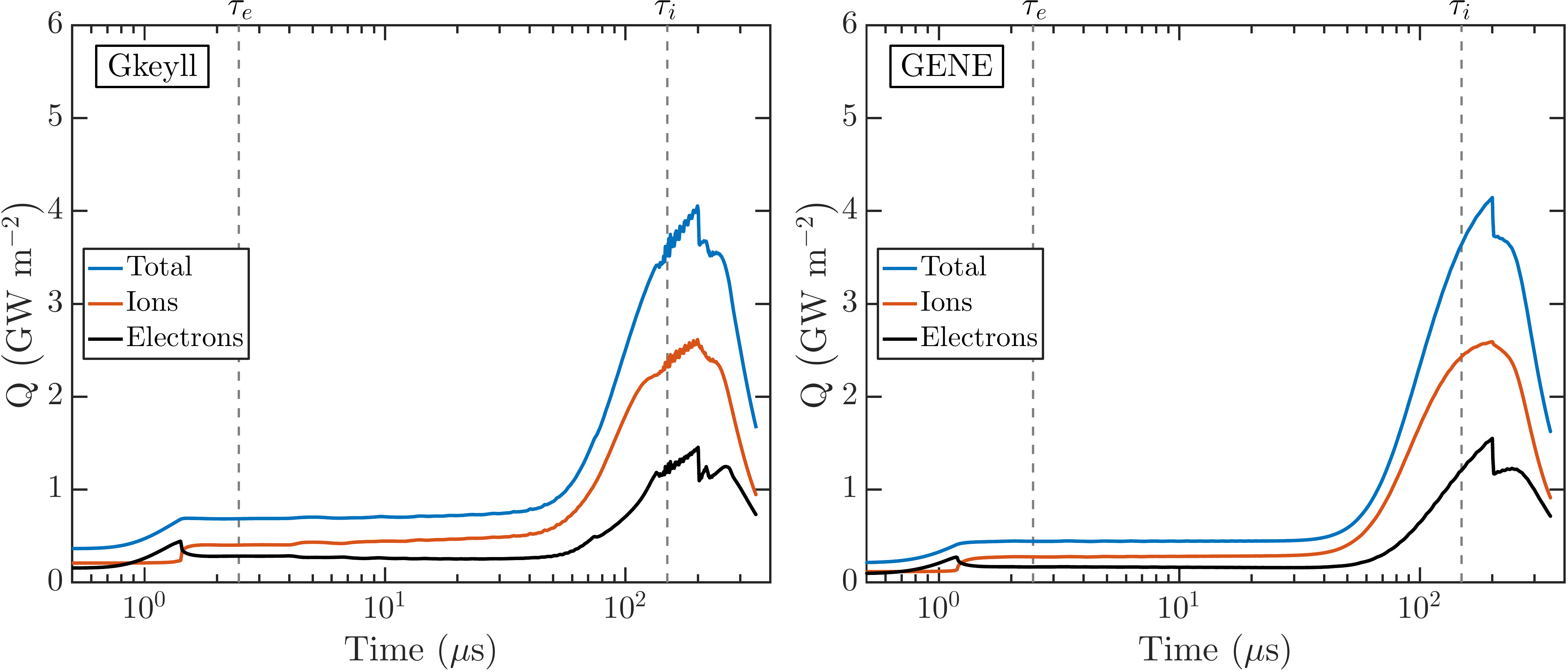}
\caption[Comparison of the parallel heat flux at the divertor
 plate versus time obtained from a Gkeyll simulation and a GENE simulation.]
  {Comparison of the parallel heat flux at the divertor
  plate versus time with drift-kinetic electrons obtained from
  a Gkeyll simulation and a GENE simulation. The electron and ion
  transit times $\tau_e$ and $\tau_i$ are indicated by the
  vertical dashed lines. Data from the GENE simulation was provided
  by and used with permission from Q.\,Pan.}
  \label{fig:sol-1d-gkeyll-gene-comp}
\end{figure}

Our gyrokinetic 1D1V test problem \citep[originally reported in][]{Shi2015}
was later implemented in a finite-volume version of the GENE code \citep{Pan2016}.
A comparison of the Gkeyll and GENE results is shown in figure~\ref{fig:sol-1d-gkeyll-gene-comp}.
Even with different numerical implementations, excellent quantitative agreement
is obtained between the two codes, although the small oscillations in the heat fluxes
on the ion-transit-time scale are absent in the GENE simulation.

\subsection{Divertor Heat Flux with an Adiabatic-Electron Model}
We have also investigated a model that includes the effect of kinetic ions
but assumes an adiabatic response for the electrons.
Specifically, the electron density takes the form
\begin{align}
  n_e(z) &= n_{e}(z_R) \exp \left(\frac{e(\phi-\phi_{sh})}{T_e}\right),
\end{align}
where $n_{e}(z_R)$ is the electron density evaluated at the domain edge.
This expression can be inverted to give another algebraic equation to determine
the potential, similar to the electrostatic gyrokinetic model with a fixed $k_\perp \rho_{\mathrm{s}0}$.
Since the time step is set by the ions, these simulations have an execution time a factor of
${\sim}\sqrt{m_i/m_e}$ faster than the gyrokinetic simulation.
This property makes the adiabatic-electron model useful as a test
case for code development and debugging.

The sheath potential $\phi_{sh}$ can be determined by assuming that $f_e$ at the
target plate is a Maxwellian with temperature $T_e$.
By using logical-sheath boundary conditions and quasineutrality, one finds
\begin{align}
  \phi_{sh} &= -\frac{T_e}{e} \log \left(\frac{\sqrt{2\pi}\Gamma_i}{n_i
 v_{\rm te}}\right),
\end{align}
where $\Gamma_i$ is the outward ion flux, and all quantities are evaluated at the domain edge.
For simplicity, we selected $T_e$ in our simulations to be the
field-line-averaged value of the ion temperature $T_i(z)$, but more accurate
models for $T_e$ could be used. 

\begin{figure}
\centering
\includegraphics[width=0.75\textwidth]{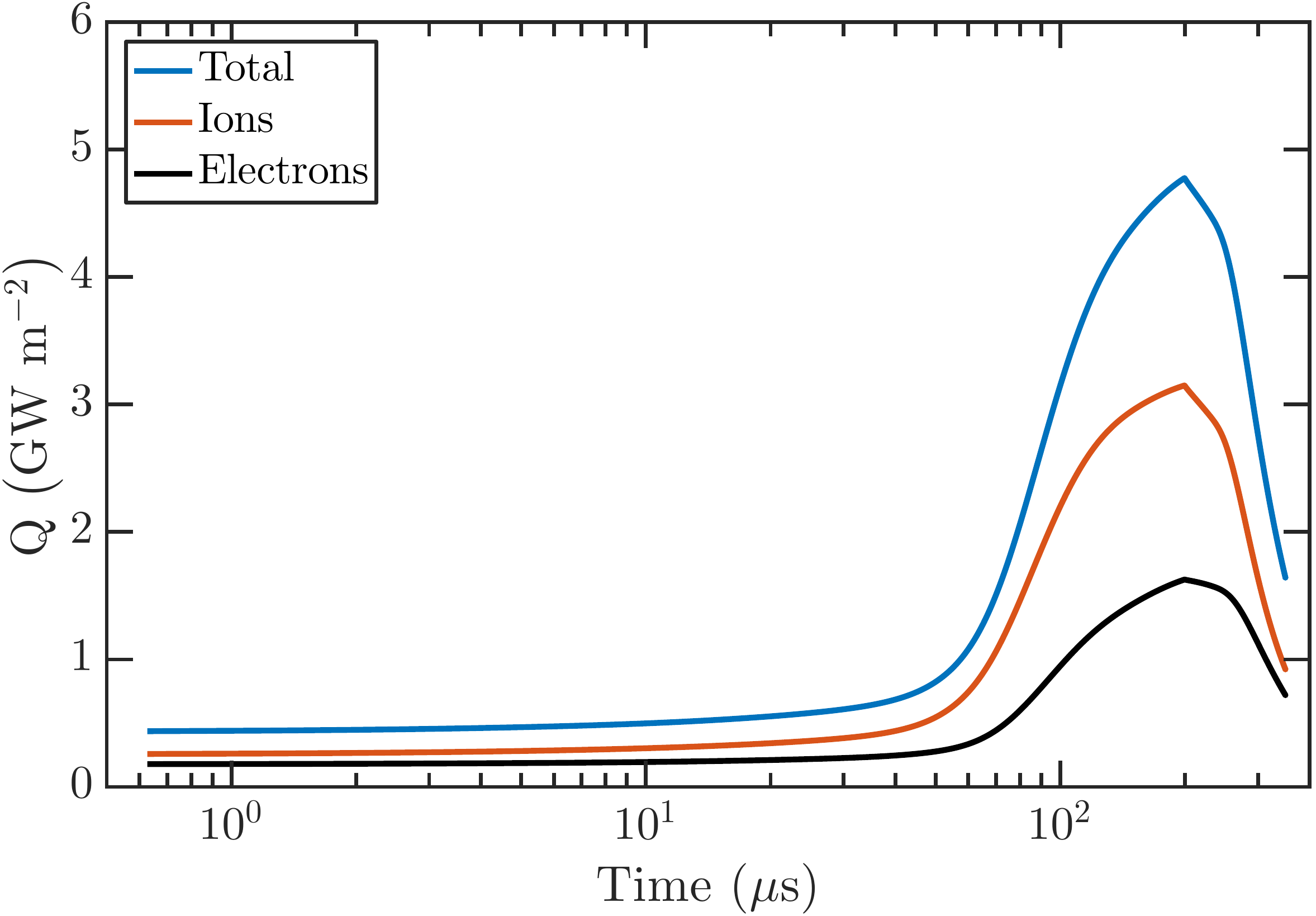}
\caption{\label{fig:heat-flux-be}Parallel heat flux at the divertor
 plate versus time from the adiabatic-electron model in the ELM heat-pulse problem.} 
\end{figure}

Figure~\ref{fig:heat-flux-be} shows the parallel heat flux on the target
plate versus time using adiabatic electrons. As expected, kinetic-electron effects
seen in figure~\ref{fig:heat-flux-es} are not resolved by this model.
When compared to a simulation using kinetic electrons, the main heat flux
at $t \approx 100$--$200$~$\mu$s is overestimated but still predicted fairly well.
The peak total heat flux is ${\approx}4.77$~GW~m$^{-2}$ instead of at ${\approx} 4.06$~GW~m$^{-2}$
from the simulation with kinetic electrons.

In this model, the electron parallel heat flux on the target plate is calculated as
\begin{align}
  Q_e &= \frac{1}{2}m_e \int_{v_c}^{\infty} \mathrm{d}v_\parallel \, f_e v_\parallel^3
  + (T_\perp - e\phi_{sh}) \int_{v_c}^{\infty} \mathrm{d}v_\parallel \, f_e v_\parallel \nonumber\\
&= (T_e + T_\perp) \int_{0}^{\infty} \mathrm{d}v_\parallel \, f_i v_\parallel.
\end{align}

\section{1D2V Model with Collisions \label{sec:1d-sol-collisions}}
\begin{figure}
\centering
\includegraphics[width=\textwidth]{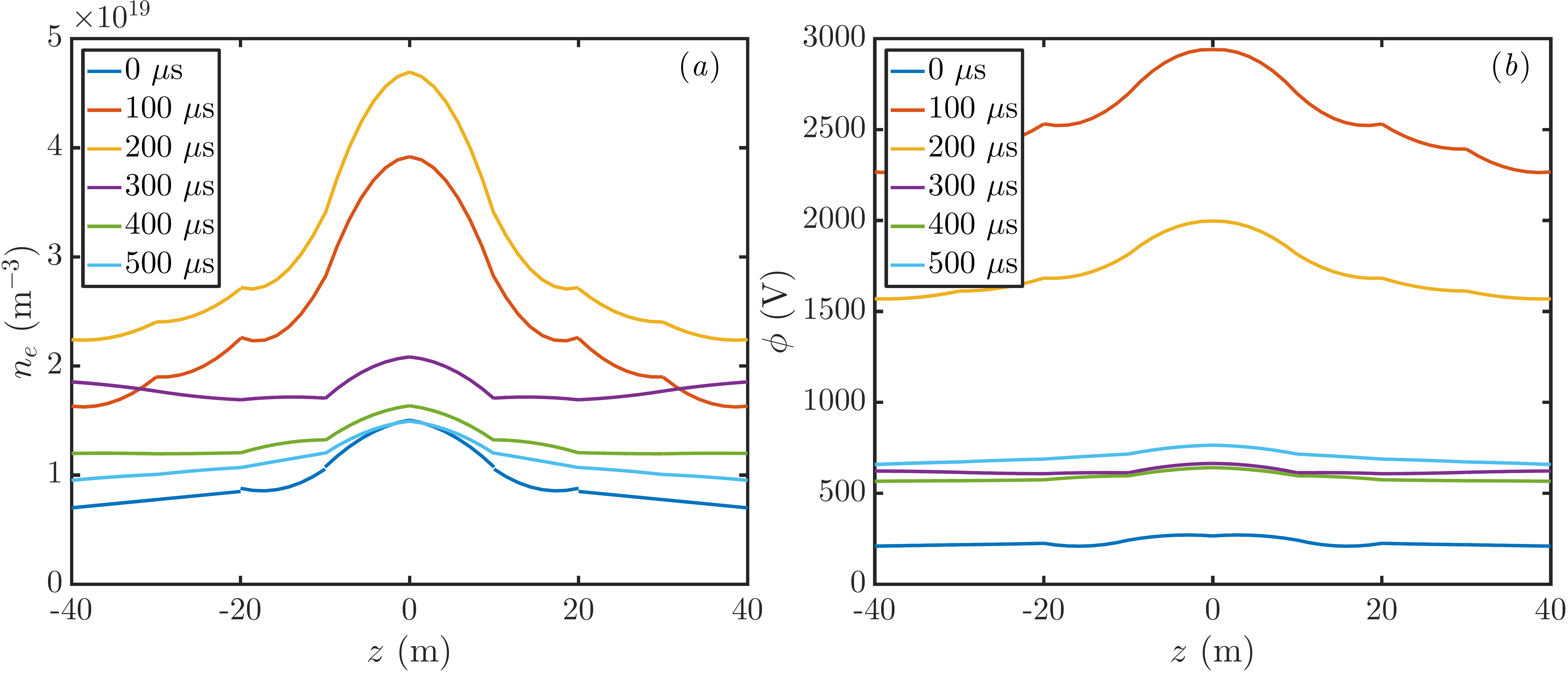}
  \caption[Snapshots of the electron-density and electrostatic-potential profiles for a 1D2V
  ELM heat-pulse simulation with collisions.]{\label{fig:1d2v_n_and_phi} Snapshots of the ($a$) electron-density
  and ($b$) electrostatic-potential profiles in $100$~$\mu$s intervals between $t=0$~$\mu$s
  and $t = 500$~$\mu$s for a 1D2V ELM heat-pulse simulation with same-species
  Lenard--Bernstein collisions.} 
\end{figure}

\begin{table}
\begin{center}
\caption[Parameters for the phase-space grid used in 1D2V ELM heat-pulse simulations.]{
Parameters for the phase-space grid used in 1D2V ELM heat-pulse simulations.
Piecewise-quadratic basis functions are used, resulting in 20 degrees of freedom per cell.}
\bigskip
\begin{tabular}{cccc}
\toprule
\textbf{Coordinate} & \textbf{Number of Cells} & \textbf{Minimum} & \textbf{Maximum} \\
\midrule
$z$ & 8 & $-L_\parallel$ & $ L_\parallel$ \\
  $v_\parallel$ & 16 & $-4 \sqrt{T_{\mathrm{ped}}/m_s} $ & $4 \sqrt{T_{\mathrm{ped}}/m_s} $ \\
  $\mu$ & 8 & 0 & $4 T_{\mathrm{ped}}/B$ \\
\bottomrule
\end{tabular}
\label{tab:1d2v_sol_grid}
\end{center}
\end{table}
So far, we have focused on 1D1V ($z$,$v_\parallel$) models in the SOL.
We added a third coordinate $\mu$, the magnetic moment, and implemented
Lenard--Bernstein same-species collisions (\ref{eq:lbCollisionOp}) for electrons and ions to
the ELM heat-pulse problem for better comparison with the 1D3V PIC results of
\citet{Havlickova2012} and to gain experience with simulations using two
velocity coordinates.
Unlike in the LAPD simulations of Chapter \ref{ch:lapd}, a model for electron--ion
collisions was not implemented for these 1D2V simulations.
The grid parameters for the 1D2V ELM heat-pulse simulations are shown in table \ref{tab:1d2v_sol_grid}.
Snapshots of the electron density and electrostatic potential profiles for a 1D2V ELM heat-pulse simulation
with collisions are shown in figure~\ref{fig:1d2v_n_and_phi}.

\begin{figure}
\centering
\includegraphics[width=\textwidth]{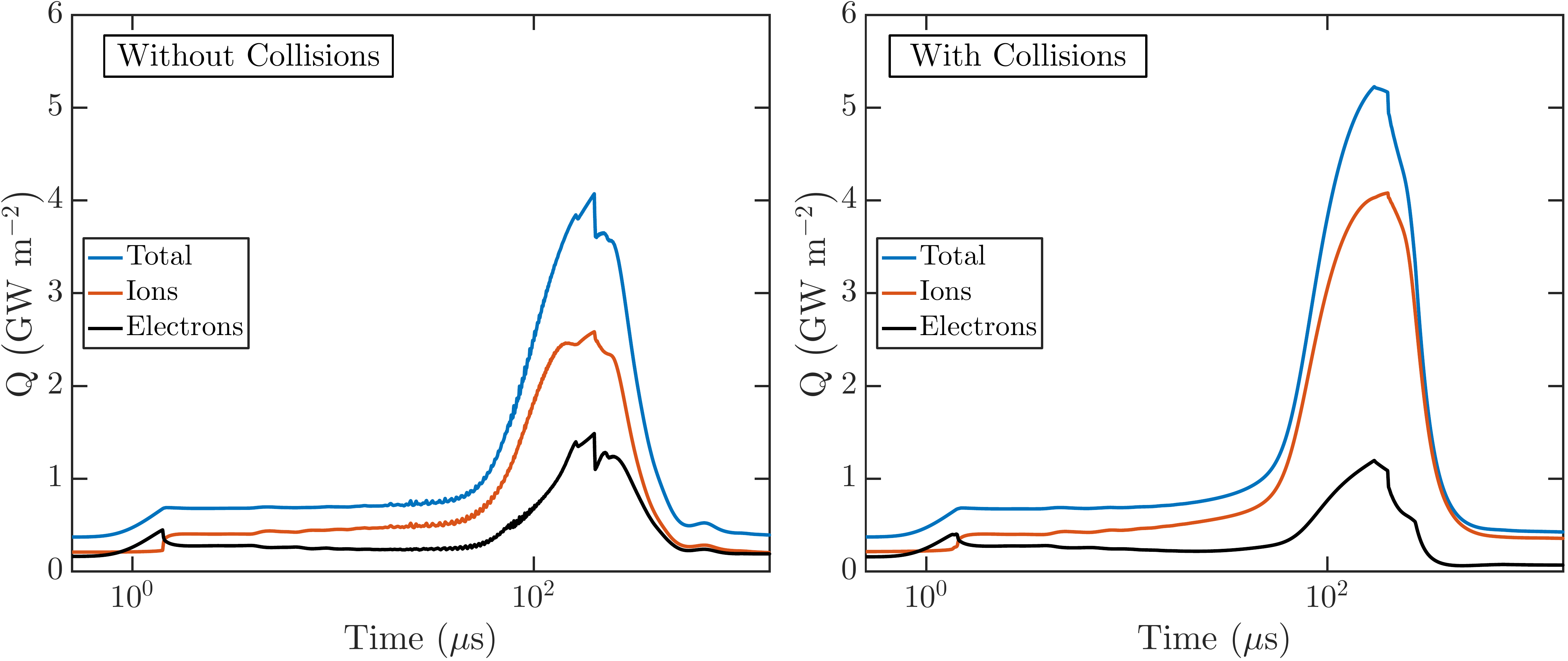}
  \caption[Comparison of the total, electron, and ion parallel heat fluxes at the divertor
  plate versus time for 1D2V cases with and without collisions.]
  {Comparison of the total, electron, and ion parallel heat fluxes
  at the divertor plate versus time for 1D2V cases with and without same-species Lenard--Bernstein collisions.}
  \label{fig:heat-flux-collision}
\end{figure}

Since $T_\perp$ is no longer a parameter of the model and can evolve through the collision operator,
we modify the calculation of the parallel heat flux to
\begin{align}
  Q_s &= \int_{v_{c,s}}^{\infty} \mathrm{d}v_\parallel \, f_s v_\parallel \left(\frac{1}{2}m_s v_\parallel^2 + \mu B \right)
  + q_s\phi_{sh} \int_{v_{c,s}}^{\infty} \mathrm{d}v_\parallel \, f_s v_\parallel.
\end{align}
The parallel temperature of the source plasma is still as described in the previous section,
and the perpendicular temperature of the source is fixed at $T_\mathrm{ped}$ even after $t=200$~$\mu$s.
Figure~(\ref{fig:heat-flux-collision}) shows a comparison of the parallel heat flux at the divertor plate
in a 1D2V case without collisions and a 1D2V case with same-species Lenard--Bernstein collisions.
Notable differences between the two cases include an ${\sim}20\%$ larger peak
total heat flux ($Q \approx 5.23$~GW~m$^{-2}$) and a slightly reduced electron heat flux.

Overall, the heat flux profiles for the case with collisions is in better agreement
with the PIC results of \citet{Havlickova2012} and \citet{Pitts2007}.
In the 1D2V case with collisions, ${\approx}44\%$ of the total energy deposited 
between $0 < t < 350$~$\mu$s is deposited before the peak in the total heat flux (compared to ${\approx}51\%$
in the 1D2V collisionless simulation).
We also find that ${\approx}19\%$ of the total energy deposited in the same time interval
is from electrons (compared to ${\approx}34\%$ in the 1D2V collisionless simulations),
which is much closer to the ${\approx}22\%$ reported by \citet{Pitts2007}.
A summary of various quantities measured in the 1D1V and 1D2V ELM-heat-pulse simulations
is given in table~\ref{tab:1d-sol-summary}.

In comparing the 1D2V simulations with and without collisions, we noticed
that the time-integrated total heat flux over long times was not the same for the two cases,
with the simulation with collisions having an ${\approx}9.9\%$ larger energy deposited
on the divertor plates.
This discrepancy is explored in more detail in Appendix~\ref{ch:1d-sol-comp},
which reveals that the difference in energies deposited on the divertor plate
is consistent with the collisionless simulations having more system energy at any given instant.

\begin{figure}
\centering
\includegraphics[width=\textwidth]{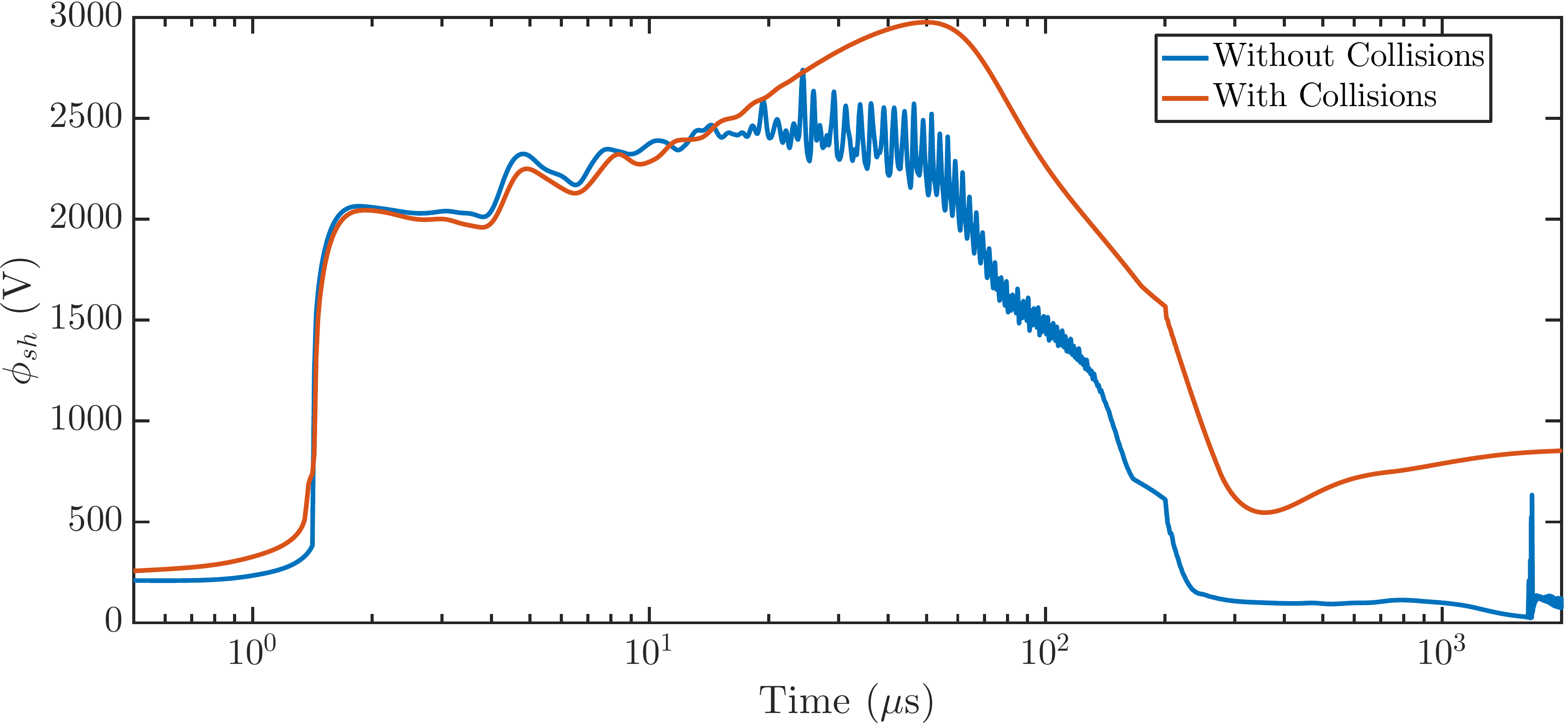}
  \caption[Comparison of the sheath-potential time history 
  between 1D2V ELM heat-pulse simulations with and without same-species Lenard--Bernstein collisions.]
  {Comparison of the sheath-potential time history 
  between 1D2V ELM heat-pulse simulations with and without same-species Lenard--Bernstein collisions.
  High-frequency oscillations in the sheath potential are not seen in the simulation with collisions.}
  \label{fig:sheath_potential} 
\end{figure}

Some differences in the distribution of heat flux between electron and ions in the
two cases can be understood by examining the sheath potential.
As seen in figure~\ref{fig:sheath_potential}, the sheath potential in the case with collisions
does not exhibit high-frequency oscillations and is generally several hundred volts larger
than the sheath potential in the case without collisions.
These trends are believed to be connected to how collisions are able to replenish high-$v_\parallel$ electrons
throughout the simulation through pitch-angle scattering.
From figure~\ref{fig:sheath_potential}, it appears that the inclusion of collisions results in a stabilizing
effect on the sheath-potential oscillations.

\begin{table}
\begin{center}
  \caption[Summary of results from various simulations of the 1D ELM-heat-pulse problem.]
  {Summary of results from various simulations of the 1D ELM-heat-pulse problem.
  The simulations included in this table are 1D1V with drift-kinetic electrons (`1D1V'), 1D1V with
  adiabatic electrons (`1D1V (A)'), 1D2V without collisions (`1D2V (NC)'), 1D2V with collisions (`1D2V (NC)'),
  and 1D3V PIC with collisions (`PIC').
  The values in the PIC column are from simulations of a 0.4~MJ $200$~$\mu$s-duration ELM
  reported in \citet{Pitts2007} and \citet{Havlickova2012}. 
  `Electron fraction of total energy' refers to the fraction of the total energy deposited on 
  the divertor plate between $0 < t < 350$~$\mu$s, and `Energy fraction before peak $Q$'
  refers to the fraction of the energy that is deposited on the divertor plate
  before the peak in the total heat flux at the divertor plate ($t \approx 200$~$\mu$s).}
  \bigskip
    \small
  \begin{tabular}{lccccc}
  \toprule
    \textbf{Quantity} & \textbf{1D1V} & \textbf{1D1V (A)} & \textbf{1D2V (NC)} & \textbf{1D2V} & \textbf{PIC}\\
  \midrule
    Peak $Q$ $\left(\mathrm{GW}~\mathrm{m}^{-2}\right)$ & 4.05 & 4.78 & 4.07 & 5.23 & 5.21\\
    Electron fraction of peak $Q$ & 0.35 & 0.34 & 0.37 & 0.23 & 0.24\\
    Electron fraction of total energy & 0.35 & 0.35 & 0.34 & 0.19 & 0.23\\
    Energy fraction before peak $Q$ & 0.51 & 0.52 & 0.52 & 0.44 & 0.27 \\
  \bottomrule
  \end{tabular}
  \label{tab:1d-sol-summary}
\end{center}
\end{table}

\subsection{A Basic Recycling Model}
When ions impact the divertor plate, some of the incident ions can reenter
plasma as cold neutrals after some interactions with the solid surface (e.g., backscattering or
desorption) and re-ionized, which is a process called \textit{recycling} \citep{Li2012}.
As long most recycled neutrals are re-ionized within the SOL and not in the core,
favorable SOL regimes with large parallel temperature gradients and low temperatures
near the targets can be achieved \citep{Stangeby2000}.
While recycling near the divertor plates (far from the last closed flux surface)
is desirable to help reduce heat loads on the divertor plates,
recycling from the main chamber wall can harm fusion performance.
The shorter distance to the last closed flux surface
from the main chamber wall when compared to the divertor makes it easier for the
recycled neutrals to penetrate into the core plasma before being ionized,
which leads to radiative cooling and contamination of the core plasma.
Experiments on Alcator C-Mod have observed significant amounts of recycling from the 
main chamber wall \citep{Umansky1998,Terry2007}.
Here, we discuss an extension to the 1D2V SOL simulations to model particle recycling only
near divertor plates.

The recycling model provides a time-dependent source of particles near the divertor plate
whose amplitude depends on the particle fluxes to the wall $\Gamma_i = \Gamma_e$
and the recycling coefficient $R$, which is between 0 and 1 and controls
the level of recycling.
The $R$ parameter in this model is an input parameter, so it is not modeled using
an empirical scaling law, atomic-physics models, or Monte Carlo calculations.
For $R=0$, we have a situation similar to the previous ELM heat-pulse simulations in which
there is no particle recycling, although we modify the midplane source
for use in a steady-state calculation so that it has no time dependence.
For $R=1$, all particles that flow to the divertor plate come back into the system, and so
there is no steady-state solution in this case.
In these simulations, we use a midplane source that is constant in time ($g(t)=1$) at with a reduced
amplitude $S_0 = 0.075 A n_{\mathrm{pred}} c_{\mathrm{s},\mathrm{ped}}/L_s$ when
compared to the midplane source used in the ELM heat-pulse simulations.
The midplane source is treated as an isotropic Maxwellian with temperature $T_S = 200$~eV.
The parameters for these 1D2V SOL recycling simulations are summarized in table \ref{tab:1d2v_sol_recycling_grid}.
\begin{table}
\begin{center}
\caption[Parameters for the phase-space grid used in 1D2V SOL recycling simulations.]{
Parameters for the phase-space grid used in 1D2V SOL recycling simulations.
Piecewise-quadratic basis functions are used, resulting in 20 degrees of freedom per cell.}
\bigskip
\begin{tabular}{cccc}
\toprule
\textbf{Coordinate} & \textbf{Number of Cells} & \textbf{Minimum} & \textbf{Maximum} \\
\midrule
$z$ & 12 & $-L_\parallel$ & $ L_\parallel$ \\
  $v_\parallel$ & 16 & $-4 \sqrt{T_S/m_s} $ & $4 \sqrt{T_S/m_s} $ \\
  $\mu$ & 8 & 0 & $4 T_S/B$ \\
\bottomrule
\end{tabular}
\label{tab:1d2v_sol_recycling_grid}
\end{center}
\end{table}

\begin{figure}
\centering
\includegraphics[width=\textwidth]{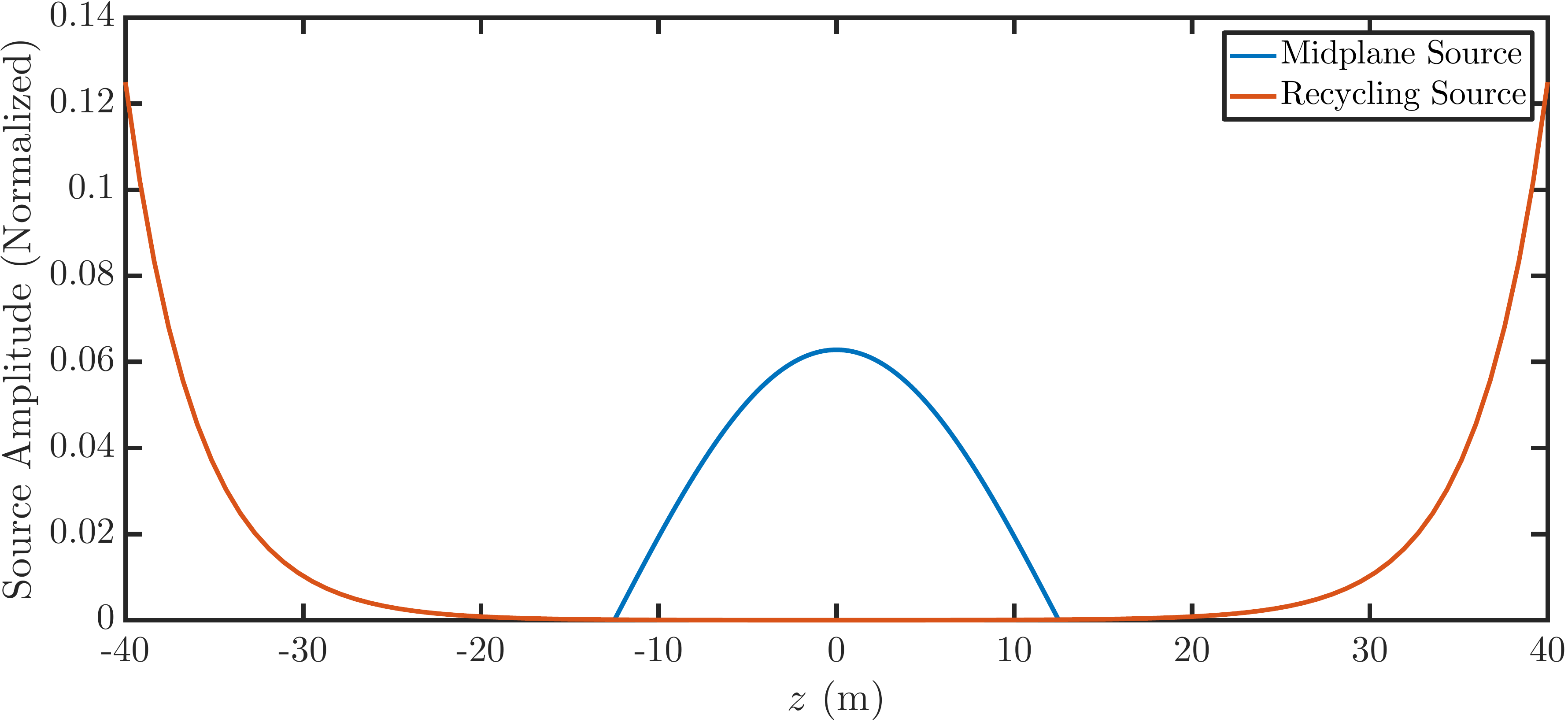}
  \caption[Source profiles for the recycling simulations.]
  {\label{fig:recycling_source} Source profiles for the recycling simulations.
  The recycling and midplane plasma sources are both normalized to a domain-integrated
  area of 1. For recycling coefficients $0 \le R < 1$, a steady-state solution exists in which the
  sources are in balance with the plasma outflow to the divertor plates.} 
\end{figure}

In diverted plasmas, the recycled neutrals are believed to be ionized near the divertor plates
\citep{Stangeby2000}.
The recycling source for electrons and ions has the following form:
\begin{equation}
  S_{R}(z,\boldsymbol{v},t) =  R\Gamma_\mathrm{tot}(t) \frac{1}{2 L_R } \frac{ e^{-(L_\parallel-|z|) / L_R } }{ 1-e^{-L_\parallel/L_R} } F_M(\boldsymbol{v}, T_R),
\end{equation}
where $\Gamma_{\mathrm{tot}}$ is the total outward flux (lower and upper surfaces in $z$),
which is the same for both electrons and ions since logical-sheath boundary conditions are used.
The function $S_{R}$ is normalized such that it contributes a source of electrons and ions
of amplitude $R \Gamma_\mathrm{tot}(t)$ into the simulation domain.
We choose the recycling-source scale length $L_R = 4$ m and the temperature of the
recycled particles $T_R = 30$~eV.
Here, $F_M(\boldsymbol{v}, T_R)$ represents an isotropic, normalized Maxwellian with temperature $T_R$.
Figure~\ref{fig:recycling_source} shows the profiles of the recycling and midplane plasma sources in the system.

\begin{figure}
\centering
\includegraphics[width=\textwidth]{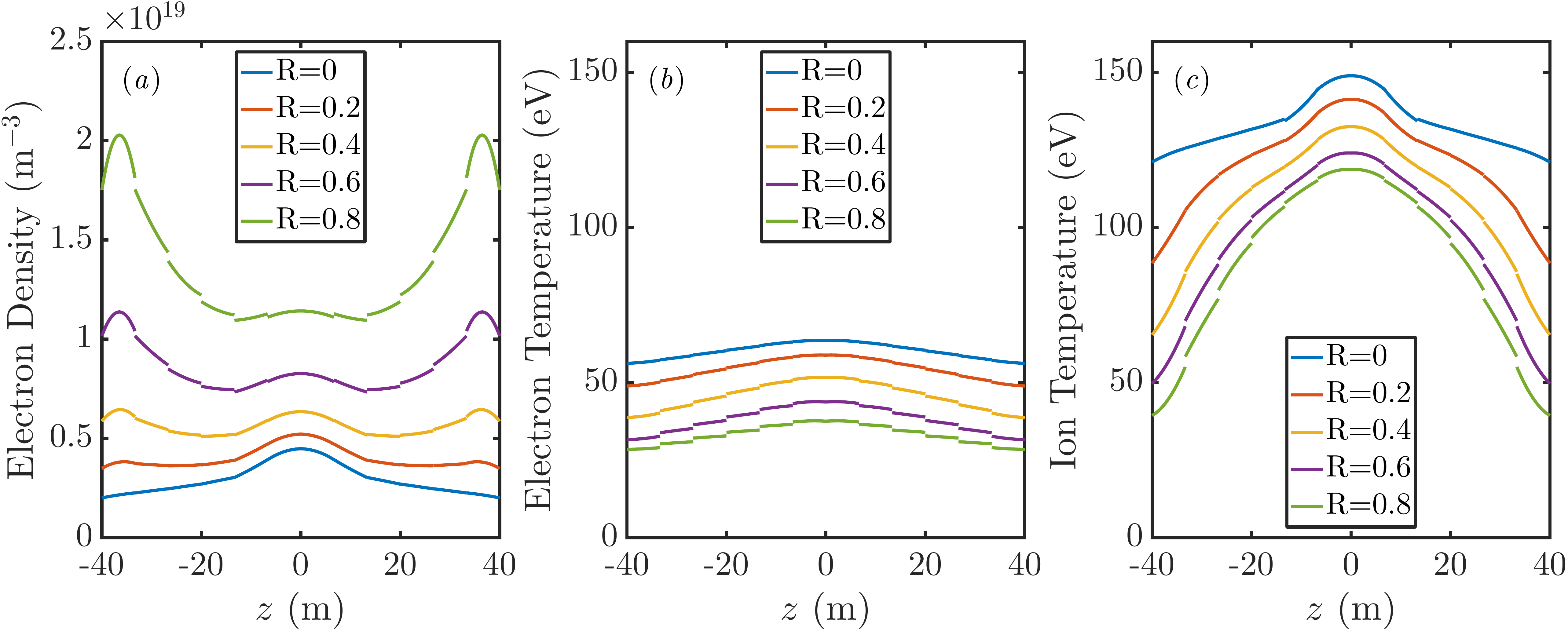}
  \caption[Steady-state profiles of the electron density,
  electron temperature, and ion temperature for different values of the recycling
  coefficient from 1D2V SOL simulations.]
  {\label{fig:recycling_steady_state} Steady-state profiles 
  of the ($a$) electron density, ($b$) electron temperature, and ($c$)
  ion temperature for different values of the recycling
  coefficient $R$ from 1D2V SOL simulations.} 
\end{figure}

Figure~\ref{fig:recycling_steady_state} shows the steady-state profiles calculated from running a set of
1D2V simulations with different values of the recycling coefficient.
The slowest simulations are run to $t = 8$ ms to reach a steady state, while
the fastest simulation only took 2.8 ms to reach a steady state.
We see that as the recycling coefficient is increased, the electron density becomes increasingly peaked
near the divertor plates (and eventually the density in front of the divertor plates becomes twice
the density of the midplane), and the electron and ion temperatures in the SOL are reduced.
Due to the rapid adiabatic-electron response, the ion temperature profile becomes much more peaked compared to
the electron temperature profile, which remains relatively flat even for large values of $R$.

\section{Conclusions\label{sec:1d-sol-conclusions}}
We have used a gyrokinetic-based model to simulate the propagation
of an ELM heat pulse along a scrape-off layer to a divertor target plate.
We have described a modification to the ion polarization term to slow
down the electrostatic shear Alfv\'{e}n wave.

Our main results include the demonstration that this gyrokinetic-based
model with logical-sheath boundary conditions is able to agree well with
Vlasov and full-orbit (non-gyrokinetic) PIC simulations, without needing
to resolve the Debye length or plasma frequency. This simplification allows the
spatial resolution to be several orders of magnitude coarser than the
electron Debye length (and the time step several orders of magnitude
larger than the plasma period) and thus leads to a much faster
calculation. Our results also confirm previous work that the
electrostatic potential in this problem varies to confine most of the
electrons on the same time scale as the ions, so the main ELM heat
deposition occurs on the slower ion-transit time scale.

Additionally, we have described a model using adiabatic electrons
that is useful for code development and debugging.
This model does not include kinetic-electron effects but runs
much faster than simulations with kinetic electrons.
Later, the original 1D1V model with drift-kinetic electrons was extended to 1D2V, and
Lenard--Bernstein collision operators for self-species collisions were added.
The 1D2V simulation with collisions obtained better quantitative agreement with the
original 1D3V PIC simulation of \citet{Pitts2007}, which also had collisions.
Some initial work was then performed on including recycling effects in the model
though source terms localized near the divertor plates with a strength
proportional to a user-specified recycling coefficient and the particle fluxes
to the divertor plates.

Since we have assumed only a single $k_\perp$ mode in these simulations to
limit the high frequency of the electrostatic shear Alfv\'{e}n wave,
future work can allow for a spectrum of $k_\perp$ modes.
For 1D electromagnetic simulations, this modification requires inverting
the $\nabla_\perp^2$ operators that appear in the gyrokinetic Poisson
equation and Amp\'ere's law.
The inclusion of magnetic fluctuations will be important when a
spectrum of very low $k_\perp$ modes is kept in order to limit the
frequency of the shear Alfv\'{e}n wave at low $k_\perp$.
These models could eventually include more
detailed effects such as secondary electron emission,
charge exchange, and radiation, and could be used to study different types
of divertor configurations.

%% file: ch-simulations-of-lapd/chapter-simulations-of-lapd.tex
\chapter{Simulations of the Large Plasma Device \label{ch:lapd}}
\begin{figure}
  \centerline{\includegraphics[width=\textwidth]{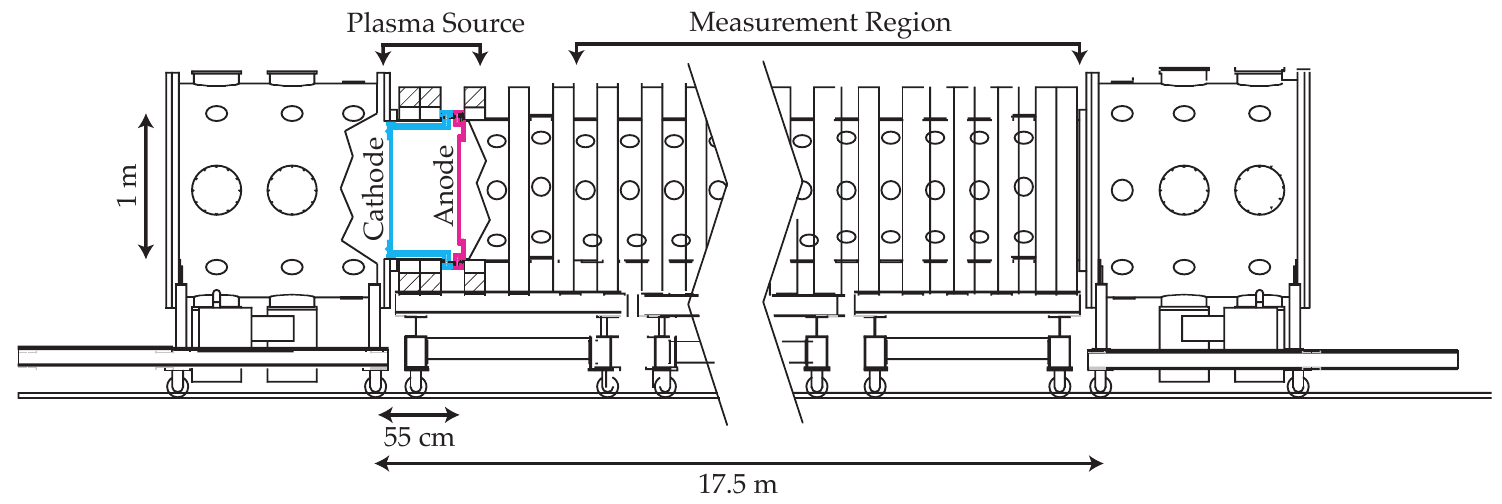}}
  \caption[Diagram showing a side view of the LAPD device.]
  {Diagram showing a side view of the LAPD device. The barium oxide (BaO) cathode
  at one end of the device is indicated in blue, which creates a plasma column that is
  18~m in length and 60~cm in radius \citep{Gekelman2016}.
  This figure is used with permission from T.\,Carter.}
\label{fig:lapd-diagram}
\end{figure}

In this chapter, we present results from gyrokinetic continuum
simulations of electrostatic plasma turbulence in the Large Plasma Device (LAPD) at UCLA \citep{Gekelman1991,Gekelman2016}
using kinetic electrons with a reduced mass ratio and a single kinetic ion species.
The LAPD is a linear device that creates a plasma column in a straight, open-field-line configuration.
Figure~\ref{fig:lapd-diagram} shows a diagram of the LAPD device.
Despite its relatively low plasma temperature, the LAPD contains some of the basic elements
of a SOL in a simplified (no X-point geometry, straight magnetic field lines, etc.), well-diagnosed
setting, making this device a useful benchmark of gyrokinetic algorithms for boundary-plasma simulation.
The LAPD plasma's relatively high collisionality also facilitates comparisons with Braginskii fluid codes,
and good agreement between the two approaches is expected.
The results presented here for the unbiased LAPD plasma were first published in \citet{Shi2017}.

We found that a major challenge in these simulations was the
severe constraint on the explicit time-step size due to the high collision frequencies.
To somewhat alleviate this time-step restriction, we used a reduced the electron--electron
and electron--ion collision frequencies by a factor of 10, although the simulations were still costly.
Additionally, we believe that the high collisionality of LAPD accentuated positivity issues
caused by the numerical implementation of the collision operators.
Nevertheless, the work presented in this chapter is an important demonstration
of the feasibility of our discontinuous Galerkin (DG) approach and helps build
confidence in the results before we discuss the more hypothetical helical SOL model in
Chapter~\ref{ch:helical-sol}.

Our work is a gyrokinetic extension of prior fluid simulations of LAPD \citep{Rogers2010,Popovich2010a},
and in particular we follow much of the same simulation set-up as in \citet{Rogers2010}.
Originally, we had hoped to make direct quantitative comparisons with the fluid simulations of \citet{Rogers2010},
but we eventually realized that the omission of a term in their momentum equation
resulted in the questionable modeling of the neutrals as having a wind velocity 
that is the same as the ion flow velocity.
These simulations are the first 5D gyrokinetic continuum simulations on open field lines
including interactions with sheath losses and are also the first 5D
gyrokinetic simulations including a sheath model of a basic laboratory plasma experiment.
LAPD experiments relevant to the simulations in this chapter are reported in
\citet{Carter2009,Carter2006,Schaffner2013,SchaffnerThesis2013}.
Relevant experimental data from LAPD are also plotted in some figures of papers presenting
fluid simulations of LAPD \citep{Popovich2010b,Fisher2015,Fisher2017,Friedman2012}.

In considering these results, the reader should keep in mind that the main purpose
of these simulations is a proof-of-principle demonstration of gyrokinetic continuum simulations
with sheath-model boundary conditions.
The simulation parameters of \citet{Rogers2010} do not appear to correspond to any
particular set of experiments performed on LAPD and result in a core $T_e \approx 3$~eV, while
most LAPD experiments report $T_e \approx 6$~eV.
A faithful set of axial boundary conditions for quantitative modeling of LAPD also requires
more investigation, as a portion of the innermost field lines terminate on the hot cathode 
at one end of the device instead of on the anode mesh (near ground) \citep{Leneman2006},
and the plasma may be detached at the other end \citep{Friedman2013}.
Neutral effects such as ion--neutral collisions can also be important to capture some 
qualitative behaviors \citep{Krasheninnikov2003,Maggs2007,Carter2009}.
Careful quantitative comparisons with LAPD experiments is deferred for future work.
Nevertheless, to satisfy the reader's curiosity, we comment on how our simulations compare
to the experiments in a qualitative sense.
With additional levels of sophistication, the gyrokinetic capability we have developed may eventually find use in
the quantitative design and prediction of LAPD experiments.

\begin{figure}
  \centerline{\includegraphics[width=\textwidth]{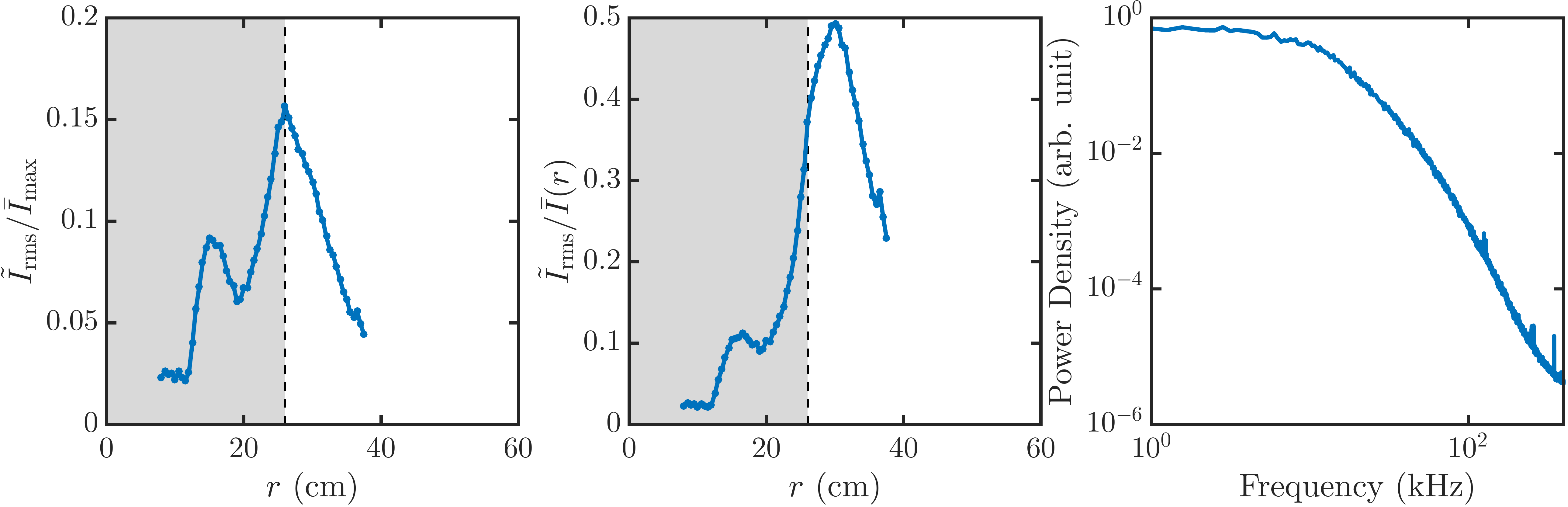}}
  \caption[Normalized r.m.s. $I_\mathrm{sat}$ fluctuation level as a function of radius and
  $I_\mathrm{sat}$ fluctuation power spectral density as measured in an experiment on LAPD.]
  {Ion-saturation-current fluctuation statistics as measured in an experiment on LAPD
  (magnetic field strength of 0.05~T).
  ($a$) The r.m.s. ion-saturation-current ($I_\mathrm{sat}$) fluctuation level as a function of radius, 
  normalized globally to the peak value of the background $I_\mathrm{sat}$ profile.
  ($b$) The r.m.s. $I_\mathrm{sat}$ fluctuation level as a function of radius,
  normalized locally to the background ion saturation current $\bar{I}_\mathrm{sat}(r)$.
  ($c$) The $I_\mathrm{sat}$-fluctuation power spectral density.
  The shaded regions in $(a)$ and ($b$) illustrate the
  core region (inside the limiter edge at $r = 26$~cm).
  The radial plot range has been extended from $r = 0$~cm to $r=60$~cm to aid in
  comparisons with figure~\ref{fig:rmsDensity}, which shows similar plots made using
  simulation data. These plots were made using triple-Langmuir-probe
  data provided by T.\,Carter and D.\,Schaffner.}
\label{fig:lapd-experiment-fluctuations}
\end{figure}

Figure~\ref{fig:lapd-experiment-fluctuations} shows ion-saturation-current ($I_\mathrm{sat}$) data measured
with a triple Langmuir probe on LAPD between $r = 8$~cm and $r=37.5$~cm in 0.5~cm increments.
Figure~\ref{fig:lapd-experiment-fluctuations}($a$) shows the profile of the root mean square (r.m.s.)
$I_\mathrm{sat}$ fluctuation level, normalized globally to the peak value of the background
$I_\mathrm{sat}$ profile.
This global normalization for profiles of fluctuation levels is often done in 
papers that include LAPD data.
Another useful normalization of the fluctuation levels is employed in
figure~\ref{fig:lapd-experiment-fluctuations}($b$), which also shows the $I_\mathrm{sat}$ fluctuation level,
but normalized locally to the background $I_\mathrm{sat}$ profile.
Figure~\ref{fig:lapd-experiment-fluctuations}($c$) shows the $I_\mathrm{sat}$-fluctuation power spectral density,
which is characteristic of the broadband turbulence that is observed in unbiased LAPD plasmas.
Most of the $I_\mathrm{sat}$-fluctuation power is seen to be concentrated at low frequencies
(${\lesssim} 5$~kHz).

\section{Simulation Parameters \label{sec:lapd-results} }
We selected the parameters for our simulations of an LAPD-like helium plasma based on those used
by \citet{Rogers2010} in a previous Braginskii-fluid-based study, with some modifications for use in a kinetic model.
These parameters are summarized in table~\ref{tab:lapd_input_parameters}.
As done by \citet{Rogers2010}, we have also used a reduced mass ratio of $m_e/m_i = 1/400$, which allows for larger time steps to be taken,
but weakens the adiabatic electron response.
These parameters are for a fairly collisional case, and the assumption that the collision
frequency is small compared to the gyrofrequency is sometimes marginal.
We have reduced the electron--electron and electron--ion collision frequencies
by a factor of 10 for these simulations,
which increases the minimum stable explicit time step size while keeping
the collisional mean free path small compared to the parallel length of the simulation box.
The rectangular simulation box (an approximation to the cylindrical LAPD plasma) has perpendicular lengths
$L_\perp = L_x = L_y = 100 \rho_{\mathrm{s}0} \approx 1.25$~m and parallel length $L_z = 1440 \rho_{\mathrm{s}0} \approx 18$~m, where
$\rho_{\mathrm{s}0} = c_{\mathrm{s}0}/\Omega_i$ and $c_{\mathrm{s}0} = \sqrt{T_{e0}/m_i}$.
The grid parameters are summarized in table~\ref{tab:lapd_grid}, and 32 degrees of freedom are 
stored in each 5D cell.
With these parameters, $T_{e,\mathrm{min}} \approx 0.91$~eV, $T_{\parallel e,\mathrm{min}} = 0.32$~eV,
and $T_{\perp e,\mathrm{min}} = 1.2$~eV.
For time stepping, the Courant number is set to 0.1.
Some typical time and length scales of interest for the LAPD plasma we simulate are provided
in table~\ref{tab:lapd_plasma_parameters}.

\begin{table}
\begin{center}
\caption[Summmary of input parameters used in LAPD simulations.]
  {Summary of input parameters used in LAPD simulations.
  The normalizations (used for the simulation domain size
  and source terms) are based on the simulations of \citet{Rogers2010}
  and are used to set the box size and source terms for the simulation.
  The ion mass is expressed in terms of the proton mass $m_p$. Note that these
  normalization values do not necessarily reflect the values observed
  in the simulations.}
  \bigskip
  \begin{tabular}{ccl}
  \toprule
  \textbf{Symbol} & \textbf{Value} & \textbf{Description} \\
  \midrule
    $T_{e0}$ & 6 eV & Electron temperature normalization \\
    $T_{i0}$ & 1 eV & Ion temperature normalization \\
    $m_i$ & $3.973 m_p$ & Mass of ion species \\
    $B$ & 0.0398 T & Background axial magnetic field strength \\
    $n_0$ & $2 \times 10^{18}$ m$^{-3}$ & Density normalization \\
    $c_\mathrm{s0}$ & ${\approx}1.2 \times 10^4$ m/s & Ion sound speed normalization \\
    $\rho_{\mathrm{s}0}$ & ${\approx}1.25$ cm & Ion sound radius normalization \\
    $\Omega_i$ & ${\approx}9.6 \times 10^5$ rad/sec & Ion gyrofrequency normalization \\
    $L_\perp$ & ${\approx}1.25$ m & Width of simulation domain in $x$ and $y$ \\
    $L_z$ & ${\approx}18$ m & Length of simulation domain in $z$ \\
  \bottomrule
  \end{tabular}
  \label{tab:lapd_input_parameters}
\end{center}
\end{table}

\begin{table}
\begin{center}
\caption[Parameters for the phase-space grid used in LAPD simulations.]
  {Parameters for the phase-space grid used in the LAPD simulations. The temperatures
  appearing in the velocity-space extents are
  $T_{i,\mathrm{grid}} = 1$~eV and $T_{e,\mathrm{grid}} = 3$~eV.
  Piecewise-linear basis functions are used, resulting in 32 degrees of freedom per cell}
  \bigskip
  \begin{tabular}{cccc}
  \toprule
  \textbf{Coordinate} & \textbf{Number of Cells} & \textbf{Minimum} & \textbf{Maximum} \\
  \midrule
  $x$ & 36 & $-50 \rho_{\mathrm{s}0}$ & $50 \rho_{\mathrm{s}0}$ \\
  $y$ & 36 & $-50 \rho_{\mathrm{s}0}$ & $50 \rho_{\mathrm{s}0}$ \\
  $z$ & 10 & $-L_z/2$ & $ L_z/2$ \\
  $v_\parallel$ & 10 & $-4 \sqrt{T_{s,\mathrm{grid}}/m_s} $ & $ 4 \sqrt{T_{s,\mathrm{grid}}/m_s} $ \\
    $\mu$ & 5 & 0 & $0.75 m_s v_{\parallel,\mathrm{max}}^2/ (2 B_0)$ \\
  \bottomrule
  \end{tabular}
  \label{tab:lapd_grid}
\end{center}
\end{table}

\begin{table}
\begin{center}
\caption[Some time and length scales of interest for characterizating the LAPD plasma.]
  {Some time and length scales of interest for characterizing the LAPD plasma, assuming
  $T_i \approx 1$~eV and $T_e \approx 3$~eV.
  The electron quantities shown in this table are
  computed using the real electron mass.
  With a reduced electron mass $m_e/m_i = 1/400$, $\tau_{ee} \approx 0.2$~$\mu$s
  (but $\lambda_{ee}$ remains unchanged since it has no mass dependence).
  The electron--electron and electron--ion collision frequencies are
  further reduced by a factor of 10 in the simulations.}
  \bigskip
  \begin{tabular}{ccl}
  \toprule
    \textbf{Symbol} & \textbf{Value} & \textbf{Description} \\
  \midrule
    $\tau_{ee}$ & 0.05 $\mu$s & Typical electron--electron collision time \\
    $\lambda_{ee}$ & 3.3 cm &  Typical mean free path for electron--electron collisions \\
    $\tau_{ii}$ & 1.1 $\mu$s &  Typical ion--ion collision time \\
    $\lambda_{ii}$ & 0.5 cm & Typical mean free path for ion--ion collisions \\
    $\rho_i$ & 0.5 cm & Typical ion gyroradius \\
    $\rho_e$ & 0.01 cm & Typical electron gyroradius \\
    $\rho_\mathrm{s}$ & 0.9 cm & Typical ion sound gyroradius \\
  \bottomrule
  \end{tabular}
  \label{tab:lapd_plasma_parameters}
\end{center}
\end{table}

These simulations were run with 648 CPU cores, taking several wall-clock days to reach a quasi-steady state.
This case is highly collisional ($\lambda_{ee}/(L_z/2) \approx 0.004$) compared
to tokamak boundary plasmas \citep[see][table 1]{Umansky2011}, and the time step is limited by
the present explicit algorithm for collisions.
An implicit algorithm for collisions is expected to reduce the cost of these simulations by a large factor.
The underlying kinetic solver parallelizes well in multiple dimensions and the execution time
is approximately linear in the number of cells.

By assuming that the magnetic field is a constant $\boldsymbol{B} = B \hat{z}$,
the 5D gyrokinetic system we solve (see (\ref{eq:gke}) and (\ref{eq:gkPB_full}))
has the following simplified form for $\mathcal{J}\boldsymbol{\Pi}$:
\begin{align}
\mathcal{J}\boldsymbol{\Pi} &= 
	\begin{pmatrix}
		0 & -1/q_s & 0 & 0 & 0 \\
		1/q_s & 0 & 0 & 0 & 0\\
    0 & 0 & 0 & B/m_s & 0 \\
    0 & 0 & -B/m_s & 0 & 0 \\
		0 & 0 & 0 & 0 & 0
	\end{pmatrix}.
\end{align}

The initial density profile for both ions and electrons is chosen to be
$n_0 A(r;c_{\mathrm{edge}}=1/20)$, where $r = \sqrt{x^2 + y^2}$ and $A(r;c_{\mathrm{edge}})$ is a function that falls from the peak value
of $1$ at $r=0$ to a constant value $c_{\mathrm{edge}}$ for $r > L_\perp/2$:
\begin{equation}
A(r;c_{\mathrm{edge}}) = 
  \begin{cases}
    (1-c_{\mathrm{edge}}) \left( 1 - \frac{r^2}{(L_\perp/2) ^2} \right)^3 + c_{\mathrm{edge}} & r < L_\perp/2, \\
    c_{\mathrm{edge}}  & \mathrm{else}.
  \end{cases}
\end{equation}
The initial electron temperature profile has the form $5.7 A(r;c_{\mathrm{edge}}=1/5)$~eV,
while the initial ion temperature profile is a uniform 1~eV.
Both electrons and ions are initialized as non-drifting Maxwellians, although
future runs could be initialized with a specified non-zero mean velocity as a function of the parallel coordinate
computed from simplified 1D models  \citep{Shi2015} to reach a quasi-steady state more quickly.
This idea is used to initialize the distribution functions in the simulations for the helical-SOL simulations
discussed in Chapter~\ref{ch:helical-sol} (see Appendix~\ref{ch:initial-helical-sol} for the calculation).

Although we expect the quasi-steady state of the system to be insensitive to the choice of initial conditions,
we found that it was important to start the simulation with a non-uniform density profile
to avoid exciting large transient potential oscillations that resulted in extremely small
restrictions being imposed on the time step (for stability, the time step is automatically adjusted
based on the maximum gyrocenter characteristic velocities).
Because the boundary conditions force $\phi$ to a constant on the side walls (see Section \ref{sec:lapd_bc}), electrons near the domain boundaries in $x$ and $y$ are quickly lost at thermal speeds
from the simulation box.
We believe that this large momentary imbalance in the electron and ion densities is the source of this stability issue.

\begin{figure}
  \centerline{\includegraphics[width=\textwidth]{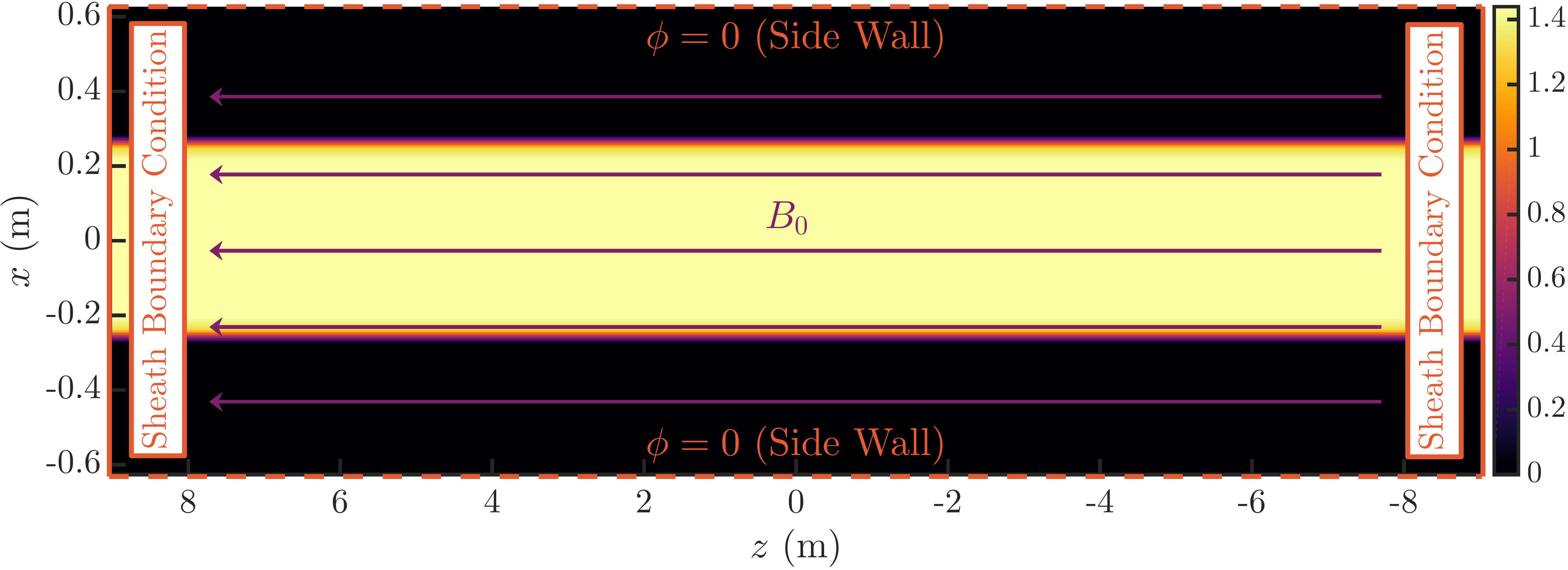}}
  \caption[LAPD-simulation density source rate in the $x$--$z$ plane.]{LAPD-simulation
  density source rate (in $10^{21}$~m$^{-3}$~s$^{-1}$) in the $x$--$z$ plane at $y = 0$ m.
  The solid orange lines on the left and right edges of the plot indicate the locations at which
  sheath-model boundary conditions are applied. The dashed orange lines on the top and bottom
  edges indicate the locations of the side walls, which we take to be grounded.
  }
\label{fig:lapd_source}
\end{figure}

\begin{figure}
  \centerline{\includegraphics[width=\textwidth]{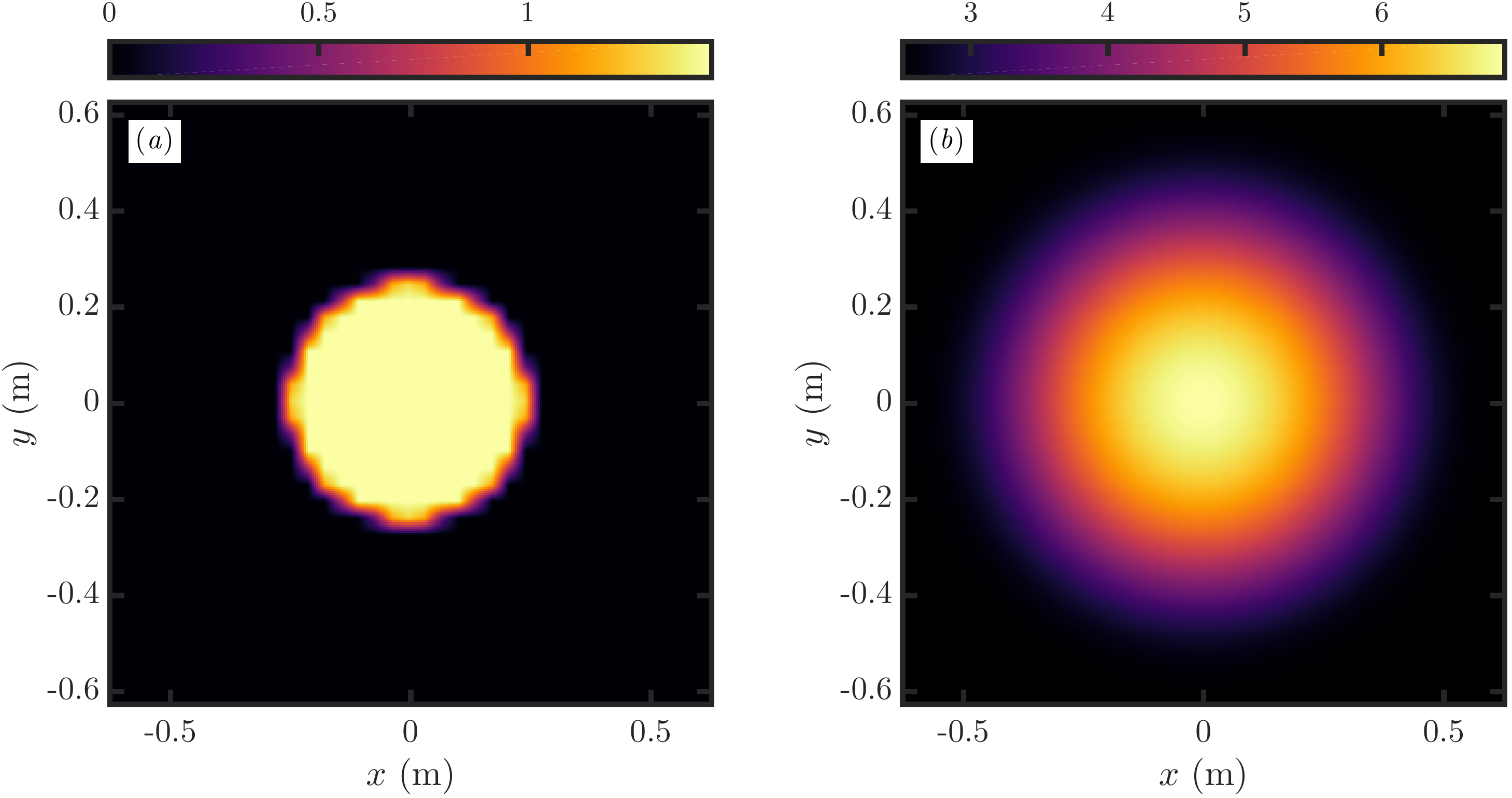}}
  \caption[LAPD-simulation density source rate and temperature of source electrons
  in the $x$--$y$ plane.]{LAPD-simulation ($a$) source density rate (in $10^{21}$~m$^{-3}$~s$^{-1}$)
  and ($b$) temperature of source electrons (in eV) in the $x$--$y$ plane.
  In this view, the side walls are located on all four sides of the square boundary.
  Note that the actual LAPD device has a circular cross section instead of the square cross section
  used in the simulation.}
\label{fig:lapd_source_xy}
\end{figure}

The electron and ion sources have the form
\begin{equation}
S_s = 
  1.08 \frac{n_0 c_{\mathrm{s}0}}{L_z} \left\{ 0.01 + 0.99 \left[ \frac{1}{2}-\frac{1}{2} \tanh\left( \frac{r-r_s}{L_s} \right) \right] \right \} F_{M,s}(v_\parallel,\mu; T_s) \label{eq:source},
\end{equation}
where $r_s = 20 \rho_{\mathrm{s}0} = 0.25$ m, $L_s = 0.5 \rho_{\mathrm{s}0} = 0.625$~cm, and
$F_{M,s}(v_\parallel,\mu; T_s)$ is a normalized non-drifting Maxwellian distribution for species $s$ with temperature $T_s$.
Figure~\ref{fig:lapd_source} shows the simulation geometry and density source rate in the $x$--$z$ plane at $y=0$~m,
and figure~\ref{fig:lapd_source_xy} shows the density source rate and temperature in the $x$--$y$ plane.
The ion source has a uniform temperature of 1~eV, while the electron source has a temperature profile given by
$6.8 A(r;c_{\mathrm{edge}}=1/2.5)$~eV.
In the actual LAPD experiment, the radius of the plasma column is controlled using
a floating-plate limiter with a variable aperture size that is situated near
the source end of the device \citep{Carter2006}.
For clarity, we will refer to the region of the simulation domain
with $r < r_s$ as the \textit{core region}, since most of the electrons and ions are
sourced there, the $r = r_s$ location as the \textit{limiter edge}, and the region with
$r > r_s$ as the \textit{edge region}.

Unlike the sources used by \citet{Rogers2010},
the sources we use model the neutrals as being ionized at zero mean velocity.
In the fluid equations of \citet{Rogers2010}, a zero-velocity plasma source would give rise to an additional
term $-S_n V_{\parallel i} / n$ on the right-hand side of the $\partial_t V_{\parallel i}$ equation,
which is kept in the more general equations of \citet{Wersal2015}.
In our simulations, electrons and ions are also sourced in the $r > r_s$ region at $1/100$th the amplitude of the central source rate
to avoid potential issues arising from zero-density regions.
While there are no high-energy electrons emitted from the cathode source
in the $r > r_s$ region in the actual LAPD device, \citet{Carter2009} have discussed
the possibility of ionization in this region by electrons in the main plasma in some
experiments with elevated edge electron temperatures.
Note that (\ref{eq:source}) does not represent the only source of energy in the system,
as the positivity-adjustment procedure also results in some energy being added to the particles at large $r$
and near the sheath entrances,
as discussed in Section \ref{sec:positivity}.

\section{Boundary Conditions and Energy Balance \label{sec:lapd_bc}}
Dirichlet boundary conditions $\phi = 0$ are used on the $x$ and $y$ boundaries for the potential solve
(taking the side walls to be grounded to the $\phi_w = 0$ end plates),
while no boundary condition on $\phi$ is required in $z$ because (\ref{eq:gkp}) contains no $z$ derivatives.
The distribution function uses zero-flux boundary conditions in $x$, $y$, $v_\parallel$, and $\mu$,
which amounts to zeroing out the interface flux evaluated on a boundary where zero-flux boundary conditions are to be applied.
This ensures that particles are not lost through the domain boundaries in $x$, $y$, $v_\parallel$, and $\mu$.
It should be noted that zero-flux boundary conditions on the $x$ and $y$ boundaries are a result of the choice of a constant $\phi$ on the
side-wall boundaries, so the $E \times B$ velocity at these boundaries is parallel to the wall.
Sheath-model boundary conditions, discussed in the previous section, are applied
to the distribution functions on the upper and lower boundaries in the $z$ direction.

To demonstrate how the choice of $\phi = 0$ affects the energy balance in our
long-wavelength gyrokinetic system with a linearized polarization term in the gyrokinetic Poisson equation,
we define the plasma thermal energy as
\begin{equation}
W_K = \int \mathrm{d}^3x \sum_s \int \mathrm{d}^3 v f_s H_0,
\end{equation}
where $H_0 = \frac{1}{2} m v_\parallel^2 + \mu B$.
Neglecting sources and collisions for simplicity, the gyrokinetic equation in a straight, constant magnetic field can be written as
\begin{equation}
\frac{\partial f_s}{\partial t} + \frac{\partial}{\partial z} \left( v_\parallel f_s \right) + \nabla \cdot \left(\boldsymbol{v}_E f_s \right)
+ \frac{\partial}{\partial v_\parallel} \left( \frac{q_s}{m_s} E_\parallel f_s \right)= 0,\label{eq:gksimple}
\end{equation}
where $E_\parallel = -\boldsymbol{b}\cdot \nabla \langle \phi \rangle$ and $\boldsymbol{v}_E = \mathbf{b} \times \nabla \langle \phi \rangle / B$.

Multiplying (\ref{eq:gksimple}) by $H_0$ and integrating over phase space,
\begin{align}
\frac{\partial W_K}{\partial t} =& - \int \mathrm{d}x \, \mathrm{d}y \sum_s \int \mathrm{d}^3 v \,
H_0 v_\parallel f_s \Big|_{z_{\mathrm{lower}}}^{z_{\mathrm{upper}}}
+ \int \mathrm{d}^3 x \sum_s \int \mathrm{d}^3 v \, v_\parallel f_s  q_s E_\parallel \nonumber \\
=& - \int \mathrm{d}x \, \mathrm{d}y \sum_s \int \mathrm{d}^3 v \, 
H_0 v_\parallel f_s \Big|_{z_{\mathrm{lower}}}^{z_{\mathrm{upper}}} +
\int \mathrm{d}^3 x \, j_\parallel E_\parallel, \label{eq:w_k}
\end{align} 
where we have used the fact that the normal component of $\boldsymbol{v}_E$ vanishes on the side walls
(since $\phi$ is a constant on the side walls)
and zero-flux boundary conditions on $f_s$ in $v_\parallel$.
The first term on the right-hand side is the parallel heat flux out to the sheaths and the second term is the parallel acceleration by the electric field,
which mediates the transfer of energy between thermal and field energies in this model (this term appears with the opposite sign in the
equation for the evolution of $E \times B$ energy).

To calculate the field energy evolution, we take the time derivative of the gyrokinetic Poisson equation (\ref{eq:gkp}),
\begin{align}
-\nabla_\perp \cdot \left( \epsilon \nabla_\perp \frac{\partial \phi}{\partial t} \right) =& \sum_s q_s \int \mathrm{d}^3 v \frac{\partial f_s}{\partial t} \nonumber \\
=& -\sum_s q_s \int \mathrm{d}^3 v \, \left[ \frac{\partial}{\partial z} \left( v_\parallel f_s \right) + \nabla \cdot \left(\boldsymbol{v}_E f_s \right)
 \right] \nonumber \\
=& -\frac{\partial j_\parallel}{\partial z} - \nabla \cdot (\boldsymbol{v}_E \sigma_g) \label{eq:gkp_dt},
\end{align}
where $\epsilon = n_{i0}^g e^2 \rho_{\mathrm{s}0}^2/T_{e0}$.
Next, we multiply (\ref{eq:gkp_dt}) by $\phi$ and integrate over space:
\begin{align}
-\int \mathrm{d}^3 x \, \phi \nabla_\perp \cdot \left( \epsilon \nabla_\perp \frac{\partial \phi}{\partial t} \right) =&
-\int \mathrm{d}^3 x \, \phi \left[ \frac{\partial j_\parallel}{\partial z} + \nabla \cdot (\boldsymbol{v}_E \sigma_g) \right] \nonumber \\
- \int \mathrm{d} \boldsymbol{S}_\perp \cdot \phi \epsilon \nabla_\perp \frac{\partial \phi}{\partial t} 
+ \frac{1}{2} \int \mathrm{d}^3 x \, \epsilon \frac{\partial \left( \nabla_\perp \phi \right)^2}{\partial t} =&
-\int \mathrm{d} x \mathrm{d} y \, \phi j_\parallel \Big|_{z_{\mathrm{lower}}}^{z_{\mathrm{upper}}} + \int \mathrm{d}^3 x \, \frac{\partial \phi}{\partial z} j_\parallel \nonumber\\
& - \int \mathrm{d} \boldsymbol{S}_\perp \cdot \phi \boldsymbol{v}_E \sigma_g + \int \mathrm{d}^3 x \, \nabla \phi \cdot \boldsymbol{v}_E \sigma_g. \label{eq:gkp_dt_energy}
\end{align}
The integral involving $\int \mathrm{d} \boldsymbol{S}_\perp$ on the right-hand side is zero because $\boldsymbol{v}_E$ has no normal component on the side walls.
By assuming that $\phi = 0$ on the side walls, the term on the left-hand side involving $\int \mathrm{d} \boldsymbol{S}_\perp \phi$ is also zero and we have
\begin{equation}
\frac{\partial W_\phi}{\partial t} = \frac{\partial}{\partial t} \left( \frac{1}{2} \int \mathrm{d}^3 x \, \epsilon \left(\nabla_\perp \phi \right)^2 \right) =
-\int \mathrm{d} x \mathrm{d} y \, \phi j_\parallel \Big|_{z_{\mathrm{lower}}}^{z_{\mathrm{upper}}} - \int \mathrm{d}^3 x \, j_\parallel E_\parallel.
\label{eq:w_phi}
\end{equation}
If the wall is biased instead of grounded, as done in a set of experiments by \citet{Carter2009},
one must retain the first term on the left-hand side of (\ref{eq:gkp_dt_energy}) in energy-balance considerations.
The second term on the right-hand side of (\ref{eq:w_phi}) is equal and opposite to the second
term on the right-hand side of (\ref{eq:w_k}), and so cancels when the two equations are added together.
The total energy is the sum of the kinetic energy $W_k$ and the field energy $W_\phi$.
Substituting the definition of $\epsilon$, this field energy can be written as
$W_\phi = \int d^3 x \, n_{i0}^g m_i v_E^2/2$, indicating that it can be interpreted as
the kinetic energy associated with the $E \times B$ motion.
(The $n_{i0}^g$ factor can be generalized to the full density $n_i^g({\boldsymbol{R}},t)$
as described in Section \ref{sec:model}, with an additional contribution to the Hamiltonian.)
The first term on the right-hand side of (\ref{eq:w_phi}) corresponds to work done on particles as they are
accelerated through the sheath. The $\phi$ in this boundary term is the potential at the $z$
boundaries of the simulation domain, where the sheath entrances are.
When $j_{\parallel}=0$ at the sheath entrance, then the energy lost by electrons as they drop
through the sheath is exactly offset by the energy gained by ions as they drop through the sheath.
If more electrons than ions are leaving through the sheath, then the net energy lost in the unresolved sheath
region contributes to an increase in the field energy.

There is also room for improvements in the side-wall boundary conditions.
Identifying the left-hand side of (\ref{eq:gkp_dt}) as $-\partial \sigma_{\mathrm{pol}}/ \partial t = \nabla \cdot \boldsymbol{j}_{\mathrm{pol}}$,
and integrating over all space,
\begin{align}
\int \mathrm{d}^3 x \, \nabla \cdot \boldsymbol{j}_\mathrm{pol} =& \int \mathrm{d} \boldsymbol{S} \cdot \boldsymbol{j}_\mathrm{pol} \nonumber \\
=& \int \mathrm{d} \boldsymbol{S}_\perp \cdot \epsilon \frac{\partial \boldsymbol{E}_\perp}{\partial t},
\end{align}
so we see that there is an ion polarization current into the side wall when the electric field
pointing into the side wall is increasing in time, which is physically reasonable.
However, if the sign of the electric-field time derivative reverses,
it is not possible to pull ions out of the side wall (where they are trapped by quantum effects, or return as neutrals),
and a boundary layer might form near the side walls.
In fusion devices, it is rare for the magnetic field to be exactly parallel to the wall,
so it could be appropriate to use a model of the Chodura magnetic pre-sheath \citep{Chodura1982}.
\citet{Geraldini2017} also recently studied a gyrokinetic approach to the magnetic pre-sheath.

The inclusion of charge-neutral source terms and number-conserving collision operators to the above analysis does not result in
additional sources of $E \times B$ energy, since they lead to the addition of terms to the right-hand side of (\ref{eq:w_phi}) of the form
\begin{equation}
-\int \mathrm{d}^3x \, \phi \sum_s q_s \int \mathrm{d}^3 v \, S_s (\boldsymbol{R}, \boldsymbol{v}, t) = 0.
\end{equation}

\section{Simulation Results \label{sec:lapd_unbiased}}
\begin{figure}
  \centerline{\includegraphics[width=\textwidth]{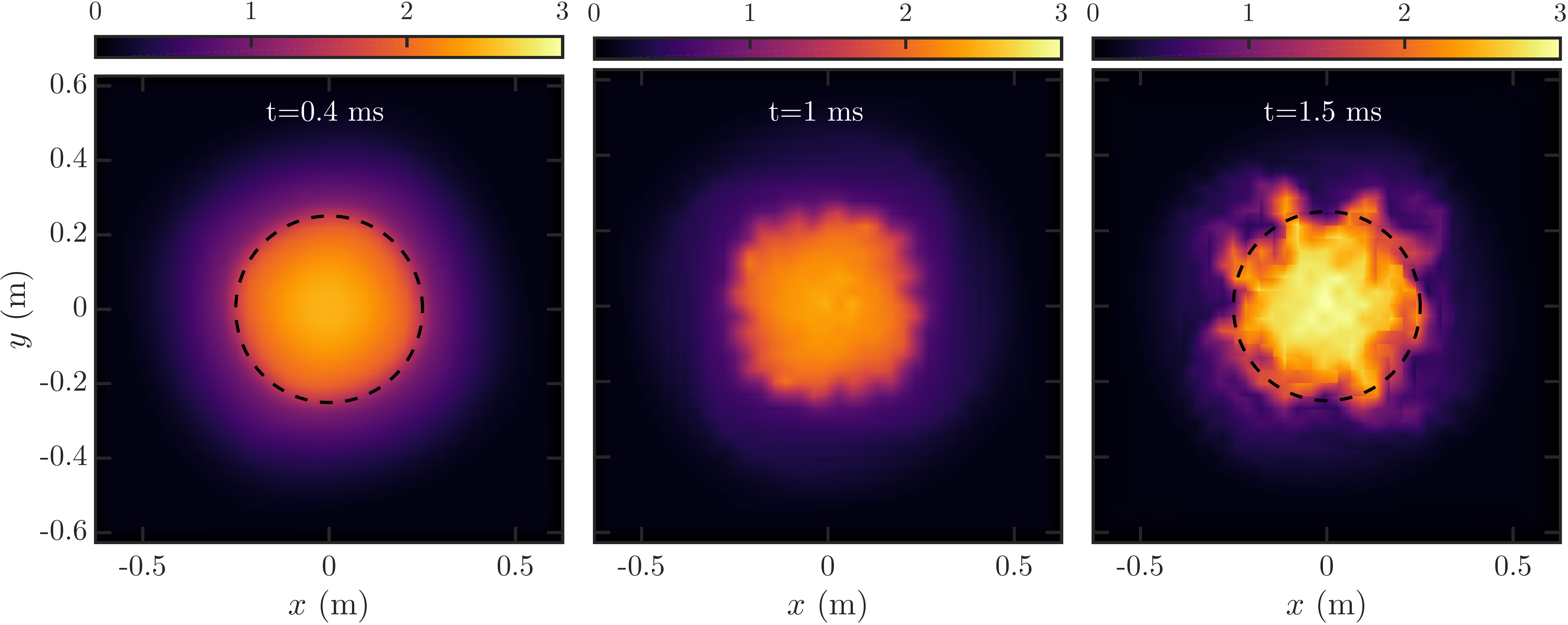}}
  \caption[Snapshots of the total electron density in the $x$--$y$ plane
  from a 5D gyrokinetic simulation of an LAPD plasma at three times.]
  {Snapshots of the total electron density (in $10^{18}$ m$^{-3}$) in the $x$--$y$ plane
  from a 5D gyrokinetic simulation of an LAPD plasma at $t=0.4$ ms, $t=1$ ms,
  and $t=1.5$ ms. The sources are concentrated inside the dashed line indicated (not shown
  on the $t=1$ ms plot), which is located at $r = r_s = 20 \rho_{\mathrm{s}0}$.
  Starting from the initial condition, the source steepens the plasma profiles until
  short-wavelength structures begin to grow. At longer times,
  larger-wavelength structures develop and remain for the rest of the simulation.
  For additional details about how this plot and ones like it
  were created, see Appendix~\ref{ch:plot-creation}.}
  \label{fig:lapd_density_evolution}
\end{figure}

In this section we present results from our gyrokinetic simulation.
Our goal here is not to argue that our simulations are a faithful model of the LAPD plasma,
but instead to demonstrate the ability to carry out gyrokinetic continuum simulations
of open-field-line plasmas in a numerically stable way and to
demonstrate a reasonable level of qualitative agreement 
by making contact with turbulence measurements from the real LAPD device
and previous Braginskii fluid simulations \citep{Ricci2010,Fisher2015,Friedman2012},
since we have used similar plasma parameters and geometry.
Starting from the initial conditions described in Section \ref{sec:lapd-results}, the electron and
ion distributions evolve for a few ion sound transit times ($\tau_s \sim (L_z/2)/c_s \approx 1.1$~ms using
$T_e = 3$~eV) until a quasi-steady state is reached, during which the
total number of particles of each species remains approximately constant.
We have found that the time discretization results in relative errors in the
energy conservation of ${\sim} 10^{-4}$.
A discussion about the energy-conservation properties of these simulations is presented
in Appendix \ref{ch:lapd-discrete-energy}.
Some snapshots of the total electron density in the transient stage are shown in
figure~\ref{fig:lapd_density_evolution}.

\begin{figure}
  \centerline{\includegraphics[width=\textwidth]{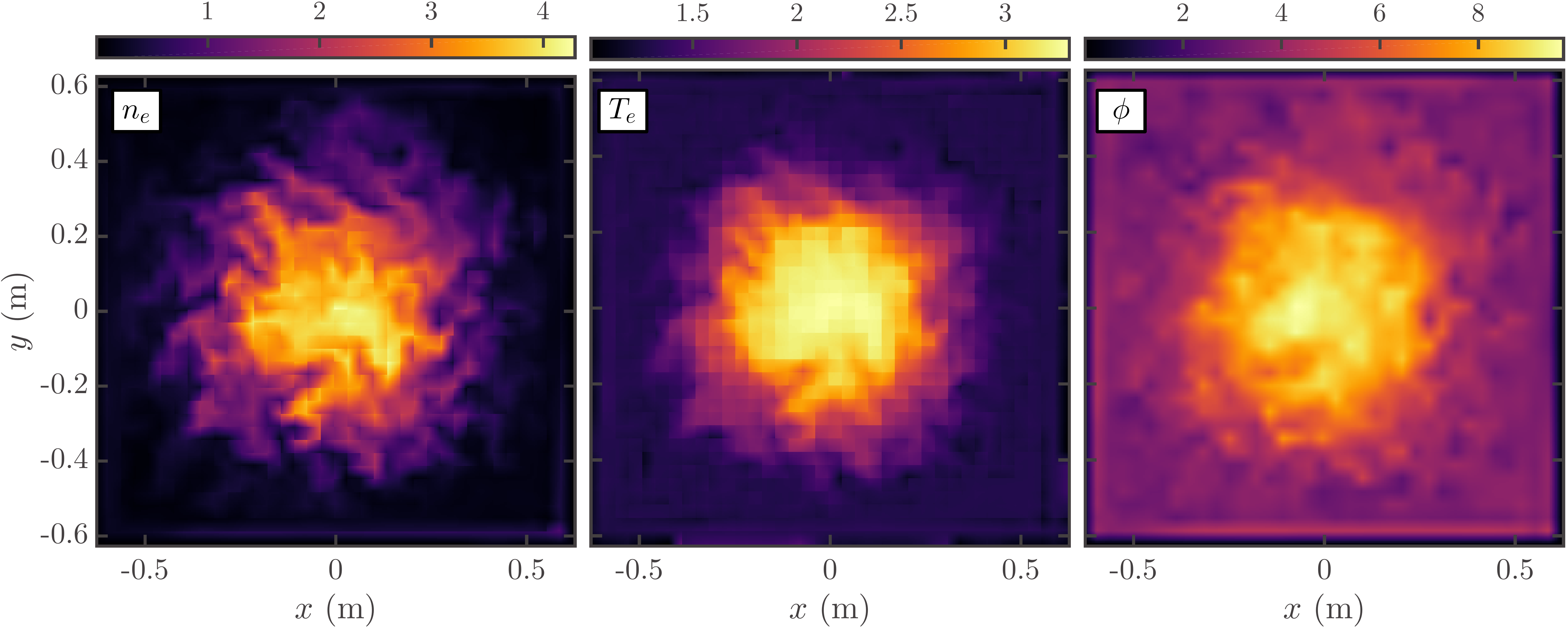}}
  \caption[Snapshots in the $x$--$y$ plane of the total electron density, electron temperature, and
  electrostatic potential from a 5D gyrokinetic simulation of an LAPD plasma.]
  {Snapshots of the total electron density $n_e$ (in $10^{18}$~m$^{-3}$),
  electron temperature $T_e$ (in eV), and electrostatic potential $\phi$ (in V) from a 5D gyrokinetic
  simulation of an LAPD plasma.
  The plots are made in center of the box at $z=0$~m.
In this simulation, a continuous source of plasma concentrated inside $r_s = 0.25$~m is transported radially outward by the turbulence
as it flows at near-sonic speeds along the magnetic field lines to the end plates, where losses are mediated by
sheath-model boundary conditions.
The plots are made in the $x$--$y$ plane perpendicular to the magnetic field in the middle of the device after a few ion transit times.}
\label{fig:lapd2dxy}
\end{figure}

\begin{figure}
  \centerline{\includegraphics[width=\textwidth]{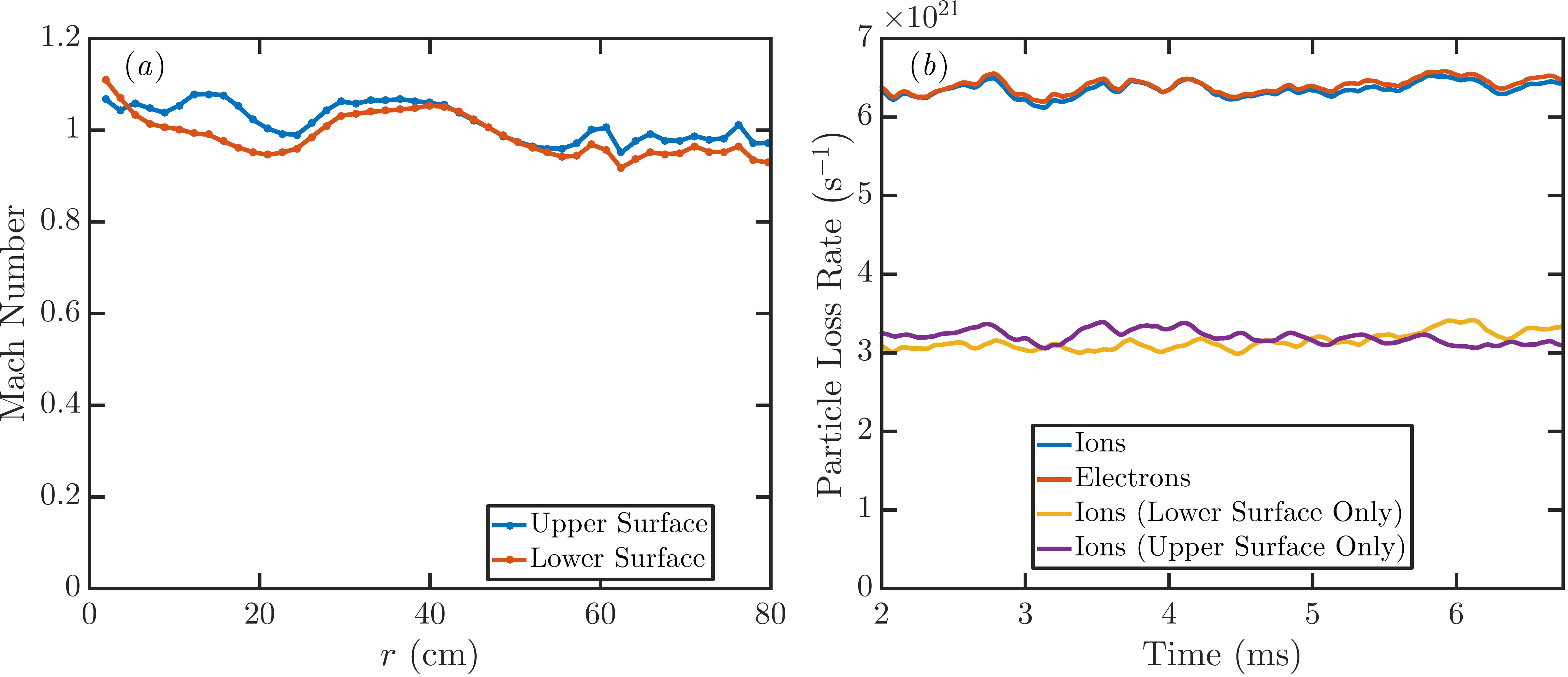}}
  \caption[Time-averaged radial profile of the ion-outflow Mach number across each parallel boundary
  and total outflow of electron and ion guiding centers across the parallel boundaries
  during a quasi-steady state.]
  {($a$) Time-averaged radial profile of the ion-outflow Mach number across each parallel boundary,
  denoted as upper and lower surfaces.
  The Mach number is defined as $u_\parallel/c_\mathrm{s}$,
  where $c_\mathrm{s} = \sqrt{(T_e + \gamma T_i)/m_i}$ and $\gamma = 5/3$.
  ($b$) Total outflow of electron and ion guiding centers across the parallel boundaries
  during a quasi-steady state (in s$^{-1}$).
  Also indicated is the outflow for ions across each surface individually.}
  \label{fig:outward_flux}
\end{figure}

As seen in LAPD experiments \citep{Schaffner2012,Schaffner2013}, we observe a weak spontaneous rotation in the
ion-diamagnetic-drift direction.
Figure~\ref{fig:lapd2dxy} shows snapshots in the perpendicular plane of the total electron density,
electron temperature, and electrostatic potential
after a few ion transit times, which are qualitatively similar to the snapshots presented from
Braginskii fluid simulations of LAPD \citep{Rogers2010,Fisher2015}.
Figure~\ref{fig:outward_flux}($a$) shows the time-averaged radial profile of the ion-outflow Mach number
across each parallel boundary, where we define the Mach number as $u_\parallel/c_\mathrm{s}$.
Since the electron mean free path is smaller than the effective parallel grid spacing,
we do not resolve the collisionless transition layer in front of the sheath in which the outflow
at the collisional sound speed ($\gamma = 5/3$) transitions to an outflow at the collisionless sound speed
($\gamma = 3$).
Therefore, we expect the sound speed at the sheath to be $c_\mathrm{s} = \sqrt{(T_e + \gamma T_i)/m_i}$ with
$\gamma = 5/3$.
Despite the fact that our sheath-model boundary conditions do not enforce outflows at the sound speed,
we see that the Mach number is very close to 1 at the sheath entrance.
Figure~\ref{fig:outward_flux}($b$) shows the total loss rate of electron and ion guiding centers
across the parallel boundaries over a period of several microseconds.
We confirm that the particle loss rate is in an approximate balance with the
particle source rate in this quasi-steady state.
We note that in addition to the particle source of approximately $5.5 \times 10^{21}$~s$^{-1}$, there is a time-varying
particle source of approximately $10^{21}$ s$^{-1}$ that comes from adding electrons and ions to the
system in order to keep the density above a floor of $2 \times 10^{12}$~m$^{-3}$ everywhere
(see Section \ref{sec:positivity}).

\begin{figure}
  \centerline{\includegraphics[width=\textwidth]{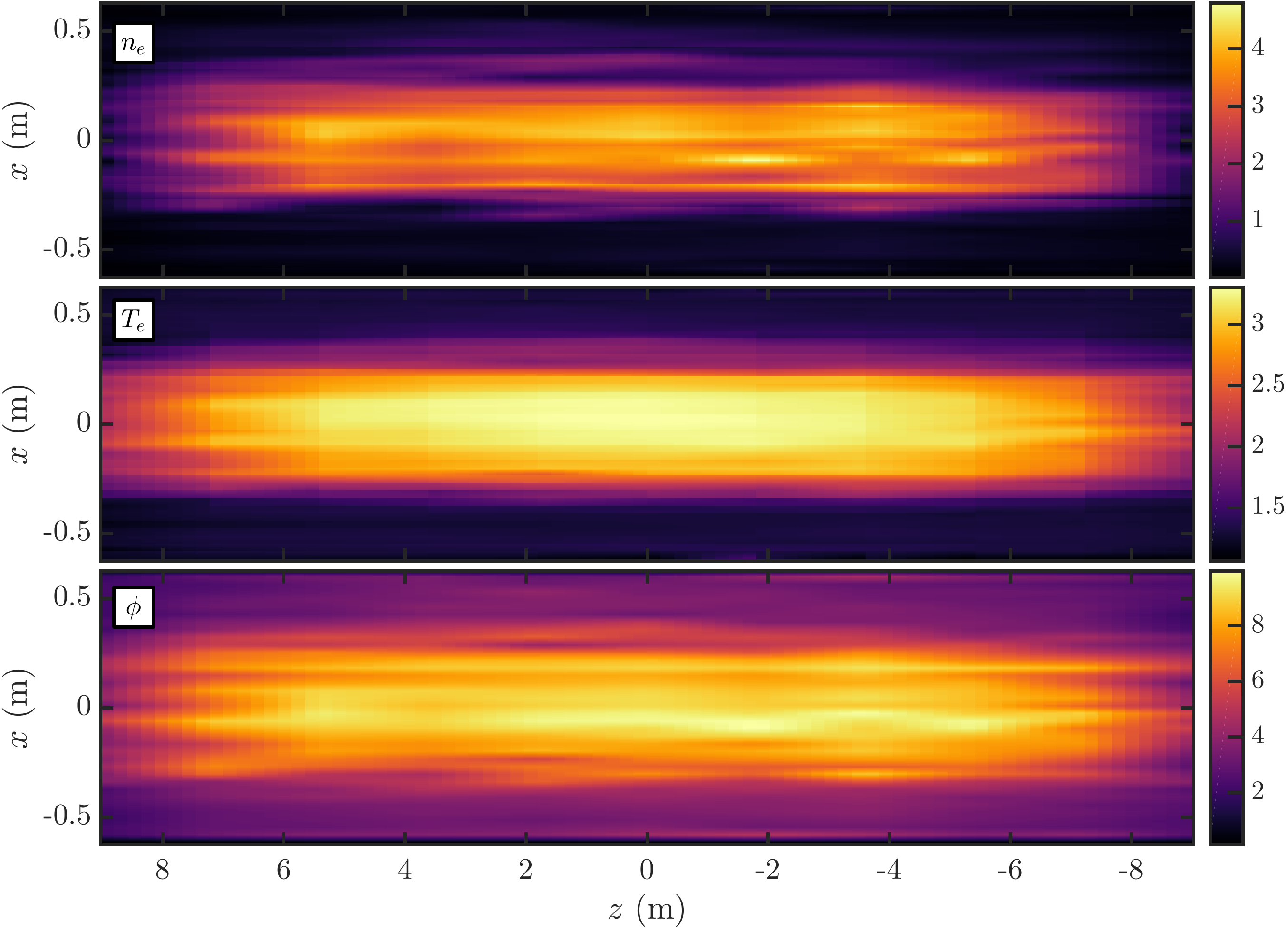}}
  \caption[Snapshots in the $x$--$z$ plane of the total electron density, electron temperature, and
  electrostatic potential from a 5D gyrokinetic simulation of an LAPD plasma.]
  {Snapshots of the total electron density $n_e$ (in $10^{18}$ m$^{-3}$),
  electron temperature $T_e$ (in eV), and electrostatic potential $\phi$ (in V) from a 5D gyrokinetic
  simulation of an LAPD plasma.
  The plots are made in the $x$--$z$ plane at $y=0$ m after a few ion transit times.}
  \label{fig:lapd2dxz}
\end{figure}

Figure~\ref{fig:lapd2dxy}($c$) shows that a boundary layer with a width of order the sound gyroradius forms in the potential
near the side walls, where the potential drops to match the boundary conditions $\phi = 0$ on the side walls.
This means that a normal sheath at the ends in $z$ with $\phi_s \sim 3 T_e$ cannot occur
very close to the side walls.
However, one can still eventually get a quasi-steady state with the electron flux,
$\sim n_e v_{te} \exp(-e \phi_s / T_e)$, of order the ion flux
because the electron density becomes very small near the side walls and the electrons become colder there.
Figure~\ref{fig:lapd2dxz} shows the same fields as in figure~\ref{fig:lapd2dxy}, but the plots are made in the $y=0$ plane
to show the parallel structure.

\begin{figure}
  \centerline{\includegraphics[width=\textwidth]{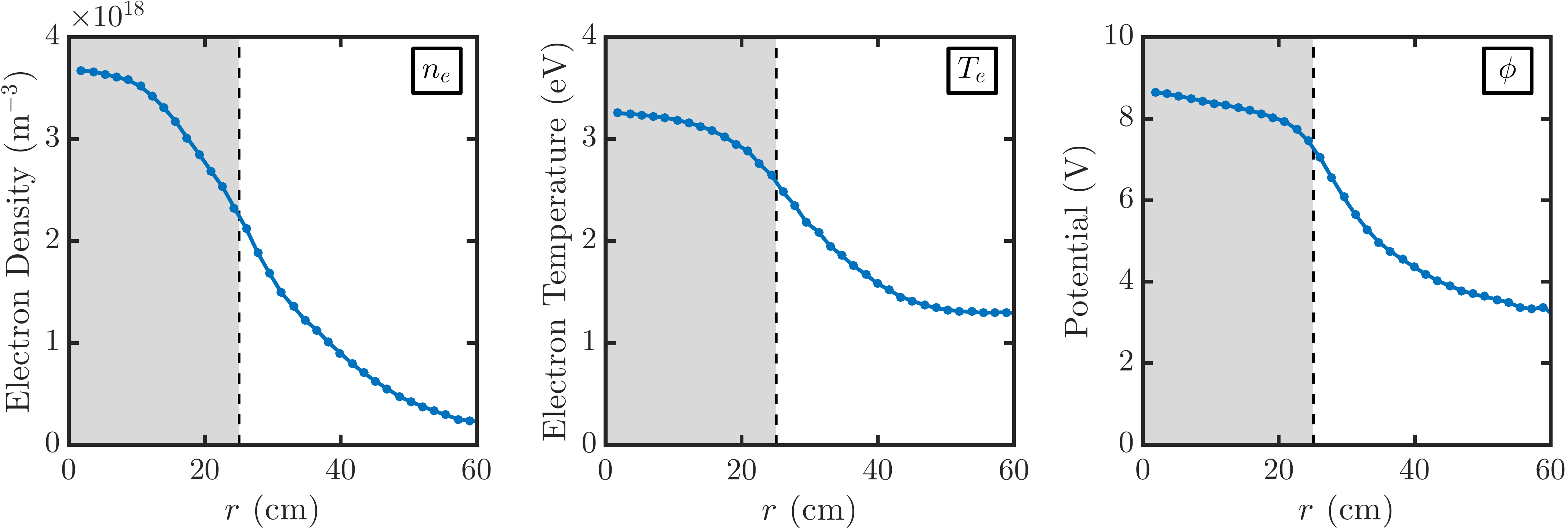}}
  \caption[Average profiles of electron density, electron temperature, and electrostatic potential as a
  function of radius from a 5D gyrokinetic simulation of an LAPD plasma.]{Average profiles of the
  electron density $n_e$, electron temperature $T_e$, and electrostatic potential $\phi$ as a function of radius.
  The fields in the region $-4$~m $<z< 4$~m are time averaged over several ion transit times
  after the simulation has reached a quasi-steady state,
  evaluated at eight equally spaced points in each cell, and then binned by radius.
  The shaded region illustrates the core region inside the limiter edge at $r=r_s$.}
  \label{fig:lapd1d}
\end{figure}

Figure~\ref{fig:lapd1d} shows the time-averaged radial profile of $n_e$, $T_e$, and $\phi$ computed
by averaging the data in the region $-4$~m $<z<4$~m.
We focus on this region since it is similar to the region in which probe measurements are taken in the LAPD,
and there is little parallel variation in this region.
Particle transport in the radial direction is especially evident in figure~\ref{fig:lapd1d} from the broadening in the $n_e$ profile.
In figure~\ref{fig:lapd1d}, the electron temperature drops off at mid-radii
but is rather flat at large $r$.
To understand this, note that there is a ${\approx}2.7$~eV residual electron source at large $r$ (see (\ref{eq:source})),
and that the observed temperature is close to the limit of the coldest temperature that can be represented on
the grid when collisions dominate and the distribution function is isotropic,
so $T_{e,\mathrm{min}} \approx T_{\perp e,\mathrm{min}} = 1.2$~eV.
Our choice of velocity-space grid is a compromise between resolving low energies and
the need to go up to significantly higher energies than the temperature of
the source (which has a maximum temperature of 6.7~eV) to represent the tail.
This will be improved in future work using a non-uniformly spaced velocity grid or exponential
basis functions, which can represent a range of electron energies much more efficiently.
We do not expect the non-vanishing $T_e$ at large $r$ to affect the results significantly
because both $n_e$ and the $n_e$ fluctuation level are small at large $r$.

\begin{figure}
  \centerline{\includegraphics[width=\textwidth]{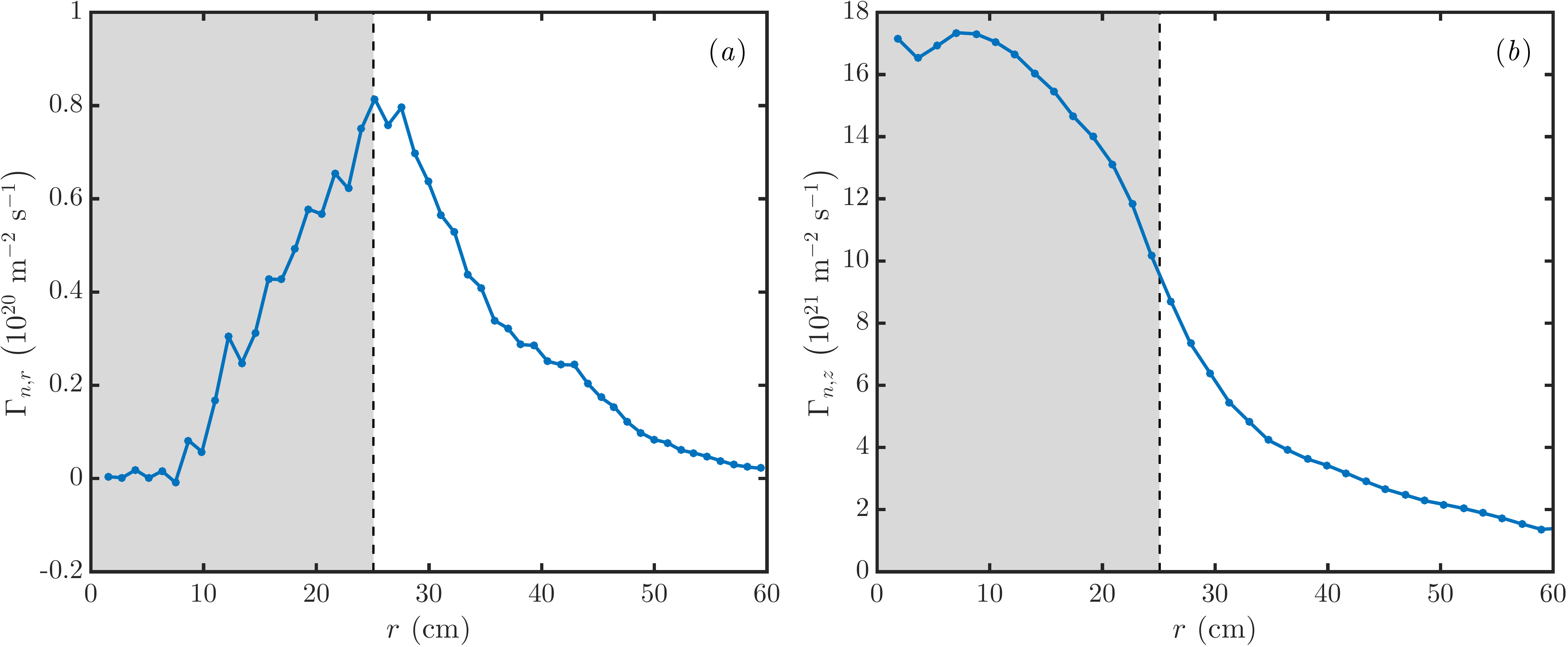}}
  \caption[Profiles of the time-averaged radial particle flux due to electrostatic fluctuations
  and total outward parallel particle flux at end plates.]
  {Profiles of the ($a$) time-averaged radial particle flux $\Gamma_{n,r}$ due to electrostatic fluctuations
  and ($b$) time-averaged total outward parallel particle flux $\Gamma_{n,z}$, which includes contributions
  from both end plates.
  The shaded region illustrates the core region inside the limiter edge at $r=r_s$.
  These profiles indicate that the parallel particle loss rate in the core is approximately
  $2.8 \times 10^{21}$~s$^{-1}$, while radial transport into the edge region
  due to turbulence is approximately $2.3 \times 10^{21}$~s$^{-1}$,
  so radial turbulent particle transport
  is comparable to parallel particle transport in these simulations (owing to the large
  ${\approx}18$~m length of the LAPD plasma column).}
  \label{fig:lapd-unbiased-flux}
\end{figure}

Figure~\ref{fig:lapd-unbiased-flux}($a$) shows the time-averaged radial $E \times B$ particle flux due to
electrostatic fluctuations, which we define as
\begin{equation}
  \Gamma_{n,r} = n_e \boldsymbol{v}_E \cdot \hat{\boldsymbol{r}}.
\end{equation}
The profile of $\Gamma_{n,r}$ is a measure of the turbulent radial particle transport in the plasma.
To get a sense of how much the turbulence broadens the profiles radially, we can compare $\Gamma_{n,r}$
to the outward parallel particle flux $\Gamma_{n,z}$, which is shown in figure~\ref{fig:lapd-unbiased-flux}($b$).
We see that the total particle outflow in the core-plasma region is 
$\int_0^{r_s} \mathrm{d}r \, 2 \pi r \Gamma_{n,z} \approx 2.8 \times 10^{21}$ s$^{-1}$.
From figure~\ref{fig:lapd-unbiased-flux}($a$), we also see that the radial transport out of the
core region is $2 \pi r_s L_z \Gamma_{n,r} \left(r=r_s \right) \approx 2.3 \times 10^{21}$ s$^{-1}$.
We conclude that rates at which particles leave the core region through
turbulent radial transport and through parallel losses at the end plates are comparable.
Note that the cross-field turbulent particle transport in linear devices is often negligible compared
to the parallel particle transport.
The large length and radius of the LAPD plasma column enable
the turbulence to have a noticeable effect on the plasma profiles.

\begin{figure}
  \centerline{\includegraphics[width=\textwidth]{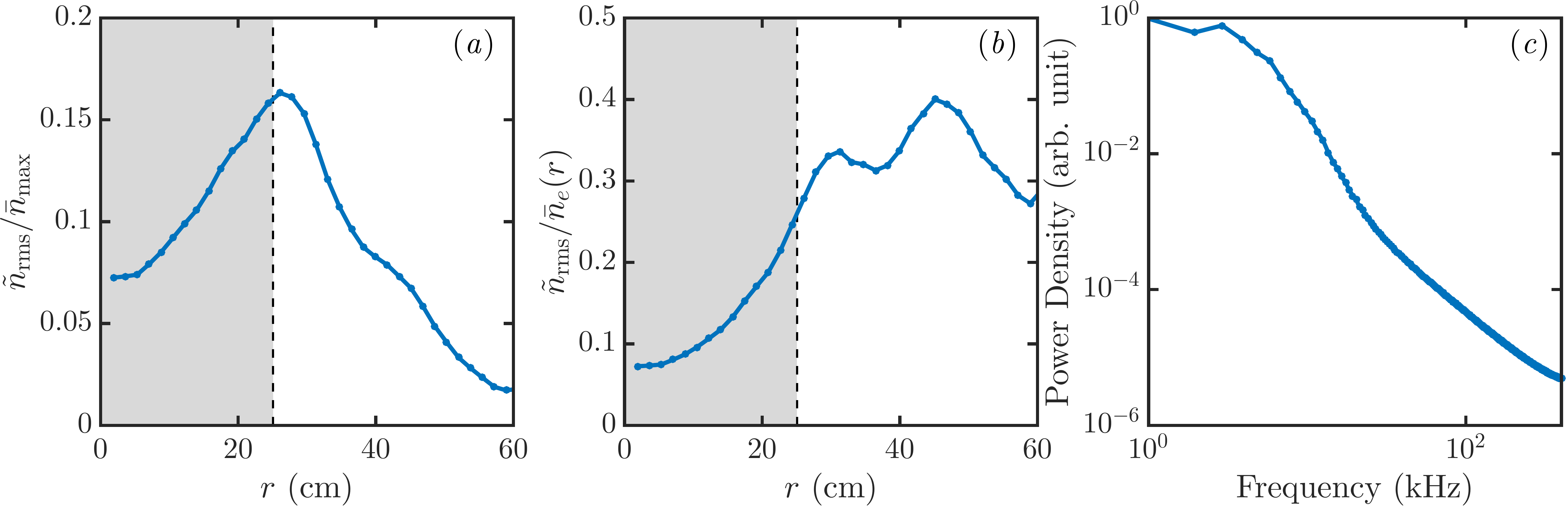}}
  \caption[Normalized r.m.s. density fluctuation level as a function of radius and
  density-fluctuation power spectral density computed from a 5D gyrokinetic simulation
  of an LAPD plasma.]
  {Density fluctuation statistics computed from a 5D gyrokinetic simulation of an LAPD plasma.
  ($a$) The r.m.s. density fluctuation level as a function of radius, 
  normalized globally to the peak background electron density $\bar{n}_\mathrm{max} \approx 3.68 \times 10^{18}$~m$^{-3}$.
  ($b$) The r.m.s. density fluctuation level as a function of radius,
  normalized locally to the background electron density $\bar{n}_e(r)$.
  ($c$) The density fluctuation power spectral density. These plots are in good
  qualitative agreement with LAPD measurements \citep{Carter2009,Popovich2010b,Friedman2012,Fisher2015,Carter2006},
  reproducing features such as the maximum in the density fluctuation level (normalized to $\bar{n}_\mathrm{max}$)
  occurring at the limiter edge and at the ${\sim}10\%$ level,
  the decay of normalized density fluctuation level both radially inward and outward
  from the peak location, and the broadband fluctuation spectrum that drops by approximately six orders of magnitude
  from ${\sim}10^3$~Hz to ${\sim}10^5$~Hz. The shaded regions in ($a$) and ($b$)
  illustrate the core region inside the limiter edge at $r=r_s$.}
\label{fig:rmsDensity}
\end{figure}

Electron-density-fluctuation profiles have also been measured in LAPD \citep{Carter2009,Popovich2010b,Friedman2012,Fisher2015,Carter2006}.
We define the density fluctuation as $\tilde{n}_e(x,y,z,t) = n_e(x,y,z,t) - \bar{n}_e(x,y,z)$,
where $\bar{n}(x,y,z)$ is computed by averaging the electron density using a 1 $\mu$s sampling interval
over a period of 1~ms.
The density fluctuation level is normalized both to the peak amplitude of $\bar{n}_e$
at $r = 0$ \citep[as done in][]{Friedman2012}
and to the local value of $\bar{n}_e(x,y,z)$ and then binned by radius to
calculate profiles of the r.m.s. density fluctuation level,
which are shown in figures~\ref{fig:rmsDensity}$(a)$ and \ref{fig:rmsDensity}$(b)$.
Similar to measurements reported in LAPD, we find that the maximum in the
density fluctuation level (normalized to $\bar{n}_\mathrm{max}$)
occurs at the limiter edge and at the ${\sim}10\%$ level.
The decay of the density fluctuation level (normalized to $\bar{n}_\mathrm{max}$)
both radially inward and outward from the peak location is also commonly observed in LAPD.
Figure~\ref{fig:rmsDensity}$(c)$ shows the power spectral density of electron-density fluctuations,
which is computed by averaging the power spectra at each node
in the region $25$~cm $<r<30$~cm and $-4$~m~$<z<4$~m.
Similar to measurements reported in LAPD, we find that the turbulence has a broadband spectrum
that drops by approximately six orders of magnitude from ${\sim}10^3$~Hz to ${\sim}10^5$~Hz.

\begin{figure} \centerline{\includegraphics[width=\textwidth]{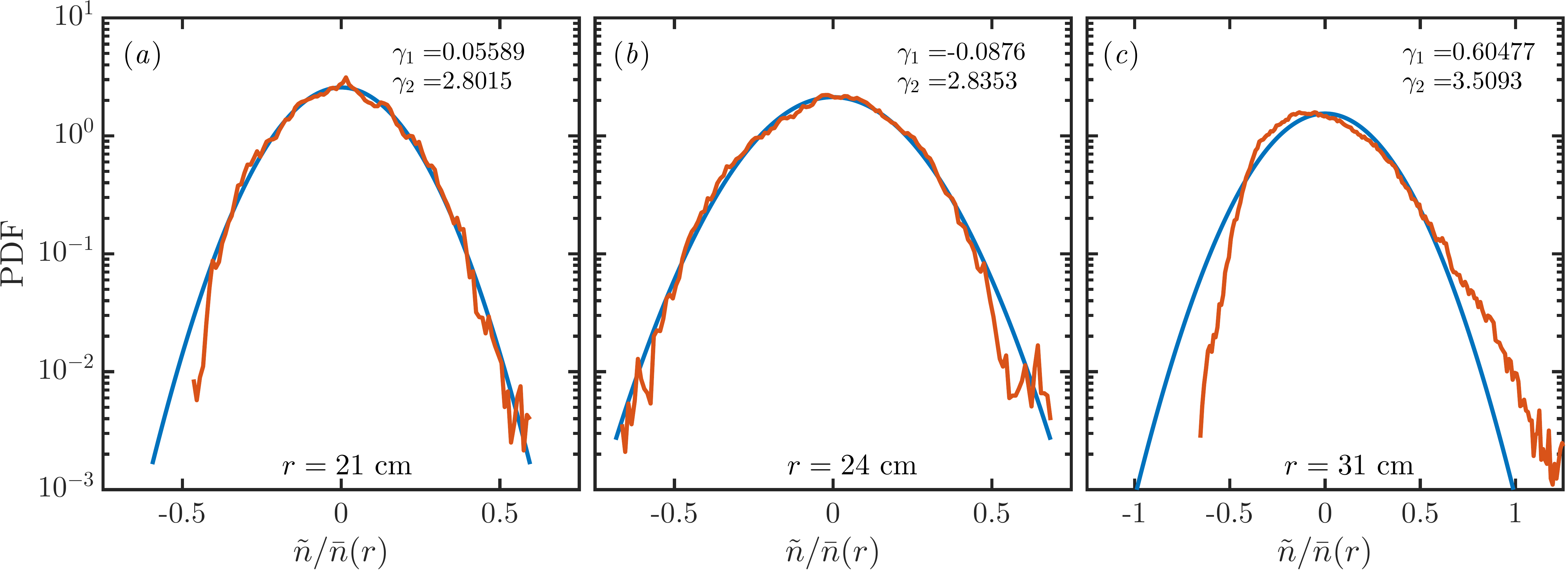}}
  \caption[Density-fluctuation-amplitude PDF at three radial locations from a 5D gyrokinetic simulation of an LAPD plasma.]
  {Density-fluctuation-amplitude PDF (in red and normalized local value of the background electron density)
  at three radial locations in the region $-4$~m~$<z<4$~m: $(a)$ 4~cm from the limiter edge
  at $r = 21$~cm, $(b)$ 1~cm from the limiter edge at $r=24$~cm, and $(c)$ a few
  centimeters outside the core region at $r = 31$~cm.
  Gaussian PDFs are shown in blue for comparison.
  Also indicated on each plot is
  the skewness $\gamma_1 = E[\tilde{n}_e^3]/\sigma^3$ and the kurtosis $\gamma_2 = E[\tilde{n}_e^4]/\sigma^4$,
  where $\sigma$ is the standard deviation of $\tilde{n}_e$ and $E[\dots]$ denotes the expected value.}
\label{fig:pdfDensity}
\end{figure}

The probability density function (PDF) of density fluctuations in LAPD has also been of interest.
\citet{Carter2006} focused on the intermittency of the density-fluctuation PDF measured at
various radial locations.
Figure~\ref{fig:pdfDensity} shows the simulation PDF at three radial locations
(using $\Delta r = 0.5$~cm wide radial intervals) in  the region $-4$~m~$<z<4$~m.
We find symmetric and near-Gaussian PDFs in the core region at $r=21$~cm,
which is  $4$~cm from the limiter edge, and at $r=24$~cm, which is
1~cm from the limiter edge.
A few centimeters outside the core region at $r=31$~cm, we find a positively skewed 
and non-Maxwellian PDF.
Here, we see that there is a particularly strong enhancement of large-amplitude
positive-density-fluctuation events.
Compared to the experimental results of \citet{Carter2006}, the PDFs we observe at the inner-two
radial locations are much closer to a Gaussian, and we do not observe
a negatively skewed PDF at the innermost radial location.

\begin{figure}
  \centering
  \includegraphics[width=0.75\linewidth]{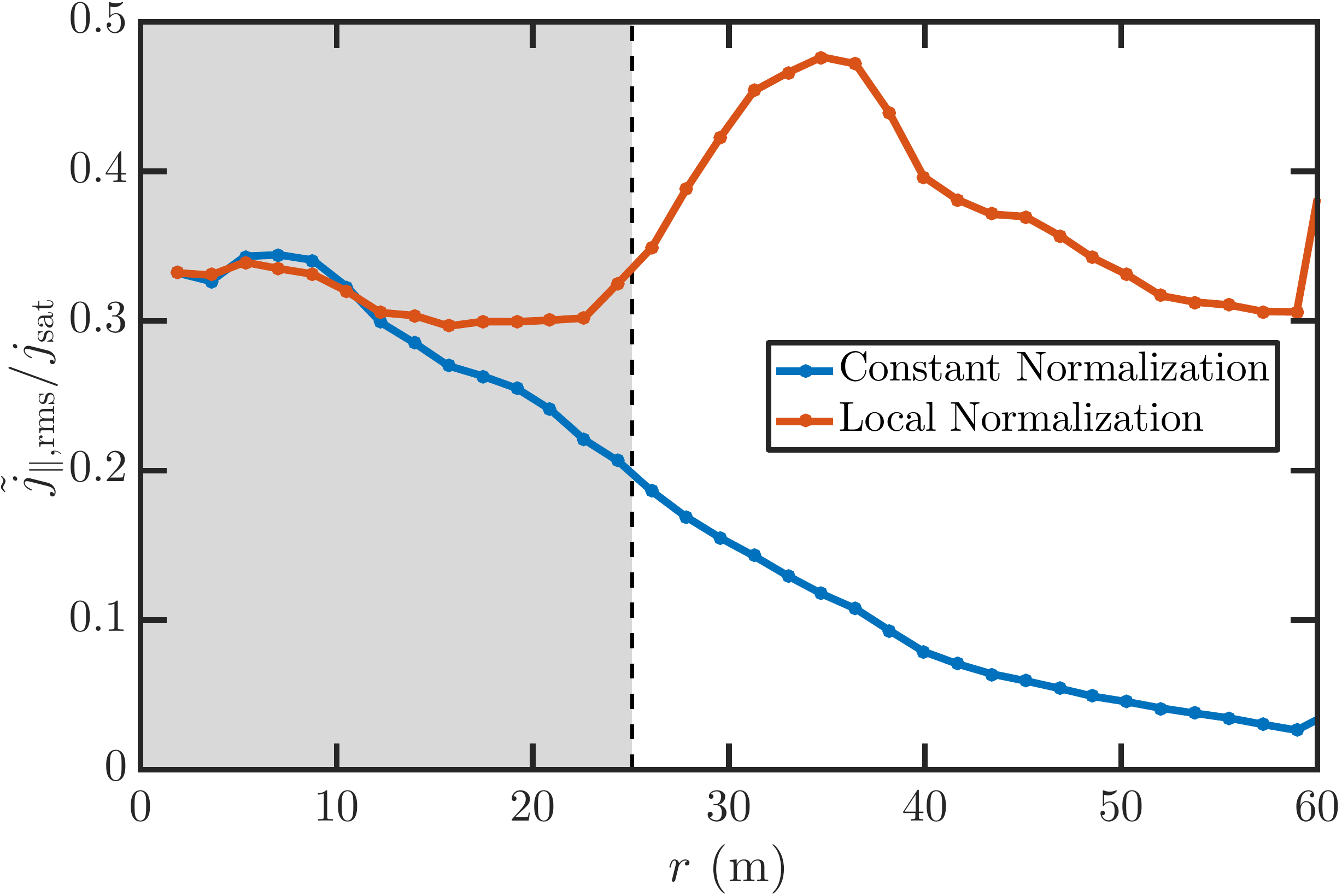}
  \caption[Radial profile of the r.m.s. current fluctuation amplitude at the sheath entrances.]
  {Radial profile of the r.m.s. current fluctuation amplitude at the sheath entrances, normalized
  to the on-axis peak value of $j_{\mathrm{sat}} = q_i n c_\mathrm{s} \approx 1300$~A~m$^{-2}$ (`constant normalization')
  and to the local value of $j_{\mathrm{sat}}$ (`local normalization').
  The large relative current fluctuation amplitudes permitted by conducting-sheath
  boundary conditions would not be present in a simulation that used insulating-sheath boundary conditions.}
\label{fig:lapd-current-fluct}
\end{figure}

Figure~\ref{fig:lapd-current-fluct} shows the r.m.s. current fluctuation level as a function of radius, measured
at the sheath entrances.
The current fluctuation amplitude is normalized to both
the on-axis peak value of $j_{\mathrm{sat}} = q_i n c_s \approx 1300$~A~m$^{-2}$ and to the 
local value of $j_{\mathrm{sat}}$.
Not shown is the mean total current at the sheath entrance, which has a peak value of approximately 100~A~m$^{-2}$.
The observed behavior is significantly different from the $j_\parallel=0$ condition that would be imposed
by insulating-sheath (logical-sheath) boundary conditions, and future work can investigate the impact of $j_\parallel = 0$ vs.
$j_\parallel \neq 0$ boundary conditions.
\citet{Thakur2013} investigated the use of both conducting and insulating end plates on the
controlled shear de-correlation experiment (CSDX), observing several changes in the turbulence characteristics.
Kelvin--Helmholtz modes driven by sheared azimuthal flows appear to be absent in
drift-reduced Braginskii fluid simulations using insulating-sheath boundary conditions \citep{Vaezi2017,Leddy2017}.

\section{Limiter-Biasing Simulations\label{sec:lapd_biasing}}
\begin{figure}
    \centering
    \subfloat{{\includegraphics[width=.4\linewidth]{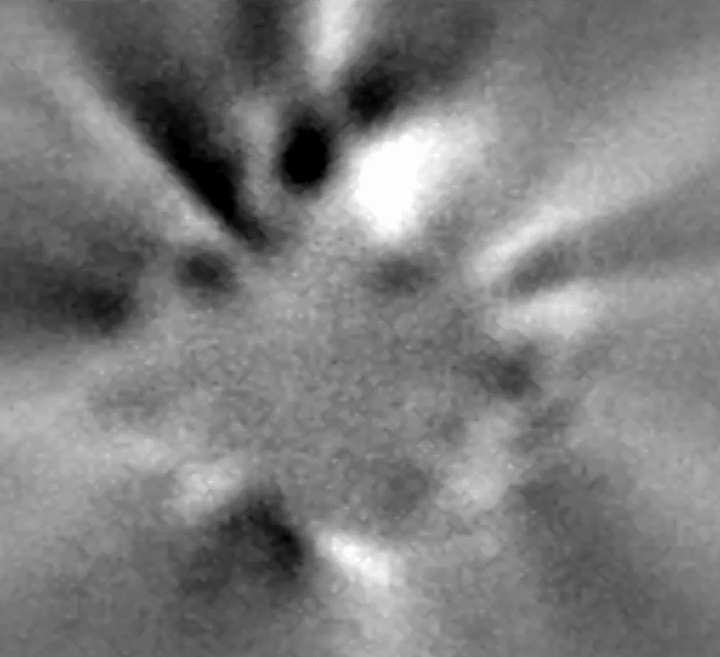} }}%
    \qquad
    \subfloat{{\includegraphics[width=.4\linewidth]{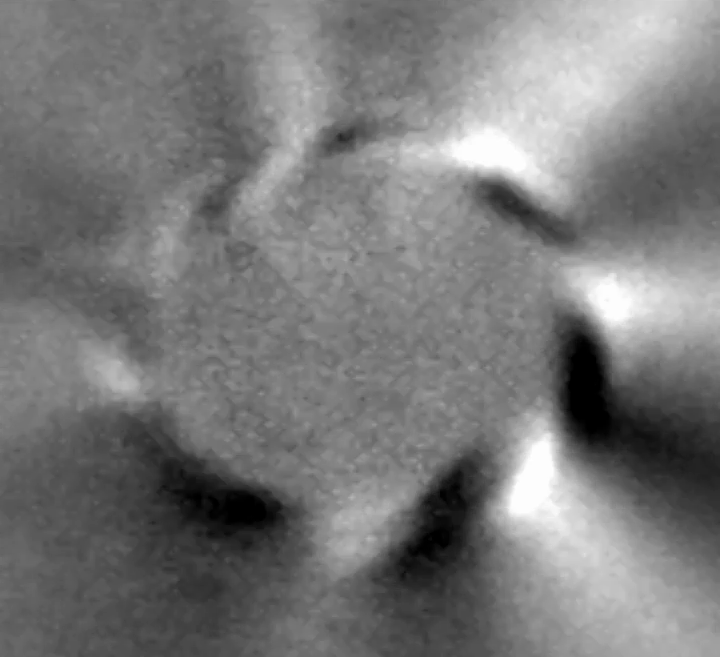} }}%
    \caption[Passive-imaging measurements of visible light 
    from a LAPD discharge before and after limiter biasing is turned on.]
    {Passive-imaging measurements of visible light 
    from a LAPD discharge at 0.05~T (left) before limiter biasing is turned on
    and (right) after limiter biasing is turned on.
    The measured light is due to emission from neutral Helium in the device.
    Qualitative changes in the turbulence are observed after limiter biasing is turned on,
    including a reversal of the plasma rotation direction for large positive bias voltages
    (relative to the source anode) and the development of a coherent mode in strongly biased cases.
    Note that the fast-framing camera, located at the end of the device downstream from the source,
    records light from the entire ${\sim}18$~m-long plasma column.
    These images are used with permission from T.\,Carter.}%
    \label{fig:lapd-visible-light}
\end{figure}

Without having to write additional \CC{} code, we can trivially modify these simulations of LAPD to
include a model of a biasable limiter.
In previous experiments on LAPD \citep{Schaffner2012,Schaffner2013,SchaffnerThesis2013},
an annular, aluminum limiter was installed that
intersected the magnetic field lines in the edge region, modifying the parallel boundary conditions
for $r > r_s$ while leaving the parallel boundary conditions in the core region ($r < r_s$) unaffected.
This limiter can be biased positive and negative relative to the anode of the plasma source, which modifies
the plasma potential in the edge region and consequently results in a $\partial_r \phi = -E_r$ profile 
that drives a poloidal $E \times B$ rotation near the limiter edge.
By using the biasable limiter to drive strong azimuthal flows and
flow shear near the limiter edge, \citet{Schaffner2012,Schaffner2013,SchaffnerThesis2013}
found that turbulent transport was suppressed.
Figure~\ref{fig:lapd-visible-light} shows visible light images of the LAPD plasma column
before and after (positive) limiter biasing in one of these experiments.
These experiments are related to tokamak experiments in the late 1980s 
in which electrically biased limiters \citep{Phillips1987} and
electrodes \citep{Taylor1989} were used to modify radial electric fields,
driving poloidal flows that consequently resulted in improvements to energy and particle confinement.

Flow shear in fusion plasmas is typically associated with a reduction in turbulence levels.
One popular theory to explain this observation is that a shear flow stretches and distorts
a turbulent eddy, shortening the eddy lifetime (eddy turnover time) by causing the eddy
to reach the eddy coherence length faster than in the no-shear-flow case \citep{Terry2000}.
This enhanced eddy decorrelation rate results in lower turbulence intensity levels.
In a fusion device, a localized region of flow shear that suppress cross-field turbulent transport
is referred to as a \textit{transport barrier}.
An edge-region transport barrier is believed to responsible for the
attractive confinement properties of the H-mode (high-confinement mode) in tokamaks \citep{Wagner1982,Wagner2007},
and advanced operational regimes for ITER rely on the formation and control of transport barriers \citep{Gormezano2007}.
A better understanding of how externally driven and spontaneously generated sheared flows
interact with turbulence in different parameter regimes is important to improving plasma confinement.

As discussed in the previous section, spontaneous azimuthal rotation of the LAPD plasma in
the ion-diamagnetic-drift (IDD) direction is observed in both simulations \citep{Shi2017,Fisher2015}
and experiments \citep{Schaffner2012,Schaffner2013,SchaffnerThesis2013} when the limiter is unbiased, which is
equivalent to setting the limiter model described in this section to $\phi_\mathrm{bias} = 0$~V.
This edge flow rotation is strongest just outside the limiter edge.
By biasing the limiter positive (relative to the anode), \citet{Schaffner2012,Schaffner2013,SchaffnerThesis2013}
observed that the plasma rotation slows down and eventually reverses direction.
Plasma rotation in the IDD direction was enhanced for negative limiter bias.
A zero-shear-flow state with increased turbulent transport was also found at an intermediate
positive limiter bias, and large flow shear in either direction was associated
with reduced levels of turbulent transport.
\citet{Schaffner2013,SchaffnerThesis2013} later compared the experimental measurements to some shear-suppression models
by using power-law fits and found limited agreement.
Only very recently were the first Braginskii fluid simulations of limiter biasing in LAPD performed
\citep{Fisher2017}.

In these simulations, we model limiter biasing by modifying the parallel boundary
condition on each end of the simulation domain in the same way for simplicity and symmetry,
and so we deviate from the LAPD experiment in this regard.
The actual LAPD experiment only has a single biasable limiter at one end of the device,
which is located 2.1~m from the cathode 1.8~m from the anode.
An electrically floating conducting end mesh is located at the other end of the device
in the experiment \citep{Schaffner2012,Gekelman2016}.
The ability to bias only one end plate while leaving the other
end plate grounded is already supported, but we have not yet investigated this case.
\citet{Fisher2017} explored both limiter-biasing configurations using the GBS code and observed
flow reversal in the electron-diamagnetic-drift (EDD) direction for large negative
bias only when both end plates were biased, which is a key feature observed in the experiments.

\begin{figure}
  \centerline{\includegraphics[width=\textwidth]{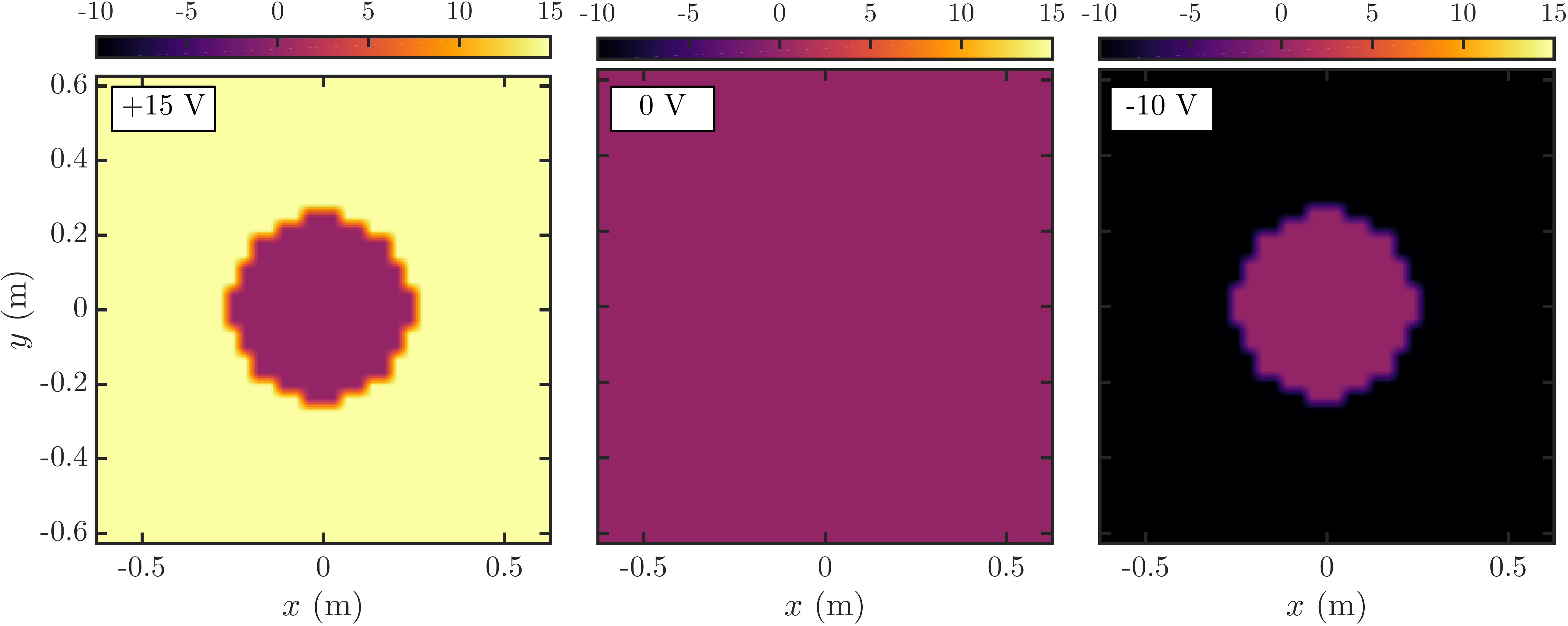}}
  \caption[The $\phi_w(x,y)$ fields used in the sheath-model boundary conditions
  to model limiters biased at three different voltages.]
  {The $\phi_w(x,y)$ fields used in the sheath-model boundary conditions to model
  limiters biased with $V_\mathrm{bias} = +15$~V, 0~V, and -10~V.
  These fields modify the sheath potential drop in the boundary conditions applied at
  each end of the simulation domain in the parallel direction.
  We assume that the limiter has an aperture of radius $r_s = 0.25$ m, so $\phi_w$ is zero for $r < r_s$ because
  these innermost field lines always terminate on a grounded end plate and not on the limiter.
  The jagged nature of the boundary between the $\phi_w = 0$ and $\phi_w = V_\mathrm{bias}$ regions
  is due to the use of a Cartesian grid.}
  \label{fig:lapd_phi_bias}
\end{figure}

\begin{figure}
  \centerline{\includegraphics[width=\textwidth]{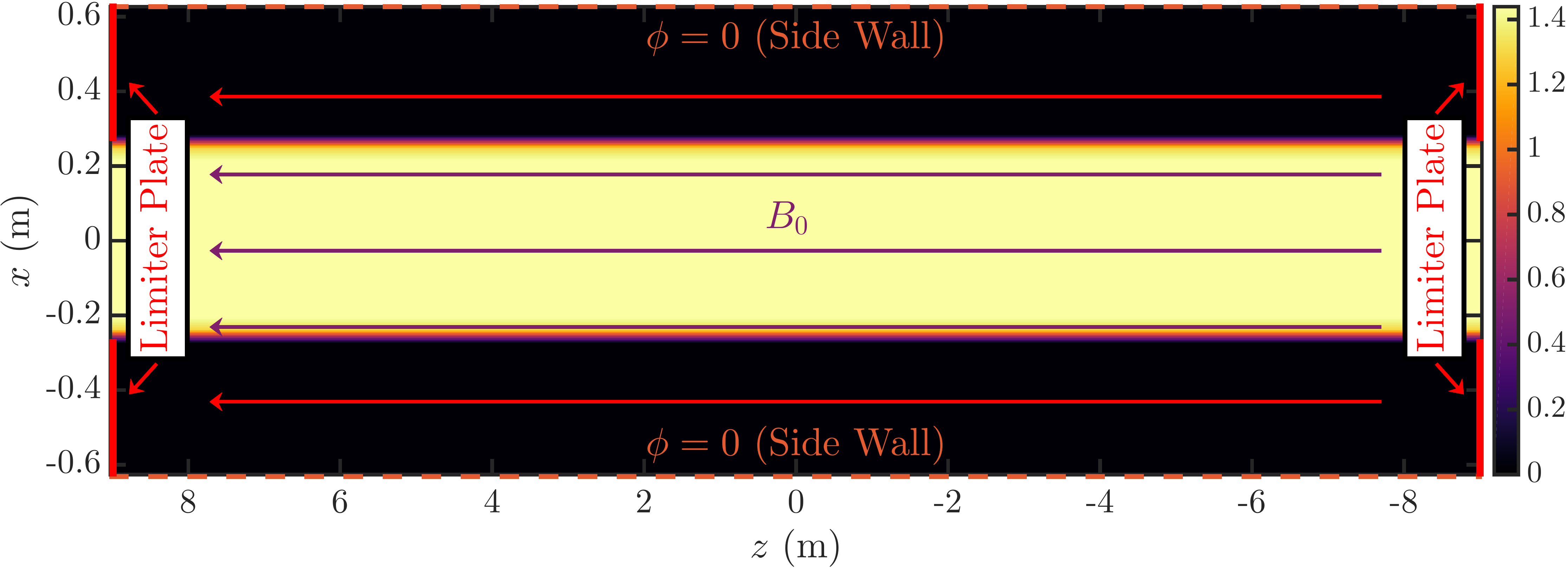}}
  \caption[LAPD-simulation density source rate in the $x$--$z$ plane with additional
  annotations to indicate biasable-limiter locations.]{LAPD-simulation
  density source rate (in 10$^{21}$~m$^{-3}$~s$^{-1}$) in the $x$--$z$ plane at $y = 0$ m with additional
  annotations to indicate biasable-limiter locations. Sheath-model boundary conditions
  are still applied at the entire upper and lower boundaries in $z$.
  The field lines in red are in contact with the biasable limiter, while the field lines in purple
  terminate on grounded end plates.}
\label{fig:lapd-bias-source}
\end{figure}

As discussed in Section \ref{sec:sheath}, the sheath potential drop at a point located
on the parallel boundaries is
\begin{equation}
  \Delta \phi(x,y) = \phi_{sh}(x,y) - \phi_w(x,y),
\end{equation}
where $\phi_{sh}$ is the potential at the sheath entrance and $\phi_w$ is the wall potential, which
was set to $0$ for the simulation reported in the previous section to model grounded end plates.
We model the effect of the limiter through the use of a non-zero $\phi_w(x,y)$ field that
has the form
\begin{equation}
  \phi_w(x,y) =
\begin{cases}
  0 & \text{if } \sqrt{x^2+y^2} < r_s,\\
  V_{\mathrm{bias}} & \text{else.}
\end{cases}
\end{equation}
Here, we investigate the use of bias voltages $V_{\mathrm{bias}} = -10$~V and $V_{\mathrm{bias}} = +15$~V,
and we compare these results to the unbiased simulation with $V_{\mathrm{bias}}=0$~V.
These two bias voltages were chosen because they resulted in large-amplitude sheared flows in the
experiment \citep{Schaffner2012,Schaffner2013,SchaffnerThesis2013}.
The corresponding $\phi_w(x,y)$ fields are shown in figure~\ref{fig:lapd_phi_bias}.
By applying a bias voltage to the limiter, we cause the electrostatic potential in the edge region
(where the plasma is on magnetic field lines that terminate on the limiter plates)
to increase or decrease relative
to the values in the unbiased case, and large $E \times B$ flows can be induced for large
enough $V_\mathrm{bias}$.
Figure~\ref{fig:lapd-bias-source} shows the locations where the limiter plates
modify the sheath-model boundary conditions in an $x$--$z$ cut of the simulation domain.
As in figure~\ref{fig:lapd_source}, the plasma source, side-wall boundary conditions, and magnetic
field direction are also indicated.

To save computational time in these simulations, we take initial conditions
from the steady state of the unbiased LAPD simulation described in Section \ref{sec:lapd_unbiased}.
Therefore, we must use the same parameters as in the unbiased LAPD simulations described in
table~\ref{tab:lapd_grid}..
The experiments performed by \citet{Schaffner2012,Schaffner2013,SchaffnerThesis2013} have a higher
peak electron temperature (8~eV), stronger magnetic field (0.1~T), and lower plasma density
($2\times 10^{18}$~m$^{-3}$).
For each of the three limiter bias voltages, we run a simulation for at least $1.5$~ms and
use data from the last $1$~ms of the simulation in our analysis.
The limiter biasing is applied instantaneously from $t=0$~s, so we do not ramp up $V_\mathrm{bias}$ from $0$~V
to the desired value over some specified duration to model the circuit response in the real experiment.
In the actual limiter-biasing experiments \citep{Schaffner2012,Schaffner2013,SchaffnerThesis2013},
the limiter was biased for ${\approx}$5~ms during the ${\approx}15$~ms discharge

\begin{figure}
  \centerline{\includegraphics[width=\textwidth]{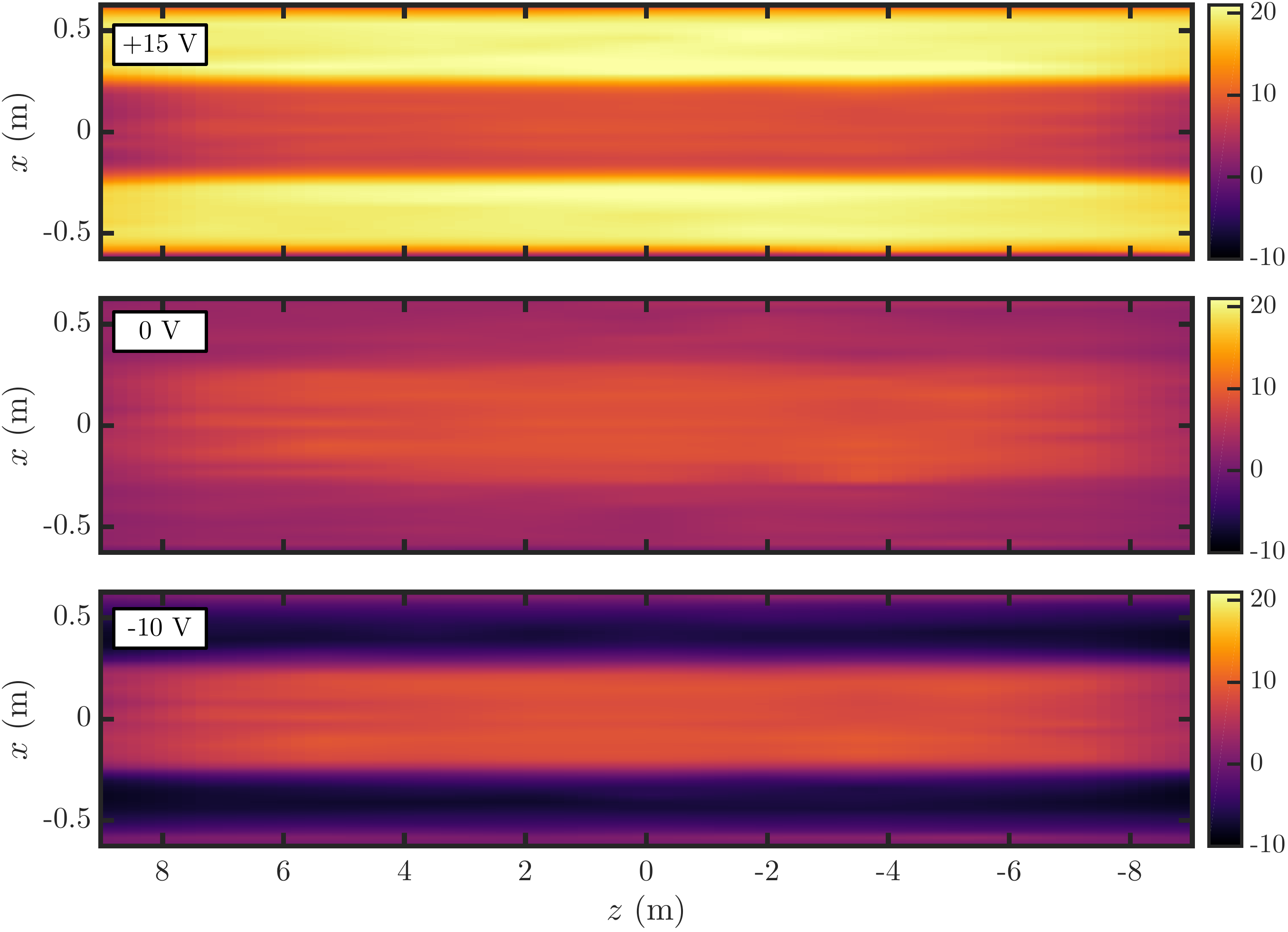}}
  \caption[Snapshots of the electrostatic potential in three LAPD simulations with
  different values of limiter biasing.]
  {Snapshots of the electrostatic potential (in V) in three LAPD simulations, which have
  +15~V limiter biasing, 0~V limiter biasing (grounded limiter), and -10~V limiter biasing.
  Limiter biasing modifies the potential in the edge region while leaving the potential in the core
  region relatively unchanged, which results in a poloidal $E \times B$ rotation
  near the limiter edge at $r_s = 20 \rho_{\mathrm{s}0} = 0.25$ m.
  All three simulations start from the same initial condition taken from an unbiased LAPD simulation
  that has reached a quasi-steady state, and the snapshots are taken at $t=100$ $\mu$s.
  The plots are made in the $x$--$z$ plane at $y=0$ m.}
  \label{fig:lapd_phi_bias_xz}
\end{figure}

Due to the rapid parallel response of the electrons, we see that the plasma potential adjusts to the limiter
biasing on a time scale on the order of the electron transit time, which is ${\approx}84$~$\mu$s
for electrons in the colder edge region ($T_e = 1.2$~eV).
Figure~\ref{fig:lapd_phi_bias_xz} shows the parallel structure of the electrostatic potential
100~$\mu$s from the start of the limiter biasing simulations.
We see that the value of the potential in the core region ($r<r_s$) is nearly the same regardless
of $V_\mathrm{bias}$, but the potential in the edge region self-consistently adjusts to stay a few
$T_e$ above $V_\mathrm{bias}$ to keep the electron and ion fluxes to the end plates approximately in balance.

\begin{figure}
  \centerline{\includegraphics[width=\textwidth]{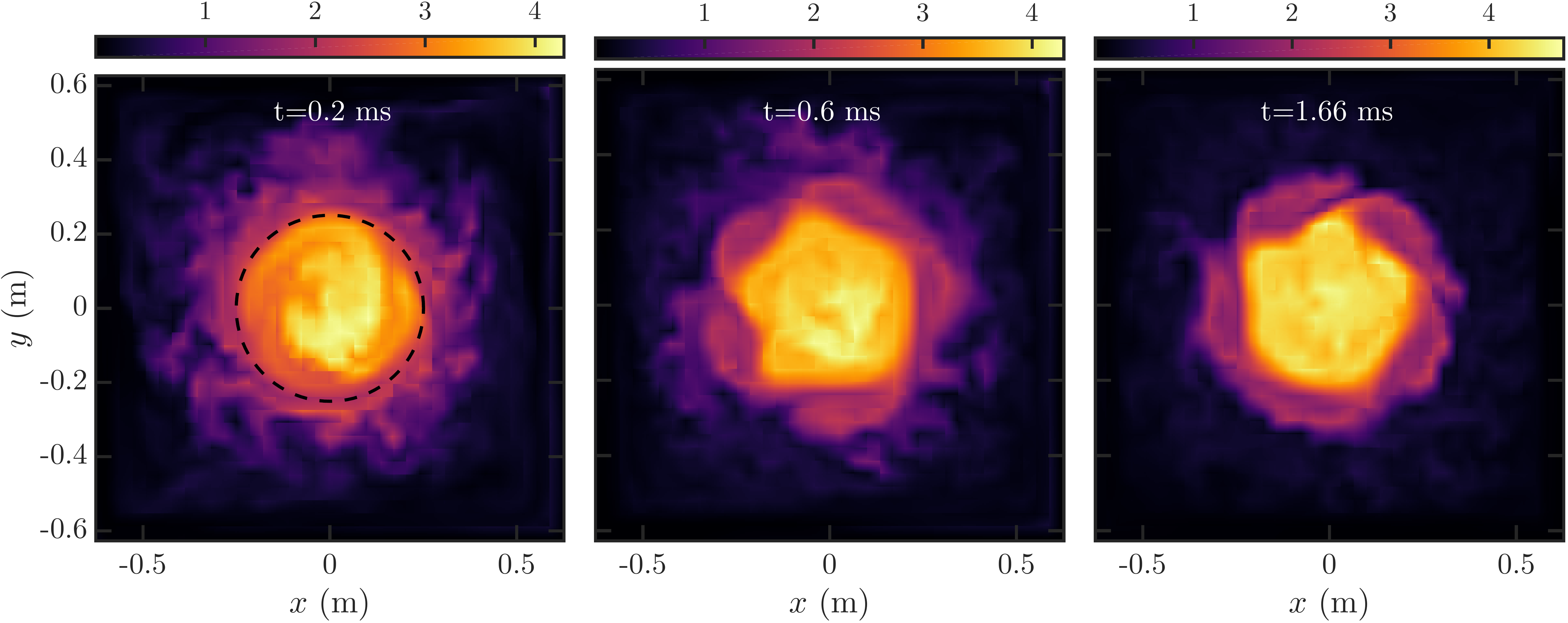}}
  \caption[Snapshots of the electron density in the $x$--$y$ plane at for a simulation
  of an LAPD plasma with +15~V limiter biasing.]
  {Snapshots of the electron density in the $x$--$y$ plane at $z=0$ m for a simulation with
  +15~V limiter biasing at $t=0.2$~ms, $t=0.6$~ms, and $t=1.66$~ms.
  The plasma immediately reverses its rotation direction and radial particle transport
  is suppressed. Eventually, a coherent mode localized to the limiter edge develops and persists for the rest of
  the simulation. The limiter edge, which is the boundary between the core and edge regions, is indicated
  by the dashed line in the $t=0.2$~ms plot. Note that each plot uses a different color scale to better
  show the features.}
  \label{fig:lapd_bias_evolution}
\end{figure}

Figure~\ref{fig:lapd_bias_evolution} shows the time evolution of the electron density viewed in the $x$--$y$
plane at $z=0$~m for the simulation with $V_\mathrm{bias} = +15$~V.
The plasma immediately reverses rotation and a strong sheared flow in the EDD direction is observed
while the density in the core rises due to a decrease in radial particle transport.
At $t=0.6$~ms, we see the emergence of a coherent mode of mode number localized to the limiter edge
with mode number $m = 5$.
At much later times, we see that this coherent mode persists, while the radial transport of density to regions
not in the immediate vicinity of the limiter edge is extremely small.
In the simulation with $V_\mathrm{bias} = -10$~V, the rotation in the IDD direction is enhanced
and a coherent mode with $m = 6$ develops.
Coherent modes were observed in cases with high shearing rates on LAPD, although
the mode number values were not described in \citet{Schaffner2012} or in \citet{Schaffner2013}.
In drift-reduced Braginskii fluid simulations, $m = 6 \pm 1$ was reported for the $V_\mathrm{bias} = +18$~V
case investigated by \citet{Fisher2017}.
\citet{Schaffner2012,SchaffnerThesis2013} proposed that this mode is caused by the Kelvin--Helmholtz instability or
the rotational-interchange instability.

\begin{figure}
  \centerline{\includegraphics[width=\textwidth]{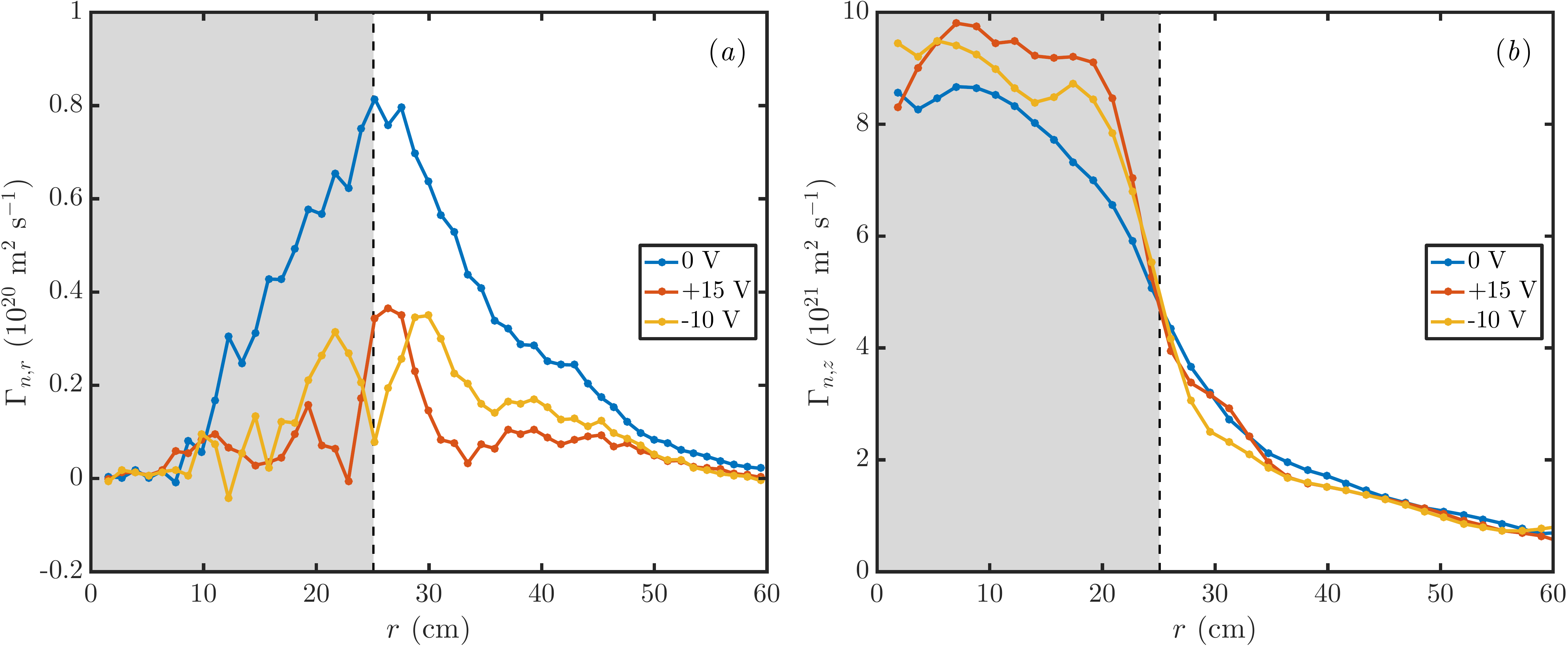}}
  \caption[Comparison of the time-averaged radial particle fluxes due to $E \times B$ fluctuations
  and the time-averaged total outward parallel particle fluxes for simulations with different values
  of limiter biasing.]
  {Comparison of the ($a$) time-averaged radial particle fluxes $\Gamma_{n,r}$ due to $E \times B$ fluctuations
  and ($b$) the time-averaged total outward parallel particle fluxes $\Gamma_{n,z}$
  for simulations with different values of limiter biasing.
  The shaded region illustrates the core region inside the limiter edge at $r=r_s$.}
  \label{fig:lapd-biased-flux}
\end{figure}

Figure~\ref{fig:lapd-biased-flux}($a$) shows a reduction in the time-averaged
radial $E \times B$ particle flux in the simulations with applied limiter biasing.
The peak radial particle flux is reduced by approximately 50\% in the biased simulations,
and the radial region over which there is significant radial particle flux is also much narrower
in the biased simulations.
Figure~\ref{fig:lapd-biased-flux}($b$) shows that the time-averaged total outward parallel particle fluxes
in the simulations with $+15$~V and $-10$~V limiter biasing are enhanced in the
core, which is consistent with the decreased radial particle transport in the biased simulations. 
These trends are also observed in LAPD experiments \citep[see][figure 3$(c)$]{Schaffner2013}.
The reduction in the cross-field turbulent fluxes result in somewhat steeper density profiles.

\begin{figure}
  \centerline{\includegraphics[width=\textwidth]{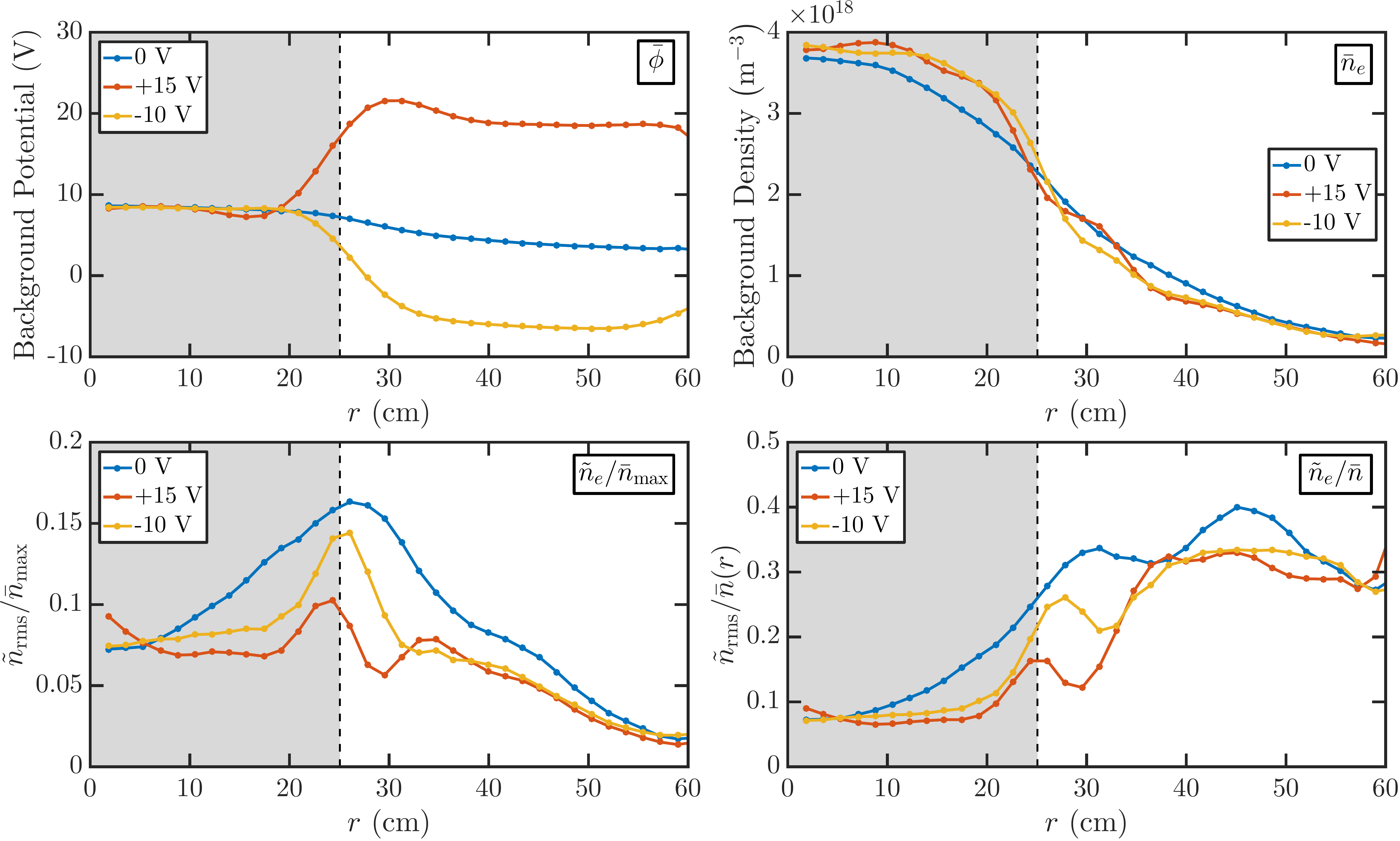}}
  \caption[Comparison of the radial profiles of the background electrostatic potential, background electron density,
  and r.m.s. electron-density fluctuation levels for three LAPD simulations
  with different values of limiter biasing.]
  {Comparison of the radial profiles of the background electrostatic potential, background electron density,
  and r.m.s. electron-density fluctuation levels (normalized to both the peak background
  electron density and to the local background electron density) for three LAPD simulations,
  which have 0~V, +15~V and -10~V limiter biasing. The core region whose field lines always
  terminate on a grounded end plate is shaded in gray, while the field lines in the unshaded edge region
  terminate on a biasable limiter.
  The background potential is significantly modified
  in the biased cases, which results in strong $E \times B$ azimuthal flows near the limiter edge.
  This sheared flow suppress radial particle transport, which results in an elevated density
  in the core region and a reduced density in the edge region. The density fluctuation levels
  are also lower in the vicinity of the limiter edge in the biased cases.}
  \label{fig:lapd_bias_comp_1d}
\end{figure}

Figure~\ref{fig:lapd_bias_comp_1d} shows radial profiles of the mean potential, mean density,
and r.m.s. electron-density fluctuation level (normalized to the peak on-axis density for the 0 V case)
for the three simulations with different values of the limiter bias voltage.
These plots are made using data in the same $-4$ m $< z <$ 4 m region that was used for
the radial profile plots in Section \ref{sec:lapd_unbiased}.
In the plot of the mean potential (figure~\ref{fig:lapd_bias_comp_1d}($a$)),
we see that the potential in the core region is nearly identical for the three cases except
near the limiter edge, where steep radial gradients in the potential develop in the $+15$ V and $-10$ V cases.
In the edge region, the gradients in the potential are small away from the limiter edge.
From this plot, we that the large, sheared azimuthal flows that are observed in the 
$+15$~V and $-10$~V cases is due to $E \times B$ motion of the plasma.
Consistent with a reduction in radial particle transport,
we see that mean electron density in the core region slightly increases, while
the density at large $r$ decreases when comparing the two biased cases with the unbiased case.
The electron-density fluctuation level is also lower for the $+15$ V and $-10$ V cases, although
the fluctuation level for all three cases peaks at the limiter edge location, which is likely
due to the coherent mode for the biased cases.
The reduction of low-frequency density fluctuation amplitudes in the biased cases
are believed to the primary factor in the reduction of turbulent radial particle flux
in the limiter biasing experiments on LAPD \citep{Schaffner2012}.

We now consider the radial-particle-flux reduction observed in the biased simulations
in more detail.
The time-averaged radial $E \times B$  particle flux $\langle \Gamma_{n,r} \rangle$
can be spectrually decomposed into the amplitudes, cross phase,
and cross coherency of density and potential fluctuations
\citep{Powers1974,Carter2009}.
We define the Fourier-transform pair
\begin{align}
  f(t) &= \frac{1}{\sqrt{2 \pi}} \int_{-\infty}^{\infty} \mathrm{d}\omega \, \hat{f}(\omega) e^{-i \omega t},\\
  \hat{f}(\omega) &= \frac{1}{\sqrt{2 \pi}} \int_{-\infty}^{\infty} \mathrm{d}t \, f(t) e^{i \omega t},
\end{align}
and the complex-valued cross-power spectrum
\begin{equation}
  P_{1,2}(\omega) = \hat{f}_1^*(\omega) \hat{f}_2 (\omega) = | P_{1,2}(\omega) | e^{i \alpha_{1,2}(\omega)}.
\end{equation}
Noting that $\boldsymbol{v}_E \cdot \hat{\boldsymbol{r}}= E_\theta/B$, we have
\begin{align}
  \int_{-\infty}^{\infty} \mathrm{d}t \, \tilde{n} \tilde{v}_E &=
    \frac{1}{B}  \int_{-\infty}^{\infty} \mathrm{d}t \, \tilde{n} \tilde{E}_\theta \\
    &= \frac{1}{B} \int_{-\infty}^{\infty} \mathrm{d}t
      \left[ \frac{1}{\sqrt{2\pi}} \int_{-\infty}^{\infty} \mathrm{d}\omega \, \hat{n}(\omega) e^{-i\omega t} \right]
      \left[ \frac{1}{\sqrt{2\pi}} \int_{-\infty}^{\infty} \mathrm{d}\omega' \, \hat{E}_\theta(\omega) e^{-i\omega' t} \right]\\
    &= \frac{1}{2\pi B} \int_{-\infty}^{\infty} \mathrm{d} \omega \, \hat{n}(\omega)
      \int_{-\infty}^{\infty} \mathrm{d} \omega' \, \hat{E}_\theta(\omega')
      \int_{-\infty}^{\infty} \mathrm{d} t \, e^{-i(\omega + \omega')t} \\
    &= \frac{1}{B} \int_{-\infty}^{\infty} \mathrm{d} \omega \, \hat{n}(\omega)
      \int_{-\infty}^{\infty} \mathrm{d} \omega' \, \hat{E}_\theta(\omega') \delta(\omega + \omega')\\
    &= \frac{1}{B} \int_{-\infty}^{\infty} \mathrm{d} \omega' \, \hat{n} (-\omega') \hat{E}_\theta (\omega') \\
    &= \frac{1}{B} \int_{-\infty}^{\infty} \mathrm{d} \omega' \, \hat{n}^* (\omega') \hat{E}_\theta (\omega'),
\end{align}
where the last equality holds because $\tilde{n}$ is real.
Furthermore, we can write
\begin{align}
  \int_{-\infty}^{\infty} \mathrm{d}t \, \tilde{n} \tilde{v}_E &=
  \frac{1}{B} \int_{0}^{\infty} \mathrm{d} \omega' \left( \hat{n}^* (\omega') \hat{E}_\theta (\omega') + \hat{n}^* (-\omega') \hat{E}_\theta (-\omega') \right)\\
  &= \frac{1}{B} \int_{0}^{\infty} \mathrm{d} \omega' \left( \hat{n}^* (\omega') \hat{E}_\theta (\omega') + \hat{n} (\omega') \hat{E}_\theta^* (\omega') \right)\\
  &= \frac{2}{B} \int_{0}^{\infty} \mathrm{d} \omega' \operatorname{Re} \left( \hat{n}^* (\omega') \hat{E}_\theta (\omega') \right) \\
  &= \frac{2}{B} \int_{0}^{\infty} \mathrm{d} \omega \operatorname{Re} \left( P_{n,E} \right) \\
  &= \frac{2}{B} \int_{0}^{\infty} \mathrm{d} \omega \left| P_{n,E}(\omega) \right| \cos \boldsymbol{(} \alpha_{nE}(\omega) \boldsymbol{)} \label{eq:particle_flux_total} \\
  &= \frac{2}{B} \int_{0}^{\infty} \mathrm{d} \omega \left| \hat{E}_\theta(\omega)\right|
    \left| \hat{n}(\omega)\right| \left| \gamma_{n,E}(\omega) \right|
    \cos \boldsymbol{(} \alpha_{n,E}(\omega) \boldsymbol{)} \label{eq:particle_flux_sep}
\end{align}
We expanded the integrand in (\ref{eq:particle_flux_total}) into
four terms in (\ref{eq:particle_flux_sep}): the amplitude spectral densities
of the electric-field and density fluctuations $\left| \hat{E}_\theta(\omega)\right|$ and
$\left| \hat{n}(\omega)\right|$, the \textit{coherence spectrum} $\left| \gamma_{n,E}(\omega) \right|$,
and the cosine of the \textit{cross phase} $\cos \boldsymbol{(} \alpha_{n,E}(\omega) \boldsymbol{)}$.

The cross-power spectrum $P_{n,E}(\omega)$ is first computed at each node as:
\begin{equation}
P_{n,E}(\omega) = \hat{n}_e^* \hat{E}_\theta,
\end{equation}
where $\hat{n}_e(\boldsymbol{R},\omega)$ and $\hat{E}_\theta(\boldsymbol{R},\omega)$ are the Fourier transforms of the time series of
$\tilde{E}_\theta (\boldsymbol{R},t)$ and $\tilde{n}_e(\boldsymbol{R},t)$.
The cross-power spectrum is then spatially averaged, and the cross phase is computed as
\begin{equation}
  \alpha_{n,E}(\omega) = \mathrm{Im} \log \boldsymbol{(} \langle P_{n,E}(\omega) \rangle \boldsymbol{)},
\end{equation}
where $\langle \dots \rangle$ denotes a spatial average in the region $20$ cm $<r<30$ cm and $-4$ m $<z<4$ m.
The coherence spectrum is defined as \citep{Powers1974}
\begin{equation}
| \gamma_{n,E}(\omega) | = \frac{| \langle P_{n,E}(\omega) \rangle | }
  { \langle P_{n,n}(\omega) \rangle^{1/2} \langle P_{E,E}(\omega) \rangle^{1/2} } \label{eq:coherence},
\end{equation}
where $P_{n,n} = \left| \hat{n}(\omega)\right|^2$ and $P_{E,E} = \left| \hat{E}_\theta(\omega)\right|^2$
are the real-valued power spectra of $\tilde{n}_e$ and $\tilde{E}_\theta$, respectively.

\begin{figure}
  \centerline{\includegraphics[width=\textwidth]{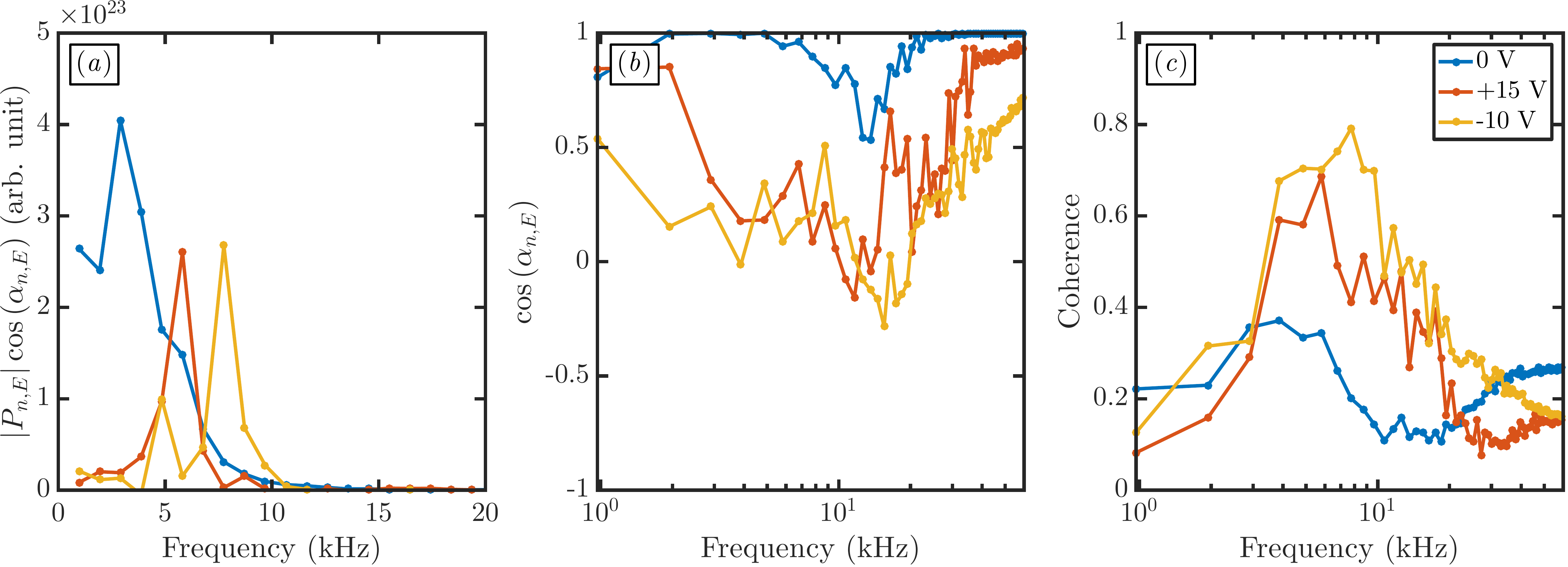}}
  \caption[Comparison of the total radial-particle-flux integrands, cosine of cross phases,
  and coherence spectra for three LAPD simulations with different values of limiter biasing.]
  {Comparison of the ($a$) total radial-particle-flux integrands (\ref{eq:particle_flux_total}),
  ($b$) cosine of cross phases, and ($c$) coherence spectra (\ref{eq:coherence}) for three LAPD simulations,
  with different values of limiter biasing. The primary cause of the reduction of radial particle flux
  in this region ($20$ cm $<r<30$ cm) in the biased simulations (see figure~\ref{fig:lapd-biased-flux})
  appears to be a reduction in $n_e$ and $E_\theta$ fluctuation amplitudes at low frequencies
  (see figure~\ref{fig:lapd-bias-power-density}), 
  as opposed to large changes in the cross phase or coherence.}
  \label{fig:lapd-bias-phase-and-coherence}
\end{figure}

Figure~\ref{fig:lapd-bias-phase-and-coherence} shows plots of the total radial-particle-flux integrand
(\ref{eq:particle_flux_total}), cosine of the cross phase, and the coherence spectrum
for the unbiased and biased simulations.
In figure~\ref{fig:lapd-bias-phase-and-coherence}$(a)$, we see that that frequencies higher than
${\sim}10$~kHz do not contribute to the integral for both unbiased and biased cases, which is due
to the exponential decay at high frequencies in the cross-power amplitude spectrum.
We see in figure~\ref{fig:lapd-bias-phase-and-coherence}$(b)$ that the cross phase in the biased cases
is less favorable for outward radial transport when compared to the cross phase in the unbiased case, while
figure~\ref{fig:lapd-bias-phase-and-coherence}$(c)$ shows that the coherence in the biased cases is more favorable for
outward radial transport.
The net effect, however, appears to be a reduction in the outward radial particle flux in the biased cases.

The results for the unbiased simulation is similar to the spectra measured in LAPD
\citep[see][figure 10]{Carter2009} at frequencies below 10~kHz,
where the fluctuation levels are the strongest as shown in the power spectrum (see figure~\ref{fig:rmsDensity}$(c)$).
At these low frequencies, \citet{Carter2009} report a cross phase that is ${\approx}0$ and a cross coherency
that is ${\approx}0.2$--$0.6$ for the unbiased case.
We note that there is a questionable feature in the low-frequency ($<10$~kHz) components
of the cross phase and coherence of the unbiased simulation
in figures~\ref{fig:lapd-bias-phase-and-coherence}($b$) and ($c$).
In a frequency window where the coherence is relatively low ($<0.5$), the cross phase
is not well defined, so we expect the corresponding cross phase in the same frequency window
to be unstable.
Instead, we see in figure~\ref{fig:lapd-bias-phase-and-coherence}($b$)
that the low-frequency cross phase is relatively flat for the unbiased case.
Similar results are obtained by using a 3 times larger time-sampling window
for the spectral decomposition.
We note that figure~10 in \citet{Carter2009} also shows a similar trend.
This feature could be due to errors in calculating $E_\theta$, both in the simulation and in the
experiment.

\begin{figure}
  \centerline{\includegraphics[width=\textwidth]{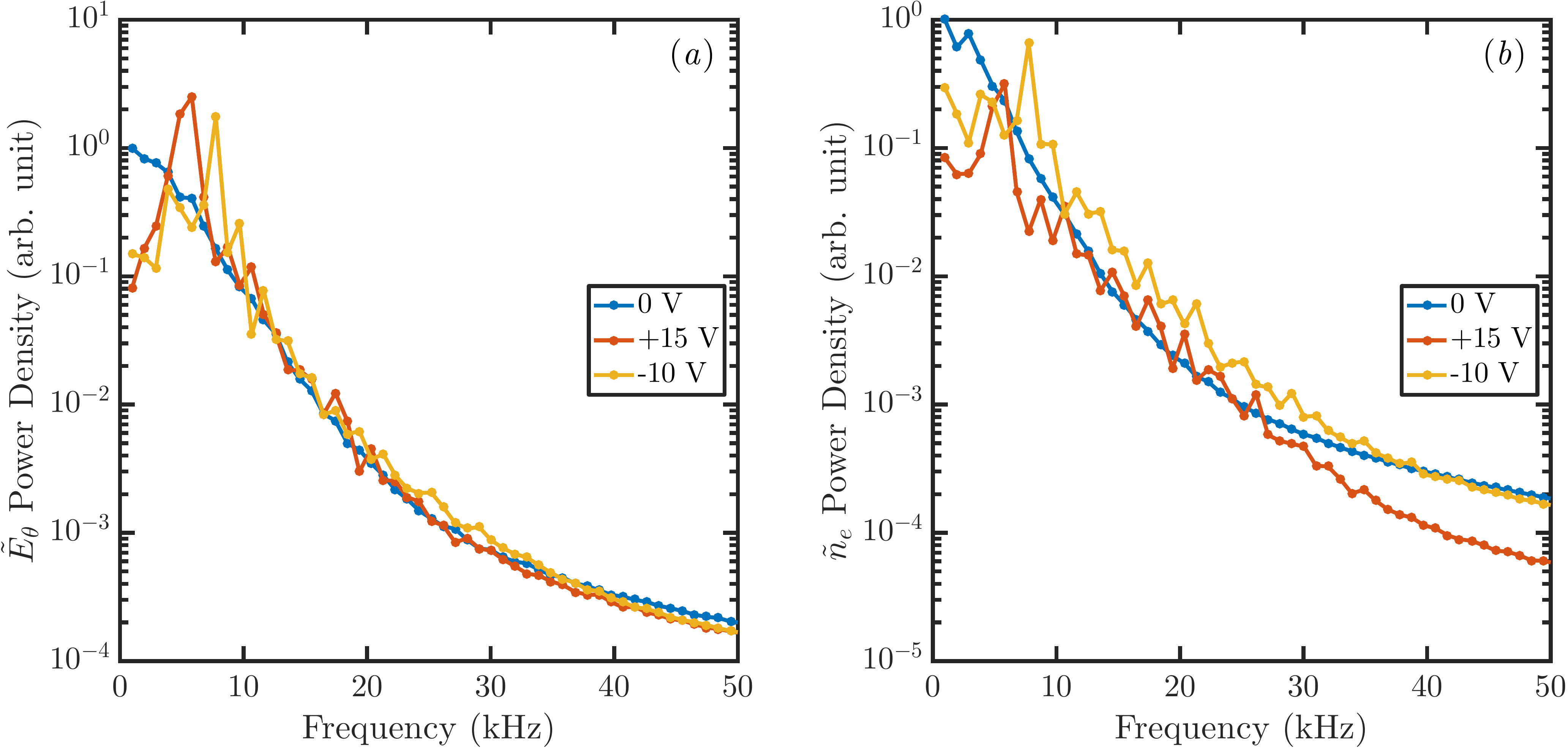}}
  \caption[Comparison of the electric-field-fluctuation power spectral densities and
  the density-fluctuation power spectral densities for three LAPD simulations
  with different values of limiter biasing.]
  {Comparison of the ($a$) electric-field-fluctuation power spectral densities and ($b$)
  the density-fluctuation power spectral densities for three LAPD simulations
  with different values of limiter biasing. In both plots, a peak from the coherent mode
  is observed in the spectra of the biased simulations (at ${\approx}5.8$~kHz for $+15$~V 
  and ${\approx}7.8$~kHz for $-10$~V).}
  \label{fig:lapd-bias-power-density}
\end{figure}

The power spectral densities of $E_\theta$ and $n_e$ fluctuations are shown in
figure~\ref{fig:lapd-bias-power-density}.
In the biased simulations, we see a peaking in the fluctuation power spectra
at ${\approx}5.8$~kHz for $+15$~V and ${\approx}7.8$~kHz for $-10$~V, which
are believed to be Kelvin--Helmholtz modes in the experiments
\citep{Schaffner2012,SchaffnerThesis2013}.
Biasing experiments on LAPD show a peak at ${\approx}12$~kHz in the
power spectral density of density fluctuations for a case in which
the limiter was biased at 13.1~V \citep[see][figure~6.9($b$)]{SchaffnerThesis2013}.

\section{Conclusions \label{sec:lapd-conclusion}}
We have presented results from the first 3D2V gyrokinetic continuum simulations of turbulence
in an open-field-line plasma.
The simulations were performed using a version of the Gkeyll code that employs an energy-conserving discontinuous
Galerkin algorithm.
We found it important to include self-species collisions in the electrons to avoid driving
high-frequency instabilities in our simulations.
Our gyrokinetic simulations are generally in good qualitative 
agreement with previous Braginskii fluid simulations of LAPD and with
previously published experimental data.

We use sheath-model boundary conditions for electrons that are a kinetic
extension of the sheath model used in past fluid simulations,
which allows self-consistent currents to flow into and out of the end plates.
In this approach, the sheath potential is determined from the gyrokinetic Poisson
equation (analogous to how the vorticity equation is used in the fluid approach
of \citet{Rogers2010}).
The ion boundary conditions used at present are the same as for the logical-sheath model,
in which ions flow out at whatever velocity they have been accelerated to at the sheath edge.
This boundary condition appears to work well for the time period of this LAPD simulation.
As discussed in Section~\ref{sec:sheath}, future work is planned to consider improved models of
a kinetic sheath, including the role of rarefaction dynamics near the sheath that may
modify the outflowing distribution function and the effective outflow Mach number.

We demonstrated in Section \ref{sec:lapd_biasing} that a biasable limiters could be modeled
through a simple modification to the sheath-model boundary conditions.
By biasing the limiter positive or negative relative to the grounded end plates,
a sheared flow in the azimuthal direction could be driven at the limiter edge.
Consistent with the experiments \citep{Schaffner2012,Schaffner2013,SchaffnerThesis2013},
we see that radial turbulent particle transport is suppressed
in the strongly biased simulations relative to the unbiased simulation.
Future work can explore more values for the limiter biasing and identify the bias voltage
that results in a zero-flow-shear state, in which the outward radial particle transport is maximized.

A number of possible modifications to the simulations could allow closer quantitative modeling of the
LAPD experiment to enable gyrokinetic simulation as a tool for planning LAPD experiments,
especially in the lower-collisionality regime $T_i \sim T_e \sim 10$~eV accessible
in the upgraded device \citep{Gekelman2016}.
In the real LAPD experiment, a cathode--anode discharge emits an energetic 40--60~eV electron beam that ionizes
the background gas along the length of the device
\citep{Leneman2006,Gekelman2016,Carter2009},
creating the plasma source that we have directly modeled in our simulations.
At present, we are ignoring the current from these energetic electrons
and modeling the anode as a regular conducting end plate.
Because the anode in the actual device is a semi-transparent mesh, there is finite pressure
on the other side of the anode from the main plasma that
can act to slow down ion outflows and thus relax the Bohm sheath criterion.

Since our simulations are kinetic, future work could include the non-Maxwellian high-energy electrons and a
model of the ionization process instead of using explicit source terms.
Future work could also investigate the primary mechanism driving the cross-field transport
observed in our simulations (such as linear drift-wave, nonlinear, or Kelvin--Helmholtz instabilities) by
analyzing the energy dynamics of the system \citep{Friedman2012,Friedman2013} and through
the use of presence/absence tests \citep{Rogers2010,Fisher2015}.
Additionally, the applicability of the current set of parallel boundary conditions
should be investigated more carefully.
It is possible that the assumption that all magnetic field lines terminate on grounded end plates
or a biasable limiter is not sufficiently accurate for modeling the LAPD experiment,
and a more realistic set of parallel boundary conditions that also allows
some field lines to terminate on the cathode or anode structures
could lead to qualitatively different results.
We also note in passing that the capability we have developed could also eventually find use
in simulating certain plasma-mass-filter concepts \citep{Ochs2017,Gueroult2016,Gueroult2014,Fetterman2011},
which direct ${\sim}10$~eV ions to be preferentially lost 
along open magnetic field lines (some of which terminate on a concentric set of 
biasable electrodes to drive plasma rotation) at different ion-mass-dependent locations.

We plan several improvements to our numerical algorithms.
The time step restriction in our LAPD simulations is currently set by the electron-electron collision frequency.
A super-time-stepping method, such as the Runge-Kutta-Legendre method \citep{Meyer2014}, or an implicit method
could significantly alleviate this restriction.
The use of non-polynomial basis functions \citep{Yuan2006} for efficient velocity-space discretization
is expected to reduce the computational cost of these simulations (by allowing for a coarser velocity-space grid)
and to preserve the positivity of the distribution function.
Future studies will also implement the full nonlinear ion polarization density in gyrokinetic Poisson equation (\ref{eq:gkp}),
which is related to removing the Boussinesq approximation in fluid models \citep{Dudson2015,Halpern2016}.
This modification requires replacing $n_{i0}^g$ in (\ref{eq:gkp}) with the full $n_i^g(\boldsymbol{R},t)$
and retaining a corresponding second-order contribution
to the Hamiltonian (\ref{eq:hamiltonian}) necessary for energy conservation \citep{Krommes2012,Krommes2013,Scott2010}

%% file: ch-simulations-of-helical-sol/chapter-simulations-of-helical-sol.tex
\chapter{Simulations of a Helical Scrape-Off-Layer Model \label{ch:helical-sol}}

The work presented in this chapter builds on on our previous work with flux-driven simulations of
open-magnetic-field-line turbulence in the Large Plasma Device \citep{Gekelman2016}
using the gyrokinetic continuum capabilities of the Gkeyll code (Chapter~\ref{ch:lapd}).
In those LAPD simulations, the magnetic field was straight and uniform, and the plasma was highly collisional,
which necessitated the use of an artificial electron-to-ion mass ratio ($m_i/m_e = 400$)
and reduced electron collision frequencies to make the simulations tractable.
Nevertheless, we found that our numerical approach based on discontinuous Galerkin methods
and sheath-model boundary conditions was stable and produced qualitatively reasonable results,
which led to the first demonstration of open-field-line turbulence with a gyrokinetic continuum code.
These restrictions are relaxed for the simulations presented in this chapter,
which now include a slightly more complex magnetic geometry.

We have added magnetic curvature and $\nabla \boldsymbol{B}$ drifts to the code and can simulate a
helical magnetic geometry approximating that in simple magnetized tori (SMTs), such as TORPEX \citep{Fasoli2006}
and Helimak \citep{Gentle2008}.
In tokamaks, probe and imaging diagnostics have revealed the existence of
intermittent coherent structures in the SOL referred to as \textit{plasma filaments} or \textit{blobs}, which
convectively transport particles, heat, momentum, and current across magnetic field lines \citep{DIppolito2011}.
Blob are characterized by densities that are much higher than local background levels,
a structure that is highly elongated along the magnetic field (much larger than the plasma radius),
and much smaller scales perpendicular to the magnetic field ${\sim}10 \rho_i$,
where $\rho_i$ is the ion gyroradius \citep{DIppolito2011,Zweben2007}.

The curvature and $\nabla \boldsymbol{B}$ forces are believed to set up a charge-separated
dipole potential structure across the blob cross-section that results in its outward radial propagation via
convective $E\times B$ transport \citep{Krasheninnikov2001,DIppolito2011}.
Finite-temperature effects of the blob can also cause spin motion if the blob is sheath-connected, which can
reduce this radial motion \citep{Myra2004}.
Numerically, blob dynamics have been studied using seeded-blob fluid simulations \citep{Angus2012,Riva2016,Walkden2015,Shanahan2016},
although recent turbulence simulations observe self-consistent blob formation \citep{Churchill2017,Ricci2013}.
In contrast to previous work on the simulation of turbulence in SMTs based on the cold-ion drift-reduced
Braginskii equations \citep{Ricci2008,Ricci2009a,Li2011},
we employ a gyrokinetic approach that allows us to investigate plasmas with
$T_i \gtrsim T_e$, which is commonly observed in the scrape-off layer (SOL) \citep{Boedo2009,Kocan2011,Kocan2012}.

Although our simulations do yet not simultaneously contain open and closed-field-line regions
\citep{Ribeiro2008,Zweben2009,Halpern2016,Dudson2017,Zhu2017,Li2017},
we believe that many basic properties of SOL turbulence and transport are reproduced in this
open-field-line model.
Additionally, the turbulence in this helical open-field-line geometry with parameters appropriate for 
a tokamak SOL has not been previously studied using a gyrokinetic PIC approach, either.
We do acknowledge, however, that gyrokinetic PIC codes that have the necessary capabilities for
the problem described in this chapter have already been developed \citep{Churchill2017,Korpilo2016}.

\begin{figure}
  \centering
  \includegraphics[width=0.3\linewidth]{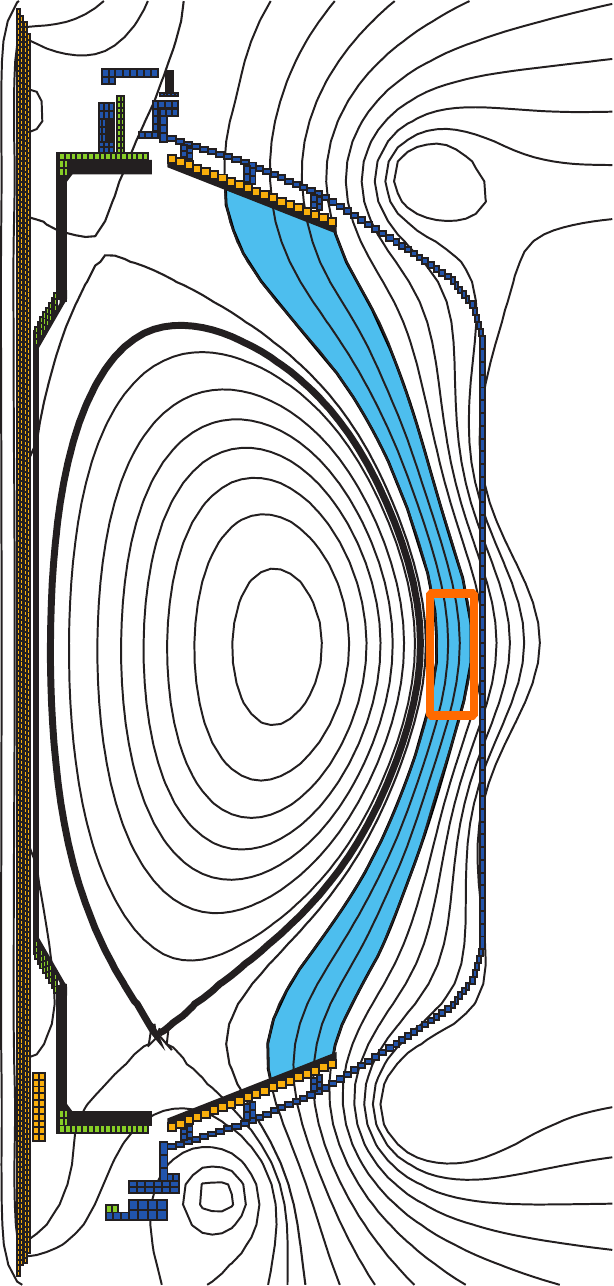}
    \caption[Illustration of the basic tokamak region approximated by the helical-SOL simulation domain.]
    {Illustration of the basic tokamak region approximated by the helical-SOL simulation domain
    (highlighted in blue).
    The SOL is modeled as a region of helical magnetic field lines that terminate on
    grounded end plates at each end, with their parallel length determined by the
    magnetic-field-line incidence angle.
    The orange box roughly indicates the projection of a flux-tube cross section onto
    this poloidal plane. This figure was adapted from figure~1.1 of \citet{StoltzfusDueck2009}.}
    \label{fig:helical_sol_cartoon}
\end{figure}

\section{Simulation Parameters}
In the non-orthogonal field-aligned geometry used in the simulation \citep{Beer1995,Hammett1993},
$z$ measures distances along field lines, $x$ is the radial coordinate, and $y$ is
constant along a field line and measures distances perpendicular to field lines.
The simulation geometry is a flux tube on the outboard side that wraps
around the torus a number of times, terminating on material surfaces at each end in $z$.
We use parameters roughly approximating a singly ionized H-mode deuterium plasma in the NSTX SOL
\citep{Zweben2015,Zweben2016}:
$n_0 = 7 \times 10^{18}$~cm$^{-3}$, $T_e \sim 30$~eV, $T_i \sim 60$~eV, $B_{\mathrm{axis}} = 0.5$~T,
$R_0 = 0.85$~m, and $a_0 = 0.5$~m.
Although we use parameters for an H-mode plasma, we do not attempt or claim to capture H-mode physics (e.g.
an edge transport barrier) in our simulations.

The simulation box has dimensions $L_x = 50 \rho_{\mathrm{s}0} \approx 14.6$~cm,
$L_y = 100 \rho_{\mathrm{s}0} \approx 29.1$~cm, $L_z = L_p/\sin \theta$,
where $L_p = 2.4$~m, $\rho_{\mathrm{s}0} = c_{\mathrm{s}0}/\Omega_i \approx 2.9$~mm, and $\theta$ is the
magnetic-field-line incidence angle.
For the results presented in this chapter, we used $\sin \theta = B_v/B_z = \left ( 0.2, 0.3, 0.6 \right)$, which
corresponds to $L_z = \left(12,8,4 \right)$~m.
The magnetic field is taken to be comprised primarily of a toroidal component with
a smaller vertical component (referred to as $B_v$), resulting in a helical-field-line geometry
that roughly approximates the tokamak SOL, as shown in figure~\ref{fig:helical_sol_cartoon}.
In this study, the magnetic-field-line incidence angle is not accounted for in the sheath boundary conditions
(no Chodura sheath \citep{Chodura1982}).
The phase-space-grid parameters are summarized in table~\ref{tab:helical_grid}.
With these parameters, $T_{e,\mathrm{min}} \approx 12.1$~eV, $T_{\parallel e,\mathrm{min}} = 4.3$~eV,
and $T_{\perp e,\mathrm{min}} = 16$~eV.
The simulation parameters and some time and length scales of interest are
listed in table~\ref{tab:helical-sol-parameters}.

\begin{figure}
  \centering
  \includegraphics[width=\linewidth]{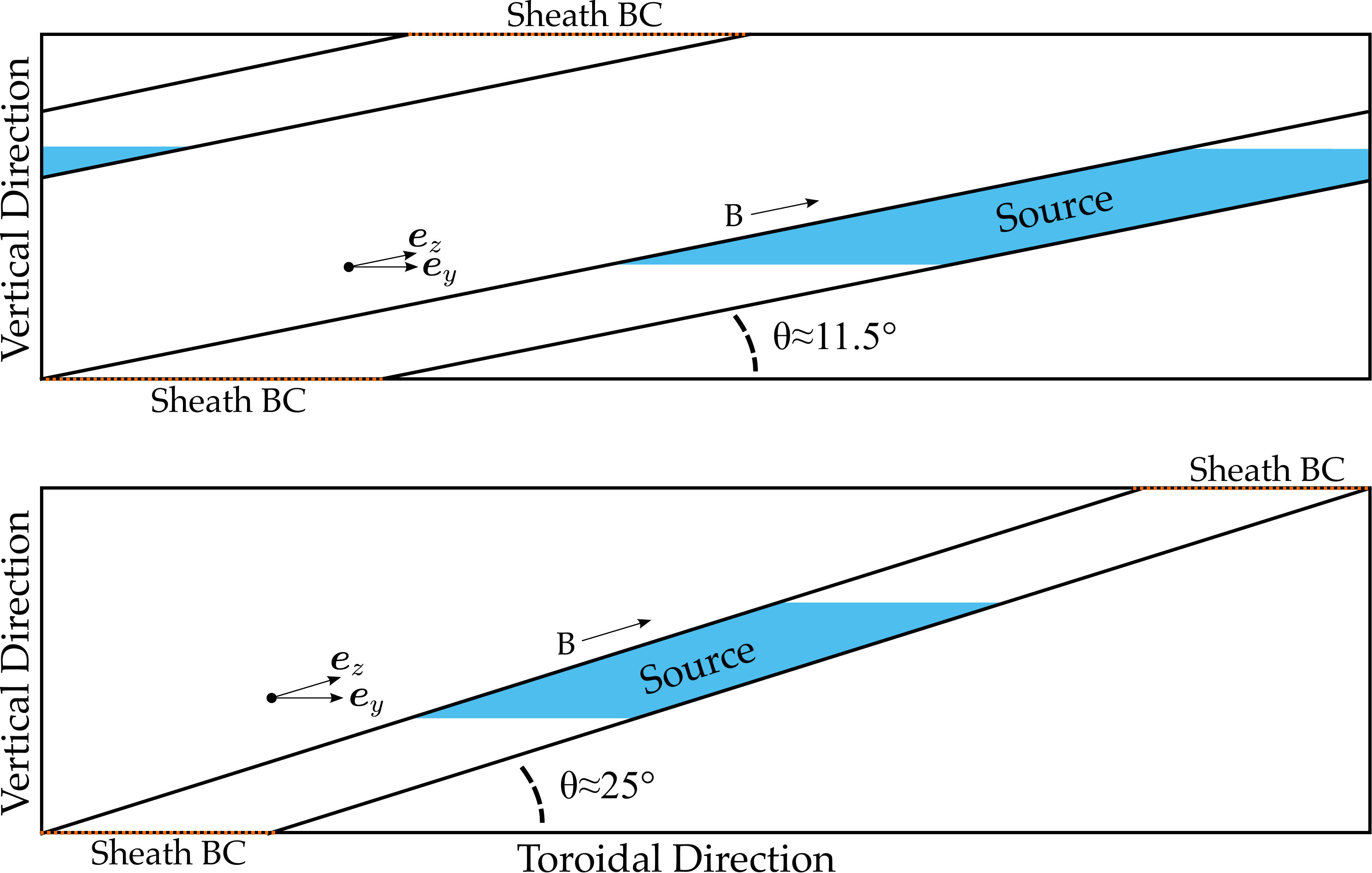}
    \caption[Illustrations of 2D simulation flux surfaces for two cases with
    different magnetic-field-line incidence angles.]
    {Illustrations of 2D simulation flux surfaces for a case with a magnetic-field-line incidence angle
    $\theta \approx 11.5^\circ$ and a case with a slightly steeper $\theta \approx 25^\circ$.
    The coordinate $y$ is constant along a field line and measures distances perpendicular
    to field lines, while the coordinate $z$ measures distances along field lines.
    Note that this coordinate system is non-orthogonal.
    The basis vectors are $\boldsymbol{e}_y = \partial_y \boldsymbol{R}$ and
    $\boldsymbol{e}_z = \partial_z \boldsymbol{R}$, where $\boldsymbol{R}$ is the
    Cartesian coordinate in space.
    Sheath-model boundary conditions are applied at each end of the flux tube in $\boldsymbol{e}_z$, and
    $\boldsymbol{e}_y$ is a periodic direction. In the steeper $\theta$ case, the flux tube covers
    a smaller fraction of the entire tokamak volume, and so the total source
    is scaled appropriately to maintain a fixed volumetric source rate.
    For additional details about the non-orthogonal coordinate system, the reader is referred to
    \citet{Beer1995,Scott1998,Hammett1993}. }
    \label{fig:helical_sol_side_cartoon}
\end{figure}

\begin{table}
\begin{center}
\caption[Summmary of simulation parameters for helical-SOL simulations.]
  {Summary of simulation parameters for helical-SOL simulations.
  These parameters are based on data for H-mode NSTX plasmas
  \citep{Zweben2015,Zweben2016}. 
  The ion mass is expressed in terms of the proton mass $m_p$.
  Also included are some time and length scales of interest,
  assuming $T_e = 25$~eV and $T_i = 40$~eV (typical values at the LCFS
  in the simulation).}
  \bigskip
  \small
  \begin{tabular}{ccl}
  \toprule
  \textbf{Symbol} & \textbf{Value} & \textbf{Description} \\
  \midrule
    $R_0$ & 0.85 m & Device major radius \\
    $a_0$ & 0.5 m & Device minor radius \\
    $T_{e,\mathrm{src}}$ & 74 eV & Electron source temperature \\
    $T_{i,\mathrm{src}}$ & 74 eV & Ion source temperature \\
    $n_0$ & $7 \times 10^{18}$~cm$^{-3}$ & Density normalization \\
    $m_i$ & $2.014 m_p$ & Mass of ion species \\
    $B_\mathrm{axis}$ & 0.5 T & On-axis magnetic field strength \\
    $B_0$ & 0.315~T & Magnetic field in middle of simulation domain \\
    $\rho_{\mathrm{s}0}$ & 2.9 mm & Ion sound radius normalization \\
    $L_p$ & 2.4 m & Poloidal distance from midplane to end plates \\
    $L_y$ & 29.1 cm & Width of simulation domain in $y$ \\
    $L_x$ & 14.6 cm & Width of simulation domain in $x$ \\
    $L_z$ & 4, 8, 12 m & Parallel length of simulation domain \\
    $B_v/B_z$ & 0.6, 0.3, 0.2 & Magnetic-field-line pitch \\
    $P_\mathrm{\mathrm{source}}$ & 270, 540, 810 kW & Total source power \\
    $S_{n,\mathrm{vol}}$ & $1.14\times 10^{23}$~m$^{-3}$~s$^{-1}$ & Volumetric source particle rate\\
    $\theta$ & 64.4$^\circ$, 30.47$^\circ$, 20.14$^\circ$ & Magnetic-field-line incidence angle \\
    $\Omega_i$ & $1.50\times 10^7$ rad/sec & Ion gyrofrequency \\
    $\tau_{ii}$ & 79~$\mu$s & Typical ion--ion collision time \\
    $\tau_{ee}$ & 0.46~$\mu$s & Typical electron--electron collision time \\
    $\lambda_{ii}$ & 3.5~m & Typical ion--ion mean free path \\
    $\lambda_{ee}$ & 0.96~m & Typical electron--electron mean free path \\
    $\rho_i$ & 2.9~mm & Typical ion gyroradius \\
    $\rho_e$ & 0.048~mm& Typical electron gyroradius \\
  \bottomrule
  \end{tabular}
  \label{tab:helical-sol-parameters}
\end{center}
\end{table}

In these equations, we neglect geometrical factors arising from a cylindrical coordinate system everywhere except
in $\boldsymbol{B}^* = \boldsymbol{B} + (B v_\parallel/\Omega_s) \nabla \times \boldsymbol{b}$,
where we make the approximation that perpendicular gradients are much stronger than
parallel gradients:
\begin{align}
  (\nabla \times \boldsymbol{b}) \cdot \nabla f(x,y,z) &= \left[ (\nabla \times \boldsymbol{b}) \cdot \nabla y \right] \frac{\partial f(x,y,z)}{\partial y}
  + \left[ (\nabla \times \boldsymbol{b}) \cdot \nabla z \right] \frac{\partial f(x,y,z)}{\partial z} \nonumber \\
  &\approx \left[ (\nabla \times \boldsymbol{b}) \cdot \boldsymbol{e}^y \right] \frac{\partial f(x,y,z)}{\partial y}.
\end{align}
Here, we assume that $(\nabla \times \boldsymbol{b}) \cdot \boldsymbol{e}^y = -1/x$, where $\boldsymbol{e}^y = \nabla y$
is a `co-basis' direction.
This type of approximation has also been employed in some fluid simulations of SMTs \citep{Ricci2009a,Ricci2010}.
We assume that $\boldsymbol{B} = B_\mathrm{axis} (R_0 / x) \boldsymbol{e}_z$. 
For the helical-SOL simulations, the 5D gyrokinetic system we solve (see (\ref{eq:gke}) and (\ref{eq:gkPB_full}))
has the following form for $\mathcal{J}\boldsymbol{\Pi}$:
\begin{align}
\mathcal{J}\boldsymbol{\Pi} &= 
	\begin{pmatrix}
		0 & -1/q_s & 0& 0 & 0 \\
		1/q_s & 0 & 0 & B_y^*/m_s & 0\\
		0 & 0 & 0 & B/m_s & 0 \\
		0 & -B_y^*/m_s & -B/m_s & 0 & 0 \\
		0 & 0 & 0 & 0 & 0
	\end{pmatrix},
  \label{eq:gkPB_helical}
\end{align}
where $B_y^* = -m_s v_\parallel / \left( q_s x \right)$.

Periodic boundary conditions are applied to both $f$ (the distribution function) and
$\phi$ (the electrostatic potential) in $y$,
the Dirichlet boundary condition $\phi = 0$ is applied in $x$, which
prevents gyrocenters from crossing the surfaces in $x$.
Sheath-model boundary conditions are applied to $f$ in $z$,
which partially reflect gyrocenters of one species and
fully absorb gyrocenters of the other species into the wall depending on the
sign of the sheath potential.
Typically, the sheath will accelerate all incident ions into the wall and reflect
the low-energy electrons back into the plasma.
As described in Section \ref{sec:lapd_sheath}, we obtain the sheath potential by
solving the gyrokinetic Poisson equation (\ref{eq:gkp}) and
evaluating $\phi$ at the sheath entrances (the surfaces of the simulation box in $z$).

\begin{table}
\begin{center}
\caption[Parameters for the phase-space grid used in helical-SOL simulations.]{
  Parameters for the phase-space grid used in the helical-SOL simulations. The parameters
  appearing in the velocity-space extents are
  $T_{i,\mathrm{grid}} = T_{e,\mathrm{grid}} = 40$ eV and $B_0 = B_{\mathrm{axis}}R_0/(R_0+a_0)$.
  Piecewise-linear basis functions are used, resulting in 32 degrees of freedom per cell}
\bigskip
\begin{tabular}{cccc}
\toprule
\textbf{Coordinate} & \textbf{Number of Cells} & \textbf{Minimum} & \textbf{Maximum} \\
\midrule
$x$ & 18 & $R_0 + a_0 - L_x/2$ & $R_0 + a_0 + L_x/2$ \\
$y$ & 36 & $-50 \rho_{\mathrm{s}0}$ & $50 \rho_{\mathrm{s}0}$ \\
  $z$ & 10 & $-L_p/\left( 2 \sin \theta \right)$ & $ L_p/\left( 2 \sin \theta\right)$ \\
$v_\parallel$ & 10 & $-4 \sqrt{T_{s,\mathrm{grid}}/m_s} $ & $ 4 \sqrt{T_{s,\mathrm{grid}}/m_s} $ \\
  $\mu$ & 5 & 0 & $0.75 m_s v_{\parallel,\mathrm{max}}^2/ (2 B_0)$ \\
\bottomrule
\end{tabular}
\label{tab:helical_grid}
\end{center}
\end{table}

\begin{figure}
  \centerline{\includegraphics[width=\textwidth]{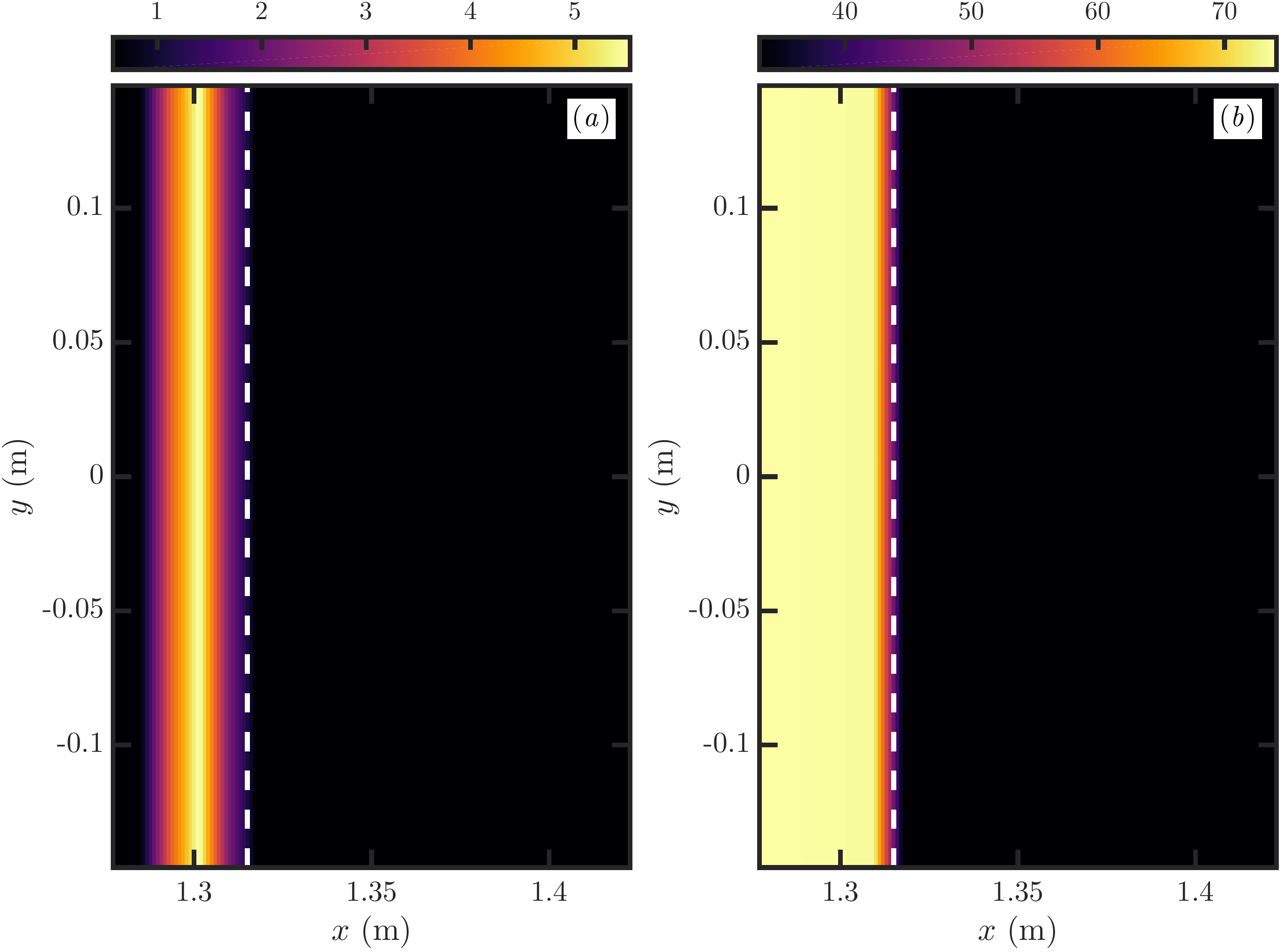}}
  \caption[Helical-SOL-simulation density source rate and temperature of
  the electron and ion sources in the perpendicular $x$--$y$ plane.]{Helical-SOL-simulation ($a$) source
  density rate (in $10^{23}$~m$^{-3}$~s$^{-1}$) and ($b$) source temperature of electrons and ions (in eV)
  in the $x$--$y$ plane at $z=0$~m. The source shown here is used
  for the $B_v/B_z = 0.6$ simulation. For other values of $B_v/B_z$, the source fueling rate
  is scaled to keep the volumetric source rate fixed.
  The dashed white line in each plot indicates the edge of the source region for comparison
  with other figures in this chapter.}
\label{fig:helical_source_xy}
\end{figure}

\begin{figure}
  \centerline{\includegraphics[width=\textwidth]{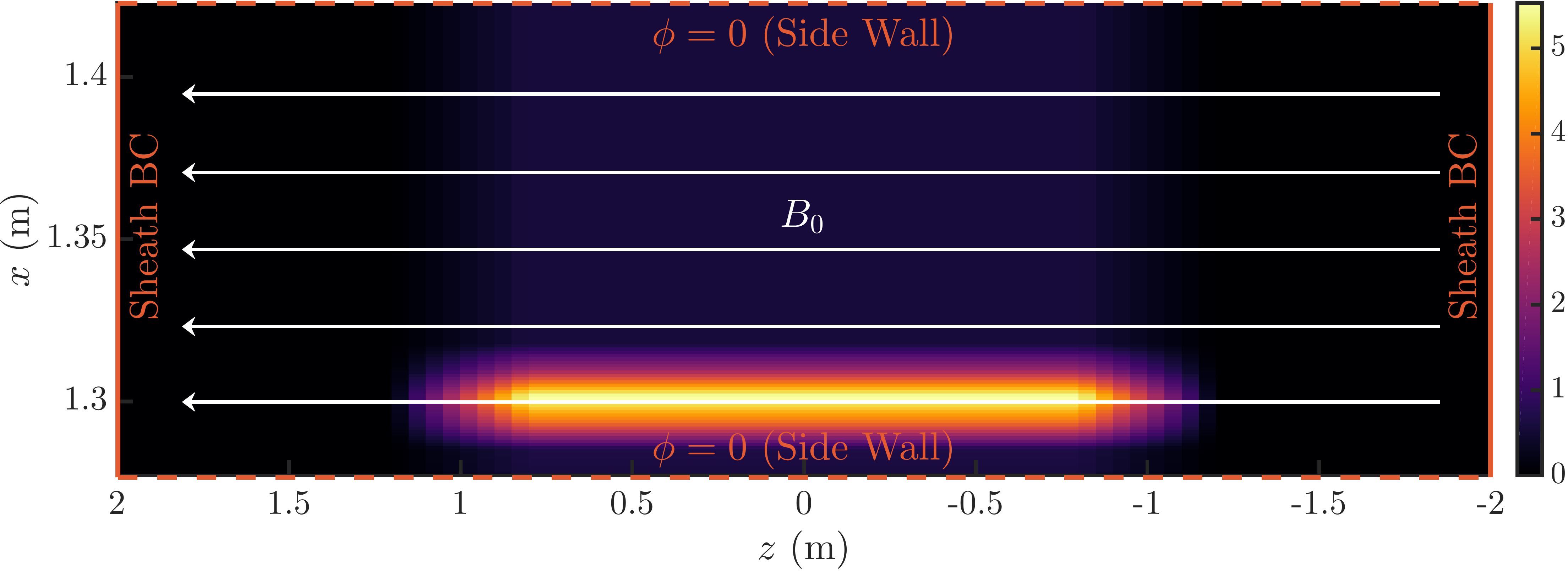}}
  \caption[Helical-SOL-simulation density source rate in the $x$--$z$ plane.]
  {Helical-SOL-simulation plasma density source rate (in 10$^{23}$~m$^{-3}$~s$^{-1}$) in the $x$--$z$ plane
  for the $B_v/B_z = 0.6$ case.
  Annotations indicate the direction of the magnetic field, side-wall boundary conditions, and
  sheath-model boundary condition locations. The source is uniform in the periodic $y$ direction.}
\label{fig:helical_source_xz}
\end{figure}

The plasma density source has the following form:
\begin{equation}
  S(x,z) = \begin{cases}
  S_0 \mathrm{max}\left( \exp\left(\frac{-(x-x_s)^2}{2 \lambda_s^2 }\right), 0.1\right) & |z| < L_z/4, \\
  0 & \mathrm{else},
  \end{cases}
  \label{eq:helical_sol_source}
\end{equation}
where $x_s = -0.05 \; \mathrm{m} + R_0 + a_0$, $\lambda_s = 5 \times 10^{-3}$~m,
and $S_0$ is chosen so that the source has total (electron plus ion) power
$P_{\mathrm{source}} = 0.27 L_z/L_{z0} $~MW, where $L_{z0} = 4$~m.
The expression for the source power comes from multiplying $P_{\mathrm{SOL}} = 5.4$~MW, the total power into the SOL,
by the fraction of the total device volume covered by the simulation box.
A floor of $0.1 S_0$ is used in the $|z| < L_z/4$ region to prevent regions of $n \ll n_0$ from developing at large $x$,
which can result in distribution-function positivity issues.
The distribution function of the sources are non-drifting Maxwellians
with a temperature profile $T_{e,\mathrm{src}} = T_{i,\mathrm{src}} = 74$~eV
for $x < x_s + 3 \lambda_s$ and $T_{e,\mathrm{src}} = T_{i,\mathrm{src}} = 33$~eV for $x \ge x_s + 3 \lambda_s$.
These choices result in an integrated source particle rate of ${\approx}9.6 \times 10^{21}$ s$^{-1}$
for the $L_z = L_{z0}$ ($B_v/B_z = 0.6$) case.
Plots of the density source rate and the source temperature of electrons and ions in the
$x$--$y$ plane at $z = 0$~m are shown in figure~\ref{fig:helical_source_xy}.
Figure~\ref{fig:helical_source_xz} shows the parallel variation of the density source rate in the $x$--$z$ plane.
The density source rate shown in these two figures are for the $B_v/B_z = 0.6$ simulation.

We cannot yet include a closed-field-line region in our simulations, so we only simulate a SOL.
The $x < x_s + 3\lambda_s$ region will be referred to as the source region in this chapter,
while the $x \ge x_s + 3\lambda_s$ region will be referred to as the SOL region.
We treat the $x = x_s + 3\lambda_s$ location as the last closed flux surface.
For these reasons, we will only comment on the dynamics in the SOL region of our simulations and neglect the
plasma behavior in the source region, which is believed to be strongly influenced by the source presence and the inner-wall
radial boundary condition.

While larger time steps can be taken in these simulations, we note that
the use of a spatially varying magnetic field also increases the computational cost for two reasons.
First, the number of Gaussian quadrature points required for the exact evaluation of
integrals is increased from $2$ to $3$ (at present, Gaussian quadrature rules for multiple
dimensions are constructed from a tensor product of the 1D Gaussian quadrature rule).
To give an example, consider the $-B_y^*/m_s$ entry in $\mathcal{J} \Pi$ (\ref{eq:gkPB_helical}).
This term results in the following integral to solve the gyrokinetic
equation after multiplying (\ref{eq:gke}) by an arbitrary test function $\psi_k$, integrating over all space
$\int \mathrm{d}\Lambda = \int \mathrm{d}^3 x \int \mathrm{d}^3 v$, and performing an integration by parts
to move a partial derivative onto $\psi_k$:
\begin{equation}
  -\int \mathrm{d}\Lambda \, \frac{\partial \psi_k}{\partial v_\parallel} \frac{B_y^*}{m_s} \frac{\partial H_s}{\partial y} f_s \nonumber .
\end{equation}
The integrand is generally of order $4p$ in $x$, with each term contributing one power of $p$.
Since we use $p=1$ (piecewise-linear basis functions) in these simulations, the Gaussian quadrature rule must be able
to integrate fourth-order polynomials in $x$ exactly, which requires 3 quadrature nodes in $x$.
For the LAPD simulations, this integral was zero because $B_y^* = 0$ for the straight-magnetic-field-line system,
so only 2 quadrature nodes in $x$ were required.

Secondly, more matrix-multiplication operations are required to solve the gyrokinetic equation (\ref{eq:gke})
because the Jacobian $\mathcal{J} = B_\parallel^* = B$ can have spatial variation.
In the LAPD simulations, the Jacobian-weighted mass matrices $\mathbb{M}$ (see (\ref{eq:1d_conservation_num}))
were the same in every cell, so every computing zone could compute and store one set of matrices
at the beginning of the computation that would be used to solve the gyrokinetic equation at each
time step.
By allowing for a spatially varying $\mathcal{J}$, it becomes memory intensive
to store all the Jacobian-weighted basis-function matrices, since a set of matrices would needed
to be stored for every variation of the Jacobian in a zone.
A separate version of the gyrokinetic equation solver was written that evaluates certain
matrix products involving the inverse of the Jacobian-weighted mass matrices on every time step, rather
than storing the matrix products, which are independent of time, in memory.

From our prior experience in simulating LAPD (Chapter \ref{ch:lapd}), we observed that uniform,
zero-velocity initial conditions can take a significant time ${\sim}{\tau_i} = (L_z/2)/v_{ti}$
to reach a quasi-steady state.
To reduce the computational cost, we decided to initialize our helical-SOL simulations from profiles 
calculated using a 1D single-fluid analysis that neglects electric fields and thermal conduction.
These initial conditions are derived and specified in Appendix \ref{ch:initial-helical-sol}.

\section{Simulation Results}
\begin{figure}
\includegraphics[width=\linewidth]{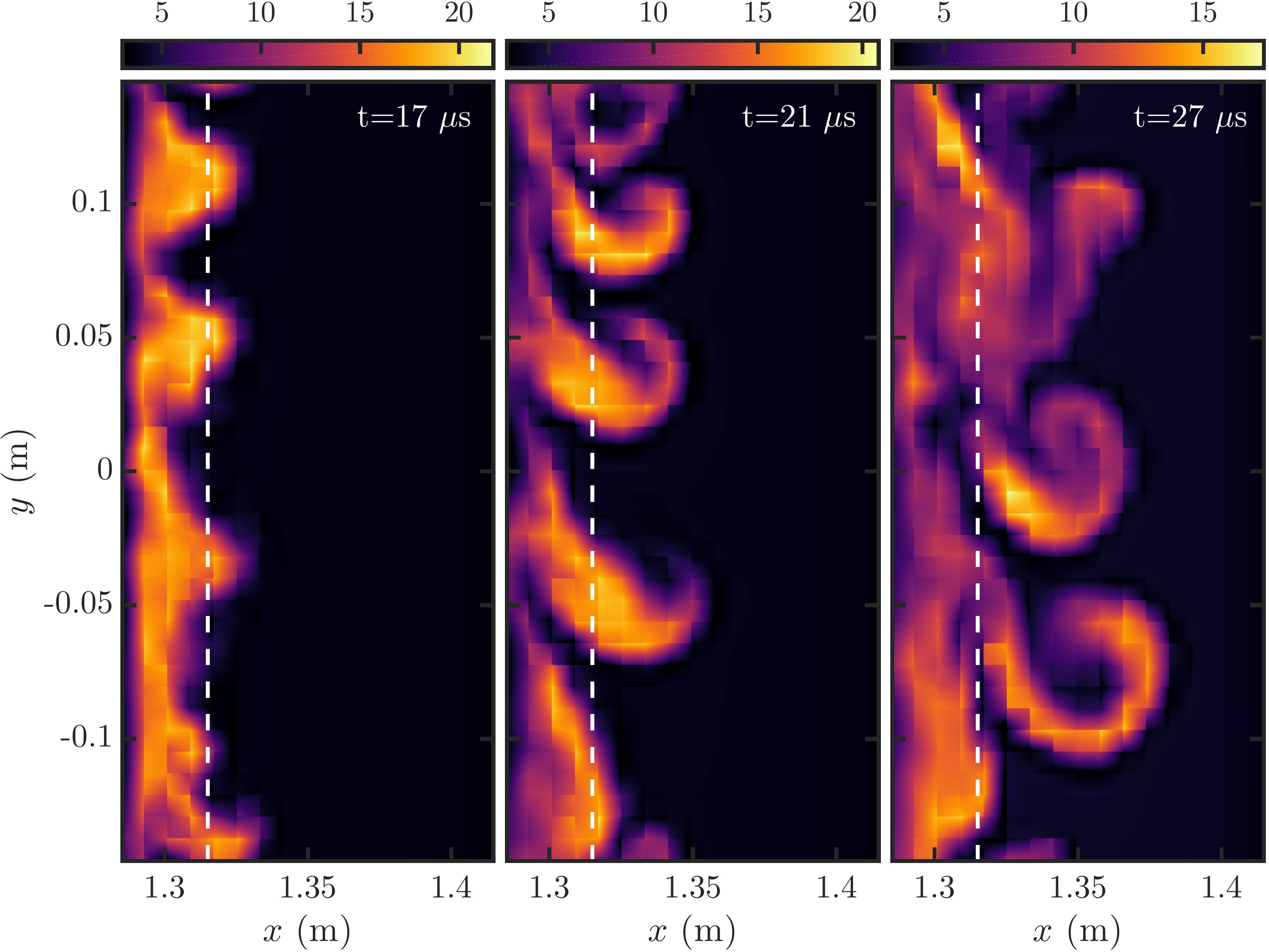}
  \caption[Snapshots of the electron density at various times near the beginning of a 5D gyrokinetic simulation
  of a helical SOL.]
  {Snapshots of the electron density (in 10$^{18}$ m$^{-3}$)
  at various times ($t=17$~$\mu$s, 21~$\mu$s, and 27~$\mu$s) near the beginning of a simulation
  in the perpendicular $x$--$y$ plane at $z=0$~m.
  This simulation has $B_v/B_z = 0.3$.
  The dashed line indicates the region in which the source is concentrated.
  Note that each plot uses a different color scale to better
  show the features.}
  \label{fig:blob_formation} 
\end{figure}

\begin{figure}
\includegraphics[width=\linewidth]{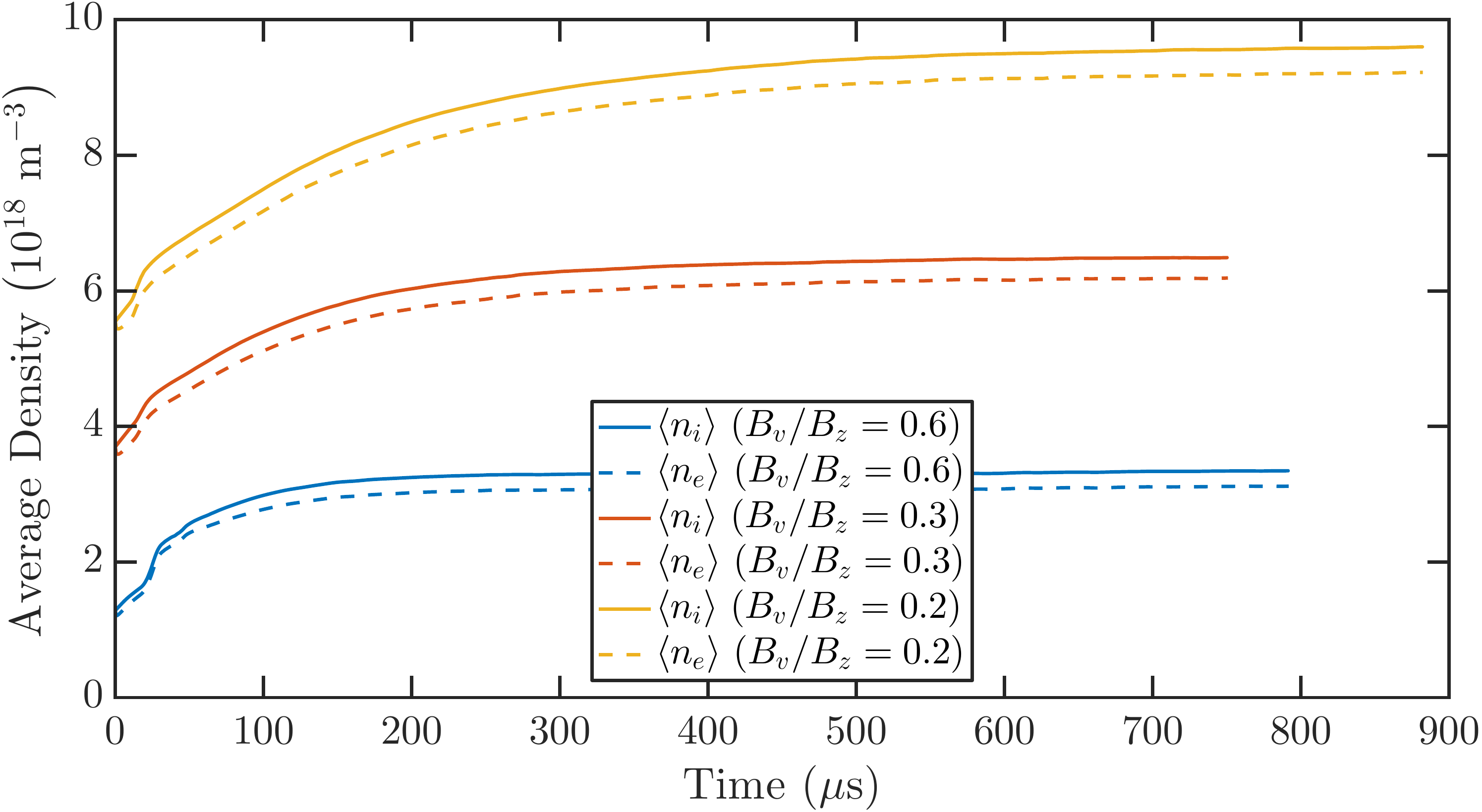}
  \caption[Average density of ions and electrons vs. time for
  three simulations with different magnetic-field-line pitches.]
  {Average density of ions (solid lines) and electrons (dashed lines) vs. time for
  three simulations with different magnetic-field-line pitches.
  Starting from an initial condition, the simulations reach a quasi-steady state
  in which the particle losses to the end plates are balanced by the plasma sources.}
  \label{fig:helical_density_vs_time}
\end{figure}

\begin{figure}
\includegraphics[width=\linewidth]{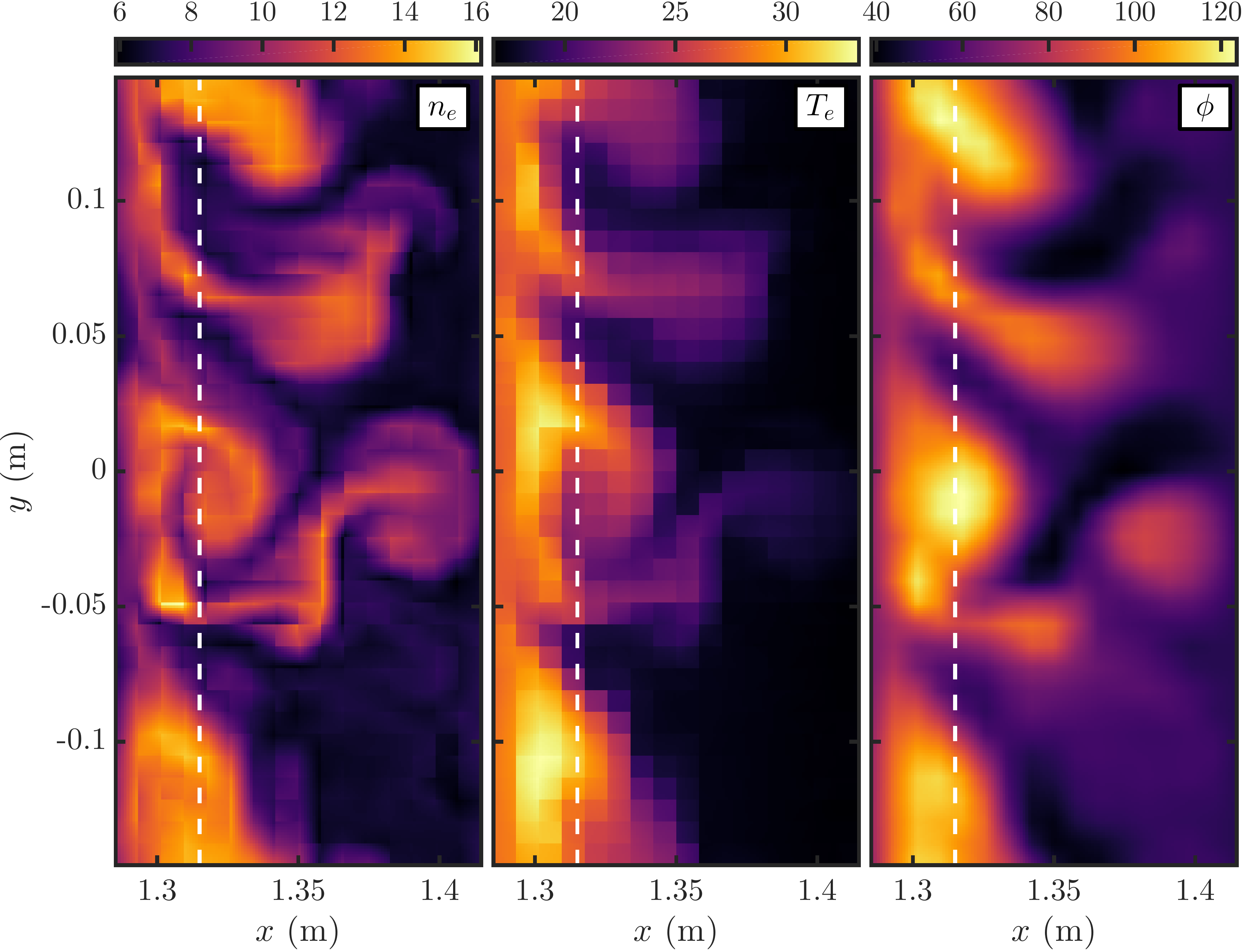}
  \caption[Snapshots of the electron density, electron temperature, and electrostatic potential from the
  quasi-steady state of a 5D gyrokinetic simulation of a helical SOL.]
  {Snapshots of the electron density (in 10$^{18}$~m$^{-3}$), electron temperature
  (in eV), and electrostatic potential (in V) in the plane perpendicular to the magnetic field
  at $z=0$~m. This plot is made at $t=625$~$\mu$s, which is after several ion
  transit times ($\tau_i \approx 46$~$\mu$s). This simulation has $B_v/B_z = 0.3$.
  The dashed line indicates the region in which the source is
  concentrated. A mushroom structure in the blob density is observed at large $x$.
  For additional details about how this plot and ones like it
  were created, see Appendix~\ref{ch:plot-creation}.}
  \label{fig:test145} 
\end{figure}

Starting from an initial condition estimated by the steady-state solution of 1D fluid equations,
the sources steepen the plasma profiles, quickly triggering curvature-driven modes that grow on a time scale
comparable to $\gamma \sim c_\mathrm{s}/\sqrt{R \lambda_p}$ (see Appendix \ref{ch:interchange} for details).
We emphasize that our system does not contain ballooning modes since there are no `good-curvature' regions.
As shown in figure~\ref{fig:blob_formation}, radially elongated structures extending far from the source region
are generated and subsequently broken up by sheared flows in the $y$ direction in the source region,
leaving radially propagating blobs.
Using the time-averaged profiles from the same $B_v/B_z = 0.3$ ($L_z = 8$~m) simulation,
we estimate $\gamma \sim 1.9 \times 10^5$~s$^{-1}$ using $\lambda_p \approx 2.9$~cm, $T_e \approx 30$~eV,
and $R = x_s = 1.3$~m.
On a time scale long compared to $\gamma^{-1}$ and $\tau_i = (L_z/2)/v_{ti}$, a quasi-steady state is reached
in which the particle losses to the end plates are balanced by the plasma sources.
Time traces of the average density in the simulation domain are shown in figure~\ref{fig:helical_density_vs_time}.
Snapshots of the electron density, electron temperature, and electrostatic potential from the quasi-steady state ($t=625$ $\mu$s)
for the $B_v/B_z = 0.3$ case are shown in figure~\ref{fig:test145}.

\begin{figure}
\includegraphics[width=\linewidth]{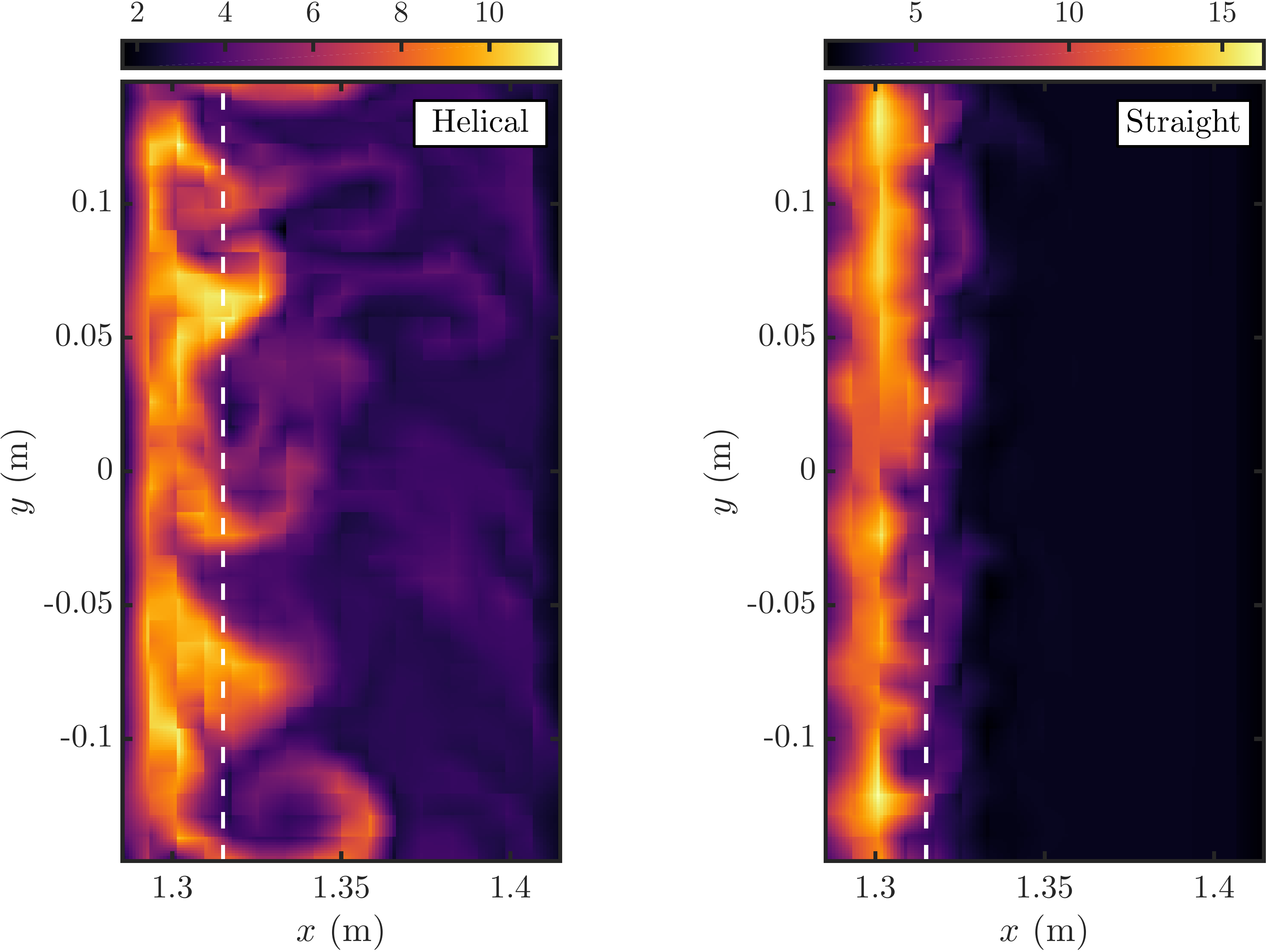}
  \caption[Comparison of an electron density snapshot between a simulation
  in a helical-magnetic-field-line geometry and a simulation in a straight-magnetic-field-line
  geometry.]
  {Comparison of an electron density snapshot (in $10^{18}$~m$^{-3}$) between ($a$) a simulation
  in a helical-magnetic-field-line geometry and ($b$) a simulation in a straight-magnetic-field-line
  geometry with $B = B_\mathrm{axis}$.
  The formation of blobs in the helical-SOL simulation results in the transport of density
  to large $x$ and a broad density profile.
  Coherent structures of elevated plasma density do not appear to convect to
  large $x$ straight-magnetic-field-line case, and so density is mostly localized to the source region.
  The plots are made in the perpendicular $x$--$y$ plane at $z = 0$~m and $t = 681$~$\mu$s.
  The dashed line indicates the region in which the source is concentrated.
  Note that each plot uses a different color scale to better show the features.}
  \label{fig:curvature_comparison}
\end{figure}

For the steepest magnetic-field-line-pitch case ($B_v/B_z = 0.6$),
we performed a second simulation with magnetic-curvature effects removed
and keeping all other parameters the same.
The resulting magnetic geometry consists only of straight magnetic field lines,
as in the LAPD simulations of Chapter~\ref{ch:lapd}, so coherent structures of elevated
plasma density cannot become polarized by curvature forces.
As shown in the electron density snapshot comparison in figure~\ref{fig:curvature_comparison},
the presence of magnetic curvature appears to have
an important role in the turbulent dynamics of the system.
When magnetic-curvature effects are removed, the radial propagation of
coherent structures appears to be significantly weakened or absent,
and most of the density is localized to the source region.

\begin{figure}
\includegraphics[width=\linewidth]{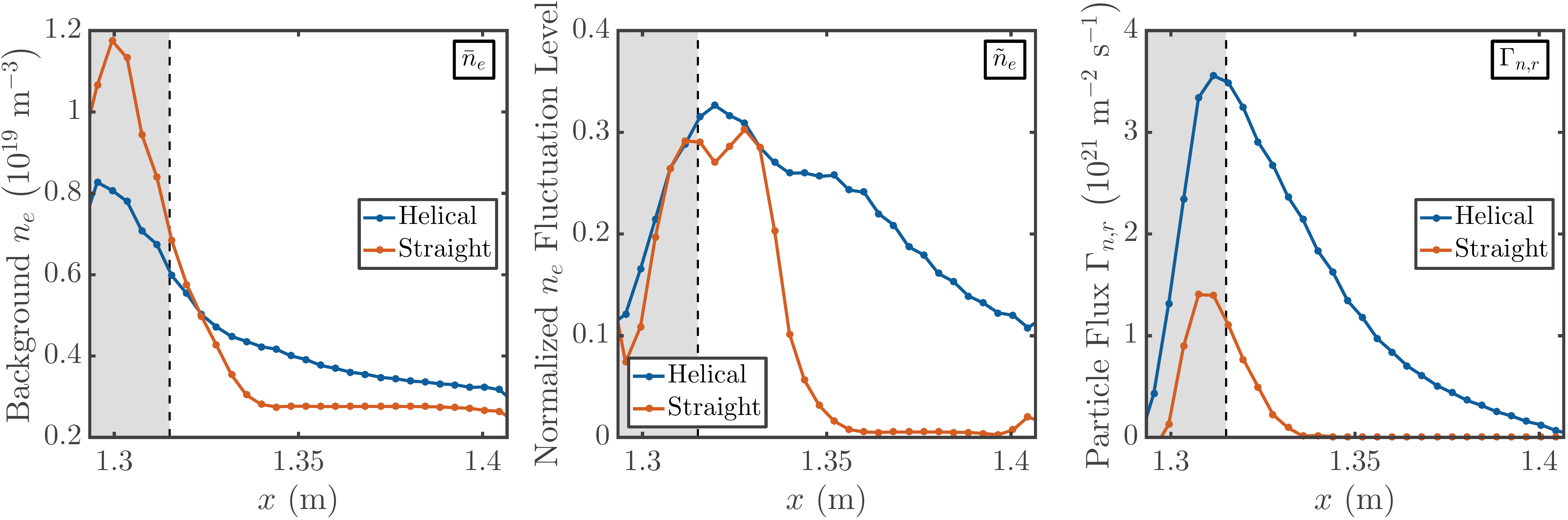}
  \caption[Radial profiles of the background electron densities,
  normalized-electron-density fluctuation levels, and radial $E \times B$ particle fluxes for
  a helical SOL simulation and a straight-field-line simulation.]
  {Radial profiles of the background electron densities (in $10^{19}$~m$^{-3}$),
  normalized-electron-density fluctuation levels, and radial $E \times B$ particle fluxes $\Gamma_{n,r}$
  (in $10^{21}$~m$^{-2}$~s$^{-1}$) for
  a helical SOL simulation and a straight-field-line simulation with $B = B_\mathrm{axis}$.
  These plots are computed using data near the midplane in the region -0.5~m$<z<0.5$~m and 
  sampled at 0.25~$\mu$s intervals over a ${\sim}400$~$\mu$s period.
  The shaded area indicates the region in which the source is concentrated.
  The background density profile in the straight-field-line simulation does not decay to 0
  at large $x$ due to the presence of a constant low-amplitude source in that region to help mitigate
  positivity issues with the distribution function.}
  \label{fig:curvature_comparisons_1d}
\end{figure}

Figure~\ref{fig:curvature_comparisons_1d} compares radial profiles of the background electron-densities,
normalized electron-density fluctuation levels, and radial $E \times B$ particle fluxes $\Gamma_{n,r}$
between these two simulations.
The radial particle flux due to electrostatic turbulence is estimated
as $\Gamma_{n,r} = \langle \tilde{n}_e \tilde{v}_r \rangle$ \citep{Zweben2007},
where $v_r = E_y/B$ and the brackets $\langle \dots \rangle_t$ indicate an average in time
over a period that is long compared to the fluctuation time scale.
The fluctuation of a time-varying quantity $A$ is denoted as $\tilde{A}$, which is related to the total $A$ as
$\tilde{A} = A - \langle A \rangle_t$.
Notable differences between these two simulations are found in all three quantities plotted.
Compared to the helical-SOL simulation, the straight-field-line simulation has a background density
profile that decays more rapidly, fluctuation levels that quickly drop to $\approx{0}\%$ outside
$x \approx 1.35$~m, and a ${\approx}2.5$ times smaller $\Gamma_{n,r}$ that also
drops to $\approx{0}$ outside $x \approx 1.34$~m.

\begin{figure}
\includegraphics[width=\linewidth]{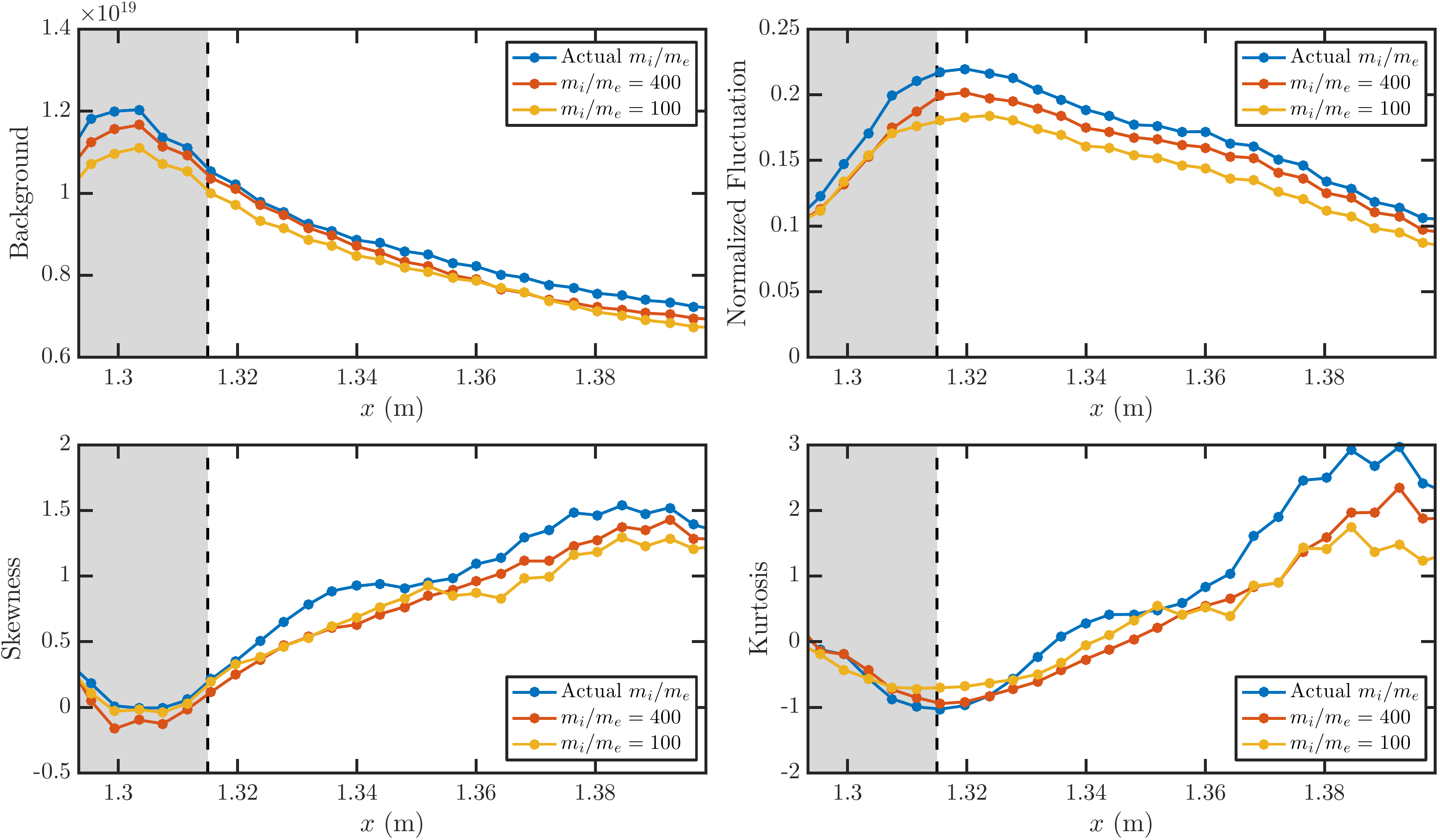}
  \caption[Comparison of electron-density profiles and fluctuation statistics for cases
  with varying electron mass.]
  {Comparison of electron-density profiles and fluctuation statistics for three cases
  with varying electron mass.
  As the mass ratio is reduced from the actual value $m_i/m_e \approx 3698$ to an artificially
  low value $m_i/m_e = 100$, the qualitative trends are unchanged, and the quantitative differences
  are small. The shaded area indicates the region in which the source is concentrated.}
  \label{fig:mass_ratio_density}
\end{figure}

We have also performed a scan of the mass ratio $m_i/m_e$ from the actual ratio of 3698 down to 100
(by increasing the electron mass), and we observed no significant quantitative or
qualitative changes in the turbulence.
The electron-density profile and fluctuation statistics are shown for a mass ratio scan in
figure~\ref{fig:mass_ratio_density}.
The mass ratio might play an important role in a different parameter regime, however.

Effects connected to $B_v \sim B_p \sim I_{\mathrm{plasma}}$ are explored by
since $\sin\theta = B_v/B_z$.
We have performed simulations at three values of magnetic-field-line pitches
$B_v/B_z = (0.2,0.3,0.6)$, which correspond to $L_z = (12,8,4)$~m and
$\theta = (20.14^\circ,30.47^\circ,64.4^\circ)$.
In the various simulations, we scale the source appropriately to maintain a fixed volumetric source rate.
In all these simulations, the source is localized to the $z\in \left[-L_z/4, L_z/4\right]$ region
to model a source with a fixed poloidal extent.
As $\theta$ is decreased, the plasma profiles are observed to become less peaked, implying
that turbulence transport in the $x$-direction increased with decreasing $\theta$.

\begin{figure}
  \centering
  \includegraphics[width=0.75\linewidth]{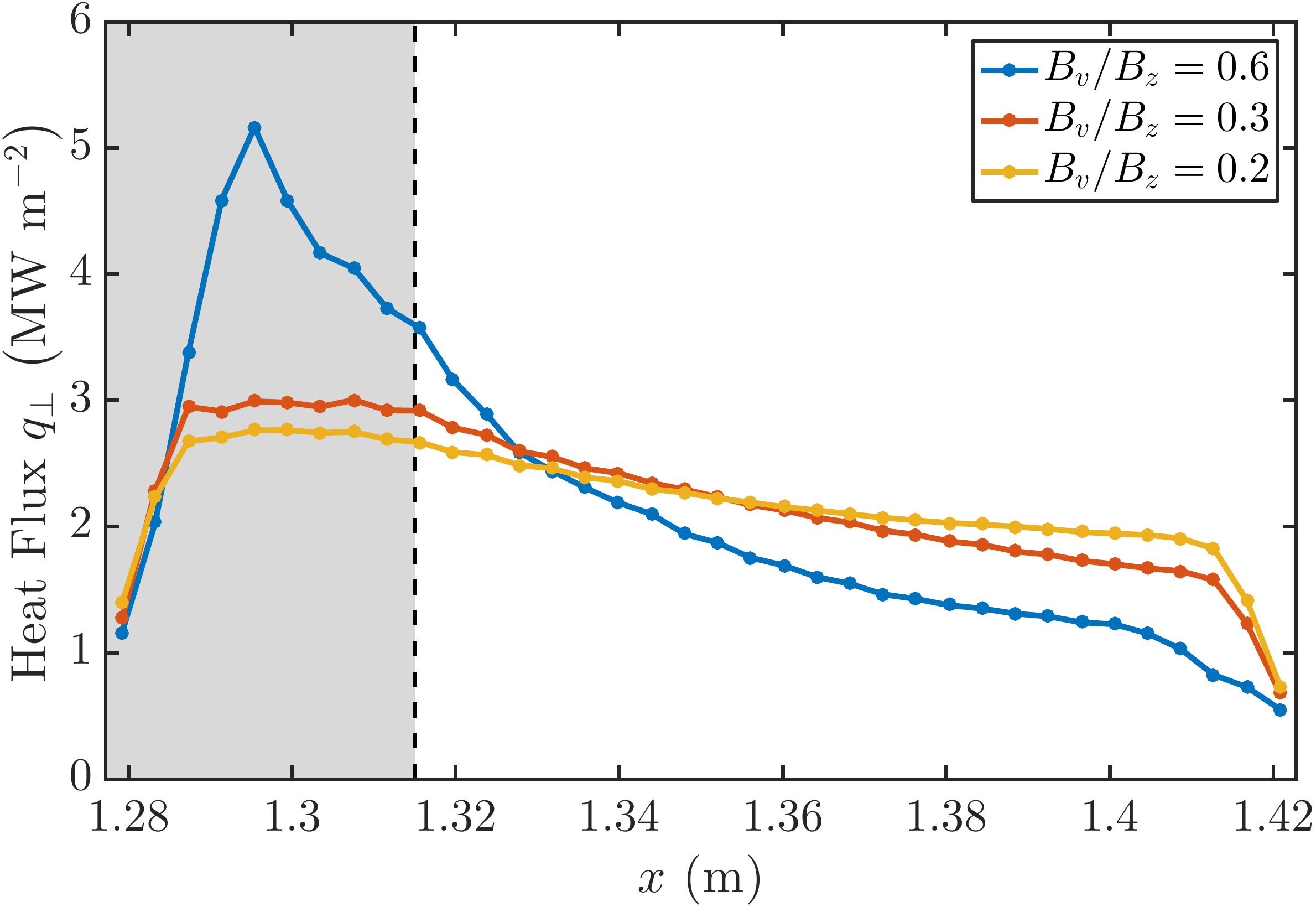}
  \caption[Comparison of the steady-state parallel heat flux normal to the divertor
  plate for cases with different magnetic-field-line pitches.]
  {Comparison of the steady-state parallel heat flux normal to the divertor plate for three cases with
  different magnetic-field-line pitches. The shaded area indicates
  the region in which the source is concentrated. The heat-flux profile is observed
  to broaden as $B_v/B_z$ is decreased.
  Since a large amount of plasma reaches the outer radial boundary, where parallel losses are enhanced due
  to the side-wall boundary conditions, the profiles in the shallower-pitch cases may exhibit
  more of an exponential fall off by increasing the box size in the radial direction.}
  \label{fig:helical-heat-flux} 
\end{figure}

We calculate the steady-state parallel heat flux $q = \sum_s \int \mathrm{d}^3 v \, H_s v_\parallel f_s$
at the sheath entrance and average $q$ in the $y$-direction to obtain a radial profile of the steady-state parallel heat flux
for each case.
To compare the heat fluxes on an equal footing, we plot the component of the parallel heat flux normal to the divertor plate $q_\perp = q \sin \theta$
in figure~\ref{fig:helical-heat-flux} .
Compared to the $B_v/B_z = 0.6$ case, the heat-flux profiles for the cases with a shallower pitch
are much broader.
This behavior is consistent with the observation in tokamaks that the SOL heat-flux width is inversely
proportional to the poloidal magnetic field (analagous to $B_v$ in this model) and the plasma
current \citep{Eich2013,Makowski2012}, although the physics reasons behind the scaling in our model
and in a tokamak SOL may be quite different.
We note that a significant amount of plasma in the smallest $\theta$ case reaches the outer radial wall,
where it is quickly lost in the parallel direction since there is no $E_\parallel$ to constrain the plasma
flows on the outer radial boundary.
Simulations with a larger domain extent in the $x$ coordinate might exhibit more of a exponential fall off
in the radial profiles than observed at these present box sizes.

\begin{figure}
  \centering
\includegraphics[width=0.75\linewidth]{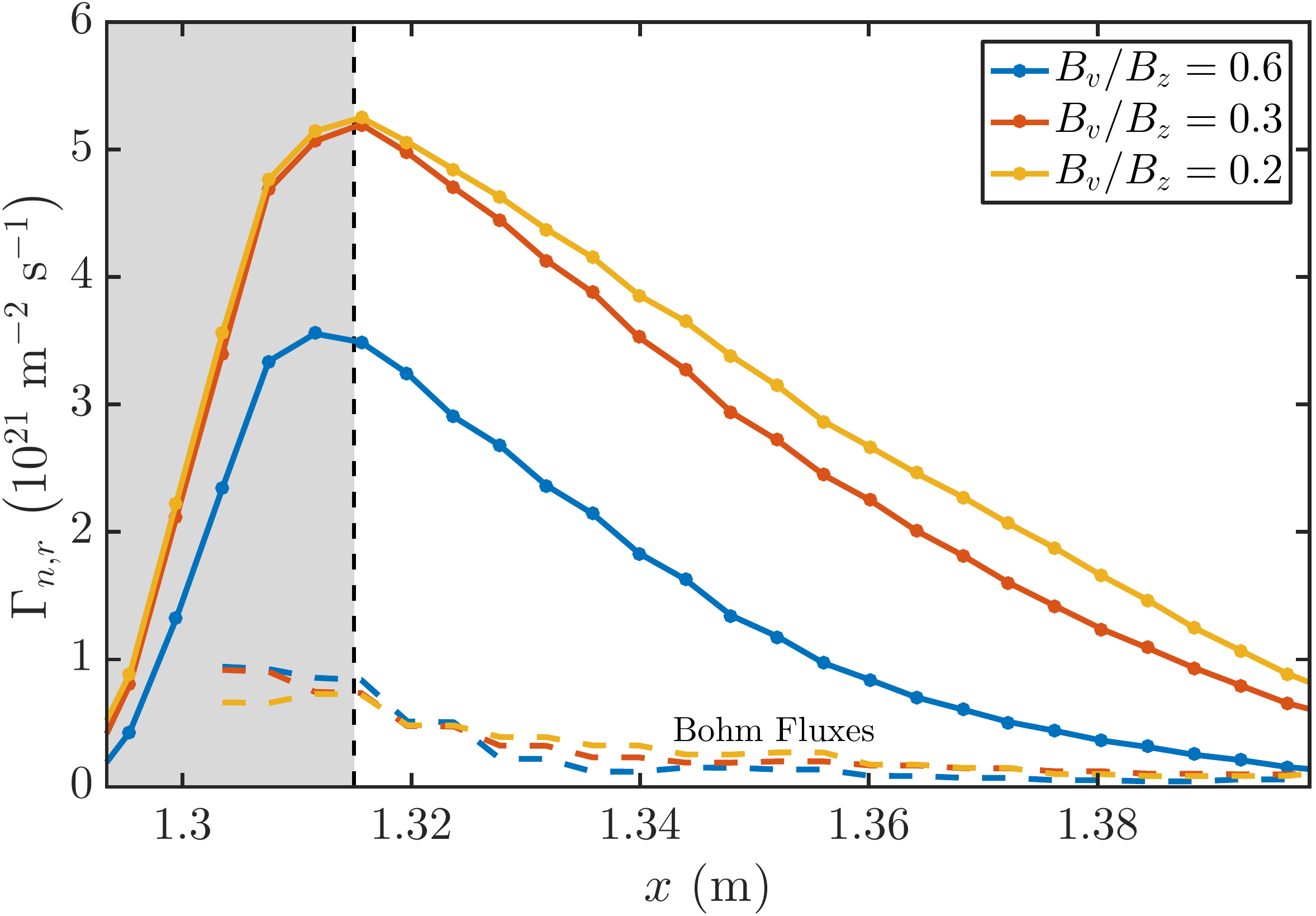}
  \caption[Comparison of the radial $E \times B$ particle flux evaluated near the midplane for cases with
  different magnetic-field-line pitches.]
{\label{fig:helical-particle-flux} Comparison of the radial $E \times B$ particle flux evaluated
  near the midplane for three cases with different magnetic-field-line pitches.
  The shaded area indicates the region in which the source is concentrated.
  The dashed lines are Bohm flux estimates for comparison.}
\end{figure}

The broad heat flux profiles in figure~\ref{fig:helical-heat-flux} can be connected to the increased
outward radial turbulent transport as $B_v/B_z$ becomes shallower.
We compute the steady state radial particle flux $\Gamma_{n,r}$ near the midplane
in the region -0.5~m$<z<0.5$~m for each value of $B_v/B_z$ and plot the $y$-averaged fluxes in
figure~\ref{fig:helical-particle-flux} (solid lines).
Since the simulation box occupies a larger fraction the device volume as $B_v/B_z$ is decreased, but
the source occupies the same fraction of the simulation box and has a fixed volumetric source rate
(see figure~\ref{fig:helical_sol_side_cartoon}),
the background density levels increase as $B_v/B_z$ decreases.
Therefore, the magnitude of the $\Gamma_{n,r}$ profiles in figure~\ref{fig:helical-particle-flux}
should not be taken alone as a measure of turbulence levels.
The $\Gamma_{n,r}$ profiles can be compared with the radial particle fluxes that result from
assuming Bohm diffusion, i.e. $\Gamma_B = D_B \partial_x n_e$, where the diffusion coefficient
$D_B = \frac{1}{16} \frac{k_B T_e}{eB}$.  In the $x > 1.36$~m region,
$\Gamma_{n,r}/\Gamma_B \approx 16$ for the $B_v/B_z = 0.2$ case,
while $\Gamma_{n,r}/\Gamma_B \approx 8$ for the $B_v/B_z = 0.6$ case.
One might expect the maximum level of turbulent transport to be around the levels set by
$D_B$, but it is important to remember that $D_B$ is a \textit{diffusive} transport estimate.
The \textit{convective} transport of blobs in these simulations appears to be responsible for
the much-higher turbulent fluxes.
Experimental data from tokamaks also suggest that the higher-than-Bohm particle transport
in the SOL is due to the non-diffusive transport of blobs \citep{Krasheninnikov2008,Zweben2007}.

\begin{figure}
\includegraphics[width=\linewidth]{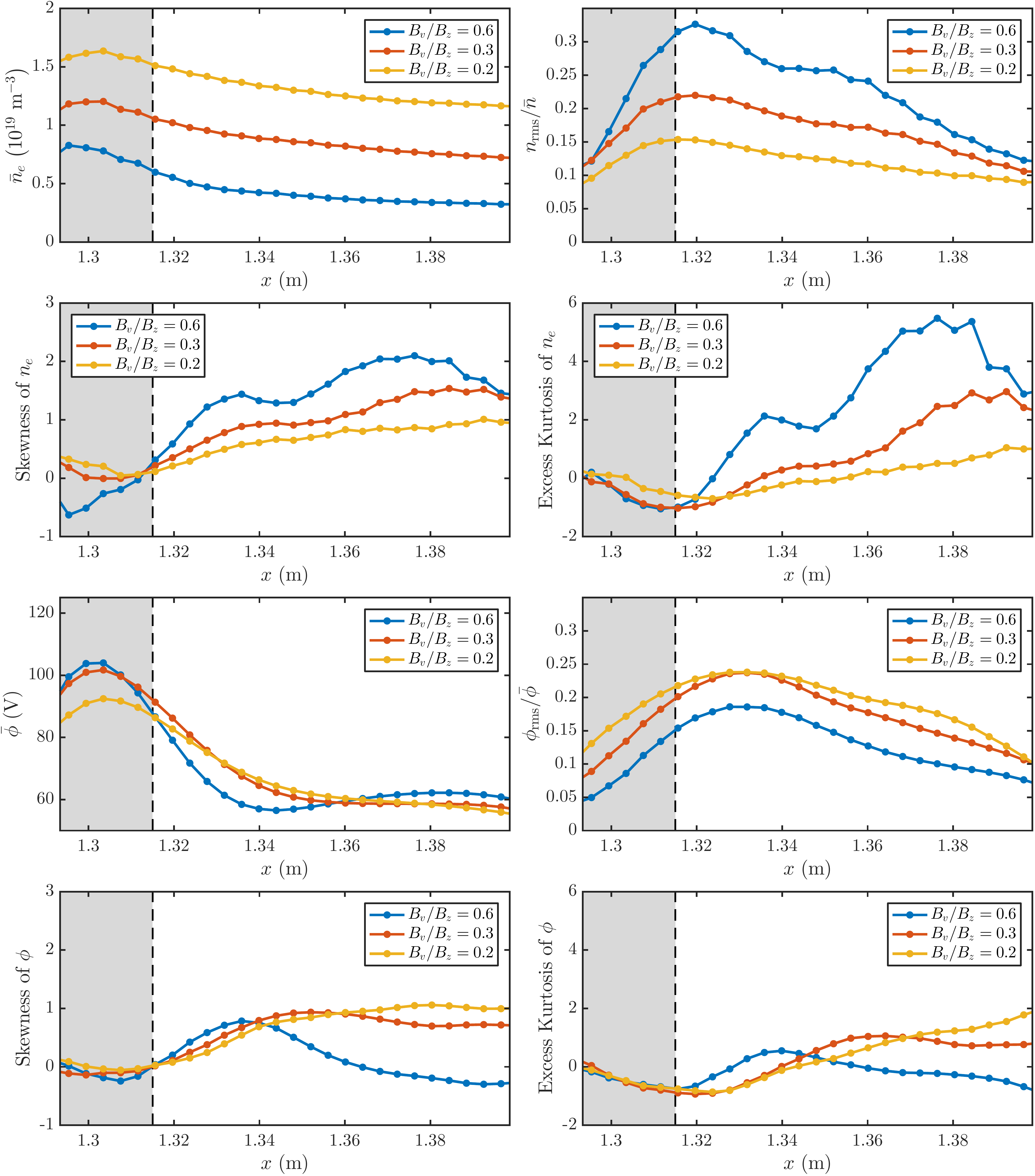}
  \caption[Comparison of the electron-density fluctuation statistics and electrostatic-potential fluctuation statistics
  computed near the midplane for cases with different magnetic-field-line pitches.]
  {Comparison of the electron-density fluctuation statistics (first two rows)
  and electrostatic-potential fluctuation statistics (bottom two rows) computed near the $z=0$~m plane for three cases with
  different magnetic-field-line pitches. The potential fluctuations are notably
  less intermittent than the density fluctuations. The shaded area indicates
  the region in which the source is concentrated.}
  \label{fig:n_and_phi_statistics} 
\end{figure}

Density fluctuation statistics are often of interest in the SOL to characterize the turbulence.
Considering again a time-varying quantity $A$, we define the skewness of $A$ as $E[\tilde{A}^3]/\sigma^3$
and the excess kurtosis of $A$ as $E[\tilde{A}^4]/\sigma^4 - 3$, where $\sigma$ is the standard
deviation of $A$ and $E[\dots]$ denotes the expected value.
Figure~\ref{fig:n_and_phi_statistics} shows the radial profiles of the normalized fluctuation level,
skewness, and excess kurtosis for electron-density fluctuations and electrostatic-potential fluctuations computed
near the $z=0$~m plane.
Unlike in Chapter~\ref{ch:lapd}, where we sometimes normalized density fluctuations by a global value,
we normalize density and potential fluctuations to their local background values in this chapter.
The positive skewness and excess kurtosis values are signatures of intermittency, which indicates
an enhancement of large-amplitude positive-density-fluctuation events and is connected to the
transport of blobs \citep{Zweben2007}.

A somewhat counter-intuitive result is the reduction of density fluctuation levels as $B_v/B_z$ is decreased,
given that figures~\ref{fig:helical-heat-flux} and \ref{fig:helical-particle-flux} indicate that turbulent
spreading is increased as $B_v/B_z$ is decreased.
The skewness and excess kurtosis plots in figure~\ref{fig:n_and_phi_statistics} indicate that
the density fluctuations become closer to a normal distribution as $B_v/B_z$ is decreased.
These trends in the density fluctuation statistics can be understood by noting that
the background density profile becomes less peaked and more uniform in the $x$-direction as $B_v/B_z$ is decreased,
so a blob that is formed in the source region propagating in the SOL has a density
that is closer to the background level, which results in lower relative fluctuation,
skewness, and excess kurtosis values when compared to the large $B_v/B_z$ case.
Additionally, the density flux is constrained by the use of a fixed volumetric source rate,
so as the background density increases with decreasing $B_v/B_z$,
the relative density fluctuation levels tend to decrease.
We also observe that the potential fluctuations are much less intermittent than the density fluctuations
at the same $B_v/B_z$.
This observation could be a real, physical effect, but we note that the fact that the temperature at large $x$
runs into the grid resolution (the lowest temperature that can be represented on the velocity grid)
could be influencing the potential fluctuation statistics in this region.
Unlike the density fluctuations, the normalized potential fluctuation levels tend to increase with decreasing
$B_v/B_z$.

\begin{figure}
\includegraphics[width=\linewidth]{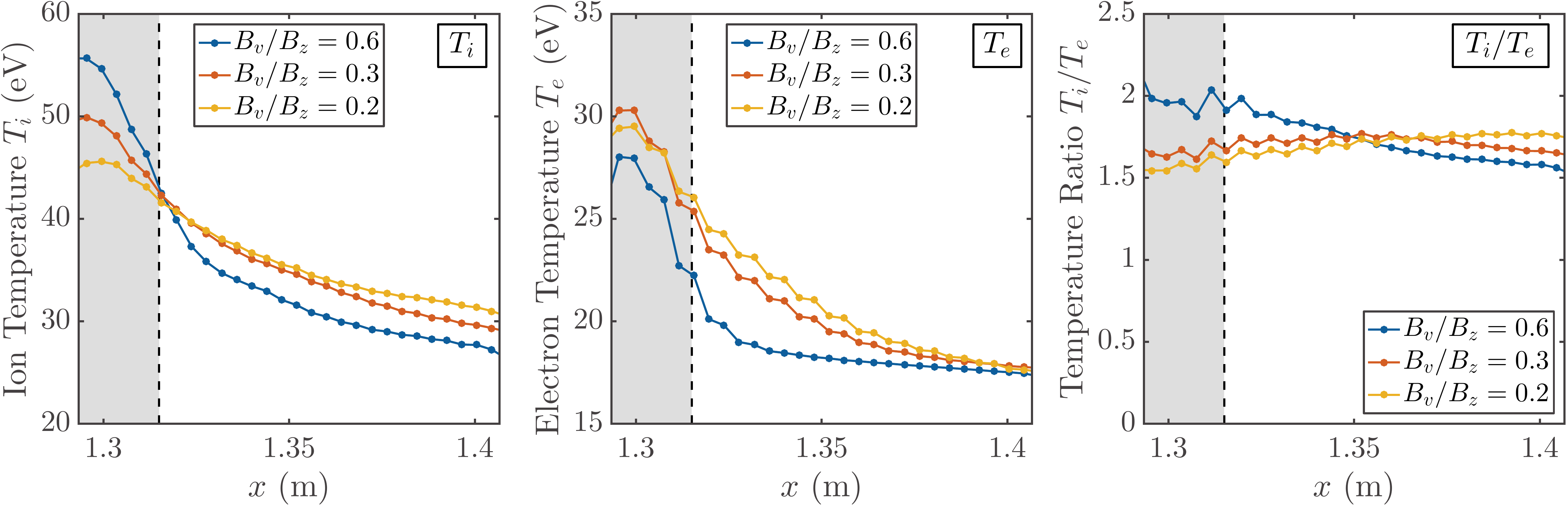}
  \caption[Radial profiles of the steady-state ion and electron temperatures
  and ion-to-electron temperature ratios near the midplane
  for cases with different magnetic-field-line pitches.]
  {Radial profiles of the steady-state ion and electron temperatures near the midplane
  and ion-to-electron temperature ratios for cases with different magnetic-field-line pitches.
  Although both electrons and ions are sourced at the same temperature, the sheath allows
  high-energy electrons to be rapidly lost from the system, resulting in lower
  electron temperatures in the SOL if collisions are not rapid enough to equilibrate 
  the two species \citep{Stangeby1990,Kocan2011}.}
  \label{fig:helical-sol-temperatures} 
\end{figure}

Figure~\ref{fig:helical-sol-temperatures} shows radial profiles of the
steady-state ion and electron temperatures and ion-to-electron
temperature ratios near the midplane for
different $B_v/B_z$.
Similar to the heat-flux profiles shown in figure~\ref{fig:helical-heat-flux},
the profiles are steepest for the case with $B_v/B_z=0.6$ and decay
more gradually in the lower-$B_v/B_z$ cases.
SOL measurements typically show that the ratio $T_i/T_e$ 
increases with radius \citep{Kocan2011}.
We see this trend in figure~\ref{fig:helical-sol-temperatures} for $B_v/B_z = 0.3$ and 0.2,
but not for $B_v/B_z = 0.6$.
This reversed trend for $B_v/B_z = 0.6$ is likely connected to the relatively flat
$T_e$ at large $x$.
In the $B_v/B_z = 0.6$ case, the low-amplitude source of ${\sim}33$~eV electrons at large
$x$ (see the form of the 
plasma source (\ref{eq:helical_sol_source})) could be setting $T_e$ in this region.
The flat $T_e$ at large $x$ could also be an artifact from the electron's running
into a floor in the temperature at large $x$ (although we note that $T_{\perp e,\mathrm{min}} = 16$~eV,
which is still a few eV lower than the $T_e$ seen in this region).
For all three simulations, $T_i/T_e$ falls in the range 1.5--2, which is within the range of
1--10 that is observed a few centimeters outside the LCFS in tokamaks \citep{Kocan2011}.

\begin{figure}
\includegraphics[width=\linewidth]{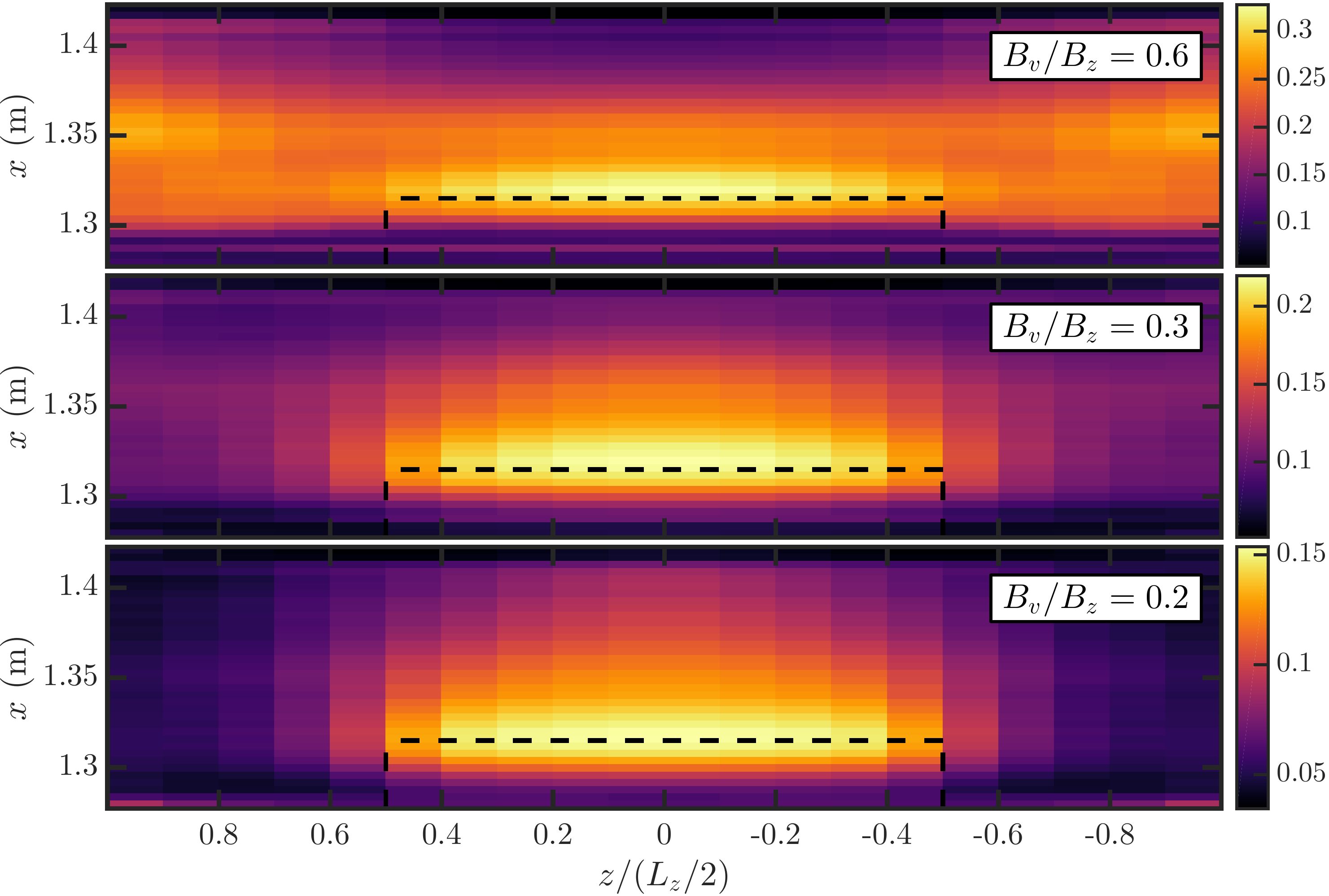}
  \caption[Comparison of the parallel structure of the normalized r.m.s. electron-density fluctuation amplitude
  for cases with different magnetic-field-line pitches.]
  {Comparison of the parallel structure of the normalized r.m.s. electron-density fluctuation amplitude
  for three cases with different magnetic-field-line pitches.
  While the density fluctuations are primarily $k_\parallel =0$ in the $B_v/B_z = 0.6$ case,
  more parallel structure is observed in the lower-$B_v/B_z$
  cases. The source region is indicated by the dashed black lines.}
  \label{fig:parallel_rms_density} 
\end{figure}

The normalized root-mean-square (r.m.s.) electron-density fluctuation level in the $x$--$z$ plane is shown
in figure~\ref{fig:parallel_rms_density}.
For all three values of $B_v/B_z$, the density fluctuation levels are the strongest in the source region
$|z| < L_z/4$.
The normalized density fluctuation levels in the $B_v/B_z = 0.6$ case are fairly uniform along the
field lines, while they tend to fall off by about a factor of 2--3 towards the sheaths in the
smaller $B_v/B_z$ cases.
This effect is likely a result of the stronger influence of the sheath on the potential as
the distance from the source to the sheath is decreased.
The instantaneous snapshots of $\tilde{n}_e$ (not shown) indicate a strong $k_\parallel =0$ component for the
largest $B_v/B_z$ cases, while more parallel structure is apparent in the smaller $B_v/B_z$ cases.

\begin{figure}
  \includegraphics[width=\linewidth]{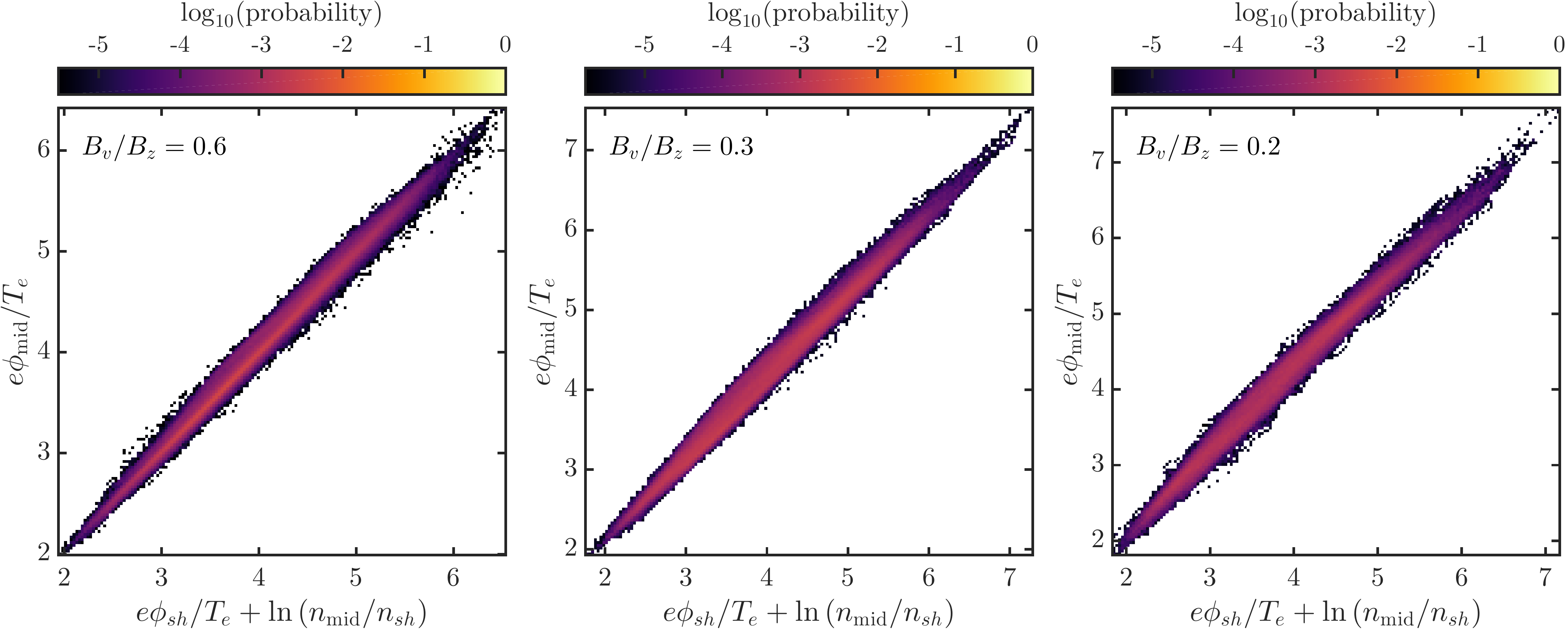}
  \caption[Comparison of the cross-coherence between $e\phi_{\mathrm{mid}}/T_e$ and
  $e\phi_{sh}/T_e + \ln \left(n_{\mathrm{mid}}/n_{sh} \right)$
  for cases with different magnetic-field-line pitches.]
  {Comparison of the cross-coherence between the midplane potential $e\phi_{\mathrm{mid}}/T_e$ and
  $e\phi_{sh}/T_e + \ln \left(n_{\mathrm{mid}}/n_{sh} \right)$ for three cases
  with different magnetic-field-line pitches $\theta$. Here, $\phi_{sh}$ is the sheath
  potential, $n_\mathrm{mid}$ is the midplane electron density, and $n_{sh}$ is the sheath
  electron density.
  These plots are created by binning ordered pairs of the two quantities sampled
  every 0.25 $\mu$s over a ${\sim} 1$ ms time interval at solution nodes falling in the region
  $1.318 \text{ m} \le x \le 1.326 \text{ m}$.
  In all three cases, the two quantities are highly correlated, which
  indicates that the electrons are strongly adiabatic (near parallel force balance).}
  \label{fig:helical_cross_coherence}
\end{figure}

The fluctuation statistics can also give information about the strength of the
electron adiabatic response for each simulation. By assuming that the electrons are isothermal
along field lines, parallel force balance satisfies
\begin{align}
  \nabla_\parallel \left(n_e e E_\parallel + P_e \right) &= 0 \\
  -n_e e E_\parallel &= T_e \nabla_\parallel n_e \\
  \frac{e \nabla_\parallel \phi}{T_e} &= \nabla_\parallel \ln n_e \\
  \frac{e \phi_\mathrm{mid}}{T_e} &= \frac{e \phi_{sh}}{T_e} + \ln \left( \frac{n_{\mathrm{min}}}{n_{sh}} \right)
  \label{eq:cross_coherence},
\end{align}
where $\phi_{sh}$ and $n_{sh}$ are the electrostatic potential and electron density evaluated
at the sheath entrances and $\phi_{\mathrm{mid}}$ and $n_{\mathrm{mid}}$ are the same quantities, but evaluated
at the midplane ($z=0$~m).
To compute the cross-coherence diagnostic \citep{Scott2005,Ribeiro2005,Mosetto2013}, ordered pairs
$\boldsymbol{(}e\phi_{\mathrm{mid}}/T_e,e\phi_{sh}/T_e + \ln \left(n_{\mathrm{mid}}/n_{sh} \right) \boldsymbol{)}$
falling in the region $1.318 \text{ m} \le x \le 1.326 \text{ m}$ (approximately
where the maximum density and potential fluctuations are) are sampled
at $1$~$\mu$s intervals over a ${\sim}1$~ms period for each simulation.
Figure~\ref{fig:helical_cross_coherence} shows the resulting plots (normalized bivariate histograms),
which all indicate a strong correlation between the two sides of (\ref{eq:cross_coherence}),
and so the electrons are strongly adiabatic, meaning that the electron distribution function
along a field line closely follows a Boltzmann distribution \citep{StoltzfusDueck2009}.
This result indicates that it might be possible to obtain similar results using an axisymmetric model (with
sheath-model boundary conditions) for the parameters considered here.

\begin{figure}
  \centering
  \includegraphics[width=\linewidth]{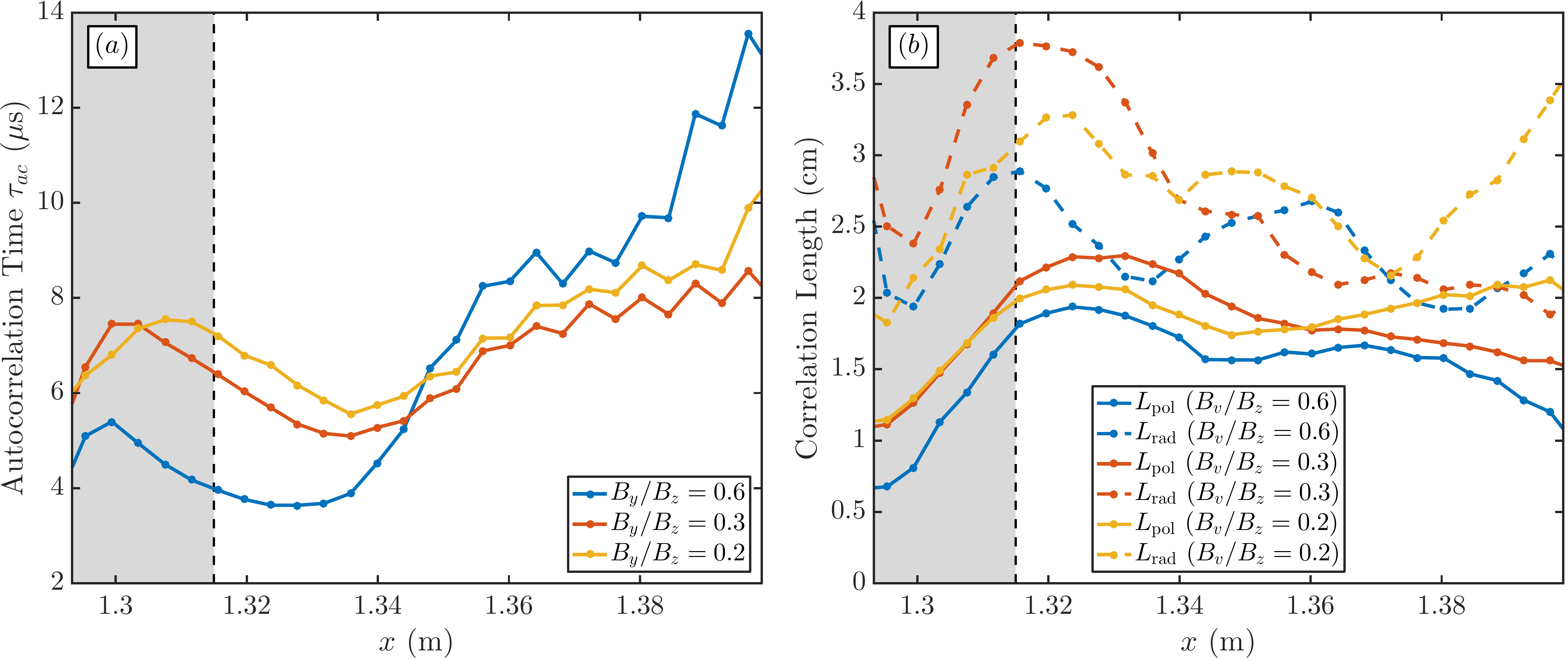}
  \caption[Radial profiles of the autocorrelation times and correlation lengths
  for cases with different magnetic-field-line pitches.]
  {Radial profiles of the ($a$) autocorrelation time and 
  ($b$) poloidal (dashed lines) and radial (solid lines) correlation lengths computed
  at the $z=0$~m plane for three cases with different
  magnetic-field-line pitches. The shaded area indicates
  the region in which the source is concentrated. $L_{\mathrm{pol}}/L_{\mathrm{rad}} \sim
  1.2$--$1.6$ is observed across the radial domain.}
  \label{fig:correlation_lengths} 
\end{figure}

Figure~\ref{fig:correlation_lengths}($a$) shows the radial profile of the
autocorrelation time $\tau_{ac}$ (computed from time traces of the density fluctuations).
In the SOL of the simulation, $\tau_{ac}$ tends to increase with radius, which is a trend observed in
to measurements on NSTX \citep[see][figure~12]{Zweben2015}.
The autocorrelation time for the $B_v/B_z = 0.2$ and $B_v/B_z=0.3$ cases is found to
vary between ${\sim}5$~$\mu$s and ${\sim}9$~$\mu$s, while the autocorrelation time
for the $B_v/B_z = 0.6$ case exhibits a larger variation in the SOL, 
with $\tau_{ac} \approx 4$~$\mu$s for $x < 1.34$~m and increasing to ${\approx}12$~$\mu$s at the
outer radial boundary.
The autocorrelation times we observe in our simulations are lower than the
$\tau_{ac} \sim 10$--$40$~$\mu$s reported by \citet{Zweben2015} for the NSTX edge and SOL,
but are well within the $\tau_{ac} \sim 2$--$20$~$\mu$s range that is typical
for edge and SOL turbulence in other tokamaks \citep{Boedo2009,Zweben2007}.

Figure~\ref{fig:correlation_lengths}($b$) shows the poloidal and radial correlation lengths ($L_{\mathrm{pol}}$ and
$L_{\mathrm{rad}}$ respectively) using the electron-density fluctuations near the $z=0$~m plane.
The correlation length at a radial location is obtained by averaging the correlation length computed at
several points in $y$.
At an individual point, the correlation length is determined from the correlation function, which is
constructed by computing the equal-time two-point autocorrelation function for density fluctuations separated
by some distance $\Delta y$ for $L_{\mathrm{pol}}$ or $\Delta x$ for $L_{\mathrm{rad}}$.
Having observed a significant wave feature in the poloidal correlation function,
we determined $L_{\mathrm{pol}}$ by fitting the poloidal correlation function to
$e^{-|\Delta y|/L_{\mathrm{pol}}} \cos( k_{\mathrm{wave} } \Delta y  )$.
The radial correlation function, which does not have a wave feature,
is computed using the full width at half maximum (FWHM) as $L_{\mathrm{rad}} = \mathrm{FWHM}/(2 \ln 2)$.

For all three values of $B_v/B_z$, we observe that the ratio
$L_{\mathrm{pol}}/L_{\mathrm{rad}}$ is between
1.2 and 1.6 for most of the radial domain, which is similar to the
$L_{\mathrm{pol}}/L_{\mathrm{rad}} \sim 1$--$2$ that is typically observed in tokamaks and stellarators
\citep{Zweben2007,Boedo2009}.
An average $L_{\mathrm{pol}}/L_{\mathrm{rad}} = 1.5 \pm 0.1$ was reported for
representative Ohmic NSTX discharges \citep{Zweben2016}, although larger ratios
$L_{\mathrm{pol}}/L_{\mathrm{rad}} \sim 3$--$4$
have been observed in some experiments \citep{Huber2005} and simulations \citep{Churchill2017}.

\begin{figure}
  \centering
  \includegraphics[width=0.75\linewidth]{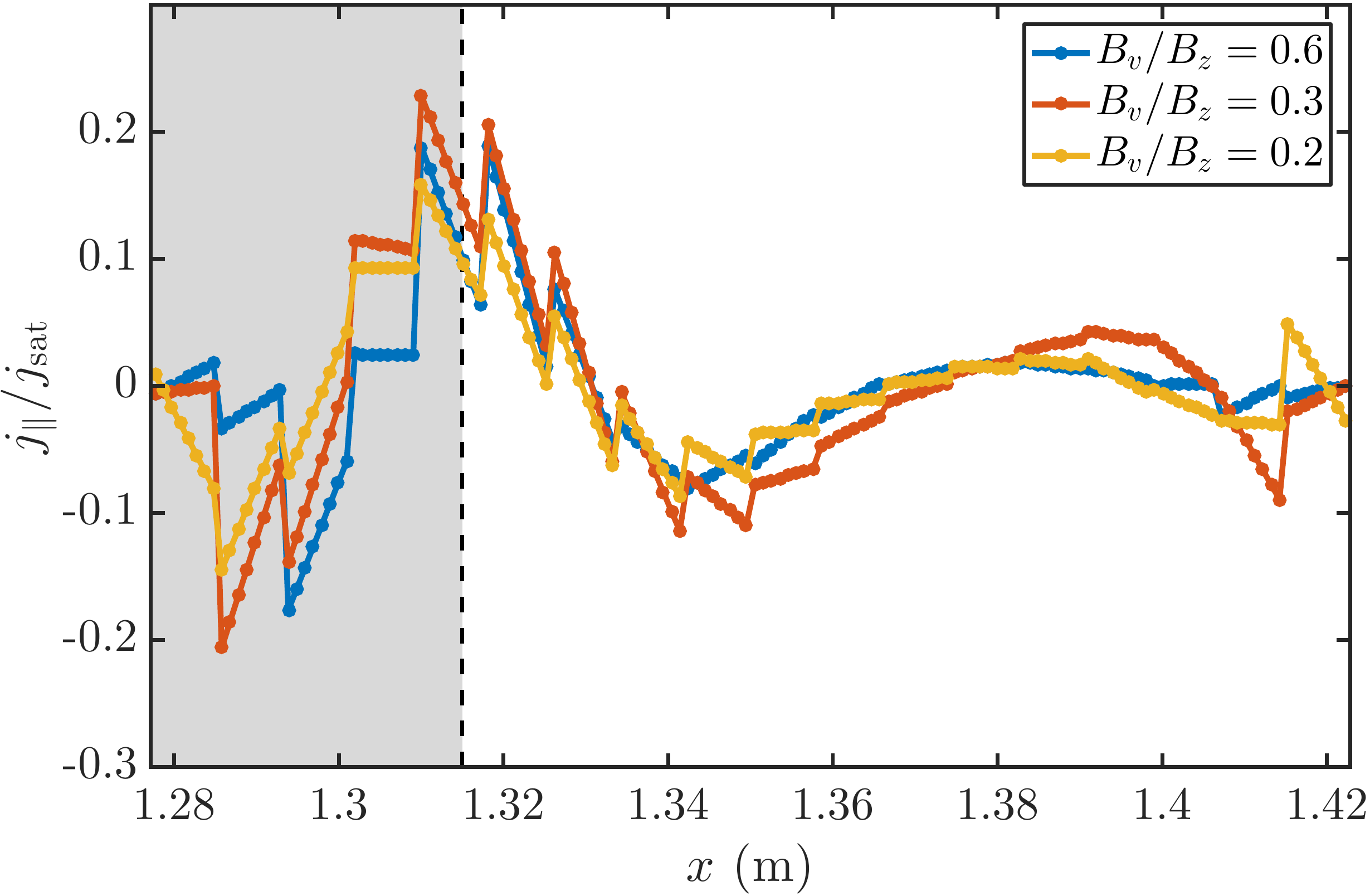}
  \caption[Radial profiles of the steady-state parallel currents into the sheaths for
  cases with different magnetic-field-line pitches.]
  {Radial profiles of the steady-state parallel currents into the sheaths for cases
  with different magnetic-field-line pitches.
  The current is normalized to the peak value of the steady-state ion saturation current
  $j_\mathrm{sat} = q_i n_i c_\mathrm{s}$ for each simulation.
  All three cases are quite quantitatively similar, featuring a large excess
  electron outflow in the source region that is balanced by a large excess
  ion outflow just outside of the source region.}
  \label{fig:helical-sol-edge-currents} 
\end{figure}

As discussed in Section \ref{sec:sheath}, there are two kinds of sheath-model boundary conditions
that are commonly used in fluid and gyrokinetic codes.
Logical-sheath boundary conditions enforce $j_\parallel = 0$ at the sheath entrances, while
current fluctuations into the sheath are permitted in conducting-sheath boundary conditions.
Figure~\ref{fig:helical-sol-edge-currents} shows the radial profiles of the steady-state
parallel current into the sheath for the three cases under consideration.
The currents have been normalized to peak steady-state 
ion saturation current $j_\mathrm{sat} = q_i n_i c_\mathrm{s}$,
where $c_\mathrm{s} = \sqrt{(T_e + \gamma T_i )/m_i}$ and $\gamma = 3$ is used because the collisionless
layer in front of the sheaths should be resolved in all three cases.
All three cases are quite quantitatively similar, and the
outward sheath currents are found to be highly symmetric in the $z$, which is consistent
with the strong adiabatic response shown in figure~\ref{fig:helical_cross_coherence}.
A large excess electron outflow (negative current) is seen in the source region,
which is compensated by a large excess ion outflow (positive current) just outside the source region.
The peak values are approximately 20\% of the ion saturation current, which motivates
future studies regarding the use of how various sheath-model boundary conditions affect
turbulence in these simulations.
In more realistic models of the SOL, the plasma source is in a region of
closed magnetic field lines, so there cannot be large electron sheath currents in the source region
in these models.
It will be interesting to explore how these profiles change when this capability is added to the code.

\section{Conclusions}
We have developed a model to investigate interchange-driven SOL turbulence in a simplified
helical-magnetic-field geometry and performed numerical simulations of the system using an
electrostatic gyrokinetic continuum code.
The blobs in our simulations appear to originate as radially elongated structures
that extend from the source region into the SOL and get broken up by sheared poloidal flows.
The blobs appear to efficiently transport plasma across the magnetic field, leading to
radial particle fluxes that are much higher than Bohm-flux estimates.
Such large-amplitude and large-scale blobs were not observed in a set of simulations
we performed without magnetic-curvature effects.
We note, however, that coherent structures with high plasma density have been
observed in linear devices with negligible magnetic curvature \citep{Antar2001,Carter2006}.
The mechanism that polarizes such coherent structures in linear devices and leads to
outward radial propagation could be due to neutral wind \citep{Krasheninnikov2003}.

We characterized the turbulence using a variety of diagnostics and found that
various quantities of interest are within the range expected for SOL turbulence in tokamaks,
such as fluctuation levels, autocorrelation times, and correlation lengths.
A summary of some quantities from our simulations is given in table~\ref{tab:helical-experiment},
which also includes experimental values from the NSTX SOL \citep{Zweben2015,Boedo2014}.
We varied the magnetic-field-line pitch in a set of simulations,
which indicated an increasing level of radial turbulent particle transport with decreasing pitch.
A cross-coherence diagnostic comparing potential fluctuations at the sheaths with those at the midplane 
indicated that all three simulations appeared to fall into a similar
turbulent regime with strongly adiabatic electrons.
The application of this model to investigate turbulence in the Helimak device \citep{Gentle2008}
is currently underway, which will also allow for comparisons with a previous GBS Braginskii
fluid simulation \citep{Li2011}.

\begin{table}
\begin{center}
\caption[Summmary of helical-SOL simulation results with comparison
  to experimental values in the SOL.]
  {Summary of helical-SOL simulation results with comparison
  to experimental values for an H-mode NSTX SOL reported in
  \citet{Zweben2015}.
  The values of $\Gamma_{n,r}$, $T_e$, and $n_e$ refer to values
  near the LCFS (whose location is not precisely known in the experiments \citep{Zweben2004}).
  Since GPI cannot be used to obtain particle fluxes, the value of $\Gamma_{n,r}$ for the NSTX case is
  taken from \citet{Boedo2014}.
  The `${\sim}$' symbol is used here to indicate that there can be large variations
  in such quantities between discharges with different parameters.
  Ion temperature measurements in the plasma boundary of NSTX were not available,
  so the value of 1--2 (seen on AUG and MAST \citep{Kocan2011}) is assumed.
  }
  \bigskip
  \begin{tabular}{ccc}
  \toprule
    \textbf{Quantity} & \textbf{Simulation Range} & \textbf{NSTX SOL} \\
  \midrule
    $\tau_{ac}$~($\mu$s) & 4--14 & 15--40  \\
    $L_\mathrm{pol}$~(cm) & 2--4 & 3--5 \\
    $L_\mathrm{rad}$~(cm) & 1--2.5 & 2--3  \\
    $\tilde{n}_\mathrm{rms}/\bar{n}$~(\%)& 10--30 & 20--100 \\
    $\Gamma_{n,r}$~$\left(10^{21}~\mathrm{m}^{-2}~\mathrm{s}^{-1} \right)$ & 3.5--5.1 & ${\sim}4$ \\
    $n_e$~$\left(10^{19}~\mathrm{cm}^{-3}\right)$ & 0.5--1.5 & ${\sim}1$ \\
    $T_e$~(eV) & 26--29 & ${\sim}29$ \\
    $T_i/T_e$ & 1.5--2 & 1--2 \\
  \bottomrule
  \end{tabular}
  \label{tab:helical-experiment}
\end{center}
\end{table}

The helical-SOL model can be extended by the addition of a closed-magnetic-field-line region (with periodic
boundary conditions in the parallel direction).
While the Gkeyll code can already perform simulations with periodicity in the parallel direction,
additional work is required to simultaneously include both open and closed-magnetic-field-line regions
in the same simulation.
The addition of good-magnetic-curvature regions and electromagnetic effects are
also important extensions that will make this model more applicable to
tokamaks.
Since our model is relatively simple compared to a realistic tokamak SOL,
the helical-SOL model could also eventually serve as a test case for the cross verification
of gyrokinetic boundary-plasma codes.
This test case might be useful for revealing major discrepancies due to different
numerical approaches, sheath-model boundary conditions, and collision operators
implemented in various codes relatively early on in the development cycle
before more significant investments are made.

%% file: ch-exponentially-weighted-basis-functions/chapter-exponentially-weighted-basis-functions.tex
\chapter{Conservative Exponentially Weighted Basis Functions\label{ch:exp-basis}}
This chapter discusses work on a conservative discontinuous Galerkin (DG) method that
employs non-polynomial approximation spaces to represent the distribution function.
As discussed in Chapter \ref{ch:models-and-numerical-methods},
the solution domain in DG methods is divided into a number of non-overlapping elements (cells)
and the numerical solution itself is represented as a linear combination of
local basis functions in each element.
One must choose a finite-element space in which the local solution is represented,
and even a different representation can be used in every element (related to p-adaptivity).
This flexibility in the mesh comes from the lack of inter-element-continuity enforcement in
the DG method.
Typically, the local approximation space is chosen to be the space of polynomials up to
a particular degree, resulting in the use of piecewise polynomial basis functions.
Piecewise polynomials may not always provide the best approximation to the solution, however.
For example, distribution functions often have Maxwellian tails that behave as $\propto \exp \left(-v^2\right)$ as
$v \rightarrow \pm \infty$, where $v$ is a velocity coordinate.
Therefore, it is of interest to examine alternative finite-element spaces for kinetic problems
to improve solution accuracy and reduce computational cost.

\citet{Yuan2006} studied DG methods using non-polynomial (trigonometric
and exponential) approximation spaces and explored methods to adjust the
approximation spaces as the solution changes over time.
They proved that non-polynomial finite element spaces satisfying
certain conditions have approximation rates similar to those of polynomial
finite element spaces of the same dimension.
Their 1D and 2D numerical results demonstrated that DG approximations based on
suitably selected non-polynomial functions 
could be much more accurate than using standard piecewise polynomials.
The authors did, however, acknowledge the challenge in efficiently identifying appropriate
approximation spaces for practical problems of interest.
Additionally, the authors did not investigate the conservation properties of the algorithm,
which we will show to be a serious issue later in this chapter.

Gyrokinetic simulations of plasma microturbulence are often computationally intensive,
requiring the calculation of distribution function in a 3D2V (three spatial
dimensions and two velocity dimensions) phase space.
It is important to pursue efficient numerical methods for 5D gyrokinetic turbulence codes because
algorithmic choices have a big impact on what problems can be simulated on current supercomputers.
Some strategies to reduce the computational cost of gyrokinetic simulations include
the use of multigrid methods, implicit-time-stepping methods, and sparse-grid methods.
For codes using grid-based (continuum) algorithms, one must be careful not to waste
grid points in regions where fine resolution of the distribution function is not necessary.
Additionally, it can be challenging to represent disparate temperatures of the same particle species
on the same phase-space grid because the velocity-space resolution must be fine enough to resolve the colder
particles, and the velocity-space extents must be large enough so that the warmer particles are still
far from the velocity-space boundaries.

\citet{Landreman2013} explored various collocation strategies for efficient velocity-space
discretization in the context of pseudospectral methods using the speed coordinate
$v$ defined on the semi-infinite domain $[0,\infty)$.
They found that calculations employing a little-known family of non-classical polynomials orthogonal
with respect to $v^\nu e^{-v^2}$ ($\nu > -1$) on the interval $v \in [0,\infty)$
\citep{Shizgal1979,Shizgal1981,Ball2002}
often yielded superior performance at both differentiating and integrating distribution functions
when compared to more commonly used collocation schemes.
These non-classical polynomials have been recently applied to the time-dependent problem of
relaxation to a Maxwellian via Fokker-Planck collisions \citep{Wilkening2015}.
While the results in \citet{Wilkening2015} were promising, the authors recognized
the additional complication arising from calculations on a dense stiffness matrix.

In this chapter, we explore the use of exponentially weighted polynomials to
represent the velocity dependence of a distribution function in problems with Fokker--Planck collisions.
We focus on the application of these ideas in the DG framework, which results in calculations with smaller
matrices due to the high degree of locality inherent to such methods.
We discuss representations of the distribution function using
exponentially weighted polynomials in Section~\ref{sec:exp-algorithm} and
show how conservation issues arise from the use of non-polynomial representations
in the standard DG method, a consequence that has not been previously discussed in the literature.
We propose a modification that allows one to use certain non-polynomial-weighted basis functions
in DG methods while respecting the conservation properties of the original equations.
Using our proposed numerical scheme, we study the 1D relaxation to a Maxwellian distribution
in Section~\ref{sec:exponential_collision_test} and the calculation of the parallel heat flux in a simplified
Spitzer--H\"arm test problem \citep{Spitzer1953} in Section~\ref{sec:exponential_heat_test}.

We note that the algorithms discussed in this chapter have not yet been implemented in the Gkeyll code.
It is possible that the exponentially weighted basis functions as they are described here
might introduce too much overhead to the calculations to be competitive with standard polynomial basis functions.
Nevertheless, we provide some evidence in this chapter that the use of non-polynomial basis functions
is an idea worth pursuing in future versions of the code because of the potentially significant savings
that it could enable.

\section{The General Algorithm \label{sec:exp-algorithm}}
We first briefly review how the standard Runge--Kutta discontinuous Galerkin (RKDG) method is applied to solving
nonlinear conservation laws of the form
\begin{align}
  \frac{\partial f}{\partial t} + \frac{\partial g(f)}{\partial v} &= 0. \label{eq:exp-hyperbolic}
\end{align}
This method uses an explicit Runge--Kutta time discretization and a
discontinuous-Galerkin space discretization.
For more information about the RKDG method, we refer the reader to Section \ref{sec:algorithms}
and the references therein.

We approximate the solution in each interval $I_j =\left[v_{j-\frac{1}{2}},v_{j+\frac{1}{2}}\right]$
by expanding in terms of $N$ local, predetermined basis functions as
\begin{align}
  f_h(v,t) &= \sum_{k=1}^{N} f_j^k(t) \psi_j^k(v),\qquad v \in I_j.
\end{align}
Typically, the basis functions span the space of polynomials up to a certain degree.
In a modal discontinuous Galerkin approach, the standard choice is to take $\psi_j^k(v)$ to
be the Legendre polynomials $P_k(v)$ defined on $I_j$.
The standard piecewise-polynomial approximation space of degree $k$ is denoted in this chapter as
\begin{align}
  V^k =& \left\{w:w|_{I_j} \in \mathrm{span}\{1,(v-v_j),\dots,(v-v_j)^k\}, v \in I_j \right\}.
\end{align}

Starting from an initial condition, the solution for $t>0$ is determined by
applying a Galerkin method to solve for $\partial f_h/\partial t$.
The weak formulation is obtained by multiplying (\ref{eq:exp-hyperbolic}) by a test function $w(v)$ and
integrating over $I_j$:
\begin{align}
  \int_{I_j}  \mathrm{d}v\, w(v) \left( \frac{\partial f_h}{\partial t} + \frac{\partial g\left(f_h\right)}{\partial v} \right) \label{eq:exp-galerkin} &= 0.
\end{align}
In the DG method, $w(v)$ is chosen to span the same space as the basis functions,
so (\ref{eq:exp-galerkin}) leads to a system of $N$ equations for $N$ unknowns $\partial{f}_j^k/\partial t$ in
each interval $I_j$.
As noted in Section \ref{sec:algorithms}, (\ref{eq:exp-galerkin}) with $w(v)$ chosen to be
the basis functions also arises from a minimization of the squared-$L^2$-norm error with respect
to $\partial f_h/\partial t$.
Integration by parts is used to move the spatial derivative from the nonlinear term $g$ onto $w(v)$:
\begin{multline}
  \int_{I_j} \mathrm{d}v \, w(v) \frac{\partial f_h}{\partial t} =
  \int_{I_j}\mathrm{d}v \frac{\partial w}{\partial v} g\left(f_h\right)   +
  - \hat{g} \left( f_h\left(v_{j+\frac{1}{2}},t \right) \right) w\left(v_{j+\frac{1}{2}}^-\right) \\
  + \hat{g} \left( f_h\left(v_{j-\frac{1}{2}},t \right) \right) w\left(v_{j-\frac{1}{2}}^+\right),
  \label{eq:exp-rkdg}
\end{multline}
where we have replaced the flux $g$ evaluated at the boundaries
by the numerical flux $\hat{g}$. The numerical flux is a single-valued function of the
left and right limits of the discontinuous solution at the boundary,
i.e. $\hat{g}\left(f_h \left(v_{j+\frac{1}{2}},t\right) \right) =
  \hat{g}\left(f_h\left(v_{j+\frac{1}{2}}^-,t\right),f_h\left(v_{j+\frac{1}{2}}^+,t\right) \right)$.
Depending on the problem, common choices are to use a centered flux or an upwind flux.
The value of the test function evaluated on the boundaries is taken from within the interval $I_j$.

If we represent $f$ using the standard piecewise-polynomial approximation space,
the numerical scheme locally conserves particle number.
The numerical scheme satisfies (\ref{eq:exp-rkdg}) with $w=1$ because piecewise constants are
in the approximation space $V^k$:
\begin{equation}
  \int_{I_j} \mathrm{d}v\, \frac{\partial f_h}{\partial t} =
  - \hat{g}\left( f_h\left(v_{j+\frac{1}{2}},t \right) \right)
  + \hat{g}\left( f_h\left(v_{j-\frac{1}{2}},t \right) \right).\label{eq:numconserv}
\end{equation}
The quantity on the left-hand side is simply the rate of change in the particle number in the interval $I_j$,
while the right-hand side is comprised of the fluxes across each boundary of cell $j$.

In this work, we are interested in approximating $f$ on each interval $I_j$
for Fokker-Planck equations using the following two non-polynomial expansions:
\begin{align}
  f_h(v_\parallel,t) &= \sum_{k=1}^{N} f_j^k(t) \psi_j^k(v_\parallel) =
  \sum_{k=1}^{N} f_j^k(t) \beta_{0,j} \exp\left(-\beta_{1,j} \frac{\left(v_\parallel-\beta_{2,j}\right)^2}{2}\right) P_k(v_\parallel), \label{eq:maxwellian}\\
  f_h(\mu,t) &= \sum_{k=1}^{N} f_j^k(t) \psi_j^k(\mu) =
  \sum_{k=1}^{N} f_j^k(t) \alpha_{0,j} \exp\left(-\alpha_{1,j} \mu \right) P_k(\mu), \label{eq:exponential}
\end{align}
where the variables $\beta_{0,j}$, $\beta_{1,j}$, $\beta_{2,j}$, $\alpha_{0,j}$, and $\alpha_{1,j}$
are free parameters of the basis functions that may also vary in time as well as in each cell.
One can recognize these forms as the standard piecewise polynomials
weighted by an exponential factor in $v_\parallel$ (the parallel velocity) or $\mu$ (the magnetic moment),
so we refer to these specific non-polynomial basis functions will be referred to as
\textit{exponentially weighted polynomials}.
For the exponentially weighted basis functions to be an attractive (and hopefully superior) alternative to
polynomial basis functions, it is important to be able to dynamically adjust
the exponential weighting factor as the solution changes in time-dependent problems.
Although our initial tests use predetermined values for the parameters of the exponential weighting factor,
we propose a method to automatically choose these parameters in Section \ref{sec:exponential-adjustment}.

\begin{figure}
  \includegraphics[width=\textwidth]{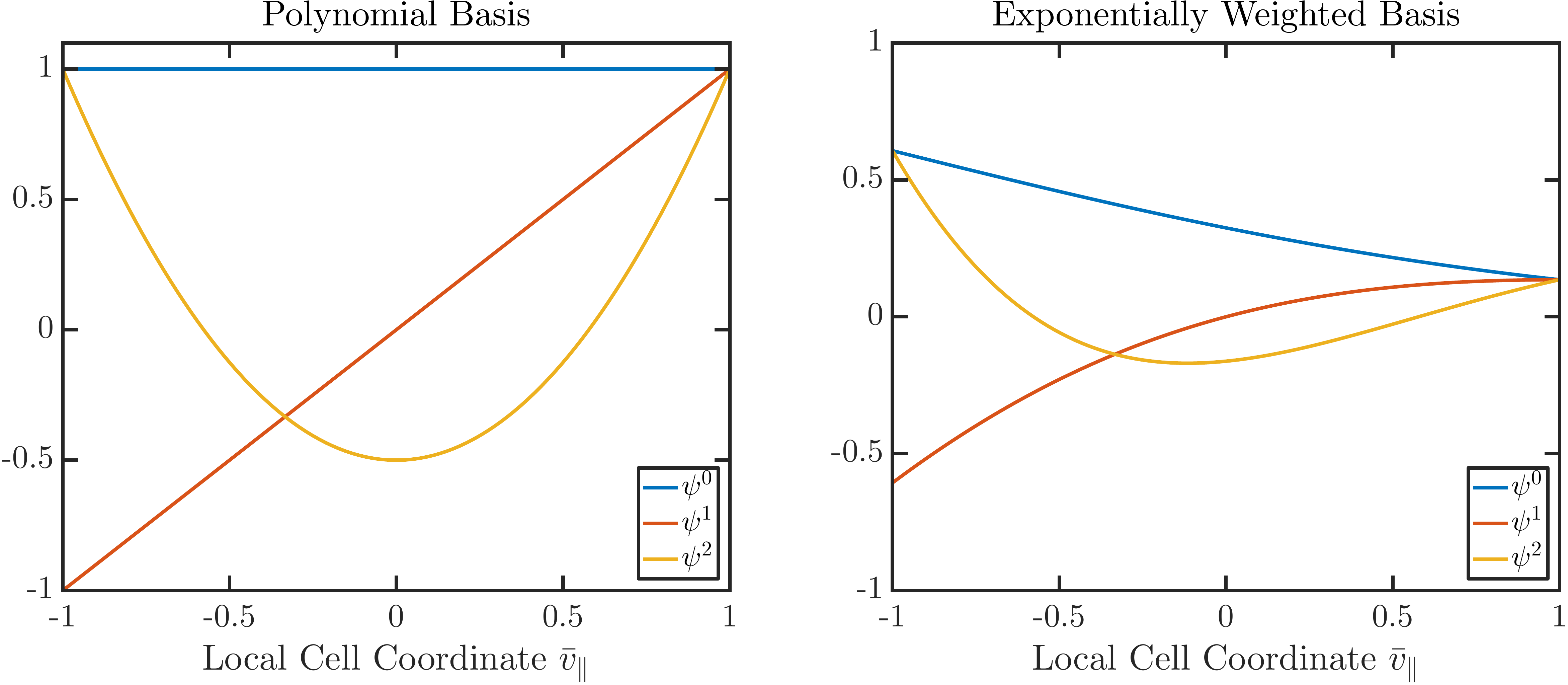}
  \caption[Comparison of standard polynomial basis functions (Legendre polynomials) with a set of exponentially
    weighted basis functions.]
    {Comparison of standard polynomial basis functions (Legendre polynomials) with a set of exponentially
    weighted basis functions (\ref{eq:maxwellian}). For the exponentially weighted basis, we use
    the parameters $\beta_{0,j} = 1$, $\beta_{1,j} = 1/4$, and $\beta_{2,j} = 1$ and plot the basis functions
    for the cell with centroid located at $v_j=4$. While the polynomial basis functions are the same
    in every cell, the exponentially weighted basis functions can vary cell-to-cell due to the weighting factor.}
  \label{fig:basisFunctionComp}
\end{figure}

Figure~\ref{fig:basisFunctionComp} shows a comparison of standard polynomial basis functions with
the exponentially weighted basis functions in (\ref{eq:maxwellian}).
Besides the advantage of potentially being a better approximation than piecewise polynomials,
these approximation spaces also permit the use of the complete velocity space domain $[0,\infty)$ in $\mu$
and $(-\infty,\infty)$ in $v_\parallel$ instead of requiring truncation of the
velocity domain at a finite and arbitrary value, which is typically chosen to be a few thermal velocities.

The approximation spaces corresponding to the representations in (\ref{eq:maxwellian}) and (\ref{eq:exponential})
are
\begin{align}
  \bar{M}^k(\beta) =& \Big\{ w:w|_{I_j} \in \mathrm{span}\big\{  \beta_0 e^{-\beta_1 (v-v_j-\beta_2)^2/2},
    \beta_0 e^{-\beta_1 (v-v_j-\beta_2)^2/2} (v-v_j),\dots,\\ \nonumber
    &\beta_0 e^{-\beta_1 (v-v_j-\beta_2)^2/2} (v-v_j)^k \big\}, v \in I_j \Big\},\\
  M^k(\alpha) =& \Big\{ w:w|_{I_j} \in \mathrm{span}\big\{ \alpha_{0,j} e^{-\alpha_{1,j}(\mu-\mu_j)},
    \alpha_{0,j} e^{-\alpha_{1,j}(\mu-\mu_j)} (\mu-\mu_j),\dots,\\ \nonumber
    & \alpha_{0,j} e^{-\alpha_{1,j}(\mu-\mu_j)} (\mu-\mu_j)^k \big\}, v \in I_j \Big\},
\end{align}
where $v_j$ and $\mu_j$ are the coordinates in $v_\parallel$ and $\mu$ of the centroid of cell $j$.

If one represents $f$ using (\ref{eq:maxwellian}) or (\ref{eq:exponential}) and uses
the standard RKDG algorithm (\ref{eq:exp-rkdg}) to solve the hyperbolic equation (\ref{eq:exp-hyperbolic}),
the resulting numerical schemes no longer conserve particle number.
We can no longer substitute $w=1$ into (\ref{eq:exp-rkdg}) because piecewise constants
are no longer contained in the approximation spaces $M^k$ and $\bar{M}^k$ used to represent $f$,
and (\ref{eq:numconserv}) is not satisfied.
This numerical scheme will be referred to as the \textit{non-conservative exponentially weighted DG method}
in the rest of this chapter.
Similarly, there are issues with the conservation of other
polynomial moments of the distribution function that were conserved
prior to discretization (such as energy or momentum).
The lack of number and energy conservation is numerically demonstrated for an example problem
in Section~\ref{sec:exponential_collision_test}.
These conservation issues are not simply of academic concern; in plasma physics,
small errors in number conservation can lead to large errors in the electric field,
and small errors in energy can lead to significant errors when integrating for long time scales.

One way to recover the conservation properties of the physical equation
is to introduce a weighting function $1/W$ into
the definition of the error that is minimized by the DG scheme such that the
weighting function cancels out the exponential weighting of the basis functions.
For $w$ in the same approximation space as $f_h$, the residual in the interval $I_j$
is now chosen to satisfy
\begin{equation}
  \int_{I_j} \mathrm{d}v \, \frac{w}{W} \left( \frac{\partial f_h}{\partial t} + \frac{\partial g\left(f_h\right)}{\partial v} \right) = 0.
\end{equation}
To be pedantic, the resulting system of equations is
\begin{multline}
  \int_{I_j} \mathrm{d}v\, \frac{w}{W} \frac{\partial f_h}{\partial t} =
  \int_{I_j} \mathrm{d}v\,  \frac{\partial}{\partial v}\left(\frac{w}{W}\right) g\left(f_h\right) 
  -\hat{g} \left( f_h\left(v_{j+\frac{1}{2}},t \right) \right)
    \frac{w \left(v_{j+\frac{1}{2}}^- \right) }{W \left(v_{j+\frac{1}{2}}^- \right)} \\
  +\hat{g} \left( f_h\left(v_{j-\frac{1}{2}},t \right) \right)
     \frac{w \left(v_{j-\frac{1}{2}}^+ \right) }{W \left(v_{j-\frac{1}{2}}^+ \right)}.
  \label{eq:weightedDG}
\end{multline}
Equation~(\ref{eq:weightedDG}) will be referred to in this chapter as the \textit{conservative
exponentially weighted DG method}.

When approximating $f$ using the forms in (\ref{eq:maxwellian}) or (\ref{eq:exponential}),
the weighting functions to restore the conservation properties are respectively
\begin{align}
  W_\mu &= \alpha_0 \exp\left(-\alpha_1 \mu \right), \label{eq:weightExp}\\
  W_{v_\parallel} &= \beta_0 \exp\left(-\beta_1 \frac{\left(v_\parallel-\beta_2\right)^2}{2}\right).
  \label{eq:weightGauss}
\end{align}
The resulting numerical scheme is considered a Petrov-Galerkin scheme,
in which the test function $w$ is not in the same approximation space as the solution $f_h$,
by interpreting $w/W$ to be the test function.

More generally, this procedure can be used to design conservative numerical schemes
that represent the solution using polynomials weighted by some non-polynomial function
(e.g., a power law) $W_j(v)$ in cell $j$ as:
\begin{align}
  f_h(v,t) &= \sum_{k=1}^N f_j^k(t) \psi_j^k(v) = \sum_{k=1}^N f_k(t) W_j(v) P_k(v)
\end{align}
If one or both velocity space extents are to be located at $\pm \infty$, then
$W(v)$ has the additional constraint that $W_j(v)P_k(v) \rightarrow 0$ as $v \rightarrow \pm \infty$ for all $k$
in the boundary cells.

For certain conservation properties to be satisfied when using a non-polynomial approximation space,
we modified the definition of the error norm minimized by the DG method by introducing the non-polynomial
weighting factor $1/W(v)$.
This means that solutions of the conservative, non-polynomial DG method will have
a higher (non-weighted) squared-$L^2$-norm error than those of the non-conservative, non-polynomial DG method.
In our numerical tests, however, we do not find this difference to be significant.

\section{1D Collision Operator Tests\label{sec:exponential_collision_test}}
In this section, we study the evolution of a 1D distribution function
represented using the exponentially weighted basis (\ref{eq:exponential}) in the presence of collisions only:
\begin{align}
  \frac{\partial f}{\partial t} &= C[f].\label{eq:collonly}
\end{align}
We choose to model collisions using the same-species Lenard--Bernstein
collision operator \citep{Lenard1958}, which has the general nonlinear form
\begin{align}
  C[f] &= \nu \frac{\partial}{\partial \boldsymbol{v}}\cdot \left[ \left(\boldsymbol{v}-\boldsymbol{u}\right)f
+ v_t^2 \frac{\partial f}{\partial \boldsymbol{v}} \right],
\end{align}
where $\nu$ is a collision frequency and the variables $\boldsymbol{u}=\boldsymbol{u}(f)$
and $v_t^2(f)$ are chosen to ensure that the collision operator conserves momentum and energy.
In addition to conserving number, momentum, and energy, this collision operator relaxes
distributions to a Maxwellian.
For simplicity, we have taken the transport coefficients that appear in the collision operator
to be constants and neglected the velocity-dependence in the collision frequency in our tests.
A complete treatment would involve the calculation of Rosenbluth potentials \citep{Rosenbluth1957}.
As an approximation to the Landau collision operator, the Lenard--Bernstein collision operator is
better suited than the Krook model to model collisions in a plasma
due to its Fokker--Planck form, which represents the dominance of small-angle scattering in plasmas.

We first investigate self-species collisions in the perpendicular
velocity coordinate $\mu$, the magnetic moment, which is equal to $m v_\perp^2/2B$
when the magnetic field is static and uniform and is commonly used as a velocity-space
coordinate in gyrokinetics.
Here, the collision operator has the form
\begin{equation}
  C[f] = \nu \frac{\partial}{\partial \mu} \left[ 2 \mu f +
    \mu_t \left(2\mu\frac{\partial f}{\partial \mu}\right) \right],\label{eq:perpcoll}
\end{equation}
where $\nu$ is taken to be a constant and $\mu_t = \int \mathrm{d} \mu \, \mu f / \int \mathrm{d}\mu \, f = \langle \mu \rangle$.
Number and energy are conserved in this system.
Total energy conservation in the exact system is obtained as
\begin{align}
  \frac{d}{dt} n\langle \mu \rangle &= \frac{d}{dt} \int_0^\infty f \mu \, \mathrm{d}\mu\nonumber\\
                                    &= \int_0^\infty \frac{\partial f}{\partial t} \mu \, \mathrm{d}\mu\nonumber\\
                                    &= -2\nu n \left( \langle \mu \rangle - \mu_t \right)\nonumber\\
                                    &= 0,
\end{align}
where $n = \int \mathrm{d}\mu \, f$.
We have also made use of the assumption that $f \rightarrow 0$ as $\mu \rightarrow \infty$ to eliminate
boundary terms.
Numerically, the drag term is calculated using an upwind flux and the 
diffusive term is calculated using the recovery-based DG method of \citet{VanLeer2005},
in which the solution and its derivatives evaluated on a boundary are found through
the recovery of a smooth solution that spans the two neighboring cells (see Section \ref{sec:collision_operator}).
We compare the performance of linear ($k=1$) polynomial basis functions with exponentially weighted
linear polynomial basis functions, both of which have two degrees of freedom per cell.

Time stepping is performed using the third-order SSP Runge--Kutta method (\ref{eq:ssp-rk3-1})--(\ref{eq:ssp-rk3-3}).
Zero-flux boundary conditions are applied at $\mu = 0$ and $\mu = \mu_{\mathrm{max}}$ for a
piecewise-polynomial representation and at $\mu = 0$ for a exponentially weighted-polynomial
representation. In the grid for exponentially weighted basis, there is a cell that extends to infinity,
so no boundary conditions are required there.
The velocity-space domain used for the polynomial case is $\mu \in [0,18\mu_t]$ and the grid used for the
exponentially weighted cases is $\mu \in [0,\infty)$, with the last cell covering the interval $[18\mu_t - \Delta \mu, \infty)$.
We use normalized values of $\mu_t = 1$ and $\nu = 1$.
The exponential weighting factors were fixed to be $\alpha_{0,j} = 1$ and $\alpha_{1,j}=1$ in every cell.

\begin{figure}
  \centering
  \includegraphics[width=0.75\textwidth]{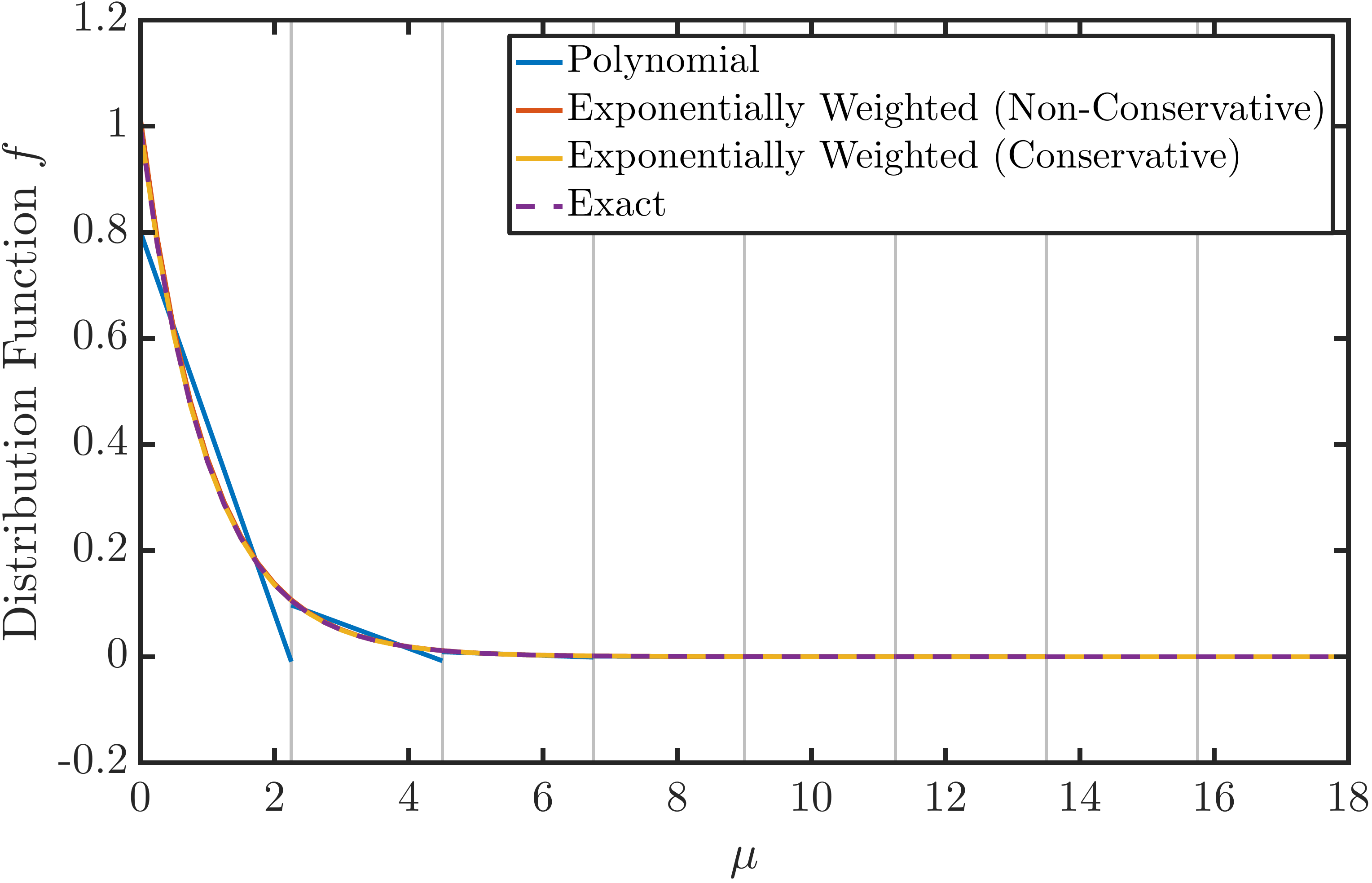}
  \caption[Comparison of the steady-state solutions to a collision operator test in $\mu$
    using three DG methods.]
    {Comparison of the steady-state solutions to a collision operator test in $\mu$
    using three DG methods.
    The methods used are standard DG with polynomial basis functions ($k=1$),
    standard DG with exponentially weighted basis functions, and
    conservative DG with exponentially weighted basis functions.
    The exact steady-state solution $f=e^{-\mu}$ is indicated in the
    dashed purple line.
    The initial condition is a top-hat distribution, and the solution to (\ref{eq:collonly}) at $t=10$ is plotted.
    The exponentially weighted solutions and the exact steady-state solution all lie on top of each other in the plot,
    while the equivalent linear-polynomial solution exhibits some negative overshoots at cell edges (indicated
    by vertical lines).}
  \label{fig:exp_distf}
\end{figure}

We set the initial condition to the projection of $f(\mu,t=0) = H(\mu-\mu_0)$ onto the respective basis functions,
where $H(\mu)$ is the Heaviside step function, and we evolve the solution to a steady state.
One can quickly verify that the exact steady-state solution to (\ref{eq:perpcoll}) is $f \propto e^{-\mu/\mu_t}$.
Figure~\ref{fig:exp_distf} compares the steady-state distribution function (plotted at $t=2$)
for various DG algorithms with the exact solution, using the same grid resolution and amount
of data ($N_\mu = 8$ and two degrees of freedom per cell) to represent the numerical solution in each case.
It is apparent that the exponentially weighted polynomials in the form of (\ref{eq:exponential})
result in a more accurate representation of the solution than standard polynomials
when compared to the exact solution.
The polynomial solution also exhibits negative overshoots at cell edges, which can cause numerical difficulties
and violates the realizability of the distribution function.
We note that despite using numerical methods that minimize different error norms,
the conservative and non-conservative exponentially weighted solutions are quite visually similar.

\begin{figure}
  \includegraphics[width=\textwidth]{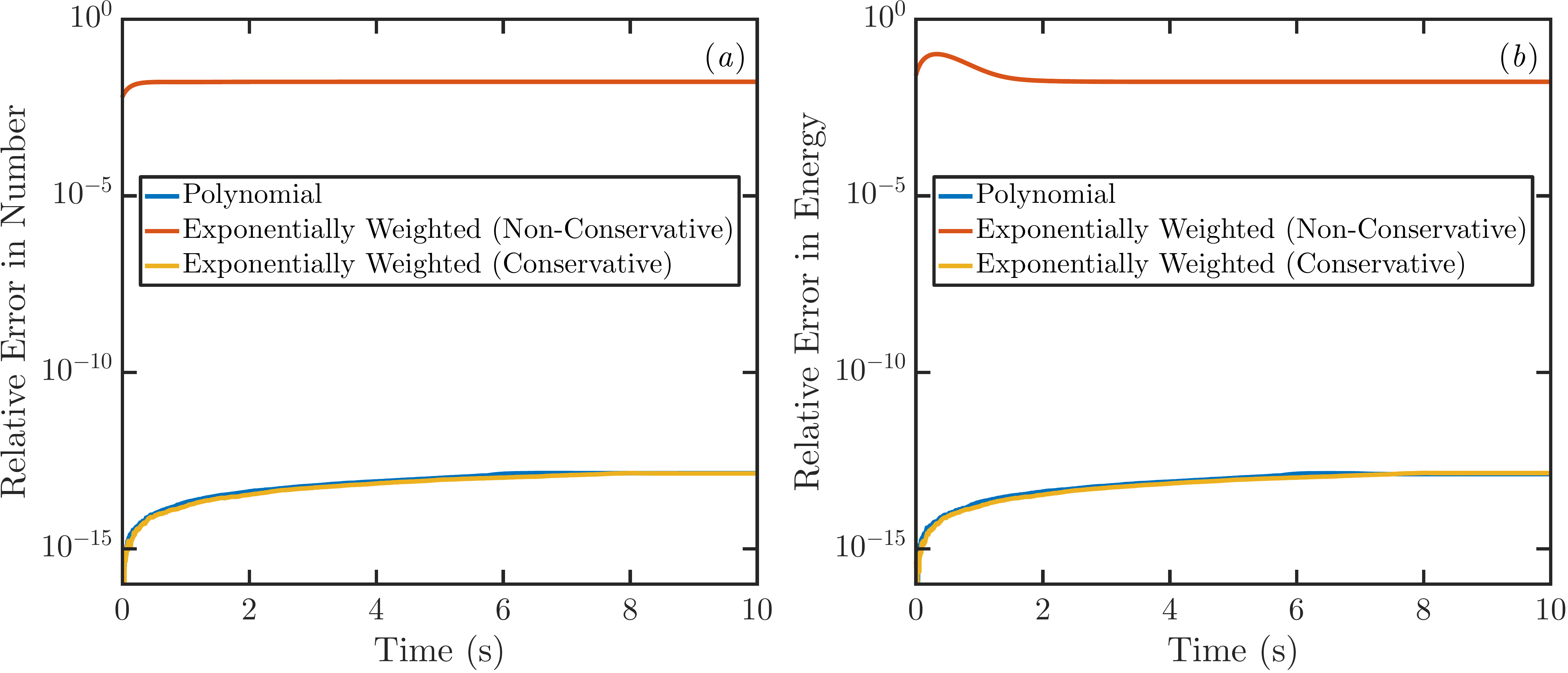}
  \caption[Comparison of the number and energy-conservation properties of various DG methods using
    polynomial and exponentially weighted basis functions for collision-operator test in $\mu$.]
    {Comparison of the ($a$) number and ($b$) energy-conservation properties of
    various DG methods using polynomial and exponentially
    weighted basis functions for a collision-operator test in $\mu$.
    The relative error is plotted for each quantity as a function of time.
    The use of exponentially weighted basis functions in the standard DG approach leads to unacceptable
    levels of error in total number and total energy, while a conservative DG method that uses
    exponentially weighted basis functions is obtained with use of an 
    appropriate weighting function as in (\ref{eq:weightedDG}).}
  \label{fig:exp_error}
\end{figure}

There are, however, serious conservation issues using the exponentially weighted 
representation of (\ref{eq:exponential}) with the standard DG algorithm (\ref{eq:exp-rkdg}).
These conservation issues are shown in figure~\ref{fig:exp_error}, which plots the relative error
in number and in energy vs. time for the three numerical methods considered.
We see that the exponentially weighted representation paired with the standard DG algorithm
does not conserve number or energy, and the degree of non-conservation is quite severe.
The piecewise-polynomial representation is able to conserve number and  energy to 
relative errors of order $\mathcal{O}(10^{-14})$,
but the exponentially weighted representation only conserves number and energy to relative errors of order
$\mathcal{O}(10^{-2})$.
While one might think that conservation errors of $\mathcal{O}(10^{-2})$ are tolerable,
small errors in charge conservation lead to large errors in the electric field in plasmas.

We verify in this numerical example that introducing a $1/W$ weighting from (\ref{eq:weightExp}) allows one to
recover the same machine-precision level of number and energy conservation as using the standard
piecewise-polynomial representation of the distribution function.
Additionally, the conservative exponentially weighted representation is a much more
accurate representation of the exact steady-state solution than the polynomial representation when compared
to the exact steady-state solution.

\begin{figure}
  \centering
  \includegraphics[width=0.75\textwidth]{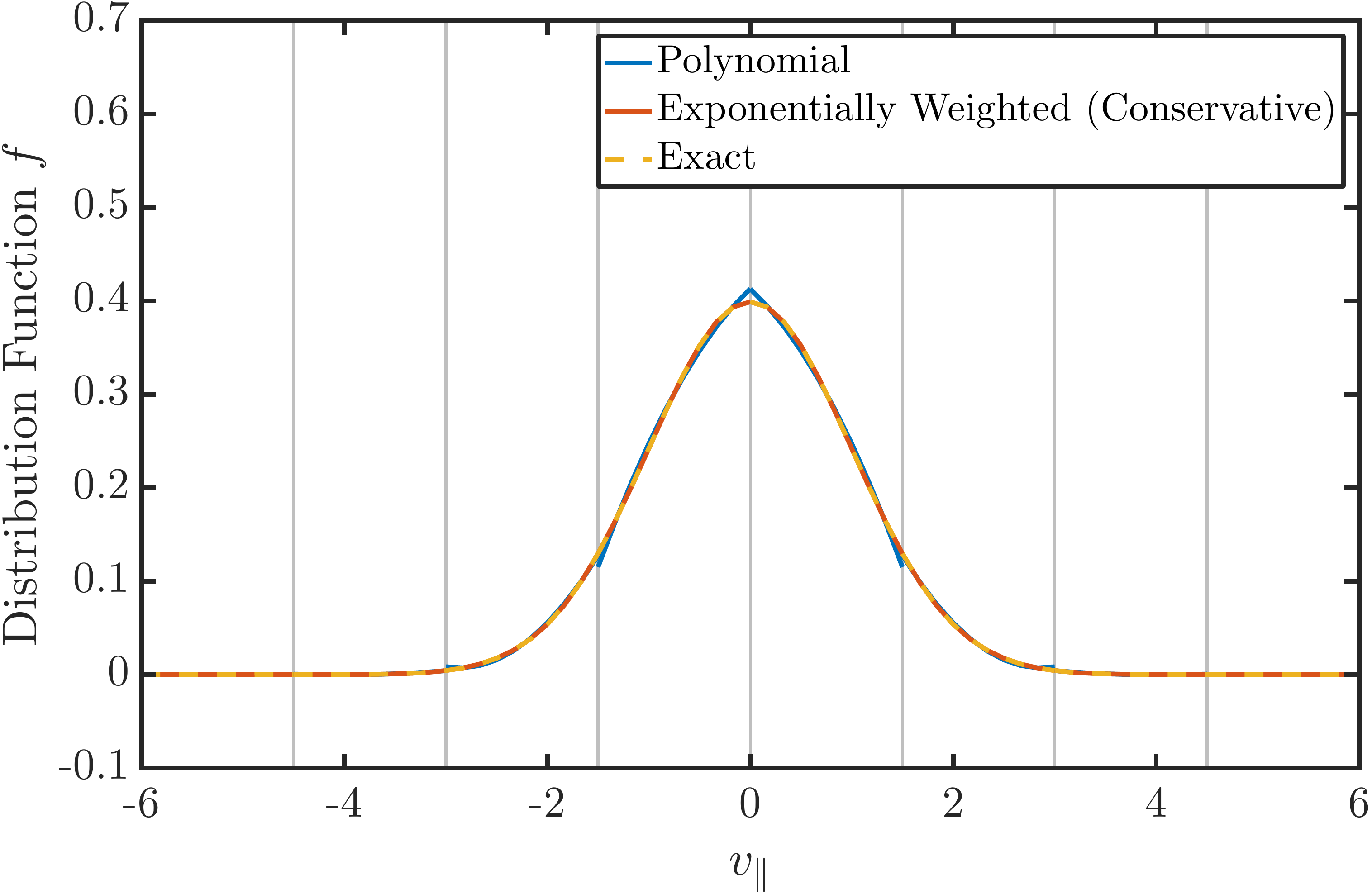}
  \caption[Comparison of the steady-state solutions to a collision operator problem in $v_\parallel$
    using two DG methods.]
    {Comparison of the steady-state solutions to a collision operator problem in $v_\parallel$
    using two DG methods.
    The two methods used are standard DG with quadratic polynomial basis functions
    and conservative DG with exponentially weighted basis functions.
    The exact steady-state solution $f=e^{-v_\parallel^2/2}/\sqrt{2\pi}$ is indicated in the
    dashed yellow line.
    The initial condition is a top-hat distribution function,
    and the solution to (\ref{eq:collonly}) at $t=10$ is plotted.}
  \label{fig:exp_distf_parallel}
\end{figure}

\begin{figure}
  \includegraphics[width=\textwidth]{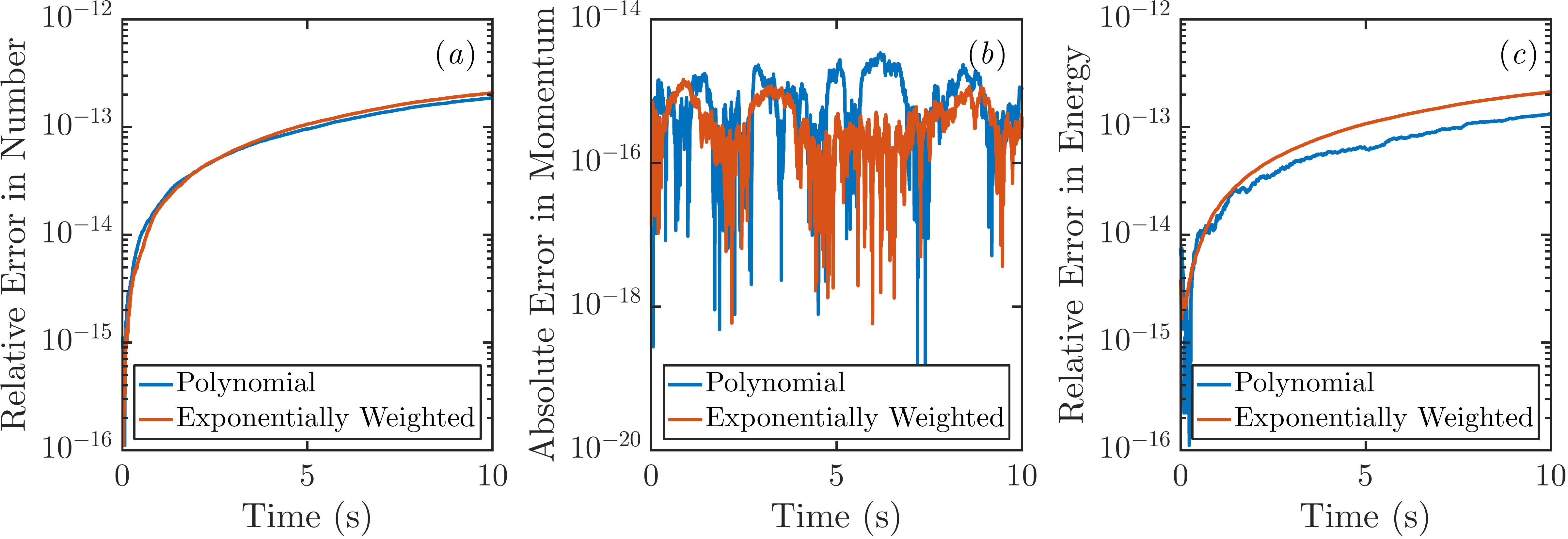}
  \caption[Comparison of the number, momentum, and energy-conservation properties of conservative DG methods using
    quadratic-polynomial and exponentially weighted basis functions for a collision-operator test in $v_\parallel$.]
    {Comparison of the number, momentum, and energy-conservation properties of conservative DG methods using
    quadratic-polynomial and exponentially
    weighted basis functions for a collision-operator test in $v_\parallel$.
    ($a$) shows the relative error in number, ($b$) shows the absolute error in momentum,
    and ($c$) shows the relative error in energy. As expected, the analytically conserved quantities are
    also conserved by these numerical methods.}
  \label{fig:exp_error_parallel}
\end{figure}

For collisions in $v_\parallel$, the collision operator is
\begin{equation}
  C[f] = \nu \frac{\partial}{\partial v_\parallel} \left(\left(v_\parallel-u_\parallel\right) f +
    v_t^2 \frac{\partial f}{\partial v_\parallel}\right) \label{eq:exp-parallelLB}.
\end{equation}
The initial condition used in these tests is the projection of
\begin{equation}
  f(v_\parallel,t=0) = \frac{H(v_\parallel + \sqrt{3} v_t) - H(v_\parallel - \sqrt{3} v_t)}{2\sqrt{3} v_t}
\end{equation}
onto the respective basis functions.
The grid has $N_{v_\parallel} = 8$ cells and $v_\parallel \in [-6 v_t, 6 v_t]$ for
the polynomial case and $v_\parallel \in (-\infty, \infty)$ for the exponentially weighted case,
with the boundary cells on the intervals $(-\infty, -6v_t + \Delta v_\parallel]$ and $[6v_t-\Delta v_\parallel,\infty)$.
We also set $v_t = 1$ and $\alpha = 1$ for these simulations.
We do not plot the results for the non-conservative DG method with exponentially weighted basis functions
for simplicity.
Figure~\ref{fig:exp_distf_parallel} shows the steady-state solutions of the standard DG method
with quadratic polynomials and the conservative DG method with exponentially weighted quadratic polynomials.
The number, momentum, and energy-conservation properties of the two DG methods are shown in
figure~\ref{fig:exp_error_parallel}.
As expected, both DG methods have similar machine-level conservation errors.

\section{A Classical-Heat-Transport Problem\label{sec:exponential_heat_test}}
Next, we benchmark the performance of exponentially weighted basis functions
using a 1D test problem in $v_\parallel$ motivated by the classical Braginskii procedure for
calculating heat conduction in a plasma (analogous to the Hilbert--Chapman--Enskog procedure
in a gas).
We solve an equation in the high-collision-frequency regime,
so the use of the exponentially weighted basis functions is appropriate.
A source term is added to the equation to drive the solution to a non-Maxwellian.

Our starting point is the following kinetic equation for the distribution function $f(z,v_\parallel,t)$
in a 2D phase space:
\begin{align}
  \frac{\partial f}{\partial t} + v_\parallel \frac{\partial f}{\partial z} &= C[f]
\end{align}
We have neglected electric and magnetic fields, but retained the behavior of small-angle
collisions that occur in a plasma through the use of a Lenard--Bernstein collision operator.
If we were modeling a gas, where large-angle scattering events are dominant, the use of a
Krook-type collision operator would be more appropriate.

The standard Chapman--Enskog-type procedure assumes that the collision frequency is large
$\nu \sim 1/\epsilon$, so the collision operator is the largest term, and expands
$f = f_0 + f_1 + \dots$.
To lowest order, $C[f_0] = 0$, so $f_0$ is a local Maxwellian of the form
\begin{align}
  f_0(z,v_\parallel) = f_M(z,v_\parallel) &= \frac{n(z)}{\sqrt{2\pi v_t^2(z)}}
  \exp \left( -\frac{\left(v_\parallel-u_\parallel\right)^2}{2 v_t^2(z)} \right).
\end{align}
The temperature $T(z)$ is related to $v_t^2(z)$ as $v_t^2(z) = T(z)/m$, where $m$ is the particle mass.

To next order, one gets
\begin{align}
  \frac{\partial f_0}{\partial t} + v_\parallel \frac{\partial f_0}{\partial z} = C\left[f_1\right].
\end{align}
We assume that there is are background temperature and density gradients in $f_0$ and that
these two parameters can be related by invoking local pressure balance
$\partial p / \partial z = \partial(nT)/\partial z = 0$.
We can then rewrite the kinetic equation as
\begin{align}
  \frac{\partial f_0}{\partial t} + \frac{v_\parallel }{L_T} \left(\frac{v_\parallel^2}{2 v_t^2} - \frac{3}{2}\right) f_0 &= C\left[f_1\right],
  \label{eq:heat1d}
\end{align}
where $L_T = T/(\partial T/\partial z)$ is the strength of the local background-temperature gradient.

For simplicity, we turn this 2D equation into a 1D equation at a single value of $z$ with a specified
input value of $L_T$.
Since the collision operator vanishes on $f_0$, we can write the right-hand side of (\ref{eq:heat1d}) as
$C[f_1] \rightarrow C[f]$, where $f = f_0 + f_1$.
In the high-collision-frequency, short-mean-free-path limit, $f_1 \ll f_0$,
we can replace $f_0$ on the left-hand side of (\ref{eq:heat1d}) with $f$.
The 1D model can then be written as
\begin{align}
  \frac{\partial f}{\partial t} &= C\left[f\right] - \frac{v_\parallel }{L_T} \left(\frac{v_\parallel^2}{2 v_t^2} - \frac{3}{2}c_1\right) f.
  \label{eq:heat1dfinal}
\end{align}
The time-dependent coefficient $c_1 = 1$ for an exact Maxwellian and must be adjusted in the simulations
to ensure that the second term on the right-hand side of (\ref{eq:heat1dfinal}) injects no momentum,
as the numerical $f_0$ can have deviations from the exact Maxwellian:
\begin{align}
  \int_{-\infty}^{\infty} \frac{v_\parallel^2}{L_T}\left( \frac{v_\parallel^2}{2v_t^2} -
    \frac{3}{2}c_1 \right) f \, \mathrm{d}v_\parallel &= 0.
\end{align}
In practice, $c_1 \approx 1$.
Since the initial condition for $f$ is chosen to have zero momentum and
there is no momentum source in the system, we set $u_\parallel = 0$ in the collision
operator (\ref{eq:exp-parallelLB}).
As in the previous section, we set $v_t = 1$,$m=1$, $n=1$, and $\nu = 1$, but one should
make $\nu$ depend on velocity in future work for a more realistic model of plasma collisions.
The initial condition $f(v_\parallel,t=0)$ is the projection of
$f_0(v_\parallel)$ onto the respective polynomial or exponentially weighted basis functions.
We also choose $T/L_T = 10^{-5}$.

The goal of this test problem is to solve (\ref{eq:heat1dfinal}) to a steady state
and calculate the heat flux $q = \int \mathrm{d} v_\parallel, m v_\parallel^3 f/2$,
which can then be compared with the analytical solution.
A prediction for the heat-flux profile $q(t)$ in the high-collision-frequency,
small-$f_1$ limit can be obtained by multiplying
(\ref{eq:heat1dfinal}) by $mv_\parallel^3/2$ (taking $c_1 = 1$) and integrating over $v_\parallel$.
The result is
\begin{align}
  q(t) &= \frac{n v_t^2}{2\nu} \frac{T}{L_T} \left( e^{-3\nu t} -1 \right)\label{eq:exp-exact-heat-flux}.
\end{align}

\begin{figure}
  \centering
  \includegraphics[width=0.75\textwidth]{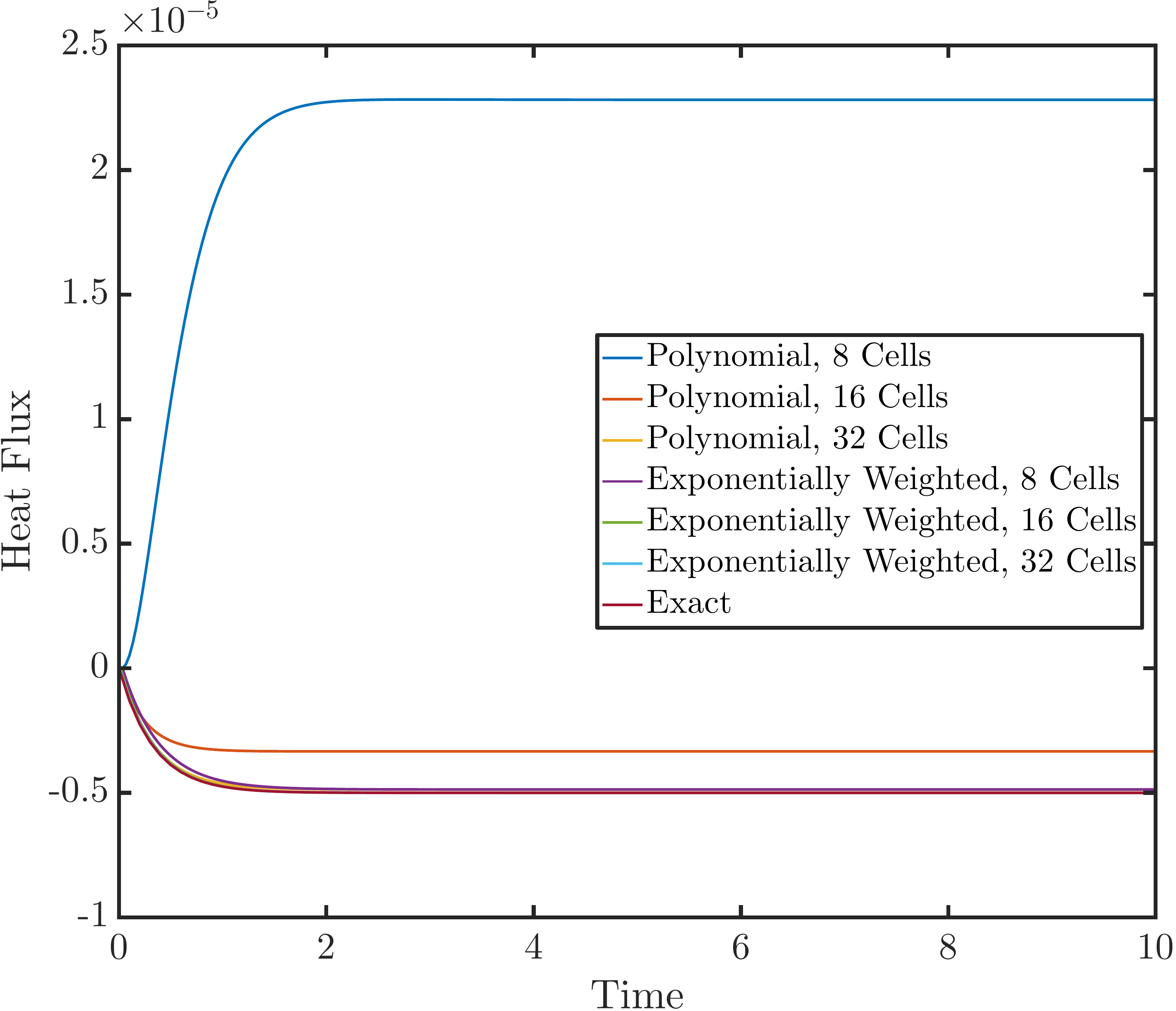}
  \caption[Time evolution of the heat flux in simulations using polynomial and exponentially weighted polynomials
  at various grid resolutions.]
  {Time evolution of the heat flux in simulations using polynomial and exponentially weighted polynomials
  at various grid resolutions. The exact solution is given in (\ref{eq:exp-exact-heat-flux}). While
  all the simulations using exponentially weighted polynomials have a $q(t)$ trace that is close to the exact
  solution, 32 cells are required for the simulation using polynomials to achieve a similar level of accuracy.
  Additionally, the heat flux has the wrong magnitude and sign for one of the polynomial cases
  performed on a coarse grid.}
  \label{fig:exp-heat-flux-time}
\end{figure}

Figure~\ref{fig:exp-heat-flux-time} shows a comparison of the heat flux $q(t)$ evolution
for simulations using linear polynomials
and exponentially weighted linear polynomials to represent the distribution function.
Here, the grid extents are kept fixed at $[-8v_t,8v_t]$ for polynomials and $(-\infty,\infty)$
for exponentially weighted polynomials, and the number of cells are varied.
All three of the exponentially weighted solutions plotted in figure~\ref{fig:exp-heat-flux-time}
are extremely close to the exact solution, while the polynomial simulation needs to use 32 cells in velocity space
to achieve similar levels of accuracy.
Additionally, the heat flux has an incorrect sign and magnitude for the lowest-resolution (8 cells) case
that uses polynomials.

\begin{figure}
  \centering
  \includegraphics[width=0.75\textwidth]{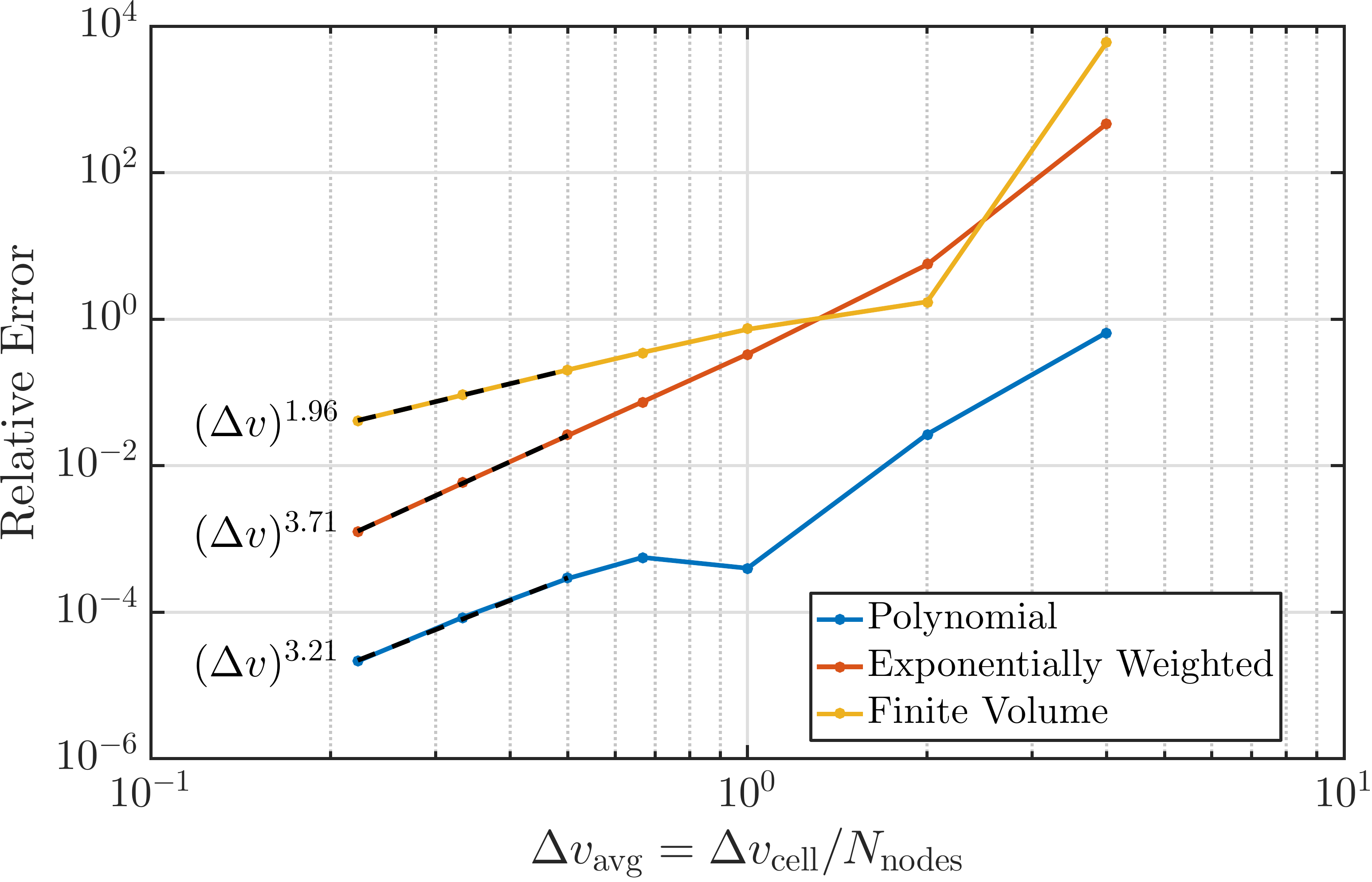}
  \caption[Relative error in the steady-state heat flux
  versus effective cell size for various numerical methods.]
  {Relative error in the steady-state heat flux (computed at $t=10$)
  versus effective cell size for various numerical methods.
  Here, $v_{\mathrm{max} = 8 v_t}$ for the polynomial and finite volume methods.
  The numerical methods plotted are
  standard DG with linear polynomials, conservative
  DG with exponentially weighted linear polynomials,
  and a standard second-order finite-volume method.}
  \label{fig:exp-heatflux}
\end{figure}

Figure~\ref{fig:exp-heatflux} shows a comparison of the relative error in the quantity
$q(t=10)$ using three numerical methods versus effective grid size.
The three numerical methods considered are standard DG with linear polynomials,
conservative DG with exponentially weighted linear polynomials,
and a standard second-order finite volume method.
Since the DG methods in figure~\ref{fig:exp-heatflux} store two pieces of data per cell
while the finite-volume method stores only one, the relative error is plotted versus
$\Delta v_{\mathrm{avg}} = \Delta v_{\mathrm{cell}} / N_{\mathrm{nodes}}$ to compare
the numerical methods on a common footing.
The velocity-space domain is
$v_\parallel \in [-8 v_t, 8v_t]$ for the standard DG and finite volume methods.
Since the outermost cells of the velocity space domain extend to $\pm \infty$,
the effective cell width for the conservative exponentially weighted DG method is computed
using only the interior cells.
The leftmost cell has domain $v_\parallel \in [-\infty, -8v_t + \Delta v_\parallel]$
and the rightmost cell has domain $v_\parallel \in [8v_t - \Delta v_\parallel, \infty]$.
For both DG methods, the total number of cells range from 4 to 72 cells.

Figure~\ref{fig:exp-heatflux} demonstrates that exponentially weighted polynomials
are ${\sim}10^2$ times more accurate than polynomials at the same resolution
in calculating the heat flux.
For the same level of error, one can use ${\sim}8$ times fewer cells in $v_\parallel$
using exponentially weighted polynomials instead of standard piecewise polynomials.
For problems in $(v_\parallel, \mu)$ coordinates used in gyrokinetics, one could potentially see
a factor of $10$ speedup using exponentially weighted basis functions, accounting some for
the additional complexity of using these exponentially weighted basis functions.

\begin{figure}
  \centering
  \includegraphics[width=0.75\textwidth]{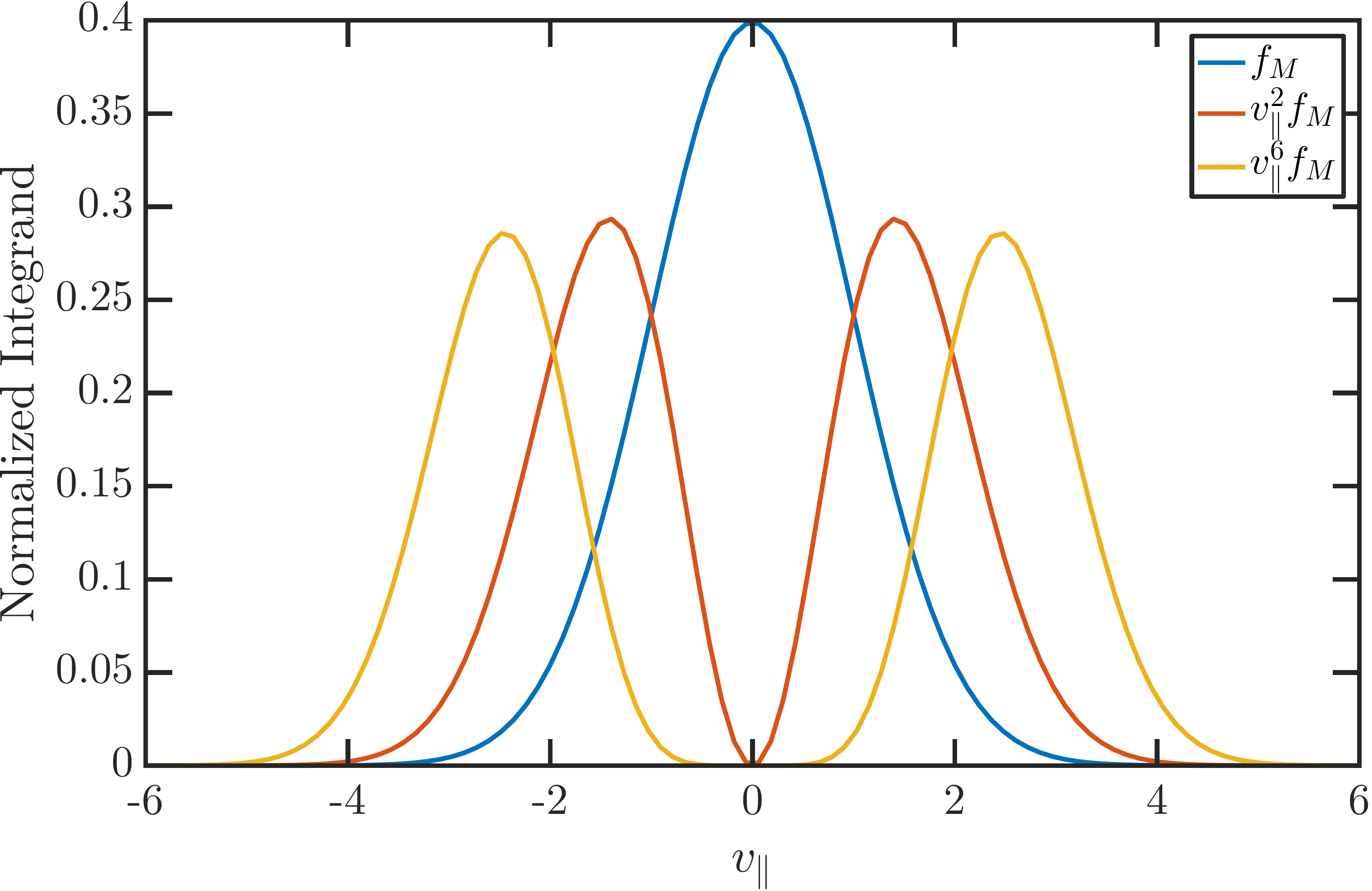}
  \caption[Illustration of the normalized integrands for the calculation of a few moments of a Maxwellian distribution.]
  {Illustration of the normalized integrands for the calculation of a few moments of a Maxwellian distribution.
  For an accurate evaluation of the heat flux, which has an integrand that scales as $v_\parallel^6 f_M$
  in the test problem considered in this section, good resolution in the tails of the distribution function
  is required. Exponentially weighted polynomials appear to be able to use fewer pieces of data to resolve
  the distribution function tails to a certain level of accuracy when compared to standard polynomials.}
  \label{fig:exp-heat-flux-integrands}
\end{figure}

We attribute the large difference between the exponentially weighted and polynomial methods
in the heat flux calculation to the need to represent the tails
of the distribution accurately, as the heat-flux integrand scales as $v_\parallel^6 f_M$ for a
velocity-independent collision frequency.
Figure~\ref{fig:exp-heat-flux-integrands} shows that the dominant contribution to the heat-flux integral
comes from the tail of the distribution function rather than the bulk.
The need to resolve the tail accurately is even greater in three velocity dimensions,
in which case the heat-flux integrand scales as $v^{11} f_M$ when
the collision frequency scales as $v^{-3}$.
If one is restricted to representing the distribution function using
piecewise polynomials, a fine resolution in the tails and a large $v_{\mathrm{max}}$ are 
both needed for the accurate evaluation of these heat-flux integrals.

\section{Adjustment of the Exponential Weighting Factor\label{sec:exponential-adjustment}}
In our tests, we selected the free parameters (the exponential weighting factor)
of our basis functions based on knowledge of the analytical solution and
required that every cell use the same values for these parameters.
In practice, these constraints would make the use of exponentially weighted basis functions
impractical for many problems of interest
in which the background temperature changes over time or the solution develops non-Maxwellian features.
As the solution in each cell changes in time, it is also important for the exponential weighting factor
in each cell to adjust so that the basis functions can continue to accurately represent the solution.

We propose a method to adjust the free parameters that
determine the approximation space defined on the interval $I_j$
based on the numerical solution in the same interval.
Essentially, we determine the new parameters such that
all of the polynomial variation in a cell is put into the exponential weighting factor.
Equivalently, we find a Maxwellian in a cell that has the same number
momentum, and energy (or just number and energy in $\mu$ coordinates)
as the solution-to-be-projected.
This procedure can be used to determine the initial values of the
basis function parameters and also to change from one approximation space to another
at the end of each time step.
To reduce the computational expense of this procedure, the adjustment of the
exponential weighting factor can be performed at regular intervals consisting of several time steps.

One advantage of this procedure is that the solution in the new approximation
space is already known once the new basis function parameters have been determined.
The methods proposed in \citet{Yuan2006} require
a stage to solve for the new parameters and another stage to project
the solution from the old approximation space to the new one.
Additionally, the 1D approximation spaces considered by \citep{Yuan2006} only had
one free parameter per cell, and the adjustment procedure proposed by the authors
does not easily generalize to multiple free parameters.

If one represents the solution using (\ref{eq:maxwellian}), the
parameters $\beta_{0,j}$, $\beta_{1,j}$, and $\beta_{2,j}$ can be found by solving
the nonlinear system using common root-finding methods:
\begin{align}
  \int_{I_j}  \mathrm{d}v_\parallel \, \beta_{0,j} \exp \left(- \beta_{1,j}\frac{(v_\parallel - \beta_{2,j})^2}{2}\right)
  &= \int_{I_j} \mathrm{d}v_\parallel \, f \\
  \int_{I_j} \mathrm{d}v_\parallel \, \beta_{0,j} \exp \left(- \beta_{1,j}\frac{(v_\parallel - \beta_{2,j})^2}{2}\right) v_\parallel
  &= \int_{I_j}\mathrm{d}v_\parallel \, f v_\parallel \\
  \int_{I_j} \mathrm{d}v_\parallel\, \beta_{0,j} \exp \left(- \beta_{1,j}\frac{(v_\parallel - \beta_{2,j})^2}{2}\right) v_\parallel^2
  &= \int_{I_j} \mathrm{d}v_\parallel\, f v_\parallel^2.
\end{align}

\begin{figure}
  \centering
  \includegraphics[width=0.75\textwidth]{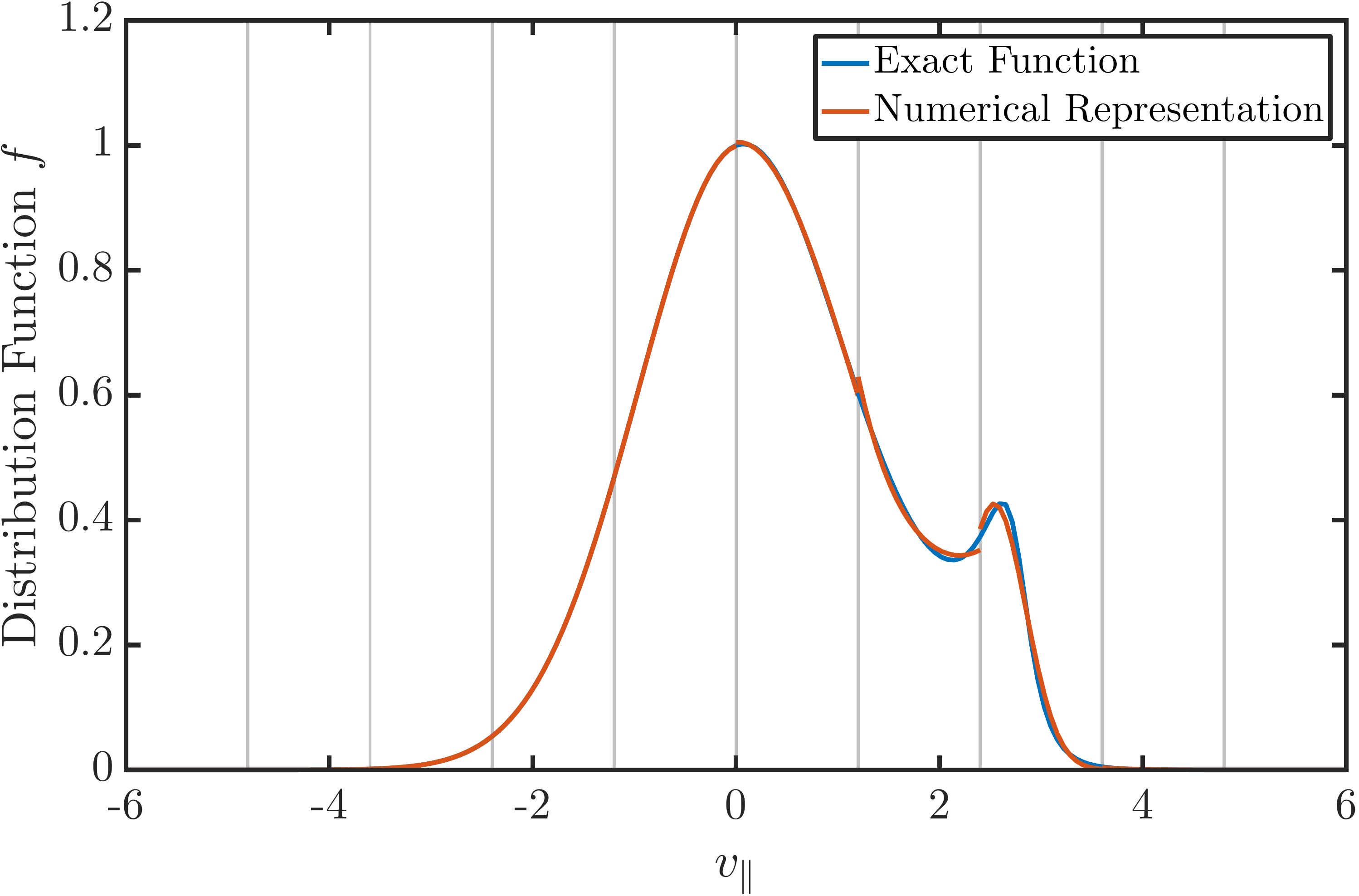}
  \caption[Calculation of an exponentially weighted approximation space from a specified initial condition.]
  {Calculation of an exponentially weighted approximation space from a specified initial condition.
  The red curve is the result of a nonlinear solve in each cell to determine the local approximation space
  such that the local solution can be represented solely as a Maxwellian (specified by three parameters).
  The blue curve is the initial condition for the calculation, given as (\ref{eq:localinit}).
  This procedure can also be employed to automatically adjust the exponentially weighted basis functions
  in a time-dependent calculation so that the basis functions can continue to provide an accurate
  representation of the solution.}
  \label{fig:localparam}
\end{figure}

Figure~\ref{fig:localparam} presents an example of applying this procedure
to calculate the initial approximation space given the analytical initial condition
\begin{align}
  f(v_\parallel) &= \exp\left(-\frac{v_\parallel^2}{2}\right) \left(1 +
\frac{0.5 v_\parallel}{(2.75-v_\parallel)^2 + 0.3^2} \right)\label{eq:localinit}.
\end{align}
This test demonstrates both the ability of exponentially weighted polynomials to
represent strongly non-Maxwellian features and the feasibility of a procedure to
determine the exponential weighting factor automatically.

The procedure for an algorithm that uses conservative exponentially weighted
polynomials for time-dependent problems can be summarized as:
\begin{enumerate}
  \item Calculate the initial parameters from the initial condition.
    This procedure determines both the initial approximation space $M_h^0$ and
    the initial numerical solution $f(t_0) \in M_h^0$.
  \item With the solution $f(t_n) \in M_h^n$ known at time $t_n$, calculate
    the preliminary numerical solution $f(t_{n+1}) \in M_h^n$ (such as by using a
    standard multi-stage Runge--Kutta method).
  \item Find the approximation space $M_h^{n+1}$ to better represent the solution
    using the same procedure that determined the initial basis function parameters.
    By construction, $f(t_{n+1}) \in M_h^{n+1}$ is also found at the same time.
    To save computational time, the adjustment procedure can be performed after
    regular intervals consisting of several time steps.
  \item Repeat Steps 2 and 3 above until the final time.
\end{enumerate}

We recognize that some or even all of the savings in computation time from
being able to use a coarser grid are lost due to the additional
complexity of using time-dependent, exponentially weighted basis functions.
Gaussian quadrature points and weights must be recomputed every time
the free parameters are adjusted, which requires evaluations of the $\erf(x)$ function.
Additionally, one must be mindful to compute the correct quadrature rule.
When using piecewise polynomials, one typically uses Gauss-Legendre quadrature for the calculation
of integrals:
\begin{align}
  \int_{-1}^{1} f(x)\,\mathrm{d}x \approx \displaystyle\sum_{i=1}^n w_i f(x_i),
\end{align}
where $n$ is the number of quadrature points, which is generally equal to the number of
local basis functions.
The quadrature rule is exact when $f(x)$ is a polynomial of degree $2n-1$ or fewer.
For conservative exponentially weighted DG, the quadrature rule must now satisfy
\begin{align}
  \int_{-1}^{1} W(x) f(x)\,\mathrm{d}x \approx \displaystyle\sum_{i=1}^n w_i' f(x_i'),
\end{align}
where $W(x)$ is the local exponentially weighted weighting function.
The quadrature rules for the weights in (\ref{eq:weightExp}) and (\ref{eq:weightGauss})
are straightforward to derive \citep{Press2007} and can be worked out as a function of the
basis function parameters in advance, since $n$ in a DG method is a small, $\mathcal{O}(1)$ number in practice.

\section{Conclusions \label{sec:exp-conclusion}}
We have developed a conservative discontinuous Galerkin method that
uses an approximation space consisting of exponentially weighted polynomials for the efficient
representation of distribution functions in the presence of collisions.
While the velocity-space domain must be truncated when using standard polynomials,
the velocity-space domain can be extended to $\pm \infty$ in the exponentially weighted polynomial approach.
The lack of number and energy conservation when one uses
exponentially weighted basis functions in the approach of \citet{Yuan2006} was demonstrated
in a test problem with a collision operator with a Fokker-Planck form.
We fixed this issue through the addition of a corresponding exponential weighting factor in the error norm
used for the DG method and showed numerically that the desired conservation properties are preserved.

We studied a Spitzer--H\"arm problem involving the accurate calculation of heat fluxes
in 1D with an additional source term to drive the solution non-Maxwellian.
We demonstrated that conservative, exponentially weighted polynomials
produced results that were ${\sim}10^2$ times more accurate than
standard piecewise polynomials at the same grid resolution.
Conservative, exponentially weighted polynomials achieved the
same level of relative error as polynomials on a ${\sim}8$ times coarser grid.
We attributed these results to the ability of the exponentially weighted representation to
better capture the variation in the distribution function tails
when compared to a piecewise-polynomial representation.
Lastly, we outlined a local procedure to determine the exponentially weighted approximation space
in each element, which adds to the computational cost of this method but makes it
much more robust and applicable to practical problems in which an accurate exponential weighting factor
cannot be determined in advance.

While we have focused on exponentially weighted polynomials in this work,
we recognize that our procedure can more generally be used with other types of non-polynomial-weighted
basis functions for applications in which exact conservation is needed or velocity-space truncation
is not desired.
For example, it might be more useful to use polynomials weighted by a power law $x^{-a}$
for cosmic-ray problems or radio-frequency-heating problems in fusion with a quasilinear operator.
We have also limited our analyses to cases in which the solution in every velocity-space element
was expanded in a similar conservative, exponentially weighted basis.
It may be advantageous to use a mixed representation consisting of standard polynomials in
the bulk of the distribution ($|v| < v_c$) and conservative exponentially weighted polynomials
in the tails $(|v| > v_c)$.
Future work could also explore efficient ways to employ a non-uniform velocity grid spacing,
as we have restricted our attention to velocity space cells of uniform width in this work.
For application to 5D gyrokinetic simulations, it is also important to generalize the use of these
1D basis functions to higher dimensions, such as $(x,y,z,v_\parallel,\mu)$.

For finite-volume or finite-difference codes, one could consider somewhat related
exponential interpolation methods instead of simple linear interpolation as is usually done.
This would be different than the Chang--Cooper algorithm \citep{Chang1970}, which looks at the ratio of the
drag and diffusion coefficients to set the degree of upwind differencing in order
to get the correct equilibrium solution.
This procedure will not give exact conservation of higher moments like energy
(for which one might add small correction terms similar to the recent work of \citet{Taitano2015}),
but it could help improve the accuracy of the results, allowing coarser grids for the same level of accuracy.

%% file: ch-conclusion/chapter-conclusion.tex
\chapter{Summary and Future Directions\label{ch:conclusion}}
\section{Summary}
This thesis presented several advances towards a gyrokinetic-continuum-simulation
capability for the boundary plasma.
The main contribution was the development of the first 
gyrokinetic continuum simulations of turbulence in straight (Chapter~\ref{ch:lapd})
and helical (Chapter~\ref{ch:helical-sol}) open-field-line plasmas.
Prior efforts to include this capability in a gyrokinetic continuum code appeared
to have run into difficulties with number conservation, energy conservation, or
sheath-boundary-condition stability.
The simulations in this thesis solved gyrokinetic equations
in an electrostatic long-wavelength (drift-kinetic) limit, which will be replaced 
with more general gyrokinetic equations in the future.
To address potential conservation issues,
we needed to evaluate, generalize, and extend existing discontinuous Galerkin (DG)
algorithms for application to the gyrokinetic system (Chapter~\ref{ch:models-and-numerical-methods}).
This work also appears to be the first application of DG methods to numerically solve a gyrokinetic system,
and we hope that the documentation of the issues that we ran into will be useful in the development
of other codes for various applications, including
gyrokinetic continuum and particle-in-cell (PIC) codes for edge and scrape-off-layer (SOL) simulations.
A key algorithm in our numerical approach is an extension of a DG algorithm for
2D incompressible flow \citep{Liu2000} to general Hamiltonian systems, which we showed
conserves number and energy.
To model the mediating effects of the plasma sheath that cannot be resolved in gyrokinetics,
we developed a kinetic analog of the conducting-wall
boundary conditions used in some fluid and gyrofluid plasma simulations that allow fluctuations
in the current to the wall, which we refer to as conducting-sheath boundary conditions.

We first implemented and evaluated DG algorithms for 1D1V (1 dimension in position space, 1 dimension
in velocity space) simulations of the parallel propagation of an edge-localized-mode (ELM)
heat pulse in the SOL using logical-sheath boundary conditions.
We showed that our simulations could recover similar results to those obtained
using a 1D1V Vlasov--Poisson code \citep{Havlickova2012} at a fraction of the computational cost (taking minutes
to run instead of fifteen hours).
Most of the savings in our simulations came from not having to resolve the restrictive time and length
scales required in fully kinetic simulations.
We later extended these simulations to 1D2V with self-species collisions modeled by a Lenard--Bernstein
collision operator, which improved the quantitative agreement with the 1D3V PIC
simulations (with collisions) of \citet{Pitts2007,Havlickova2012}.

Having developed 1D1V and 1D2V models of the ELM heat-pulse problem, we then extended our
kinetic equation solver to 2D2V, which we used to perform
simulations of electron-temperature-gradient-driven (ETG)
turbulence (not discussed in this thesis). For those simulations, we verified using parameter scans of the
background temperature gradient in a periodic slab geometry that
the correct linear growth rates were recovered and that nonlinearly saturated turbulent states were reached.
Some initial 5D (3D2V) simulations of ETG turbulence were also performed to investigate basic code stability.

After we developed a 5D gyrokinetic solver (and A.\,Hakim parallelized his gyrokinetic-Poisson-equation
solver),
we generalized the logical-sheath boundary conditions that were implemented in our 1D1V and 1D2V simulations
to 3D2V. We encountered stability issues with these boundary conditions in our simulations of the
Large Plasma Device (LAPD), which could be thoroughly documented in a future paper.
We found better success by generalizing a set of conducting-wall boundary conditions used in some
prior fluid and gyrofluid simulations of open-field-line plasmas, which became the present
set of sheath-model boundary conditions used in our model.
The final barrier in the LAPD simulations was a way to deal with issues concerning the positivity of the
distribution function, since this property was not automatically guaranteed in our algorithms.
Our current solution to maintain the positivity was discussed in Section~\ref{sec:positivity},
which is not a fully satisfactory solution but appears to work well for current applications.
The resulting simulations of LAPD turbulence appear qualitatively reasonable when compared to
experimental data from the device and drift-reduced Braginskii fluid simulations from the Global
Braginskii Solver (GBS) code (Chapter~\ref{ch:lapd}).
While our LAPD simulations were intended as a way to gain confidence in our model and algorithms
in a simplified, well-diagnosed open-field-line geometry, the gyrokinetic-simulation capability we have developed
can eventually lead to improved modeling of LAPD experiments with the future addition of
more accurate treatments of the plasma source and boundary conditions.

We later developed simulations of a helical SOL in an all-bad-curvature slab (Chapter~\ref{ch:helical-sol})
using parameters for a National Spherical Torus Experiment (NSTX) SOL,
which is much less collisional than the plasmas in LAPD.
Because of the additional interchange-instability mechanism, the radial turbulent transport in these simulations
is much stronger and the overall turbulence character is qualitatively distinct
when compared to our simulations of turbulence in LAPD.
Initial analysis suggests that these simulations are in a sheath-connected regime, and so some
qualitative features of these simulations might be well reproduced in isothermal 2D models.
The helical SOL simulations are currently being applied to simulate turbulence in the Helimak device.
This model will be used as a starting point for additional levels of sophistication on the path towards
realistic boundary-plasma simulations spanning the SOL and the confined edge.

Finally, we developed a conservative DG method that employs exponentially weighted basis functions to
discretize the velocity-space dependence of the distribution function (Chapter~\ref{ch:exp-basis}).
DG methods can be extended to use non-polynomial basis functions in a straightforward manner \citep{Yuan2006},
although such basis functions are rarely used in practice.
We found that \citet{Yuan2006} did not recognize potential conservation issues in using
non-polynomial basis functions in a standard DG method,
and we developed a Petrov-Galerkin approach that allows the use
of discontinuous exponentially weighted basis functions and also conserves number, momentum, and energy.
In a simple but non-trivial heat-flux benchmark, we demonstrated the potential savings of our
conservative exponentially weighted
DG method in reducing velocity-space resolution requirements when compared to the standard DG approach that
uses polynomials to represent the solution.
Generalization of this method to higher dimensions is left to future work.

\section{Future Directions}
We have made a number of simplifications to the simulations presented in this thesis,
and it is important to increase the sophistication of the models and improve the design and
implementation of the numerical methods used so that the code can eventually be applied to
study relevant physics issues on present-day and future tokamaks.
Of course, high priority should be placed in adding finite-Larmor-radius effects (e.g. gyroaveraging
in the gyrokinetic Poisson equation and in the gyrokinetic equation).
It will be interesting to explore how the turbulence in the helical-SOL model changes
with the inclusion of finite-Larmor-radius effects when compared to our existing simulations for
the long-wavelength (drift-kinetic) gyrokinetic system.
Here, we discuss in more detail some other less-obvious priorities for near-term future work.

\noindent
\hangindent=15pt \textbf{Distribution-function positivity.}
Since the numerical algorithm we use to solve the gyrokinetic equation does not
automatically preserve the positivity of the distribution functions
(either in a cell-average sense or everywhere within a cell),
we currently apply a positivity-adjustment procedure (Section~\ref{sec:positivity})
at the end of every intermediate Runge--Kutta substage
to remove the negative-valued portions of the distribution functions.
This procedure usually results in a non-negligible source of particles and energy (approximately 10--20\% of the
fixed plasma sources).
While the simulations discussed in this thesis appear to produce qualitatively reasonable results,
it is highly desirable to modify the algorithms used in the code so that the extra source of particles and energy
is greatly reduced or even eliminated.
The current correction procedure is overly conservative because it prevents a piecewise-linear
representation $f(x) = f_0 + f_1 (x-x_j)$ from going negative anywhere within a cell,
which restricts $|f_1|/f_0 < 1$, while there are physically realizable positive functions
that have the same moments even if $|f_1|/f_0$ is somewhat larger than this.
We are pursuing ideas to relax this constraint, which are related to the exponentially weighted
basis-function approach explored in Chapter~\ref{ch:exp-basis}.
These ideas are also related to positivity-preserving limiters and positivity-preserving fluxes
used to preserve the positivity of cell-averaged quantities \citep{Zhang2017,Zhang2010,Zhang2011b}.
Additional correction steps will be needed if the distribution function is
required be positive everywhere within a cell, which is stricter than requiring that cell averages are non-negative.
We also note that there exists a large body of literature on positivity-preserving algorithms for finite-volume
and DG methods \citep[for example, see][]{Rossmanith2011}.

\noindent
\hangindent=15pt \textbf{Code optimization.}
Although the simulations presented in this thesis were all performed using a reasonable
amount of computational resources, it is important to honestly compare the computational cost of
similar gyrokinetic simulations using different numerical methods.
Specifically, a collaborative effort should be made among existing gyrokinetic codes under
development for boundary plasma simulation to benchmark code performance on an agreed-upon test case.
While we acknowledge that timing comparisons can be a sensitive topic, 
a major reason why this project was originally undertaken was to explore the potential of advanced continuum
methods to provide cheaper gyrokinetic simulations of edge and SOL turbulence.
Additionally, it is in the interests of the fusion-research community to know if one code requires
orders of magnitude more resources to produce the same result as other codes to identify areas
in need of algorithmic improvements.
At present, the underlying DG algorithm is neither fully modal nor fully nodal in its implementation
\citep[see][Section~6.6]{Durran2010}, and
consequently is unable to realize the full computational advantages of either approach.
For example, the DG solution is represented using nodal basis functions, but the solution nodes are
different from the Gauss--Legendre quadrature nodes.
Therefore, interpolation matrices are required to evaluate the solution at quadrature nodes
whenever numerical evaluation of integrals over the volume or a surface of an element is required.
Future versions of the code should commit to implementing algorithms consistently in a
nodal or modal sense.
It is also crucial for future version of the code to exploit the well-known adaptive $h$ (element size)
and $p$ (basis function degree) grid refinement and coarsening 
features of the DG method \citep{Remacle2003} to dynamically adjust the grid so that
memory can be used efficiently.
The current approach uses a uniformly spaced grid with the same number of basis functions in each
cell, which almost certainly wastes some data by using many pieces of information
to resolve the distribution function in locations where such resolution is unneeded.

\noindent
\hangindent=15pt \textbf{Open and closed-magnetic-field-line regions.}
As a first step towards increasing the complexity of the magnetic geometry,
one should add to the code the ability to include 
a confined edge region consisting of closed magnetic field lines and a SOL region consisting of
open magnetic field lines in the same simulation \citep{Ribeiro2008,Halpern2016,Dudson2017}.
The two regions are presumably distinguished only by the parallel boundary conditions on the
distribution functions and potential.
Periodic boundary conditions must be applied to the distribution function and potential on
closed field lines and sheath-model boundary conditions are applied to the distribution function
on open field lines.
As in the purely open-field-line simulations, no parallel boundary conditions are required for the potential
in the open-field-line region.
This capability could enable studies concerning the effects of ion drift orbit excursions
from the edge into the SOL on edge rotation \citep{StoltzfusDueck2012}, radial-electric-field profiles,
and SOL profiles.

\noindent
\hangindent=15pt \textbf{Electromagnetic effects.}
One of the potential advantages of continuum methods over PIC methods discussed in Section~\ref{sec:intro-gk-codes}
is the relative ease with which numerical problems in the implementation of 
electromagnetic effects can be handled (specifically, magnetic perturbtions arising from the parallel
magnetic potential~$A_\parallel$).
Electromagnetic simulations can be still be 
challenging in some regimes for physics reasons regardless of the numerical approach taken.
The root of this advantage appears to be the lack of sampling noise in continuum methods, which avoids
the well-known Amp\'{e}re's law cancellation problem \citep{Hatzky2007} as long as the numerical integrations
for the solution of the parallel magnetic potential are performed consistently\footnote{This solution was
first presented by G.\,Hammett and F.\,Jenko in presentations at the Plasma Microturbulence
Project meeting at General Atomics on July 25, 2001 \citep[see][footnote~13]{Chen2003}.}
\citep{Dannert2004}.
While the direct cross-field transport of particles and heat due to magnetic fluctuations
is generally negligible when compared to the cross-field transport due to the $E \times B$ drift
\citep[see references in][]{Zweben2007}, magnetic induction slows the electron parallel dynamics,
weakening the adiabatic electron response.
A discussion of the dimensionless parameters to characterize the relative importance of
collisional, inertial, and inductive processes in strenghtening the non-adiabatic electron response
can be found in \citet{Scott2003,Scott2007,Ribeiro2008}.
For typical edge-plasma parameters, electromagnetic effects lead to stronger
turbulence in simulations \citep{Scott2007,Scott1997,Scott2003,Scott2010b}.
In addition to the physics studies it would enable, a demonstration that
electromagnetic effects can be handled in a stable and efficient
(i.e., without increasing the computational cost by an order of magnitude or larger) manner
for open-field-line gyrokinetic-turbulence simulations would be a significant achievement.

\noindent
\hangindent=15pt \textbf{Sheath-model boundary conditions.}
As discussed in Section~\ref{sec:sheath}, the only boundary conditions that are applied to the ions
at the sheath entrance are zero-inflow boundary conditions.
Otherwise, ions freely flow out of the domain, and there is no mechanism in the ion
boundary conditions to ensure that the Bohm sheath criterion is satisfied.
While there might not be much difference between ion outflow at thermal velocities
instead of at the ion sound speed for typical SOL parameters because $T_i$ is typically
a few times larger than $T_e$, this issue might be more of an issue in achieving quantitatively correct
simulations of basic-plasma-physics experiments.
In XGCa gyrokinetic simulations of a DIII-D SOL \citep{Churchill2016},
the authors observed subsonic ion flows at the sheath entrance with a Mach number typically 0.4--0.7,
but argued that the standard Bohm-sheath-criterion result was inapplicable.
Future work should develop a set of sheath-model boundary conditions for ions that includes the effects of
the rarefaction fan \citep{Munz1994} that accelerates ions to sonic outflow speeds if the Bohm sheath criterion
is not satisfied.
Such an improvement would benefit both continuum and PIC codes that use sheath-model boundary conditions.

\noindent
Other important additions to the code that are only mentioned in passing here include
a Rosenbluth--Fokker--Planck \citep{Rosenbluth1957,Taitano2015} or Fokker--Planck--Landau collision operator,
realistic magnetic geometry (including the X-point), neutral and impurity-species modeling,
and atomic-physics modeling.
As discussed in Chapter~\ref{ch:intro}, the fusion-energy community has recognized a need for
a first-principles gyrokinetic-simulation capability
for the boundary plasma \citep{Ricci2015b,Boedo2009,Cohen2008}.
We hope that our contributions will accelerate the development of predictive
gyrokinetic continuum codes for modeling the boundary plasma of and improving the performance
of future devices.

%% file: ch-appendicies/notes-on-plotting.tex
\chapter{\label{ch:plot-creation}Plot Creation}
Several 1D and 2D plots are presented in this thesis to illustrate simulation results.
No data interpolation (estimation) is used, such as when plotting DG solutions versus position.
This is because DG methods expand the solution in terms of predetermined basis functions,
a topic which is covered in Chapter~\ref{ch:models-and-numerical-methods}.
With the degrees of freedom known in a cell, the solution can be evaluated exactly
at an arbitrary number of points within the same cell.
For plotting scalar data versus two position coordinates. e.g. the electron
density in the $x$--$y$ plane, we generally evaluate the solution on an
$8 \times 8$ grid of equally spaced points in each cell.

\begin{figure}
  \centerline{\includegraphics[width=0.75\textwidth]{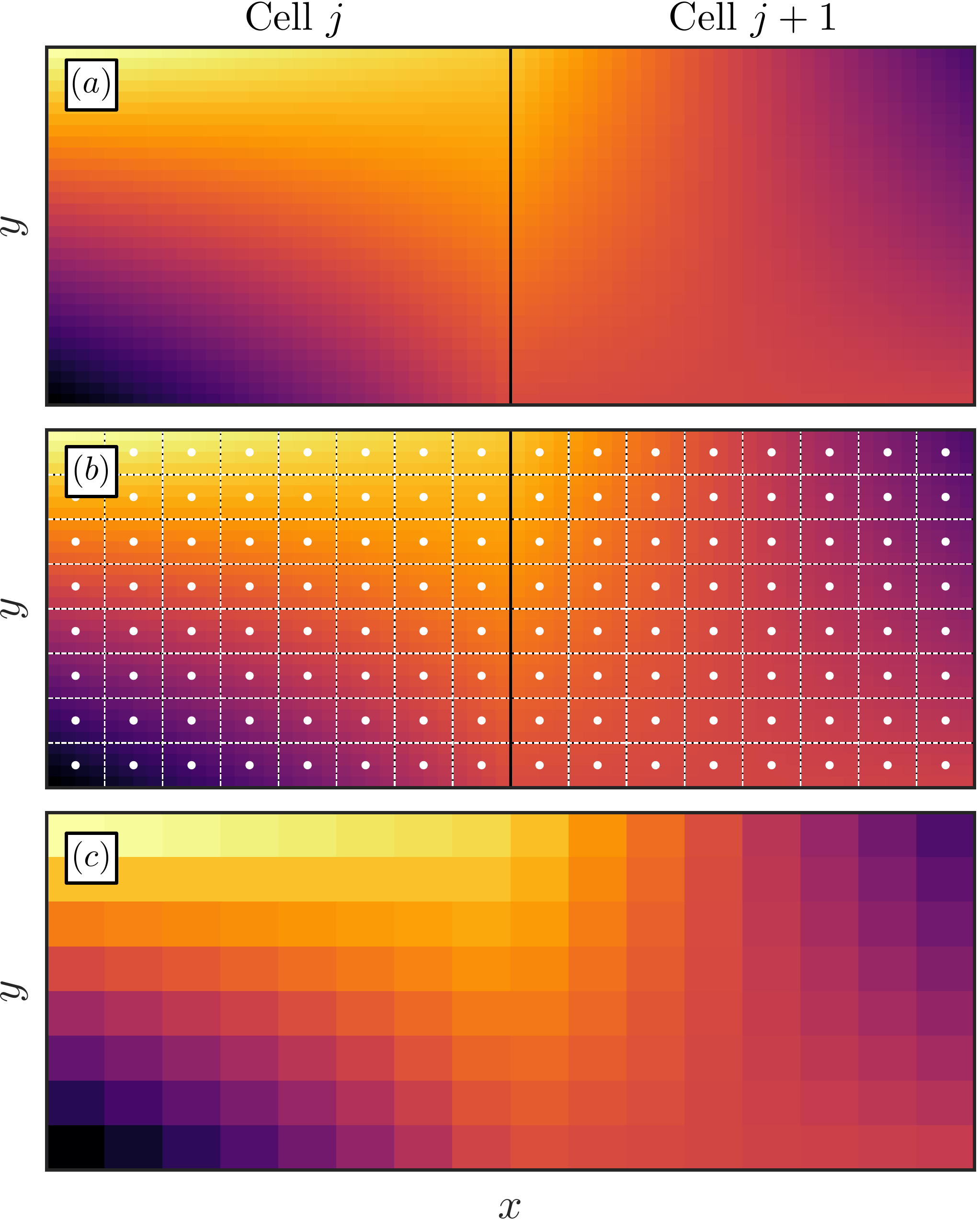}}
  \caption[Illustration of the image-creation procedure for a DG solution.]
  {Illustration of the image-creation procedure for a DG solution.
  ($a$) The complete DG solution in two neighboring cells is shown.
  ($b$) According to the desired resolution for the image that will be created,
  each cell is divided up into a number of subcells such that 
  each subcell corresponds to a single pixel in the final image.
  The value of each pixel is determined by evaluating the DG solution at the 
  center of each subcell (white markers) and mapped to a color according to a color map.
  ($c$) The final image is created from the subcell-gridded data using standard plotting routines.
  }
  \label{fig:appendix-plotting-explained}
\end{figure}

Figure~\ref{fig:appendix-plotting-explained} illustrates the procedure through which
a 2D plot of a DG solution is created for visualization in post processing.
We start with the full DG solution in figure~\ref{fig:appendix-plotting-explained}($a$),
where the full solution in two cells sharing a common boundary are shown.
Our goal is to create a matrix from this data that can be used to create an image,
in which the value of row $i$ and column $j$ in the matrix specifies the color
of the pixel at row $i$ and column $j$ in the image.

Having decided on the resolution of the image we want to generate, we divide up each cell
into a number of subcells, which is shown in figure~\ref{fig:appendix-plotting-explained}($b$).
In this example, we divide up each cell into a finer grid of $8 \times 8$ cells.
Each subcell will each correspond to a single pixel in the final image.
The value of the DG solution in the center of each subcell is evaluated without approximation
using knowledge of the basis functions and stored into a matrix.
Each value of this matrix is then assigned a particular color to each pixel according to a
color map.
With the image matrix filled out, we can use standard plotting packages to generate
an image from this data.
In Matlab, we use the \texttt{imagesc} function to create the image.
The end result of the plotting procedure is shown in figure~\ref{fig:appendix-plotting-explained}($c$).

%% file: ch-appendicies/1d-sol-differences.tex
\chapter{\label{ch:1d-sol-comp}Additional Comparisons Between ELM-Heat-Pulse Simulations}
\begin{figure}
\centering
\includegraphics[width=\textwidth]{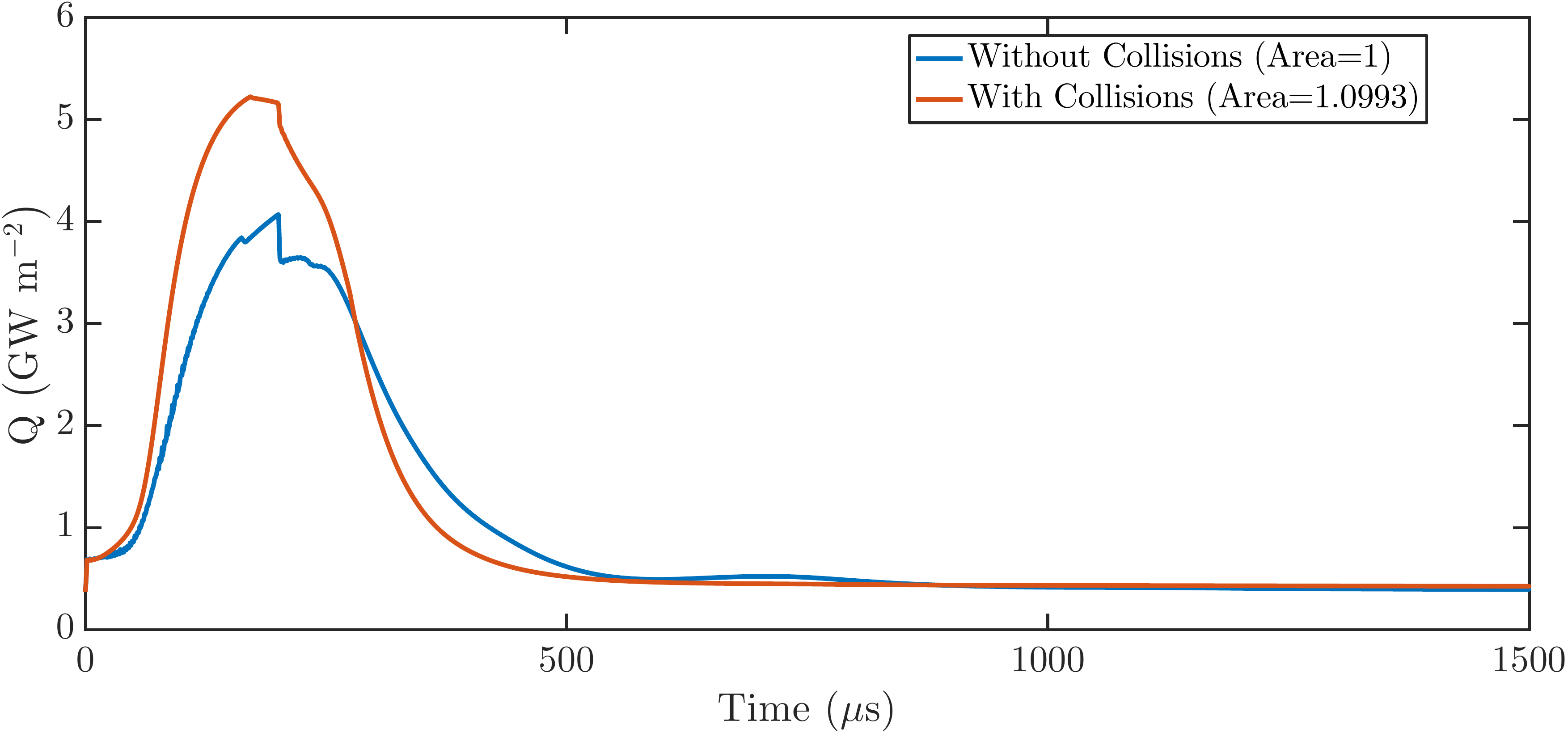}
  \caption[Comparison of the total parallel heat flux at the divertor
  plate versus time for 1D2V ELM-heat-pulse simulations with and without collisions.]
  {Comparison of the total parallel heat flux at the divertor
  plate versus time for 1D2V ELM-heat-pulse simulations with and without same-species Lenard--Bernstein collisions.
  By integrating the area under each curve and normalizing the area to the smaller value,
  we find that the time-integrated total heat flux for the case with collisions is ${\approx}9.9\%$ larger.}
  \label{fig:1d2v-total-heat-flux}
\end{figure}

In Section~\ref{sec:1d-sol-collisions}, we noted that the time-integrated total heat flux
for the 1D2V ELM-heat-pulse simulation with collisions 
is ${\approx}9.9\%$ larger (over a 1.5~ms integration window), as shown in figure~\ref{fig:1d2v-total-heat-flux}.
Here, we explore the reason for this discrepancy of approximately 0.34~MJ~m$^{-2}$.

We investigate the energy balance of the system to explain the differences in the
time-integrated parallel heat fluxes between the cases with and without collisions.
In the collisionless case, our system is described by the system
\begin{align}
\frac{\partial f_e}{\partial t} + v_\parallel \frac{\partial f_e}{\partial z} +
  \frac{q_e}{m_e} E_\parallel \frac{\partial f_e}{\partial v_\parallel} &= S_e(z,\boldsymbol{v},t),\\
\frac{\partial f_i}{\partial t} + v_\parallel \frac{\partial f_i}{\partial z} +
  \frac{q_i}{m_i} E_\parallel \frac{\partial f_i}{\partial v_\parallel} -
  \frac{e^2 k_{\perp 0}^2 \rho_{\mathrm{s}0}^2 }{m_i T_{e0}} \delta \phi E_\parallel \frac{\partial f_i}{\partial v_\parallel}
  &= S_i(z,\boldsymbol{v},t) ,\\
  n_i \left(k_{\perp 0}\rho_{\mathrm{s}0}\right)^2 \frac{e^2 \delta \phi}{T_{e0}} &= \sum_s q_s n_s .\label{eq:1d2vSimpleKinetic}
\end{align}

We consider the time evolution of the total energy:
\begin{align}
\frac{\mathrm{d} W_{\mathrm{tot}}}{\mathrm{d}t} &= \int \mathrm{d} \Lambda
\sum_s \left( f_s \frac{\partial H_s}{\partial t} + \frac{\partial f_s}{\partial t} H_s \right),
\end{align}
where $\int \mathrm{d}\Lambda = \int \mathrm{d}z \int \mathrm{d}^3 \boldsymbol{v} $.
The electron and ion Hamiltonians are
\begin{align}
H_e &= \frac{1}{2} m_e v_\parallel^2 + \mu B - e \delta \phi, \\
H_i &= \frac{1}{2} m_i v_\parallel^2 + \mu B + e \delta \phi - 
  \frac{1}{2} \frac{e^2}{T_{e0}} \left(k_{\perp 0}\rho_{s0}\right)^2 \delta \phi^2,
\end{align}
where $k_{\perp 0} \rho_{s0} = 0.2$ and the second-order Hamiltonian in the electrons has been
neglected since electrons are much less massive than the ions.
Recall that the second-order term in the Hamiltonian was constructed
so that $\int \mathrm{d}\Lambda \sum_s f_s \partial_t H_s = 0$ in Section \ref{subsec:1dsolenergy}.
Therefore, we need to calculate $\int \mathrm{d}\Lambda \sum_s H_s \partial_t f_s$,
which is not going to be zero when we include the sheath losses:
\begin{align}
\int \mathrm{d}\Lambda \, H_s \frac{\partial f_s}{\partial t} &=
  - \int \mathrm{d} \Lambda \, H_s v_\parallel \frac{\partial f_s}{\partial z} +
  \int \mathrm{d}\Lambda \, H_s \frac{1}{m_s} \frac{\partial H_s}{\partial z} \frac{\partial f_s}{\partial v_\parallel}
  + \int \mathrm{d}\Lambda \, H_s S_s. \label{eq:1d2vEnergyBalanceInterm}
\end{align}
The second term on the right-hand side of (\ref{eq:1d2vEnergyBalanceInterm}) can be integrated by parts to get
\begin{align}
\int \mathrm{d}\Lambda \, H_s \frac{1}{m_s} \frac{\partial H_s}{\partial z} \frac{\partial f_s}{\partial v_\parallel} &=
\left. \int \mathrm{d}z \int d^2 v_\perp \frac{1}{m_s} H_s \frac{\partial H_s}{\partial z} f_s \right|_{-v_{\parallel,\mathrm{max}}}^{v_{\parallel,\mathrm{max}}}
- \int \mathrm{d}\Lambda \frac{\partial H_s}{\partial v_\parallel} \frac{1}{m_s} \frac{\partial H_s}{\partial z} f_s \\
&= - \int \mathrm{d}\Lambda \frac{\partial H_s}{\partial z} v_\parallel f_s,
\end{align}
where we have used the zero-flux boundary condition in $v_\parallel$.
This term then combines with the first term on the right-hand side of (\ref{eq:1d2vEnergyBalanceInterm})
as $\int \mathrm{d}\Lambda \, v_\parallel \partial_z\left ( H_s f_s \right)$.
Therefore, the energy-evolution equation is
\begin{align}
\frac{\mathrm{d} W_{\mathrm{tot}}}{\mathrm{d}t} &=  -\left.\int \mathrm{d}^3\boldsymbol{v} \sum_s v_\parallel H_s f_s \right|_{z=-L_\parallel}^{z=L_\parallel}
+ \int \mathrm{d} \Lambda \sum_s H_s S_s.
\end{align}

We can simplify this result further by noting that the source terms have the property
$\int \mathrm{d}^3 \boldsymbol{v} S_i = \int \mathrm{d}^3\boldsymbol{v} S_e$
and using the logical sheath boundary conditions:
\begin{multline}
  \frac{\mathrm{d} W_{\mathrm{tot}}}{\mathrm{d}t} =
  -\left.\int \mathrm{d}^3\boldsymbol{v} \sum_s v_\parallel H_{s,0} f_s \right|_{z=-L_\parallel}^{z=L_\parallel}
  + \left.\int \mathrm{d}^3 \boldsymbol{v} \, v_\parallel H_{i,2} f_i \right|_{z=-L_\parallel}^{z=L_\parallel}\\
  + \int \mathrm{d}\Lambda \sum_s H_{s,0} S_s + \int \mathrm{d}\Lambda \, H_{i,2} S_i \label{eq:1d2vEnergyBalance},
\end{multline}
where $H_{s,0} = \frac{1}{2}m_s v_\parallel^2 + \mu B$ and $H_{i,2} = e^2 (k_\perp \rho_{\mathrm{s}0})^2 \delta \phi^2 / (2 T_{e0})$.
The first term on the right-hand side of (\ref{eq:1d2vEnergyBalance}) is the total parallel heat flux,
which we already have plotted in figure~\ref{fig:heat-flux-collision}.
The second term on the right-hand side of (\ref{eq:1d2vEnergyBalance}) is proportional to the outgoing
ion particle flux $\Gamma_i$, so we must compute a diagnostic for
\begin{align}
  H_{i,2}(t) = \left.\frac{1}{2} \frac{e^2}{T_{e0}} \left(k_{\perp 0}\rho_{s0}\right)^2
    \delta \phi^2 \Gamma_i(z) \right|_{z=-L_\parallel}^{z=L_\parallel} \label{eq:h2flux}.
\end{align}
The third term on the right-hand side of (\ref{eq:1d2vEnergyBalance}) will be the same for the simulations
with and without collisions, as it is time independent.
We must also keep a record of the fourth term on the right-hand side of (\ref{eq:1d2vEnergyBalance})
and $W_\mathrm{measured}(t) = \sum_s \int \mathrm{d}\Lambda\,H_s f_s$ to quantify the error in energy conservation
in the simulation.
In this case, we define the energy error as
$| \Delta W_{\mathrm{measured}} - \Delta W_{\mathrm{predicted}}|/\Delta W_{\mathrm{predicted}}$,
where
\begin{equation}
  \Delta W_{\mathrm{predicted}} = \int_0^{t_{\mathrm{end}}} \mathrm{d}t \frac{ \mathrm{d} W_{\mathrm{tot}} }{ \mathrm{d}t },
\end{equation}
and use (\ref{eq:1d2vEnergyBalance}) to calculate the integrand at the end of each time step.
We use a trapezoid rule to evaluate the integral in time for $\Delta W_{\mathrm{predicted}}$
and take $t_{\mathrm{end}} = 2$ ms in our tests.

\begin{figure}
\centering
\includegraphics[width=\textwidth]{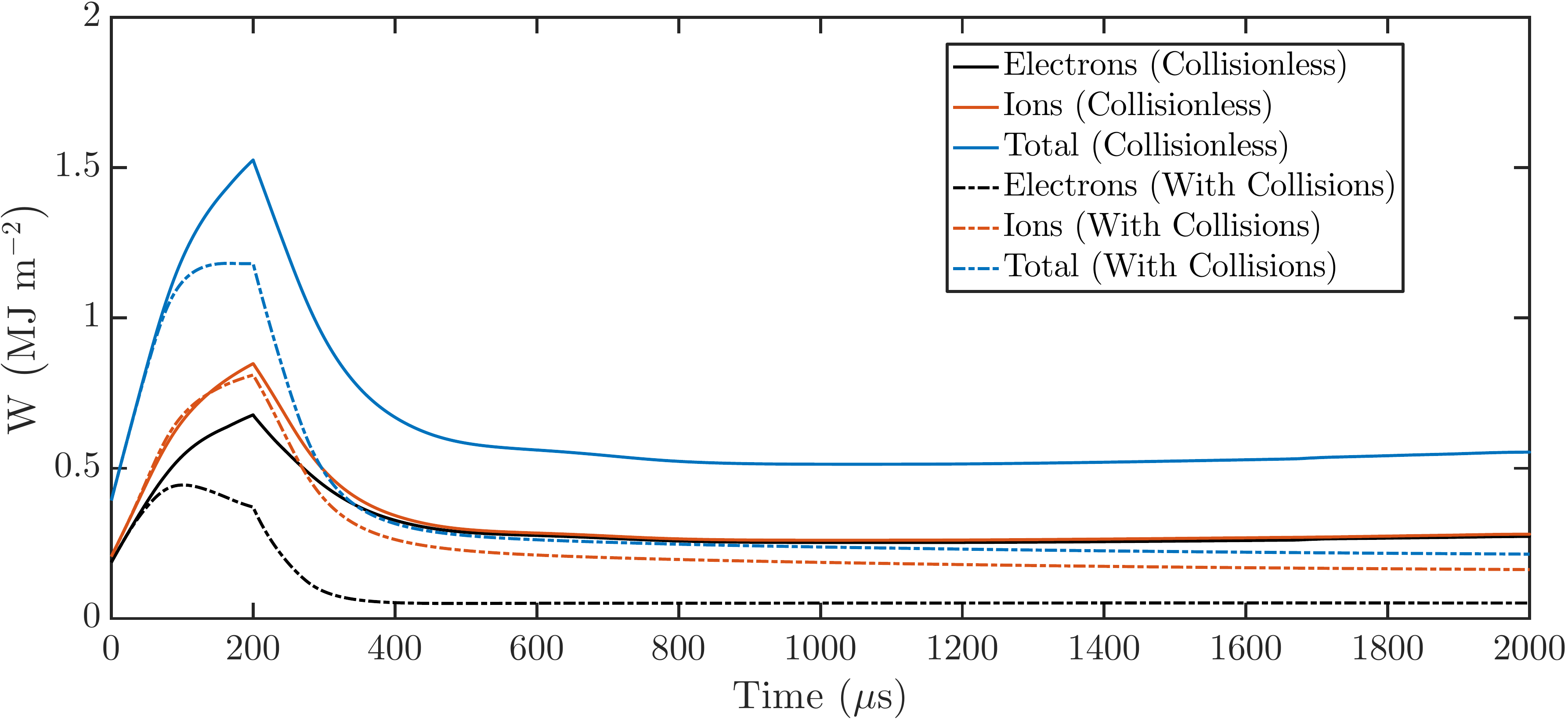}
  \caption[Comparison of the total energy and electron and ion contributions 
  between 1D2V ELM-heat-pulse simulations with and without same-species Lenard--Bernstein collisions.]
  {Comparison of the total energy and electron and ion contributions 
  between 1D2V ELM-heat-pulse simulations with and without same-species Lenard--Bernstein collisions.
  Starting from the same initial condition, the collisionless simulation has more than twice the total energy
  of the simulation with collisions at $t = 2$~ms. This plot helps explain the difference in the
  total heat flux in figures~\ref{fig:heat-flux-collision} and \ref{fig:1d2v-total-heat-flux},
  where the time-integrated total heat flux is observed to be larger in the simulation with collisions.} 
  \label{fig:1d2v_total_energy} 
\end{figure}

By comparing the energy diagnostics from the two simulations, we find that the difference in the time-integrated
heat flux is explained by difference in $W_{\mathrm{measured}}(t)$.
Figure~\ref{fig:1d2v_total_energy} shows the time traces of the total energy $W_{\mathrm{measured}}(t)$
measured in the two simulations, as well as the individual contributions from electrons and ions to
$W_{\mathrm{measured}}(t)$.
The 1D2V ELM-heat-pulse simulation without collisions simply has a larger system energy $W_{\mathrm{measured}}$
at any given instant than the simulation with collisions, which results in a lower time-integrated
heat flux for the collisionless simulation.
The two simulations both start with $W_{\mathrm{measured}} \approx 0.39$ MJ/m$^2$, but
the collisionless simulation has $W_{\mathrm{measured}(t_{\mathrm{end}}} \approx 0.55$~MJ~m$^{-2}$
and the simulation with collisions has $W_{\mathrm{measured}(t_{\mathrm{end}}} \approx 0.21$~MJ~m$^{-2}$.
Additionally, the energy error was found to be 0.04\% for the two simulations,
so energy is well conserved.

%% file: ch-appendicies/lapd-discrete-energy.tex
\chapter{\label{ch:lapd-discrete-energy}Numerical Energy Conservation in LAPD Simulations}
Here, we investigate the energy-conservation properties of our LAPD simulations
discussed in Chapter \ref{ch:lapd}.
As discussed in Section \ref{sec:algorithms}, the space discretization exactly conserves energy,
but the SSP-RK3 time integrator introduces energy conservation errors.
Therefore, it is interesting to quantify the energy non-conservation that results from the
time discretization.
The evolution of total energy $W = W_k + W_\phi$ (with perfect spatial and temporal energy conservation)
can be obtained by adding together (\ref{eq:w_k}) and (\ref{eq:w_phi}):
\begin{align}
\frac{\partial W}{\partial t} =& -P_{\mathrm{loss}} + P_{\mathrm{source}}, \label{eq:totalEnergy} \\
P_{\mathrm{loss}} =& - \int \mathrm{d}x \, \mathrm{d}y \sum_s \int \mathrm{d}^3 v \,
H_s v_\parallel f_s \Big|_{z_{\mathrm{lower}}}^{z_{\mathrm{upper}}}, \label{eq:pLoss} \\
P_{\mathrm{source}} =& \int \mathrm{d}^3 x \sum_s \int \mathrm{d}^3 v \, H_s S_s =
  \int \mathrm{d}^3 x \sum_s \int \mathrm{d}^3 v \, H_0 S_s, \label{eq:pSource}
\end{align}
where we have assumed a charge-neutral source for the second equality in (\ref{eq:pSource}).
Although we also have a non-negligible source of energy from the positivity-adjustment procedure, this
source of energy can also be measured and accounted for using diagnostics, i.e. by computing the total energy before and after the
positivity-adjustment procedure, which is applied to the distribution functions at the end of each intermediate
Runge--Kutta stage.
By taking into account the details of the multi-stage SSP-RK3 time integrator, we can derive a simple expression
for the expected change in the total energy of the system after a total time step of size $\Delta t$ that involves
a combination of power loss and source terms from each substage.

For clarity, the SSP-RK3 algorithm to advance an equation of the form $\partial_t f = \xi(f,\phi)$ from
$f(t^n) = f^n$ to $f(t^{n+1})$, where $t^{n+1} = t^n + \Delta t$, is written as \citep{Gottlieb2001,Peterson2013}
\begin{align}
  f^* =& f^n + \Delta t \xi \left(f^n, \phi^n \right), \label{eq:rk3-1} \\
  f' =& \frac{3}{4} f^n + \frac{1}{4} \left[ f^* + \Delta t \xi(f^*,\phi^*) \right], \\
  f^{n+1} =& \frac{1}{3} f^n + \frac{2}{3}\left[f' + \Delta t \xi(f',\phi') \right]. \label{eq:rk3-3}
\end{align}

The positivity-adjustment procedure is applied to $f^*$, $f'$, and $f^{n+1}$ before it is used in the $\xi(f,\phi)$ operator
of the subsequent stage, so we denote the extra energy added to the electrons and ions at the end of
each substage as $W_{\mathrm{pos}}^*$, $W_{\mathrm{pos}}'$, and $W_{\mathrm{pos}}^{n+1}$.
The SSP-RK3 algorithm (\ref{eq:rk3-1})--(\ref{eq:rk3-3}) can be combined as
\begin{equation}
f^{n+1} = f^n + \Delta t \left( \frac{1}{6} \xi (f^n,\phi^n) + \frac{1}{6} \xi (f',\phi') + \frac{2}{3} \xi (f^*, \phi^*) \right) \label{eq:rk3combined}
\end{equation}

Using (\ref{eq:totalEnergy}), we notice that the energy change associated with a term like $\xi(f^n,\phi^n)$ is
\begin{equation}
\int \mathrm{d}^3 x \sum_s \int \mathrm{d}^3 v \, H_s^n \xi(f_s^n,\phi^n) = -P_{\mathrm{loss}}^n + P_{\mathrm{source}}^n,
\end{equation}
where the superscript $n$ on $P_{\mathrm{loss}}$ and $P_{\mathrm{source}}$ indicates that (\ref{eq:pLoss}) and (\ref{eq:pSource})
are to be evaluated with $H^n$ and $f^n$.
By multiplying (\ref{eq:rk3combined}) by $H^{n+1}$, integrating over phase space, and summing over both species,
the energy change in the system after the total time step can be written as
\begin{align}
P_{\mathrm{total}} =& \frac{W^{n+1}-W^n}{\Delta t} = \left( \frac{1}{6} P_{\mathrm{loss}}^n + \frac{1}{6}P_{\mathrm{loss}}^* + \frac{2}{3}P_{\mathrm{loss}}' \right) \nonumber\\
& + \left( \frac{1}{6} P_{\mathrm{source}}^n + \frac{1}{6}P_{\mathrm{source}}^* + \frac{2}{3}P_{\mathrm{source}}'  \right) \nonumber\\
& + \frac{1}{\Delta t} \left( \frac{1}{6} W_{\mathrm{pos}}^* + \frac{2}{3} W_{\mathrm{pos}}' + W_{\mathrm{pos}}^{n+1}\right) + P_{\mathrm{err}}, \label{eq:energyEvolution}\\
W^{n} =& \int \mathrm{d}^3 x \int \mathrm{d}^3 v \left(H^n - \frac{1}{2}q_s \phi^{n} \right) f^n, \\
W^{n+1} =& \int \mathrm{d}^3 x \int \mathrm{d}^3 v \left(H^{n+1} - \frac{1}{2}q_s \phi^{n+1} \right) f^{n+1},
\end{align}
where $P_{\mathrm{err}}$ is a measure of the energy conservation error $\propto \left( \partial_t W \right)^4 \left(\Delta t\right)^3$
resulting from the time discretization scheme.
While it has a complicated expression, it can be tracked in the code by measuring all the other terms in
(\ref{eq:energyEvolution}) using diagnostics.

\begin{figure}
  \centering
  \includegraphics[width=0.75\linewidth]{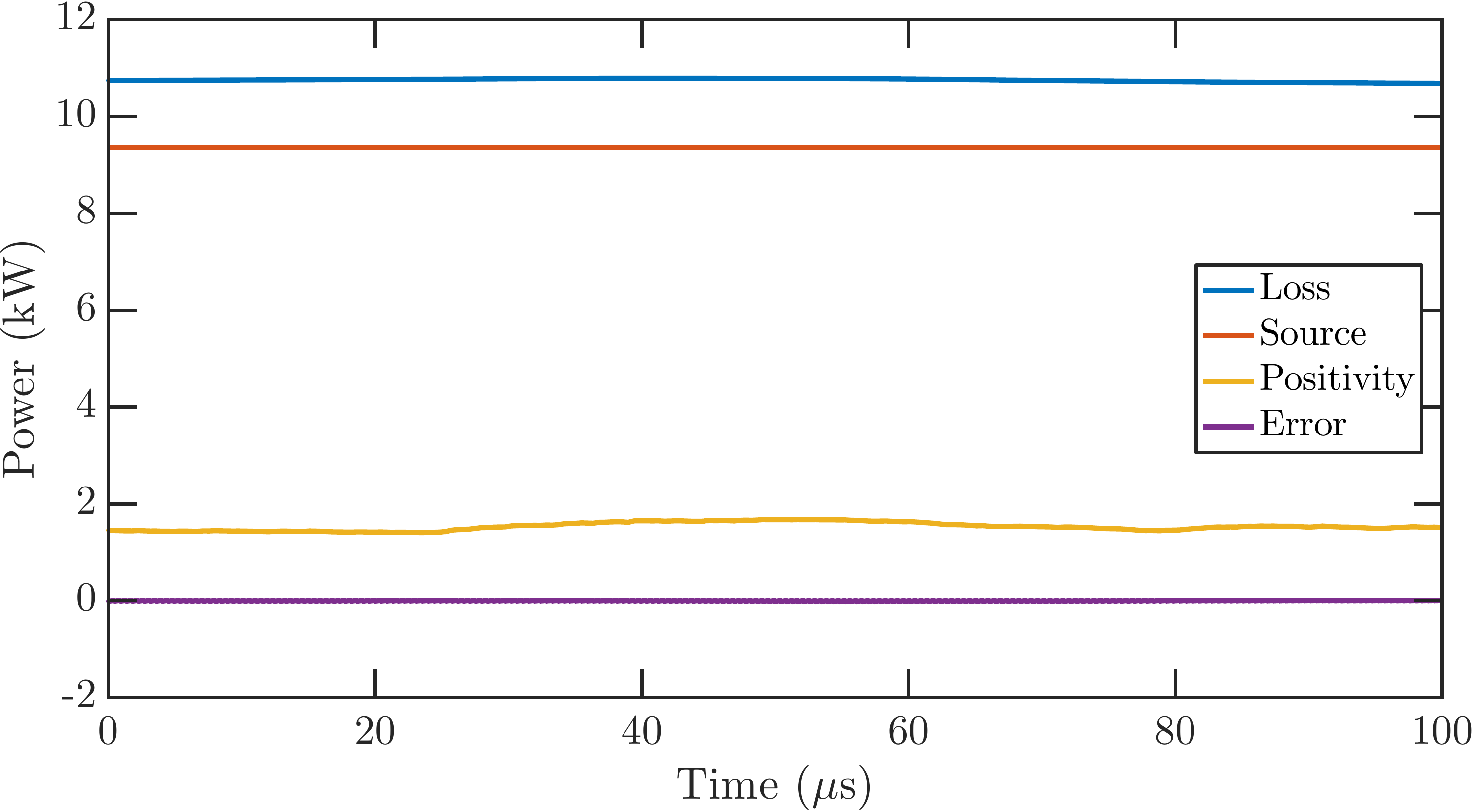}
  \caption[Time traces of diagnostics tracking power sources and sinks in a LAPD simulation.]
  {Time traces of diagnostics tracking power sources, power sinks, and power error over a 0.1 ms period
  in a LAPD simulation.
  The power error, which is defined in (\ref{eq:energyEvolution}) and arises from the time discretization,
  fluctuates in amplitude between 1 and 6~W (${\sim}10^{-4}$ relative error).
  This plot indicates that the energy conservation error
  introduced by the time discretization is extremely low.}
\label{fig:lapd-energy-trace}
\end{figure}

Figure~\ref{fig:lapd-energy-trace} shows the time traces of the terms appearing in the power balance
(\ref{eq:energyEvolution}) over a $0.1$ ms period of
the simulation in a quasi-steady turbulent state, or approximately $2 \times 10^4$ time steps.
Also plotted in figure~\ref{fig:lapd-energy-trace} is $P_{\mathrm{err}}$,
which is found to vary in magnitude between $1$--$6$ W (${\sim}10^{-4}$ relative error),
so the energy conservation error introduced by the time discretization is extremely low.

\begin{figure}
  \centering
  \includegraphics[width=\linewidth]{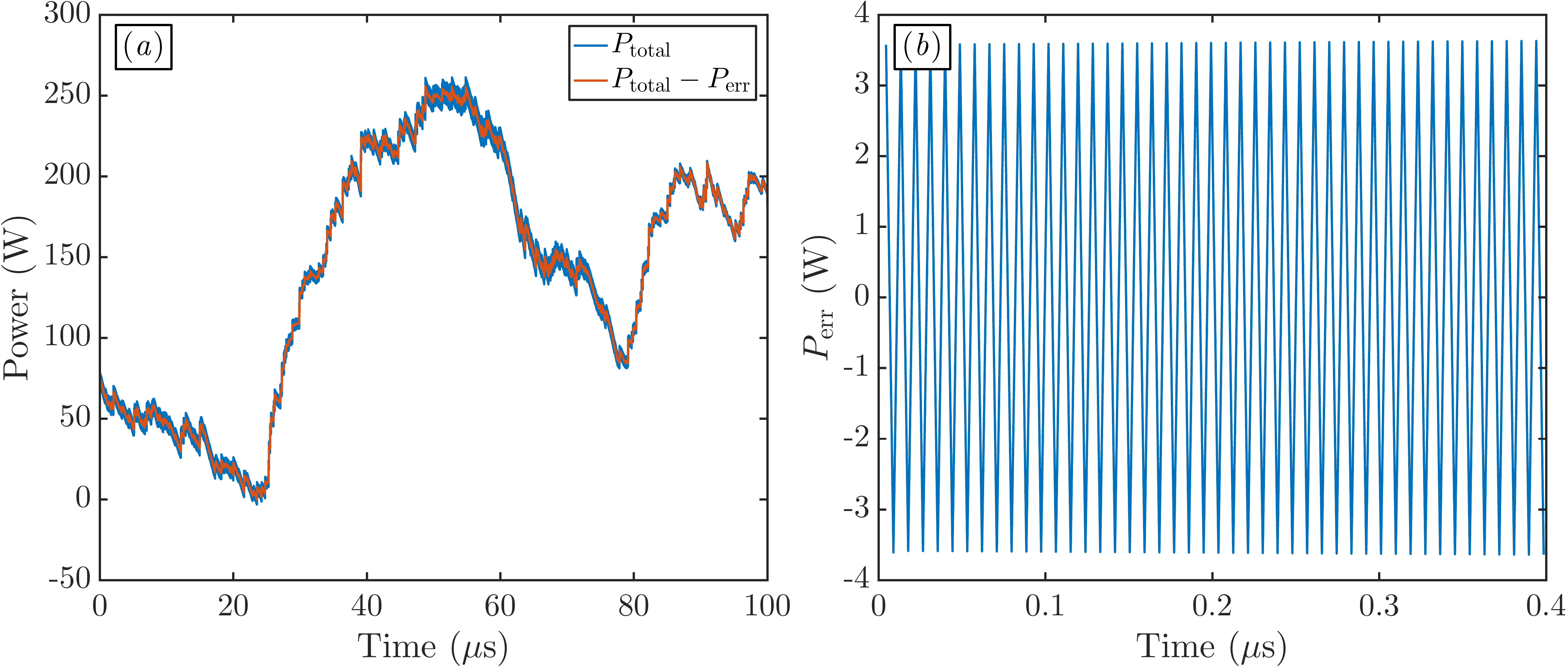}
  \caption[Time traces of power-error diagnostics in a LAPD simulation.]
  {Time traces power-error diagnostics in a LAPD simulation.
  ($a$) $P_\mathrm{err}$ is computed by taking the difference between 
  the sum of power sources and sinks in the system ($P_\mathrm{total}-P_\mathrm{err}$ in
  (\ref{eq:energyEvolution})) is compared with numerical net power $P_\mathrm{total}$, which is
  computed by taking the difference in the total system energy before and after a time step.
  ($b$) From the two curves in $(a)$, we can quantify the error in power that arises from
  the time discretization. Note that the time interval has been reduced to show the oscillations
  in $P_\mathrm{err}$. Over a larger 0.1~ms time interval, $P_\mathrm{err}$
  fluctuates in amplitude between 1 and 6~W.
  }
\label{fig:lapd-energy-trace-dual}
\end{figure}

%% file: ch-appendicies/helical-sol-initial-conditions.tex
\chapter{Initial Conditions for Helical-SOL Simulations\label{ch:initial-helical-sol}}
Here, we derived the initial conditions used in the helical-SOL simulations of Chapter~\ref{ch:helical-sol}.
The results can also be used to provide initial conditions for the LAPD simulations of Chapter~\ref{ch:lapd}.
The calculation presented here is based on notes by G.~Hammett. 

We consider a problem in which a uniform mass source $S_\rho$ and energy source $S_E$
is continuously active in the region $|z| < L_s/2$.
This fluid flows out to perfectly absorbing boundaries at $|z| = L_z/2$.
We treat the plasma as a single fluid with mass density $\rho \approx n_e m_i$ and
energy density $(3/2)n_e(T_e+T_i)$, so $S_\rho = m_i S_n$ and $S_E = (3/2) T_{\mathrm{src}} S_n$,
where $S_n$ is the electron and ion particle source rate and
$T_\mathrm{src} = T_{e,\mathrm{src}} + T_{i,\mathrm{src}}$ is the effective single-fluid source temperature.
This system is described by the steady-state fluid equations
\begin{align}
  0 &= -\frac{\partial}{\partial z} \left(\rho u \right) + S_\rho, \\
  0 &= -\frac{\partial}{\partial z} \left(\rho u^2 + p \right), \\
  0 &= -\frac{\partial}{\partial z} \left(\frac{1}{2} \rho u^3 + \frac{5}{2} p u \right)  + S_E,
\end{align}
where $u$ is the fluid velocity, $\rho$ is the mass density, and $p$ is the pressure.
We treat the source as having no mean flow in the $z$ direction.

We integrate these equations from $z=0$ to an arbitrary position $z < L_s/2$
and use the boundary condition $u(z=0) = 0$ to get
\begin{align}
  \rho u &= S_\rho, \\
  \rho u^2 + p &= p_0, \\
  \frac{1}{2} \rho u^3 + \frac{5}{2} p u &= z S_E,
\end{align}
where $p_0 \equiv p(z=0)$.
The first two equations can be solved for $\rho$ and $p$ respectively, and we obtain
a quadratic equation for $u(z)$ by substituting these expressions into the last equation.
The solution to this system is
\begin{align}
  p(z) &= \frac{3 p_0 \mp \sqrt{25 p_0^2 - 32 z^2 S_\rho S_E}}{8}, \label{eq:1d_fluid_pressure} \\
  u(z) &= \frac{5 p_0 \pm \sqrt{25 p_0^2 - 32 z^2 S_\rho S_E}}{8 S_\rho z}, \\
  \rho(z) &= \frac{z S_p}{u}.
\end{align}
Since the pressure cannot be negative, the only physical solution for small $z$
is the negative branch for $u(z)$ and the positive branch for $p(z)$.
The central pressure $p_0$ is determined by the boundary conditions at $|z| = L_z/2$.
A steady-state solution at a perfectly absorbing wall 
requires $\mathcal{M} \ge 1$ at the wall \citep{Munz1994}, where the Mach number 
$\mathcal{M}(z) \equiv u(z)/c_\mathrm{s}(z) = u(z)/\sqrt{(5/3) p(z) / \rho(z)}$.
We see that
\begin{equation}
  \mathcal{M}(z)^2 = \frac{\rho(z) u(z)^2}{(5/3) p(z)^2} = \frac{3}{5} \frac{p_0 - p}{p}.
\end{equation}
The maximum value of $\mathcal{M}(z)$ occurs at the $z$ that minimizes $p(z)$.
This value $z_\mathrm{max}$ turns out to be the $z$ that makes the
radicand in (\ref{eq:1d_fluid_pressure}) zero, so we find that 
\begin{align}
  z_\mathrm{max}^2 &= \frac{25}{32} \frac{p_0^2}{S_p S_E}, \\
  p(z_\mathrm{max}) &= \frac{3}{8} p_0, \\
  \mathcal{M}(z_\mathrm{max}) &= 1.
\end{align}
This result means that the outflow requirement must be $\mathcal{M} = 1$
and provides a constraint on the value of $p_0$ such that $z_\mathrm{max} = L_s/2$:
\begin{equation}
  p_0 = \frac{L_s}{2} \sqrt{ \frac{32}{25} S_\rho S_E}.
\end{equation}
Using this expression for $p_0$, we have the following profiles in the source
region $0 < |z| < L_s/2$:
\begin{align}
  p(z) &= p_0 \left( \frac{ 3 + 5 \sqrt{1-z^2 / \left( L_s/2 \right)^2} }{8} \right), \\
  u(z) &= \frac{\sqrt{3}}{2} \sqrt{ \frac{T_\mathrm{src}}{m_i} }
   \left( \frac{ 1 - \sqrt{1-z^2 / \left( L_s/2 \right)^2} }{ z/\left( L_s/2 \right) } \right), \label{eq:1d_fluid_u} \\
  \rho(z) &= \frac{16 S_\rho^2}{5 p_0} \left( \frac{L_s}{2} \right)^2
    \left( \frac{1 + \sqrt{1-z^2/\left( L_s/2 \right)^2 } }{2} \right).
\end{align}
In order to use these profiles to initialize a Maxwellian initial condition for a kinetic
simulation, we note that these profiles correspond to density ($n=\rho/m_i$) and temperature
($T = m_i p/\rho$) profiles in the source region given by
\begin{align}
  T(z) &= \frac{3}{5} T_\mathrm{src} \left(
    \frac{3 + 5 \sqrt{1-z^2 / \left( L_s/2 \right)^2} }
    {4 + 4 \sqrt{1-z^2 / \left( L_s/2 \right)^2} }
    \right), \label{eq:1d_fluid_temp} \\
  n(z) &= \frac{4\sqrt{5}}{3} \frac{(L_s/2) S_n}{c_\mathrm{ss}} 
    \left( \frac{1 + \sqrt{1-z^2/\left( L_s/2 \right)^2 } }{2} \right), \label{eq:1d_fluid_density}
\end{align}
where $c_\mathrm{ss} = \sqrt{(5/3) T_\mathrm{src}/m_i}$.
In the source-free regions $z>L_s/2$ or $z<L_s/2$, $n(z)$, $T(z)$, and $u(z)$ are all constant and equal to
the value that their respective profiles evaluated at the corresponding edge of the source region
at $z=L_s/2$ or $z=-L_s/2$.
The 1D equilibrium profiles (\ref{eq:1d_fluid_u}), (\ref{eq:1d_fluid_temp}), and (\ref{eq:1d_fluid_density}),
the density source in the helical-SOL simulations (\ref{eq:helical_sol_source}),
and the temperature profiles of the electron and ion sources are used to
generate spatially varying initial conditions in $(x,y,z)$.

One could go further by calculating the ion guiding-center density profile that
gives the desired equilibrium potential $\phi(x,y,z)$ when the gyrokinetic Poisson equation is solved.
For now, we simply set $n_i^g(x,y,z) = n_e(x,y,z)$ and initialize with $\phi = 0$,
as we did in the LAPD simulations.

%% file: ch-appendicies/interchange.tex
\chapter{Estimates of Interchange Instability\label{ch:interchange}}
Here, we estimate the frequencies and growth rates of interchange-like instabilities that may be relevant
for the helical SOL simulations discussed in Chapter~\ref{ch:helical-sol}.
The notes here are based on discussions with T.~Stoltzfus-Dueck.
For simplicity, we look for modes with $k_\parallel = 0$, take ions to be cold ($T_i = 0$) and
singly charged, and take electrons to be isothermal.

\section{Basic Interchange Instability}
We start with the system of equations in 2D describing the electron density,
ion gyrocenter density, and electrostatic potential:
\begin{align}
  \frac{\partial \tilde{n}_e}{\partial t} + \frac{1}{B_0} \left\{\phi, \tilde{n}_e + n_{e0} \right\}
  + \frac{1}{e}\mathcal{K} \left(\tilde{n}_e T_{e0} - n_{e0} e \phi \right) &= 0, \label{eq:ic-basic-elc} \\
  \frac{\partial \tilde{n}_i}{\partial t} + \frac{1}{B_0} \left\{\phi, \tilde{n}_i + n_{i0} \right\}
  - \frac{1}{e} \mathcal{K}\left(n_{i0} e \phi \right) &= 0, \label{eq:ic-basic-ion} \\
  -n_{i0} m_i \frac{1}{B^2} \nabla_\perp^2 \phi &= e\left(\tilde{n}_i - \tilde{n}_e\right), \label{eq:ic-qn}
\end{align}
where the densities have been split into background $n_{s0}$ and fluctuating $\tilde{n}_s$ components
and the Poisson bracket for $E \times B$ advection is
$\left\{ f,g \right\} = (\partial_x f)(\partial_y g) - (\partial_y f)(\partial_x g)$.
The curvature operator is $\mathcal{K} = \mathcal{K}^x \partial_x + \mathcal{K}^y \partial_y$,
where $\mathcal{K}^x = -(2/B) \boldsymbol{b} \times \nabla \ln B \cdot \nabla x$
and $\mathcal{K}^y = -(2/B) \boldsymbol{b} \times \nabla \ln B \cdot \nabla y$.
The term $\mathcal{K}(\tilde{n}_e T_{e0})/e$ represents the advection of density by the electron curvature drift.
Since we have assumed that the ions are singly charged, $n_{i0} = n_{e0} = n_0$.

First, we take a time derivative of the quasineutrality equation (\ref{eq:ic-qn}) and substitute
(\ref{eq:ic-basic-elc}) and (\ref{eq:ic-basic-ion}):
\begin{align}
  -n_0 m_i \frac{1}{B^2} \nabla_\perp^2 \frac{\partial \phi}{\partial t} =& 
  e\left( -\frac{1}{B_0} \left\{\phi, \tilde{n}_i + n_0 \right\} + \frac{1}{e} \mathcal{K}\left(n_0 e \phi \right) \right) \\ \nonumber
  & -e \left(-\frac{1}{B_0} \left\{\phi, \tilde{n}_e + n_0 \right\} -\frac{1}{e}\mathcal{K} \left(\tilde{n}_e T_{e0} - n_0 e \phi \right) \right) \\
  n_0 m_i  \frac{1}{B^2} \nabla_\perp^2 \frac{\partial \phi}{\partial t} =&
  \frac{e}{B_0} \left\{\phi, \left( \tilde{n}_i + n_0 \right) - \left(\tilde{n}_e + n_0\right) \right\}
  + \mathcal{K} \left(-n_0 e \phi - \tilde{n}_e T_{e0} + n_0 e \phi \right) \\
   n_0 m_i  \frac{1}{B^2} \nabla_\perp^2 \frac{\partial \phi}{\partial t} =&
  -n_0 m_i \frac{1}{B_0 B^2} \left\{\phi, \nabla_\perp^2 \phi \right\}
  - \mathcal{K} \left( \tilde{n}_e T_{e0} \right)
\end{align}

Next, we linearize the equations by taking $\tilde{n}_e / n_0 \ll 1$ and $e\phi/T_{e0} \ll 1$.
Additionally, we assume that $\nabla n_0 = -n_0 \hat{x}/L_n$ to write
\begin{gather}
  \frac{\partial \tilde{n}_e}{\partial t} + \frac{1}{B_0} \frac{\partial \phi}{\partial y} \frac{n_0}{L_n}
  + \frac{1}{e} \mathcal{K}^x \frac{\partial}{\partial x}\left( \tilde{n}_e T_{e0} - n_0 e \phi \right)
  + \frac{1}{e} \mathcal{K}^y \frac{\partial}{\partial y}\left( \tilde{n}_e T_{e0} - n_0 e \phi \right) = 0,\\
  n_0 m_i \frac{1}{B^2} \nabla_\perp^2 \frac{\partial \phi}{\partial t} = 
  -\mathcal{K}^x \frac{\partial}{\partial x} \left(\tilde{n}_e T_{e0} \right)
  -\mathcal{K}^y \frac{\partial}{\partial y} \left(\tilde{n}_e T_{e0} \right).
\end{gather}
For the helical-SOL geometry, we set $\mathcal{K}^x = 0$ because we assume that
$\nabla x \parallel \nabla B$ and $\mathcal{K}^y \sim 2/(BR)$.
Next, we look at a single Fourier component of the fluctuations by taking $\tilde{n}_e = \hat{n}_e \exp \left(i k_y y - i \omega t\right)$
and $\tilde{\phi} = \hat{\phi} \exp \left(i k_y y - i \omega t\right)$ to solve for $\omega$ as a function of $k_y$:
\begin{gather}
  \omega \hat{n}_e - \frac{k_y}{B_0} \hat{\phi} \frac{n_0}{L_n}
  - \frac{k_y}{e} \mathcal{K}^y \left( \hat{n}_e T_{e0} - n_0 e \hat{\phi} \right) = 0,\\
  - \omega n_0 m_i \frac{1}{B^2} k_\perp^2 \hat{\phi} = 
   k_y \mathcal{K}^y \hat{n}_e T_{e0}.
\end{gather}
These equations can be rearranged as
\begin{gather}
  \left( \omega  - k_y \frac{T_{e0} \mathcal{K}^y }{e}  \right) \frac{\hat{n}_e}{n_0}
  = \left( k_y \frac{T_{e0}}{e B_0 L_n} - k_y \frac{T_{e0}\mathcal{K}^y}{e} \right) \frac{e \hat{\phi} }{T_{e0}}, \label{eq:ic-ne-fluct-basic}\\
  \frac{e \hat{\phi}}{T_{e0}} = 
   -\frac{k_y}{\omega k_\perp^2} \frac{T_{e0} \mathcal{K}^y}{e} \frac{e^2 B^2}{m_i T_{e0}} \frac{\hat{n}_e}{n_0} \label{eq:ic-phi-basic}.
\end{gather}
To simplify the notation, we define the velocities
\begin{align}
  v_{de} &\equiv \frac{T_{e0} \mathcal{K}^y}{e} \sim 2 \frac{\rho_\mathrm{s}}{R_0} c_\mathrm{s} \label{eq:ic-vde} \\
  v_{*e} &\equiv \frac{T_{e0}}{e B_0 L_n} = \frac{\rho_\mathrm{s}}{L_n} c_\mathrm{s}, \label{eq:ic-vstare}
\end{align}
where $c_\mathrm{s}^2 = T_{e0}/m_i$ and $\rho_\mathrm{s}^2 = c_\mathrm{s}^2 / \Omega_{ci}^2 = m_i T_{e0}/(e^2 B^2)$.
Using these definitions and substituting (\ref{eq:ic-phi-basic}) into (\ref{eq:ic-ne-fluct-basic}),
we get
\begin{equation}
  \left(\omega - k_y v_{de} \right) = - k_y^2 \left(v_{*e} - v_{de} \right) \frac{v_{de}}{\omega k_\perp^2 \rho_\mathrm{s}^2 }
\end{equation}
We make the additional assumption that $v_{de}/v_{*e} \sim L_n/R_0 \ll 1$,
so $v_{*e} - v_{de} \approx v_{*e}$.
The mode frequency $\omega$ then satisfies the quadratic equation
\begin{equation}
  \omega^2 - k_y v_{de} \omega + \frac{k_y^2 v_{de} v_{*e}}{k_\perp^2 \rho_\mathrm{s}^2} = 0.
\end{equation}
By assuming that $k_\perp \rho_\mathrm{s} \lesssim 1$, we see that the term linear in $\omega$
is small compared to at least one of the other terms, since
\begin{equation}
  \frac{k_y v_{de} \omega}{\omega^2} \frac{k_y v_{de} \omega}{ \frac{k_y^2 v_{de} v_{*e}}{k_\perp^2 \rho_\mathrm{s}^2} }
  = k_\perp^2 \rho_\mathrm{s}^2 \frac{v_{de}}{v_{*e}} \ll 1.
\end{equation}
Therefore, we neglect this term and obtain the following result for $\omega$:
\begin{equation}
  \omega \approx \pm i \frac{|k_y|}{k_\perp} \frac{\sqrt{v_{*e} v_{de}}}{\rho_\mathrm{s}}
  \sim \pm i \frac{|k_y|}{k_\perp} \frac{\sqrt{2} c_s}{\sqrt{R_0 L_n} }.
\end{equation}
The growth rate is maximized for $k_x \rightarrow 0$, in which case we get the standard interchange estimate
that the mode grows with a rate ${\sim} c_\mathrm{s}/\sqrt{R_0 L_n}$.

\section{Addition of Sheath Effects}
Next, we attempt to incorporate sheath effects to refine our estimate for the interchange-mode growth rate.
We assume that the electron temperature is constant along a field line and that the ions are cold.
The sheath boundary conditions used in the helical-SOL simulations permit fluctuations in the parallel
current at the sheath entrance, while the steady-state parallel current is assumed to be zero.
Taking $f_e$ to be a Maxwellian, the total parallel electron current into the sheath is
\begin{align}
  j_{\parallel e} = -e \int_{v_c}^{\infty} \mathrm{d} v_\parallel \, v_\parallel \frac{n_e}{\sqrt{2\pi v_{te}^2}}
  e^{-v_\parallel^2 / 2v_{te}^2 }
    &= -\frac{n_e e}{\sqrt{2\pi v_{te}^2}} \int_{v_c}^{\infty} \mathrm{d} v_\parallel \frac{\partial}{\partial v_\parallel} \left( -v_{te}^2 e^{-v_\parallel^2 / 2v_{te}^2 }  \right) \\
    &= - \frac{n_e e v_{te} }{\sqrt{2\pi}} e^{-v_c^2 / 2v_{te}^2 },
\end{align}
where $v_c^2 = 2 e \phi/m_e$.

In the cold-ion limit, we assume that the ions are accelerated to $c_\mathrm{s}$ at the sheath entrance
(as required by the Bohm sheath criterion $u_{\parallel i} \ge c_\mathrm{s}$),
so the parallel ion current into the sheath is simply taken to be $n_i e c_\mathrm{s}$.
The zeroth-order potential comes from solving $j_{\parallel i} + j_{\parallel e} = 0$:
\begin{align}
  \frac{n_{e} e v_{te} }{\sqrt{2\pi}} e^{- e \phi_0 / T_{e0} } &= n_{i} e c_\mathrm{s} \\
  \phi_0 &= \frac{T_{e0}}{e} \ln \left( \frac{n_e v_{te}}{\sqrt{2\pi} n_i c_\mathrm{s}}  \right)
\end{align}
The individual fluctuating parallel currents at the sheath entrance are
\begin{align}
  \tilde{j}_{\parallel e} &= - \frac{n_{e0} e v_{te} }{\sqrt{2 \pi}} e^{-e\phi_0/T_{e0}}
  \left( \frac{\tilde{n_e}}{n_{e0}} + \frac{1}{2} \frac{\tilde{T}_e}{T_{e0}} -
  \frac{e \tilde{\phi}}{T_{e0}} + \frac{e \phi_0}{T_{e0}} \frac{\tilde{T}_e}{T_{e0}} \right),
  \label{eq:ic-j-par-e} \\
  \tilde{j}_{\parallel i} &= n_{i0} e c_\mathrm{s} \frac{\tilde{n}_i}{n_{i0} }.
\end{align}
The total fluctuating parallel current can be written as
\begin{align}
  \tilde{j}_\parallel &= \tilde{j}_{\parallel i} + \tilde{j}_{\parallel e} \\
  &= n_{e0} e c_\mathrm{s} \left( \frac{\tilde{n}_i - \tilde{n}_e}{n_{i0}}
  + \frac{e \tilde{\phi}}{T_{e0}} - \frac{1}{2} \frac{\tilde{T}_e}{T_{e0}} - \frac{e \phi_0}{T_{e0}} \frac{\tilde{T}_e}{T_{e0}} \right) \\
  &= n_{e0} e c_\mathrm{s} \left( -m_i n_{i0} \frac{1}{eB^2} \nabla_\perp^2 \tilde{\phi}
  + \frac{e \tilde{\phi}}{T_{e0}} - \left(\frac{1}{2} + \frac{e \phi_0}{T_{e0}} \right) \frac{\tilde{T}_e}{T_{e0}} \right) \\
  &= n_{e0} e c_\mathrm{s} \left(  \left(1 - \rho_\mathrm{s}^2 \nabla_\perp^2 \right) \frac{e \tilde{\phi}}{T_{e0}}
  - \left[\frac{1}{2} + \ln \left( \frac{n_e v_{te}}{\sqrt{2\pi} n_i c_\mathrm{s}}  \right) \right] \frac{\tilde{T}_e}{T_{e0}} \right),
  \label{eq:ic-j-par-refined}
\end{align}
where the quasineutrality condition (\ref{eq:ic-qn}) was used to write $\tilde{n}_i - \tilde{n}_e$
in terms of the fluctuating potential $\tilde{\phi}$.
In (\ref{eq:ic-j-par-refined}), the $\rho_\mathrm{s}^2 \nabla_\perp^2 e \tilde{\phi}/T_{e0}$ correction should
be discarded, as it arises from the neglect of parallel ion-polarization-density flow.

Going back to the equation describing the evolution of the electron density,
we restore the $\nabla_\parallel \left( n_e u_{\parallel e} \right)$ term in (\ref{eq:ic-basic-elc}) that was originally
neglected through the assumption of $k_\parallel = 0$:
\begin{equation}
  \frac{\partial \tilde{n}_e}{\partial t} + \frac{1}{B_0} \left\{\phi, \tilde{n}_e + n_{e0} \right\}
  + \frac{1}{e}\mathcal{K} \left(\tilde{n}_e T_{e0} - n_{e0} e \phi \right) = - \nabla_\parallel \left(n_e u_{\parallel e} \right).
\end{equation}
After linearization, we have
\begin{equation}
  \frac{\partial \tilde{n}_e}{\partial t} + \frac{1}{B_0} \frac{\partial \phi}{\partial y} \frac{n_0}{L_n}
  + \frac{1}{e} \mathcal{K}^x \frac{\partial}{\partial x}\left( \tilde{n}_e T_{e0} - n_0 e \phi \right)
  + \frac{1}{e} \mathcal{K}^y \frac{\partial}{\partial y}\left( \tilde{n}_e T_{e0} - n_0 e \phi \right) = 
  -\nabla_\parallel \left(n_0 \tilde{u}_{\parallel e} \right).
\end{equation}
We can relate $\nabla_\parallel \left( n_0 \tilde{u}_{\parallel e} \right)$
to fluctuations in the current outflow at the sheaths (located at $\pm L_z/2$) by
assuming that $\tilde{j}_\parallel$ is linear in $z$
\begin{equation}
  -\nabla_\parallel \left( n_0 \tilde{u}_{\parallel e} \right) \approx \frac{1}{e}
  \frac{2}{L_z} \left. \tilde{j}_\parallel \right|_{sh}
\end{equation}
We make the additional approximation $\tilde{j}_{\parallel} \approx \tilde{j}_{\parallel e}$
by assuming that the parallel response of the ions is too slow to contribute to the current fluctuations
at typical instability timescales.
As a first step, we also neglect electron temperature fluctuations, so we set
$\tilde{T}_e = 0$ in (\ref{eq:ic-j-par-e}) and get
\begin{equation}
  \left. \tilde{j}_\parallel \right|_{sh} = - n_0 e c_\mathrm{s} \frac{e \tilde{\phi}}{T_{e0}}.
\end{equation}

Keeping the equation for the ion gyrocenter density (\ref{eq:ic-basic-ion}) unchanged,
the linearized system that includes sheath effects is
\begin{gather}
  \frac{\partial \tilde{n}_e}{\partial t} + \frac{1}{B_0} \frac{\partial \phi}{\partial y} \frac{n_0}{L_n}
  + \frac{1}{e} \mathcal{K}^y \frac{\partial}{\partial y}\left( \tilde{n}_e T_{e0} - n_0 e \phi \right) = 
    \frac{1}{e} \frac{2}{L_z} \left. \tilde{j}_\parallel \right|_{sh},\\
  n_0 m_i \frac{1}{B^2} \nabla_\perp^2 \frac{\partial \phi}{\partial t} = 
  -\mathcal{K}^y \frac{\partial}{\partial y} \left(\tilde{n}_e T_{e0} \right)
  + \frac{2}{L_z} \left. \tilde{j}_\parallel \right|_{sh}.
\end{gather}

As before, we look at a single Fourier component
$\tilde{n}_e = \hat{n}_e \exp \left(i k_y y - i \omega t\right)$
and $\tilde{\phi} = \hat{\phi} \exp \left(i k_y y - i \omega t\right)$:
\begin{gather}
  \left( \omega  - k_y \frac{T_{e0} \mathcal{K}^y }{e}  \right) \frac{\hat{n}_e}{n_0}
  = \left[ k_y \left( \frac{T_{e0}}{e B_0 L_n} - 
      \frac{T_{e0}\mathcal{K}^y}{e} \right) 
      + i \frac{2}{L_z} c_{\mathrm{s}} \right] \frac{e \hat{\phi} }{T_{e0}}, \\
  i \omega k_\perp^2 n_0 m_i \frac{1}{B^2} \hat{\phi}
  = -\mathcal{K}^y ik_y \hat{n}_e T_{e0} + \frac{2}{L_z} n_0 e c_\mathrm{s} \frac{e \hat{\phi}}{T_{e0}}.
\end{gather}
Using the definitions for $v_{de}$ (\ref{eq:ic-vde}) and $v_{*e}$ (\ref{eq:ic-vstare})
and defining a new frequency $\omega_t = 2 c_\mathrm{s}/L_z$, the system becomes
\begin{gather}
  \left( \omega  - k_y v_{de} \right) \frac{\hat{n}_e}{n_0}
  = \left[ k_y (v_{*e} - v_{de}) 
      + i \omega_t \right] \frac{e \hat{\phi} }{T_{e0}}, \\
  \left( \omega k_\perp^2 \rho_\mathrm{s}^2 + i \omega_t \right) \frac{e \hat{\phi}}{T_{e0}}
  = -k_y v_{de} \frac{\hat{n}_e}{n_0}.
\end{gather}
These equations are again combined to give a quadratic equation for $\omega$:
\begin{equation}
  \omega^2 + \left(i\frac{\omega_t}{k_\perp^2 \rho_\mathrm{s}^2} - k_y v_{de} \right)\omega 
  + \frac{k_y^2 v_{de} v_{*e}}{k_\perp^2 \rho_\mathrm{s}^2} \approx 0, \label{eq:ic-sheath-quadratic}
\end{equation}
where we have again taken $v_{de}/v_{*e} \sim L_n/R_0 \ll 1$.
As before, we can neglect the $k_y v_{de} \omega$ term because it is small compared
to one or both of the constant and quadratic terms.
We similarly compare the new sheath-contribution term to the other two terms:
\begin{equation}
  \frac{ \left( \omega \frac{\omega_t}{k_\perp^2 \rho_\mathrm{s}^2} \right)^2}{\omega^2 \frac{k_y^2 v_{de} v_{*e}}{k_\perp^2 \rho_\mathrm{s}^2} }
  = \frac{\omega_t^2}{k_\perp^2 \rho_\mathrm{s}^2 k_y^2 v_{de} v_{*e}}
  \sim \frac{4 c_\mathrm{s}^2}{L_z^2} \frac{1}{k_\perp^2 \rho_\mathrm{s}^2 k_y^2} \frac{R_0 L_n}{2 \rho_\mathrm{s}^2 c_\mathrm{s}^2}
  = \frac{2}{k_y^2 k_\perp^2 \rho_\mathrm{s}^4} \frac{L_n}{L_z} \frac{R_0}{L_z} \equiv \epsilon_t.
\end{equation}
Here, $L_n/L_z \ll 1$, $R_0/L_z \sim \mathcal{O}(1)$, and $k_y^2 k_\perp^2 \rho_\mathrm{s}^4$ is typically small,
so $\epsilon_t$ can be big or small compared to 1.

The solution to the quadratic equation (\ref{eq:ic-sheath-quadratic}) is
\begin{equation}
  \omega = -i \frac{\omega_t}{2 k_\perp^2 \rho_\mathrm{s}^2}
    \pm \frac{i}{2} \sqrt{ \frac{\omega_t^2}{k_\perp^4 \rho_\mathrm{s}^4}
      + \frac{4 k_y^2 v_{de} v_{*e}}{k_\perp^2 \rho_\mathrm{s}^2} },
\end{equation}
so the sheath current has a stabilizing effect on the growth rate when compared to the result of the
previous section.
In the weak-sheath-current limit ($\epsilon_t \ll 1$),
\begin{equation}
\omega = -i \frac{\omega_t}{2 k_\perp^2 \rho_\mathrm{s}^2}
  \pm i \frac{|k_y|}{k_\perp \rho_\mathrm{s}}\sqrt{v_{de} v_{*e}} \sqrt{1 + \frac{\epsilon_t}{4}}
  \approx \pm i \frac{|k_y|}{k_\perp} \frac{\sqrt{2} c_\mathrm{s}}{\sqrt{R_0 L_n}} - i \frac{\omega_t}{2 k_\perp^2 \rho_\mathrm{s}^2},
\end{equation}
so the growth rate of the mode has been slightly reduced.
In the strong-sheath-current limit ($\epsilon_t \gg 1$),
\begin{equation}
  \omega = -i \frac{\omega_t}{2 k_\perp^2 \rho_\mathrm{s}^2}
    \pm i \frac{\omega_t}{2 k_\perp^2 \rho_\mathrm{s}^2} \sqrt{1 + \frac{4}{\epsilon_t}}.
\end{equation}
Choosing the upper branch, the solution is
\begin{equation}
  \omega \approx i \frac{k_y^2 v_{de} v_{*e}}{\omega_t} = i \frac{1}{\sqrt{\epsilon_t}} \frac{|k_y|}{k_\perp} \frac{\sqrt{2} c_\mathrm{s}}{\sqrt{R_0 L_n}},
\end{equation}
so the growth rate can be reduced by a large factor if $\epsilon_t \gg 1$ (e.g. at low $k_\perp$).